\documentclass[]{akari2017}

\usepackage{here}
\usepackage{graphicx}
\usepackage{sidecap}
\usepackage{hyperref}

\shorttitle{Consistent dust and gas models for protoplanetary disks}
\shortauthors{P.\ Woitke, et al.}

\begin{document}

\title{Consistent dust and gas models for protoplanetary disks\\[2mm]
        III. Models for selected objects from the FP7 DIANA$^\star$
        project} 

\correspondingauthor{Peter Woitke 
(\href{mailto:pw31@st-and.ac.uk}{pw31@st-and.ac.uk}),
\ \ main DIANA website at\ \ \url{https://dianaproject.wp.st-andrews.ac.uk}\\[1mm]
$^\star$: {\footnotesize EU FP7-SPACE-2011 project 284405
  ``DiscAnalysis'' (Analysis and Modelling of Multi-wavelength 
  Observational Data from Protoplanetary Discs)}
}

\author{P.\ Woitke}
\affiliation{SUPA, School of Physics \& Astronomy, University of 
  St Andrews, North Haugh, KY16 9SS, UK, and
  Centre for Exoplanet Science, University of St Andrews, St Andrews, UK}
\author{I.\ Kamp}
\affiliation{Kapteyn Astronomical Institute, University of Groningen, 
  The Netherlands}
\author{S.\ Antonellini}
\affiliation{Kapteyn Astronomical Institute, University of Groningen, 
  The Netherlands, and Astrophysics Research Centre, School of
  Mathematics and Physics, Queen’s University Belfast, University
  Road, Belfast BT7 1NN, UK}
\author{F.\ Anthonioz}
\affiliation{Universit\'e Grenoble-Alpes, CNRS Institut de
  Plan\'etologie et d’Astrophyisque (IPAG) F-38000 Grenoble, France}
\author{C.\ Baldovin-Saveedra}
\affiliation{University of Vienna, Department of Astrophysics,  
  T{\"u}rkenschanzstrasse 17, A-1180, Vienna, Austria}
\author{A.\ Carmona} 
\affiliation{IRAP, Universit\'e de Toulouse, CNRS, UPS, Toulouse, France}
\author{O.\ Dionatos} 
\affiliation{University of Vienna, Department of Astrophysics,  
  T{\"u}rkenschanzstrasse 17, A-1180, Vienna, Austria}
\author{C.\ Dominik}
\affiliation{Astronomical institute Anton Pannekoek, University of 
  Amsterdam, Science Park 904, 1098 XH, Amsterdam, The Netherlands}
\author{J.\ Greaves} 
\affiliation{School of Physics and Astronomy, Cardiff University, 
  4 The Parade, Cardiff CF24 3AA, UK}
\author{M.\ G\"{u}del}
\affiliation{University of Vienna, Department of Astrophysics,  
  T{\"u}rkenschanzstrasse 17, A-1180, Vienna, Austria}
\author{J.\ D.\ Ilee} 
\affiliation{Institute of Astronomy, University of Cambridge,
  Madingley Road, Cambridge CB3 0HA, UK}
\author{A.\ Liebhardt}
\affiliation{University of Vienna, Department of Astrophysics,  
  T{\"u}rkenschanzstrasse 17, A-1180, Vienna, Austria}
\author{F.\ Menard}
\affiliation{Universit\'e Grenoble-Alpes, CNRS Institut de 
  Plan\'etologie et d’Astrophyisque (IPAG) F-38000 Grenoble, France}
\author{M.\ Min} 
\affiliation{SRON Netherlands Institute for Space Research, 
  Sorbonnelaan 2, 3584 CA Utrecht, and Astronomical institute Anton
  Pannekoek, University of Amsterdam, Science Park 904, 1098 XH, 
  Amsterdam, The Netherlands}
\author{C.\ Pinte}
\affiliation{Universit\'e Grenoble-Alpes, CNRS Institut de 
  Plan\'etologie et d'Astrophyisque (IPAG) F-38000 Grenoble, France, 
  and Monash Centre for Astrophysics (MoCA) and School of Physics and 
  Astronomy, Monash University, Clayton Vic 3800, Australia}
\author{C.\ Rab} 
\affiliation{Kapteyn Astronomical Institute, University of Groningen, 
  The Netherlands}
\author{L.\ Rigon}
\affiliation{SUPA, School of Physics \& Astronomy, University of 
  St Andrews, North Haugh, KY16 9SS, UK}
\author{W.\-F.\ Thi} 
\affiliation{Max Planck Institute for Extraterrestrial Physics, 
  Giessenbachstrasse, 85741 Garching, Germany}
\author{N. Thureau} 
\affiliation{SUPA, School of Physics \& Astronomy, University of 
  St Andrews, North Haugh, KY16 9SS, UK}
\author{L.\ B.\ F.\ M.\ Waters} 
\affiliation{SRON Netherlands Institute for Space Research, 
  Sorbonnelaan 2, 3584 CA Utrecht, and Astronomical institute Anton 
  Pannekoek, University of Amsterdam, Science Park 904, 1098 XH, 
  Amsterdam, The Netherlands}


\begin{abstract}
\noindent
The European FP7 project DIANA has performed a coherent analysis of a
large set of observational data of protoplanetary disks by means of
thermo-chemical disk models. The collected data include
extinction-corrected stellar UV and X-ray input spectra (as seen by
the disk), photometric fluxes, low and high resolution spectra,
interferometric data, emission line fluxes, line velocity profiles and
line maps, which probe the dust, PAHs and the gas in these objects.
We define and apply a standardized modelling procedure to fit these
data by state-of-the-art modeling codes ({\sf ProDiMo, MCFOST,
  MCMax}), solving continuum and line radiative transfer, disk
chemistry, and the heating \& cooling balance for both the gas and the
dust. 3D diagnostic radiative transfer tools (e.g.\ {\sf FLiTs}) are
eventually used to predict all available observations from the same
disk model, the {\sf DIANA}-standard model.  Our aim is to determine
the physical parameters of the disks, such as total gas and dust
masses, the dust properties, the disk shape, and the chemical
structure in these disks. We allow for up to two radial disk zones to
obtain our best-fitting models that have about 20 free parameters.
This approach is novel and unique in its completeness and level of
consistency.  It allows us to break some of the degeneracies arising
from pure SED modeling.  In this paper, we present the results from
pure SED fitting for 27 objects and from the all inclusive {\sf
  DIANA}-standard models for 14 objects.  Our analysis shows a number
of Herbig Ae and T\,Tauri stars with very cold and massive outer disks
which are situated at least partly in the shadow of a tall and
gas-rich inner disk. The disk masses derived are often in excess to
previously published values, since these disks are partially optically
thick even at millimeter wavelength and so cold that they emit less
than in the Rayleigh-Jeans limit. We fit most infrared to millimeter
emission line fluxes within a factor better than 3, simultaneously
with SED, PAH features and radial brightness profiles extracted from
images at various wavelengths. However, some line fluxes may deviate
by a larger factor, and sometimes we find puzzling data which the
models cannot reproduce. Some of these issues are probably caused by
foreground cloud absorption or object variability. Our data
collection, the fitted physical disk parameters as well as the full
model output are available to the community through an online database
(\url{http://www.univie.ac.at/diana}).
\end{abstract}

\keywords{protoplanetary disks; 
          astronomical databases: miscellaneous;
          radiative transfer;
          astrochemistry;
          line: formation;
          molecular processes\vspace*{-3mm}}

\setcounter{page}{1}


\vspace*{-8mm}
\section{Introduction} 
\label{sec:intro} 

\noindent The European FP7-SPACE project {\sf
  DIANA}\footnote{\url{https://dianaproject.wp.st-andrews.ac.uk}}
analyzed multi-wavelength and multi-kind observational data about
protoplanetary disks by using a standardized modeling approach, in
order to learn more about the physico-chemical state of the
birthplaces of extra-solar planets, their evolution, and the
pre-conditions for planet formation. In order to place our efforts
into context, we first review the state-of-the-art of fitting disk
observations by modeling.
 
Previous studies have applied a wealth of different disk modeling
approaches and fitting techniques, often tailored towards one
particular object or a fresh set of observations from a particular new
instrument for a couple of disk sources. The approaches can be divided
in order of increasing level of self-consistency.
\begin{enumerate}
\item Retrieval modeling of a few selected observables using radiative
  transfer (RT) techniques based on simple parametric disk models. A
  single model typically runs faster than a few CPU-min, such that
  $\chi^2$ minimization, e.g. in form of genetic algorithms, and
  sometimes Monte Carlo Markov Chain (MCMC) techniques can be
  applied. \\[-5mm]
\item Multi-stage modeling, using RT-modeling of continuum
  observations first to determine the physical disk structure, before
  chemical and line-transfer modeling is applied to compare to line
  observations, \\[-5mm]
\item Analysis of large sets of multi-wavelengths observables using
  forward modeling of consistent RT and chemistry. These models are
  usually so expensive that many authors do not claim to have fitted
  all observations, but are rather seeking for a broad agreement with
  the data, in order to discuss several modeling options and new
  physical or chemical assumptions that work best to obtain that
  agreement.
\end{enumerate}
The methodical differences in these individual disk modeling and
fitting works are unfortunately so substantial that it is very
difficult to cross-compare the results, for example the disk
structures obtained, the disk masses determined, or the evolutionary
trends deduced. This is a key goal of the homogeneous DIANA modeling
approach presented in this paper. Some selected previous disk fitting
studies are exemplified in the following, ordered by increasing level
of self-consistency and complexity of the physics and chemistry
applied.

\subsection{Continuum radiative transfer modeling (dust and PAHs)}

\noindent Full 2D/3D continuum radiative transfer (RT) techniques have
been applied to model Spectral Energy Distributions (SEDs), for
example \citep{Andrews2007,Andrews2013}. More recently, SED fitting
has been extended to include continuum visibilities and/or images. For example,
\citet{Pinte2008} used the MCMC method to fit the SED and
multi-wavelength continuum images of IM\,Lup with MCFOST.
\citet{Wolff2017a} performed a grid search and used the MCMC method to
fit the SED and scattered light images of ESO-H$\alpha$\,569.
\citet{Muro-Arena2018} fitted the SED of HD\,163296 in combination
with scattered light images (VLT/SPHERE) and thermal emission (ALMA). 
\citet{Maaskant2013} fitted SEDs and Q-band images, using a stepwise
procedure, for a small number of Herbig~Ae sources with MCMax.
\citet{Maaskant2014} have modeled dust and PAHs in a couple of
Herbig~Ae transition disks, aiming at determining the properties of
PAHs in disk gaps using the disk models of their earlier work
\citep{Maaskant2013,Acke2010}. An example for recent 3D continuum
modeling is \citet{Min2016a}, who used a genetic fitting algorithm for
HD\,142527. They have not applied the MCMC algorithm but used the
history of their fitting method to identify some parameter
degeneracies and to provide rough estimates of the uncertainties in
parameter determination. Further examples to 3D disk SED fitting are
given in \citep{Price2018} and \citep{Pinte2018}.

These continuum modeling approaches are usually multi-$\lambda$ and
based on parametric disk structures, with or without hydrostatic
equilibrium.  The studies provide constraints on the disk dust
structure, dust temperature, and the dust properties, but do not allow
for much conclusions about the gas.  If MCMC methods are used, these
studies provide the credibility intervals for the determination of the
various input parameters, and thus allow for a proper assessment of
the quality and uniqueness of the fits.

\subsection{Simplified chemical models}

\noindent As an extension, models have been developed where the
density and dust temperature structure is taken from full 2D dust RT
models, but a parametric prescription is used for the molecular
concentration, without computing any chemical rates in
detail. Examples are \citet{Williams2014dk}, \citet{Boneberg2016},
\citet{Isella2016}, and \citet{Pinte2018}. The molecular abundance in
the absence of photo-dissociation and freeze-out is a free parameter
in these models, and usually $T_{\rm gas} = T_{\rm dust}$ is
assumed. The molecular abundance is then switched to zero, or to a
very small value, where one of the above mentioned chemical desruction
processes is thought to be dominant. This approach can be
multi-wavelength, but usually concentrates on a series of lines from a
single molecule, in most cases CO and its isotopologues observed at
(sub-)mm wavelengths.

Such models are still fast and allow for MCMC approaches. However,
they lack the chemical insight to explain why some molecules are
confined to certain regions. Also, such models can only be applied to
predict the lines of dominant molecules like CO, which contain
the majority of the respective elements. The concentrations of other
molecules like HCO$^+$ may not be so straightforward to guess, 
and so their line emission regions can be different.

\subsection{Atomic/molecular line emission models}

\noindent Some approaches employ pure radiative transfer techniques to
fit line observations (fluxes, radial profiles, resolved line images
or visibilities). These line radiative models are based on parametric
column densities and a given radial temperature law, but without
detailed heating/cooling balance or chemical rates.  Dust continuum
radiative transfer modeling can be part of these models.  The best fits
are derived via retrieval methods to determine the column density and
temperature profile parameters, using for example power-law prescriptions.

The Chemistry in Disks (CID) project \citep[][and subsequent
  papers]{Dutrey2007} is a good example. Dutrey et al.\ focused on
N$_2$H$^+$ lines from DM\,Tau, Lk\,Ca\,15 and MWC\,480. The results of
their fitting are radial profiles of molecular column densities and
temperature. The main goal of this approach is to invert the line
observations, as directly as possible, to determine the desired disk
properties such as chemical abundances, column densities and
temperatures, but without a detailed physical or chemical disk model
that results in those structures. The main physics included is the line
radiative transfer. Thanks to the simplicity of these models, a wide
parameter space can be explored using $\chi^2$-minimization. The
approach is often applied to spatially resolved mm-data, where the
dust continuum radiative transfer is less crucial, for example
\citep{Teague2015,Oeberg2015}. Formal errors on the parameters can be
derived from the $\chi^2$-minimization.

These results are then interpreted in the context of generic
astrochemical disk models. \citet{Teague2015} presented a disk model
for DM\,Tau to discuss the HCO$^+$ and DCO$^+$ sub-mm line
observations. The authors used a combination of $\chi^2$
minimization and MCMC fitting in visibility space to derive disk
geometry parameters such as inner/outer radius, and inclination as
well as physical parameters such as scale height, temperature and
surface density power laws for each molecule independently. The
authors subsequently use more physical disk models to explore the
radial gradient in deuteration in the disk. The stellar parameters of
DM\,Tau, an ISM-like UV radiation field and the accretion rate are
taken from the literature to build a 1+1D steady state $\alpha$-disk
model. On top of the physical disk structure, the authors solve
time-dependent chemistry using a large gas-grain chemical network
including CR, UV and X-ray reactions. A similar approach is used by
\citet{Semenov2018}. In both cases, restricted disk parameters are
varied to interpret the radial molecular column density profiles and
to learn about disk ionization or elemental depletion.
 

At mid- and far-IR wavelengths, the continuum becomes non-negligible,
thus requiring a combination of the above approach with dust RT,
e.g.\ \citep{Banzatti2012,Pontoppidan2014b}. \citet{Zhang2013} used a
detailed physical dust structure, but parametric molecular
abundance/column densities. On top of the manually fitted dust RT model
\citep[RADMC,][]{Dullemond2004}, the authors computed water lines over
a wide wavelength range (mid- to far-IR) and discussed the water ice
line for the transitional disk around TW\,Hya in the context of
Spitzer and Herschel data. \citep{Blevins2016} used a similar approach
to model Spitzer and Herschel water lines in four primordial disks.
Similar techniques are also applied for near-infrared CO ro-vibrational 
lines, for example \citep{Carmona2017}. 

The resulting disk structures of such approaches can be quite
degenerate (dust structure, temperature, column density, line emitting
region) if unresolved data is used like Spitzer and Herschel line
fluxes and SEDs. The situation improves if a large wavelength range of
lines/multiple species are used and/or spatial information is
available. However, as far as we know, detailed fitting strategies and
an evaluation of the goodness of such fits have never been attempted.

\subsection{Pure chemical models}

\noindent This approach uses a proper chemical model on top of a fixed
disk structure, i.e.\ the physical properties like densities,
temperatures and radiation fields are calculated once and then
fixed. Those quantities are either estimated or taken from a dust
radiative transfer code. Thus, the gas chemistry has no mutual
influence on the physical properties in the disk, including its
temperature structure or dust settling. This approach is used, for
example, to interpret molecular column density profiles derived from
observations (see e.g.\ CID papers cited above). In those cases, the
authors do not fit observations with a chemical model, but rather vary
some chemical parameters and discuss what matches the observations
best or what is missing, {with the intention to improve astro-chemical
  networks in general.  Some of these works may not use detailed UV
  propertries of the star in combination of UV disk radiative
  transfer, or may not be consistent with the observed SED}.

Some works go beyond this approach by using dust structures consistent
with continuum observations and tailored for specific targets.
\citet{Cleeves2015a} fitted the SED of TW\,Hya using existing dust
models and carrying out TORUS RT models \citep{Harries2000}. The gas
mass is calibrated using a parametric gas temperature profile (based
on $T_{\rm dust}$ and the local FUV radiation field) and the observed
HD flux. The disk structure is then fixed and the authors explore
cosmic ray (CR) and X-ray processes to fit the radial emission
profiles of various mm-lines (mainly ions) using LIME
\citep{Brinch2011}. A similar approach is used for IM\,Lup
\citep{Cleeves2016}. Here, a chain of dust RT, X-ray and UV RT models
is executed. A parametric gas temperature (based on $T_{\rm dust}$ and
the local FUV radiation field) is used to calculate the molecular gas
distribution based on chemical models.
LIME is used to produce CO channel maps for various levels of
CO depletion and external UV radiation fields. A $\chi^2$-minimization
strategy is applied to find the best match to the observed ALMA channel
maps, but without MCMC algorithm to determine the errors.

The computation times of such chemical models are generally orders of
magnitude higher than those of continuum RT models. Hence MCMC or
other exhaustive $\chi^2$-minimization strategies are generally
avoided, as these would require at least hundreds of thousands of such
models. This makes it difficult to evaluate in how far the results are
degenerate. The conclusions drawn from such approaches are therefore
often limited to a specific goal or question in the respective
study. \citet{Vasyunin2008} did a sensitivity analysis for the
chemistry. The errorbars given in the CID papers \citep[][and
  subsequent papers]{Dutrey2007} are based on those results.

\subsection{Radiation thermo-chemical models -- consistent dust and gas models}

\noindent This approach calculates self-consistently the dust
temperature, gas temperature, chemical abundances, and optionally the
vertical disk structure. Such models include a dust RT module,
a chemistry module, a heating/cooling module, and some post-processing
tools to derive for example visibilities, images, line profiles and
channel maps. These codes include most of the aspects mentioned
before, but not necessarily as sophisticated as used in the individual
chemical models illustrated above. Examples of such codes are ProDiMo
\citep{Woitke2009,Kamp2010,Thi2011,Aresu2011,Rab2018}, DALI
\citep{Bruderer2009,Bruderer2012,Bruderer2013}, the \citet{Gorti2011}
code, and the \citet{Du2014} code. In this approach, a small chemical
network is often used that is sufficient to predict the abundances of
the main coolants and observed simple molecules, for example no
isotope chemistry, no surface chemistry except adsorption and
desorption, and steady-state chemistry. The focus is to determine the
physical properties of disks, especially their radial/vertical
structure. They are also a critical test-bed/virtual laboratory for
our understanding of the complex coupling between radiation/energetic
particles (X-ray, UV, cosmic rays, stellar particles), dust particles
and gas.

\citet{Gorti2011} modeled the disk around TW\,Hya in the context of a
large set of observed line fluxes (e.g.\ forbidden optical lines such
as [S\,{\sc ii}], [O\,{\sc i}], near- and mid-IR lines as well as
sub-mm CO and HCO$^+$ lines). This disk model and derivations thereof
are used also in subsequent studies
\citep[e.g.][]{Bergin2013,Favre2013}. \citet{Gorti2011} compiled a
detailed input spectrum using stellar parameters from the literature
to select a suitable stellar atmosphere model for the photospheric
spectrum, completed by a FUSE spectrum and XMM-Newton X-ray data and
the Lyman\,$\alpha$ luminosity. The dust model and the surface density
distribution is a simplified version of a previous study by
\citet{Calvet2002} that matched the SED and 7mm images. It remains
unclear, however, whether the "simplified dust model" still fits the
SED and the images. The authors then vary the gas surface density on
top of the dust until they match the optical to sub-mm line fluxes. No
additional effort to consistently fit SED and lines was reported.

DALI was used to fit the CO ladder of HD\,100546 and a series of
fine-structure lines from neutral/ionized carbon and neutral oxygen
\citep{Bruderer2012} by varying a limited set of disk parameters
(e.g.\ dust opacities, outer disk radius, carbon abundance, and the
gas-to-dust mass ratio). In this case, no effort was made to fit the
continuum observables such as SED and/or images. \citet{Kama2016} used
DALI to model HD\,100546 and TW\,Hya, performing hand-fitting of the
SED by varying a limited set of disk parameters (dust and gas
depletion in gaps, dust surface density distribution, disk scale
height, flaring angle, tapering off, dust settling). In addition 
to the previously mentioned lines, they also included C$_2$H lines and line
profiles of CO and [C{\sc i}]. Typically of the order of 100 models
were explored per source. DALI has also been used to interpret ALMA
observations of disks, in particular the gas and dust
surface density distribution of transition disks, for example 
by \citet{vanderMarel2016} and \citet{Fedele2017}.

\citet{Du2015} modeled TW\,Hya for a selection of gas emission lines
from mid-IR to mm (fine-structure lines, CO isotopologues, water, OH,
and HD). They showed that their constructed dust model matches the SED
and sub-mm image, but they do not attempt to fit the sub-mm
visibilities. They fitted the line observations by adjusting the 
carbon and oxygen abundances, either considered to be 
ISM-like or modified, with a genetic algorithm. The
results of these two models are then discussed in the context of the
observations, but no detailed gas line fitting is attempted.

\citet[][ET\,Cha]{Woitke2011}, \citet[][HD\,163296]{Tilling2012},
\citet[][HD\,141569A]{Thi2014} and \citet[][HD\,135344B]{Carmona2014}
provide examples of ProDiMo\,+\,MCFOST disk fitting. For example,
\citet{Woitke2011} employed a genetic algorithm to find the best
parameter combination (11 parameters) to fit a wide range of
observables: SED, Spitzer spectrum, [O\,{\sc i}]\,6300~\AA, near-IR
H$_2$, far-IR Herschel atomic and molecular lines (partly upper
limits), and CO 3-2. In this case, the confidence intervals of the determined
model parameters are estimated by a-posteriori variation of single
parameters around the $\chi^2$-minimum.

\subsection{Grid-approach}

\noindent We list this approach here mainly for completeness. Its use
for fitting disk observations of individual targets is quite limited
due to the large number of free parameters in disk modeling, which
allows for just a few values per parameter to span several orders of
magnitude.  Often a sub-selection of parameters must be made to study
more specific questions.  Diagnostic methods derived from such grids
have to be evaluated critically in the context of the non-varied
parameters and model simplifications.  Examples for this approach are
the DENT grid \citep{Woitke2010,Pinte2010,Kamp2011} for SEDs, mid- to
far-IR and sub(mm) lines, \citet{Williams2014dk} and
\citet{Miotello2016} for (sub)mm CO isotopologues lines, and
\citet{Du2017} for water lines.  This approach is mainly driven by the
endeavor to understand the predicted changes of observables as
selected parameters are varied systematically. Ultimately, such
trends can possibly be inverted to devise new diagnostic tools for
observations.

\subsection{The DIANA approach}

\noindent The bottom line of the above summary of published disk
modeling works is that full radiation thermo-chemical models, where
all disk shape, dust and gas parameters have been commonly varied to
obtain the best fit of line and continuum data, have not yet been 
applied to more than a single object. Fitting gas line observations is
usually performed on top of a given disk dust structure. Disk modeling
assumptions vary significantly between papers, making it virtually
impossible to cross-compare the derived physical disk properties, even
if those papers come from the same group.

This is where our approach is new and makes a difference. The
ambitious goal of the DIANA project was to perform a coherent analysis
of {\em all} available continuum and line observational data for a
statistically relevant sample of protoplanetary disks. Our approach is
based on a clearly defined succession of three modeling steps: (i) to
fit the stellar and irradiation properties of the central stars; (ii)
to apply state-of-the-art 2D disk modeling software {\sf
  ProDiMo} \citep{Woitke2009a, Kamp2010, Thi2011}, {\sf MCFOST}
\citep{Pinte2006,Pinte2009} and {\sf MCMax} \citep{Min2009}, with a
fixed set of physical and chemical assumptions, to simultaneously fit
the disk shape, dust opacity and gas parameters of all objects; and
(iii) to use various post-processing radiative transfer tools,
including {\sf FLiTs} \citep[][written by M.\ Min]{Woitke2018} to
compute spectra and images that can be compared to the available
observational data. Contrary to many earlier efforts, our physical and
chemical modeling assumptions are not changed as we apply them to
different objects.  The simultaneous gas and dust modeling is
designed to be as self-consistent as possible to cover the following
feedback mechanisms:
\begin{itemize}
\item Changing the dust properties means to change the internal disk
  temperature structure, and to change the ways in which UV photons 
  penetrate the disk, which is of ample importance for the 
  photo-chemistry, freeze-out, and line formation.
\item Changing the gas properties affects dust settling. Disks with
  strong line emission may require a flaring gas structure, which can
  be different from the dust flaring if settling is taken into account
  in a physical way.
\item Changing or adding an inner disk, to fit some near-IR
  observations, will put the outer disk into a shadow casted by the
  inner disk, which changes the physico-chemical properties
  of the outer disks and related mm-observations.
\end{itemize}
These are just a few examples. Exploiting these feedback mechanisms
can help to break certain degeneracies as known, for example, from
pure SED-fitting.  Our data collection is available in a public
database \citep[DIOD,][]{Dionatos2018}, which includes photometric
fluxes, low and high-resolution spectroscopy, line and visibility
data, from X-rays to centimeter wavelengths, and respective meta-data
such as references. The database is online at
\url{http://www.univie.ac.at/diana}, together with our fitted stellar
and disk properties and detailed modeling results, which are also
available in an easy to use format at
\url{http://www-star.st-and.ac.uk/~pw31/DIANA/SEDfit}. This makes our
work completely transparent and reproducible. The predictive power of
these models can be tested against new observations, for example
unexplored molecules, other wavelength ranges or new instruments.  Our
results do not only contain the fitted observations, but we also
provide predictions for a large suite of other possible observations
(continuum and lines), which are computed for all our targets in the
same way.  The long-term purpose of our disk modeling efforts is
\begin{itemize}
\item to determine the disk masses, the disk geometry and shape, and
  the internal gas and dust properties (i.e.\ the dust and gas density
  distribution in the radial and vertical direction) for a large
  sample of well-studied protoplanetary disks,
\item to prepare cross-comparisons between individual objects, by
  applying standardized modeling assumptions and identical modeling
  techniques to each object,
\item to offer our disk modeling results, including the disk internal
  physico-chemical structure and a large variety of predicted
  observations to the community via a web-based interface,
\item to provide all relevant information and input files to ensure
  that all individual models can be reproduced, also by researchers
  from the wider community.
\end{itemize}
With our open policy to offer our modeling results to the community,
we hope to stimulate future research in neighboring research areas,
such as hydrodynamical disk modeling and planet formation theories.

\section{Target selection and stellar properties}
\label{sec:Targets}

\noindent At the beginning of the project, a full {\sf DIANA}
targetlist with 85 individual protoplanetary disks was compiled from
well-studied Herbig~Ae stars, class II T\,Tauri stars, and
transitional disk objects,
covering spectral types B9 to M3. The selection of objects was
motivated by the availability and overlap of multi-wavelength and
multi-kind, line and continuum data. However, additional criteria have
been applied as well, for example the exclusion of strongly variable
objects, where the data from different instruments would probe
different phases, and the exclusion of multiple or embedded sources,
where the observations are often confused by foreground/background
clouds or companions in the field, which is a problem in particular
when using data from instruments with different fields of view. We do
not claim that this target list is an unbiased sample. The full {\sf
  DIANA} targetlist was then prioritized and a subset thereof was
identified and put forward to detailed disk modeling. The modeling
was executed by different members of the team, but was not completed
within the run-time of {\sf DIANA} for all objects.  The completed
list of objects is shown in Table~\ref{tab:stellar}, together with the
results of our first modeling step, which is the determination of the
stellar parameters and UV and X-ray irradiation properties.

\begin{table}
\caption{Stellar parameters, and UV and X-ray irradiation properties, for 27 protoplanetary disks.}
\label{tab:stellar}
\def\z{$\hspace*{-1mm}$}
\def\zz{$\hspace*{-2mm}$}
\def\zzz{$\hspace*{-3mm}$}
\hspace*{-25mm}
\resizebox{194mm}{!}{
\begin{tabular}{c|ccc|ccc|c|llll}
\hline
&&&&&&&&&\\[-2.0ex]
object & \z SpTyp$^{(1)}$\zzz
       & $d$\,[pc]\z 
       & $A_V^{(15)}$
       & $T_{\rm eff}\rm[K]$\z 
       & \z$L_\star\rm[L_\odot]^{(15)}$\zz
       & \z$M_\star\rm[M_\odot]^{(1)}$\zz
       & \z age\,[Myr]$^{(1)}$\zz
       & $L_{\rm UV,1}^{(2)}$\zz & $L_{\rm UV,2}^{(3)}$\zz  
       & $L_{\rm X,1}^{(4)}$\zz & $L_{\rm X,2}^{(5)}$\zz\\ 
&&&&&&&&&\\[-2.0ex]
\hline
&&&&&&&&&\\[-2.0ex]
HD\,100546 & B9$^{(7)}$& 103 & 0.22 & 10470 & 30.5   & 2.5 & $>\!4.8^{(7)}$& 8.0     & 1.6(-2) & 4.9(-5) & 2.0(-5)\\
HD\,97048  & B9$^{(7)}$& 171 & 1.28 & 10000 & 39.4   & 2.5 & $>\!4.8^{(7)}$& 7.2     & 1.9(-2) & 2.1(-5) & 1.4(-5)\\
HD\,95881  & B9$^{(7)}$& 171 & 0.89 &  9900 & 34.3   & 2.5 & $>\!4.8^{(7)}$& 4.9     & 8.0(-2) 
                                                                          & 2.0(-5)$^{(11)}$\zz & 1.3(-5)$^{(11)}$\zz\zz\zz\\
AB\,Aur    & B9$^{(6)}$& 144 & 0.42 &  9550 & 42.1  & 2.5  & $>\!4.5^{(6)}$& 4.0     & 9.6(-3) & 2.3(-4) & 2.6(-5)\\
HD\,163296 & A1       & 119 & 0.48 &  9000 & 34.7   & 2.47 & 4.6          & 2.1     & 1.8(-2) & 1.5(-4) & 4.4(-5)\\
49\,Cet    & A2       & 59.4& 0.00 &  8770 & 16.8   & 2.0  & 9.8          & 1.0     & 1.7(-4) & 2.6(-4) & 5.3(-5)\\
MWC\,480   & A5       & 137 & 0.16 &  8250 & 13.7   & 1.97 & 11           & 5.6(-1) & 3.8(-3) & 1.5(-4) & 2.5(-5)\\
HD\,169142 & A7       & 145 & 0.06 &  7800 &  9.8   & 1.8  & 13           & 2.2(-1) & 1.6(-5) & 4.8(-5) & 1.4(-6)\\
HD\,142666 & F1$^{(12)}$\zz& 116 & 0.81 &  7050 &  6.3   & 1.6  & $>\!13^{(12)}$ & 3.7(-2)$^{(10)}$\zz & 5.6(-9)$^{(10)}$\zz
                                                                                                  & 1.6(-4) & 1.1(-5)\\
\zz HD\,135344B\z& F3 & 140 & 0.40 &  6620 &  7.6   & 1.65 & 12           & 3.2(-2) & 6.3(-3) & 2.4(-4) & 5.3(-5)\\
V\,1149\,Sco & F9     & 145 & 0.71 &  6080 &  2.82  & 1.28 & 19           & 5.1(-2) & 1.4(-2) & 3.7(-4) & 2.8(-5)\\
Lk\,Ca\,15 & K5$^{(16)}$ &140 &1.7  &  4730 &  1.2   & 1.0  & $\approx\!2^{(16)}$  
                                                                          & 5.1(-2) & 6.3(-3) & 5.5(-4) & 1.7(-4)\\
\zzz USco\,J1604-2130\z& K4& 145& 1.0& 4550& 0.76   & 1.2  & 10           
                           & 4.0(-3)$^{(17)}$\zz & 3.1(-4)$^{(17)}$\zz & 2.6(-4)$^{(18)}$\zz & 5.3(-5)$^{(18)}$\zz\zz\zz \\
RY\,Lup    & K4       & 185 & 0.29 &  4420 &  2.84  & 1.38 & 3.0          & 2.4(-3) & 1.5(-4) & 4.3(-3) & 3.6(-4)\\
CI\,Tau    & K6       & 140 & 1.77 &  4200 &  0.92  & 0.90 & 2.8          & 2.0(-3) & 8.7(-5) & 5.0(-5) & 1.0(-5)\\
TW\,Cha    & K6       & 160 & 1.61 &  4110 &  0.594 & 1.0  & 4.3          & 7.2(-2) & 4.4(-3) & 3.4(-4) & 1.0(-4)\\
RU\,Lup    & K7       & 150 & 0.00 &  4060 &  1.35  & 1.15 & 1.2          & 1.4(-2) & 9.0(-4) & 7.1(-4) & 3.4(-4)\\
AA\,Tau    & K7       & 140 & 0.99 &  4010 &  0.78  & 0.85 & 2.3          & 2.3(-2) & 5.8(-3) & 1.1(-3) & 3.2(-4)\\
TW\,Hya    & K7       & 51  & 0.20 &  4000 &  0.242 & 0.75 & 13           & 1.1(-2) & 4.2(-4) & 7.7(-4) & 7.0(-5)\\
GM\,Aur    & K7       & 140 & 0.30 &  4000 &  0.6   & 0.7  & 2.6          & 6.6(-3) & 2.8(-3) & 7.0(-4) & 1.2(-4)\\
BP\,Tau    & K7       & 140 & 0.57 &  3950 &  0.89  & 0.65 & 1.6          & 1.3(-2) & 1.1(-3) & 5.9(-4) & 2.5(-4)\\ 
DF\,Tau$^{(14)}$ & K7  & 140 & 1.27 &  3900 &  2.46  & 1.17 & $\approx\!2.2^{(14)}$ 
                                                                          & 3.6(-1) & 2.9(-1) & $-^{(13)}$ & $-^{(13)}$ \\
DO\,Tau    & M0       & 140 & 2.6  &  3800 &  0.92  & 0.52 & 1.1          & 1.3(-1) & 2.7(-2) & 1.1(-4) & 4.1(-5)\\
DM\,Tau    & M0       & 140 & 0.55 &  3780 &  0.232 & 0.53 & 6.0          & 7.0(-3) & 6.3(-4) & 8.4(-4) & 2.9(-4)\\
CY\,Tau    & M1       & 140 & 0.10 &  3640 &  0.359 & 0.43 & 2.2          & 7.3(-4) & 7.1(-5) & 2.1(-5) & 6.9(-6)\\
FT\,Tau    & M3       & 140 & 1.09 &  3400 &  0.295 & 0.3  & 1.9          & 5.2(-3)$^{(8)}$\zz & 8.4(-4)$^{(8)}$\zz 
                                                                              & 2.3(-5)$^{(9)}$ & 7.0(-6)$^{(9)}$\zz\zz\zz\\
RECX\,15   & M3       & 94.3& 0.65 &  3400 &  0.091 & 0.28 & 6.5          & 6.3(-3) & 4.0(-4) & 1.7(-5) & 8.2(-6)\\[1mm]
\hline
\end{tabular}}\\[2mm]
\hspace*{4mm}\resizebox{160mm}{!}{\parbox{175mm}{The table shows
    spectral type, distance $d$, interstellar extinction $A_V$, 
    effective temperature $T_{\rm eff}$, stellar luminosity $L_\star$,
    stellar mass $M_\star$, age, and UV and X-ray luminosities
    without extinction, i.\,e.\ as seen by the disk. Numbers written
    $A(-B)$ mean $A\times 10^{-B}$.  The UV and X-ray luminosities 
    are listed in units of $[L_\odot]$.\\[0.5mm]
$^{(1)}$: spectral types, ages and stellar masses are consistent with evolutionary
    tracks for solar-metallicity pre-main sequence stars by\\ 
    \hspace*{5mm} \citet{Siess2000}, using $T_{\rm eff}$ \& $L_\star$ as input,\\
$^{(2)}$: FUV luminosity from 91.2 to 205\,nm, as seen by the disk,\\
$^{(3)}$: hard FUV luminosity from 91.2 to 111\,nm, as seen by the disk,\\
$^{(4)}$: X-ray luminosity for photon energies $>\!0.1\,$keV, as seen by the disk,\\
$^{(5)}$: hard X-ray luminosity from 1\,keV to 10\,keV, as seen by the disk,\\
$^{(6)}$: no matching track, values from closest point at $T_{\rm eff}\!=\!9650\,$K 
         and $L_\star\!=\!42\rm\,L_{\odot}$,\\ 
$^{(7)}$: no matching track, values from closest point at $T_{\rm eff}\!=\!10000\,$K 
         and $L_\star\!=\!42\rm\,L_{\odot}$,\\ 
$^{(8)}$: no UV data, model uses an UV-powerlaw with $f_{\rm UV}\!=\!0.025$ 
         and $p_{\rm UV}\!=\!0.2$
         \citep[see][App.~A for explanations]{Woitke2016},\\
$^{(9)}$: no detailed X-ray data available, model uses a
         bremsstrahlungs-spectrum with $L_X\!=8.8\times 10^{28}$\,erg/s and
         $T_X\!=\!20\,$MK, based on \\ 
         \hspace*{5mm} archival XMM survey data
         (M.~G{\"u}del, priv.\,comm.),\\
$^{(10)}$: ``low-UV state'' model, where a purely photospheric spectrum is assumed,\\
$^{(11)}$: no X-ray data available, X-ray data taken from HD\,97048,\\
$^{(12)}$: no matching track, values from closest point at $T_{\rm eff}\!=\!7050\,$K
         and $L_\star\!=\!7\rm\,L_{\odot}$,\\
$^{(13)}$: no X-ray data available,\\
$^{(14)}$: resolved binary, 2$\times$ spectral type M1, luminosities $0.69\,L_\odot$ and $0.56\,L_\odot$, 
          separation $0.094''\approx 13\,$AU \citep{Hillenbrand2004},\\
$^{(15)}$: derived from fitting our UV, photometric optical and X-ray data, see Sect.~\ref{sec:step1},\\
$^{(16)}$: no matching track, values taken from \citep{Drabek2016,Kraus2012},\\
$^{(17)}$: no UV data, model uses $f_{\rm UV}\!=\!0.01$ and $p_{\rm UV}\!=\!2$ 
          \citep[see][App.~A for explanations]{Woitke2016},\\
$^{(18)}$: no X-ray data, model uses $L_X\!=\!10^{30}$erg/s and $T_X\!=\!20\,$MK 
          \citep[see][App.~A for explanations]{Woitke2016}.}}
\end{table}

\newpage
\section{Methods}
\label{sec:Methods}

\subsection{Modeling step 1:\ \ fitting the stellar parameters}
\label{sec:step1}

\noindent The first step of our modeling procedure is to determine the
stellar parameters (stellar mass $M_\star$, stellar luminosity
$L_\star$ and effective temperature $T_{\rm eff}$), as well as the
interstellar extinction $A_V$, and the incident spectrum of UV and
X-ray photons as irradiated by the star onto the disk.  These
properties are essential to setup the subsequent disk models.  The
method we have used for all objects is explained in \citet[][see
  Appendix A therein]{Woitke2016}, assuming that these parts of the
spectrum are entirely produced by the central star, without the
disk. We hence neglect scattering of optical and UV photons by the
disk surface in this modeling step. The method cannot be applied to
edge-on sources where the disk is (partly) in the line of sight
towards the star.  However, we can check this later, when absorption
and scattering by the disk is included, and can adjust in this case.
We use a large collection of optical and near-IR photometry points in
combination with low-resolution UV spectra, UV photometry points, and
X-ray measurements. 

We fit the photospheric part of each dataset by standard {\sf PHOENIX}
stellar atmosphere model spectra \citep{Brott2005}, with solar
abundances $Z\!=\!1$, after applying a standard reddening law 
\citep{Fitzpatrick1999} according to interstellar extinction $A_V$
and reddening parameter $R_V$. A standard value of $R_V\!=\!3.1$ is applied to
all stars if not stated otherwise. All photometric data in
magnitudes have been converted to Jansky ($F_{\rm filter}^{\,\rm obs}$)
based on instrument filter functions and zero-point data kindly provided
by Pieter Degroote (priv.\,comm.). The stellar model is then compared
to those data, depending on detector type, as
\begin{eqnarray}
  \mbox{CCD-detectors:} && \quad\displaystyle
  F_{\rm\,filter}^{\,\rm mod} = 
    \frac{\int \frac{1}{\lambda}\;F_\nu^{\,\rm mod}\;t_{\rm\,filter}(\lambda)\,d\lambda} 
         {\int \frac{1}{\lambda}\;t_{\rm\,filter}(\lambda)\,d\lambda} \\
  \mbox{BOL-detectors:} && \quad\displaystyle
  F_{\rm\,filter}^{\,\rm mod} = 
    \frac{\int \frac{1}{\lambda^2}\;F_\nu^{\,\rm mod}\;t_{\rm\,filter}(\lambda)\,d\lambda} 
         {\int \frac{1}{\lambda^2}\;t_{\rm\,filter}(\lambda)\,d\lambda} \ ,
  \label{filter_flux}
\end{eqnarray}
where $t_{\rm filter}(\lambda)$ are the filter transmission functions
and $F_\nu^{\,\rm mod}$\,[Jy] is the high-resolution model spectrum,
assuming that CCD detectors measure photon counts and bolometers
measure photon energy. The fit quality of the model is then determined
with respect to all selected photometric data points $i\!=\!1\,...\,I$,
within wavelength interval $[\lambda_1,\lambda_2]$ as
\begin{equation}
  \chi^2 = \frac{1}{I}\sum_{i=1}^I \left\{\begin{array}{ll}
  \displaystyle\left(\frac{\log\big(F_i^{\,\rm mod}/F_i^{\,\rm obs}\big)}
                {\sigma_i^{\,\rm obs}/F_i^{\,\rm obs}}\right)^2 
  & \quad\mbox{if\ \ } F_i^{\,\rm obs}>3\,\sigma_i^{\,\rm obs}\\[4mm]
  \displaystyle\left(\frac{F_i^{\,\rm mod}}{3\,\sigma_i^{\,\rm obs}}\right)^2 
  & \quad\mbox{otherwise}\end{array}\right.      \ ,
  \label{eq:chi}
\end{equation}
where $\sigma_i^{\,\rm obs}$ are the measurement errors. 
The selection of photometric data and wavelength fit range
was made manually for each object. Typical choices are 
$400-600$\,nm to $2-3\,\mu$m for T\,Tauri stars and 
$150-250$\,nm to $1-2\,\mu$m for Herbig\,Ae stars, depending 
on the observed level of non-photospheric emission in the UV and IR.

We have used the $(1,12)$-evolutionary strategy of
\citet{Rechenberg1994} to fit our model parameters $P_j$
(here $L_\star,T_{\rm eff},A_V$) to the data by minimizing $\chi^2$
\begin{eqnarray}
  P_{g,j}^k     &=& P_{g,j}^{\,0} + r\,\delta_g\,\Delta P_j    \quad(k=1\,...\,12) \\
  P_{g+1,j}^{\,0} &=& P_{g,j}^{\,k_{\rm best}} \\
  \delta_{g+1} &=& \left\{\begin{array}{ll}
                  \delta_g\,/\,1.4    & \quad\mbox{if}\ \ N_{\rm better}\!=\!0\\[1mm]
                  \delta_g            & \quad\mbox{if}\ \ 1\!\leq\!N_{\rm better}\!\leq\!2\\[1mm] 
                  \delta_g \times 1.4 & \quad\mbox{if}\ \ N_{\rm better}\!>\!2
                  \end{array}\right.
\end{eqnarray}
where $P_{g,j}^{\,0}$ are the parameter values of the parent of generation
$g$, $\Delta P_j$ are user-set search widths ($\Delta P_j\!=\!0$ means
to freeze the value of parameter $j$), $\delta_g$ is
the stepsize, and
\begin{equation}
  r = \sqrt{-2\ln(1-z_1)} \sin(2\pi\,z_2)
\end{equation}
are normal-distributed random numbers with mean value $\langle
r\rangle\!=0$ and standard deviation $\langle r^2\rangle=1$. They are
created from pairs of uniformly distributed pseudo-random numbers
$0\!\leq\!z_1,z_2\!<\!1$.  $k_{\rm best}$ is the index of the child
with lowest $\chi^2$ and $N_{\rm better}$ is the number of children
with $\chi^2$ better than their parent. In order
to escape local minima, it is important in this strategy to always
accept the best child as new parent, even if the fit quality of the
child is worse than the fit quality of its parent.  In numerous tests,
we found that this genetic fitting algorithm is very robust and
reliable, even if the models are noisy as will become relevant in the next
modeling step where Monte-Carlo radiative transfer techniques are
applied.

Since most of our sources are well-studied with high-resolution
spectroscopy, we used distance $d$ and effective temperatures $T_{\rm
  eff}$-values from the literature\footnote{The setup of all our models was
executed before the first GAIA data release.}.  Once $T_{\rm eff}$ and $L_\star$
are determined, we involve pre-main sequence stellar evolutionary
models \citep{Siess2000} to determine $M_\star$, the spectral type and
the age of the star. Based on those results, the stellar radius
$R_\star$ and the surface gravity $\log\,g$ can be computed which are
then used in the next iteration step to better select our photospheric
spectra (which depend on $T_{\rm eff}$ and $\log\,g$). For given $d$
and $T_{\rm eff}$, this iteration is found to converge very quickly to
a unique solution. Certain combinations of $d$ and $T_{\rm eff}$ found
in the literature, however, needed to be rejected this way, because
the procedure described above resulted in an impossible location in
the Hertzsprung-Russel diagram. 

A large collection of UV low- and high-resolution archival data was
collected from different instruments (IUE, FUSE, STIS, COS, ACS), and
then collated, averaged and successively re-binned until statistically
relevant data was obtained, using the method of weighted means
described in \citet{Valenti2000,Valenti2003}. The details are
described in \citep[][see Fig.~3 therein]{Dionatos2018}. These spectra
were then de-reddened according to our $A_V$ to obtain the stellar
UV-spectra as seen by the disk. These UV input spectra are included in
our database DIOD. The de-reddened data is used to replace the
photospheric model in the UV, and possible gaps between UV spectral data and
photospheric model spectrum are filled by powerlaws.

X-ray data was collected from XMM-Newton and Chandra.  A physically
detailed X-ray emission model was fitted to these observations
\citep{Dionatos2018}, from which we extracted a high-resolution X-ray
emission spectrum as seen by the disk by not computing the last
modeling step, namely the reduction of the X-ray fluxes by extinction.
These X-ray emission spectra are also available in our database
DIOD. As a side result, the modeling of the X-ray data provided 
estimates of the hydrogen column densities towards the sources, 
which is useful to verify our results for $A_V$.
 
The stellar properties, and in particular the assumed visual
extinction $A_V$, have a profound influence on the disk modeling
results. The stellar parameters must be carefully adjusted and checked
against UV and optical data to make sure that this part of the
spectrum is properly reproduced by the model. A blind application of
published stellar parameters can lead to substantial inconsistencies.
If $A_V$ is overestimated, for example, one needs to assume larger
values for $L_\star$, which would then make the disk warmer and
brighter in the infrared and beyond. A more substantial de-reddening
would result in a stronger UV spectrum as seen by the disk, causing
stronger emission lines, etc.  The resulting stellar properties of our
target objects are listed in Table~\ref{tab:stellar} for 27 objects.
The photometric and UV data are visualized in Fig.~\ref{fig:SEDs}.

\subsection{Modeling step 2:\ \ SED fitting}
\label{sec:SED}
 
\noindent The second step of our modeling pipeline is to fit the
spectral energy distribution (SED) of our targets including all
photometric data points and low-resolution spectra (ISO/SWS and LWS,
Spitzer/IRS, Herschel/PACS and SPIRE) from near-IR to millimeter
wavelengths. The data partly contains mid-IR PAH emission features
which we aim to fit as well. Our model is composed of a central star,
with parameters fixed by the previous modeling step, surrounded by
an axi-symmetric dusty disk seen under inclination angle $i$, which is
taken from the literature. Our physical assumptions about the gas,
dust particles and PAHs in the protoplanetary disk are detailed in
\citep[][section 3 therein]{Woitke2016}. We briefly summarize these
assumptions here
\begin{itemize}
\item passive disk model, i.e.\ no internal heating of the dust by
  viscous processes,
\item up to two radial disk zones, optionally with a gap in-between,
\item prescribed gas column density as function of radius in each disk
  zone, using a radial power-law with a tapered outer edge,
\item fixed gas-to-dust mass ratio in each zone,
\item parametric gas scale height as function of radius in each zone,
  using a radial powerlaw, 
\item dust settling according to \citet{Dubrulle1995}, with typically
  100 size-bins,
\item we apply the {\sf DIANA} standard dust opacities for disks,
  based on a power-law dust size distribution, an effective mixture of
  laboratory silicate and amorphous carbon, porosity, and a
  distribution of hollow spheres (DHS), see \citet{Min2016} and
  \citet{Woitke2016}, and
\item simplified PAH absorption and re-emission optionally included,
  see \citet[][Section 3.8]{Woitke2016}.
\end{itemize}
These disk models are run by means of our fast Monte Carlo radiative
transfer tools MCFOST and/or MCMax. The number of free parameters to
fit are
\begin{enumerate}
\item Inner and outer radius of each zone $R_{\rm in}$ and $R_{\rm
  out}$.  In the outermost zone, the outer radius is exchanged by the 
  tapering-off radius $R_{\rm tap}$, and the disk is radially extended 
  until the total hydrogen nuclei column density reaches the tiny value of 
  $10^{20}\rm\,cm^{-2}$.
\item The disk gas mass $M_{\rm disk}$ and the column density powerlaw index
  $\epsilon$ in each zone. In case of the outermost disk zone, there is in addition
  the tapering-off power-index $\gamma$.
\item The dust-to-gas ratio in each zone.
\item The gas scale height $H_0$ at some reference radius and the
  flaring index $\beta$ in each disk zone.
\item The assumed strength of turbulence in the disk $\alpha_{\rm
  settle}$ counteracting dust settling. This value
  is treated as a global parameter throughout all disk zones.
\item The minimum and maximum radius ($a_{\rm min}$,
  $a_{\rm max}$) and the powerlaw index of the size distribution 
  function $a_{\rm pow}$ of the grains in each zone
\item The volume fraction of amorphous carbon (amC) in the grains, treated
  as global parameter throughout all disk zones. The other two dust volume 
  contributors $\rm Mg_{0.7}Fe_{\rm 0.3}SiO_3$ (60\%) and porosity (25\%) are
  scaled to reach 100\% altogether. The values in brackets are
  for default choice $\rm amC\!=\!15\%$. The maximum hollow sphere
  volume fraction is fixed at 80\% (not used for fitting).
\item The PAH abundance with respect to interstellar standard $f_{\rm
  PAH}$ in each zone, and the ratio of charged PAHs (global parameter)
  in case the PAHs are included in the fit. We fix the kind of PAHs to
  circumcoronene with 54 carbon and 18 hydrogen atoms in all disk
  models.
\end{enumerate}
Altogether, we have hence $2+3+1+2+1+3+1=13$ free parameters for a
single-zone SED model without PAHs, 15 free parameters for a
single-zone model including PAHs, 21 free parameters for a two-zone
model without PAHs, and 24 free parameters for a two-zone model
including PAHs. In practice, however, the actual number of these
parameters is smaller. It is more or less impossible, for example, to
determine the radial extension of a disk by SED-fitting. Therefore,
$R_{\rm out}$ and $\gamma$ are rather estimated or values are taken
from the literature, but are not varied during the SED-fitting
stage. The same is true for $\epsilon$ in the outer disk which we
would rather fix to a default value of one, because it has very little
influence of the SED.  In consideration of two-zone disks, some
parameters may be chosen to be global, i.e.\ not zone-dependent. All
these operational decisions are left to the modeler's responsibility,
in consideration of known facts about the object under discussion.
Our general strategy was to first try a single-zone disk model, and
only if that resulted in a poor fit, we needed to repeat the fitting
exercise with a two-zone model. In cases where the object is
well-known to have a gap (pre-ALMA era), the disk was treated by a
two-zone model in the first place. Table~\ref{tab:SEDfit} summarizes
these choices in terms of the number of radial disk zones assumed,
whether PAH properties have been part of the fitting or not (only
attempted when PAH features are detected), and what was the total
number of free parameters (average is $13\,\pm\,3$). Small numbers
of parameters and/or generations indicate that we had a good
SED-fitting model to start with from previous works. Large
$\chi$-values indicate that incompatible observational data was used
for the fit (some data points might not agree with others within the
errorbars) rather than a failed fit.  This list demonstrates that a
fully automated SED-fitting is impossible. We need to decide which
disk parameters can be fitted by the available observations, and which
can't, and here human interference is unavoidable.

\begin{table}
\caption{Type of SED-fitting model, number of free parameters, 
  generations completed and models calculated, and final fit
  quality. Objects marked with ``--'' have no detections of
  PAH features, so PAHs are not included in the radiative transfer.
  ``n.a.''$=$ not recorded.}
\label{tab:SEDfit}
\centering
\def\z{$\hspace*{-1mm}$} 
\def\zz{$\hspace*{-2mm}$}
\hspace*{-17mm}\resizebox{166mm}{!}{
\begin{tabular}{c|cc|cc|cc|c}
\hline
object & \#\,disk zones\z & \z PAHs fitted\,? & \#\,free
parameters\z & \z\#\,data points & \#\,generations\z & \z\#\,models & final $\chi$\\
\hline
HD\,100546       & 2 & yes & 16 & 120 & 632 & 7584 & 1.85\\
HD\,97048        & 2 & yes & 16 & 119 & 1124& 13488& 1.65\\
HD\,95881        & 2 & yes & 14 &  69 & 114 & 1368 & 1.67\\
AB\,Aur          & 2 & yes & 13 & 140 & 533 & 6396 & 3.37\\
HD\,163296       & 2 & yes & 15 & 116 & 887 & 10644& 1.91\\
49\,Cet          & 2 & --  & 14 &  65 & n.a.& n.a. & 2.43\\
MWC\,480         & 1 & yes & 10 &  80 & 713 & 8556 & 2.30\\
HD\,169142       & 2 & yes & 15 & 126 & 1039& 12468& 2.41\\
HD\,142666       & 2 & yes & 19 &  80 & 1401& 16812& 1.91\\
HD\,135344B      & 2 & yes & 15 &  58 & 141 & 1692 & 3.33\\
V\,1149\,Sco     & 2 & yes & 17 &  72 & 665 & 7980 & 2.40\\
Lk\,Ca\,15       & 2 & --  & 14 &  65 & 1006& 12072& 2.05\\
USco\,J1604-2130 & 2 & --  & 15 &  45 & 703 & 8436 & 2.58\\
RY\,Lup          & 2 & --  & 15 &  47 & 601 & 7212 & 3.42\\
CI\,Tau          & 2 & yes & 13 &  61 & 682 & 7448 & 2.18\\
TW\,Cha          & 1 & --  & 9  &  47 & n.a.& n.a. & 2.01\\
RU\,Lup          & 1 & --  & 8  &  63 & 172 & 2064 & 2.78\\
AA\,Tau          & 1 & --  & 9  &  49 & n.a.& n.a. & 2.67\\
TW\,Hya          & 2 & --  & 13 &  63 & 3031& 36372& 2.35\\
GM\,Aur          & 2 & --  & 14 &  72 & 1004& 12048& 2.97\\
BP\,Tau          & 1 & yes & 9  &  60 & 343 & 4116 & 2.11\\
DF\,Tau          & 1 & --  & 9  &  64 & n.a.& n.a. & 3.20\\
DO\,Tau          & 1 & --  & 9  &  63 & 456 & 5472 & 1.94\\
DM\,Tau          & 2 & --  & 13 &  64 & n.a.& n.a. & 2.43\\
CY\,Tau          & 2 & yes & 14 &  65 & 333 & 3996 & 2.41\\
FT\,Tau          & 1 & --  & 9  &  51 & 179 & 2148 & 3.92\\
RECX\,15         & 1 & yes & 8  &  56 & 1018& 12216& 1.99\\
\hline
\end{tabular}}
\end{table}

To save computational time, we have converted all low-resolution
spectra into small sets of monochromatic points and added those to the
photometric data (see Fig.~\ref{fig:SEDs}). The spectral fluxes are
then only computed for these wavelengths by radiative transfer. These
model fluxes are always a bit noisy due to the application of MC
methods, which produces noise both in the temperature determination
phase and flux calculation phase. The fit quality 
of an SED-fitting model $\chi^2$ is computed according to
Eq.~(\ref{eq:chi}), but we first calculate $\chi^2$ separately in spectral
windows, for example $[0,0.3]\mu$m, $[0.3,1]\mu$m, $[1,3]\mu$m, etc., and
then average those results. This procedure makes sure that all spectral
regions have an equal influence of the fit quality, even if the 
distribution of measurement points is unbalanced in wavelength space.

\begin{figure}
  \hspace*{-7mm}
  \begin{tabular}{rr}
  \hspace*{-5mm}\includegraphics[width=87mm,height=74mm]{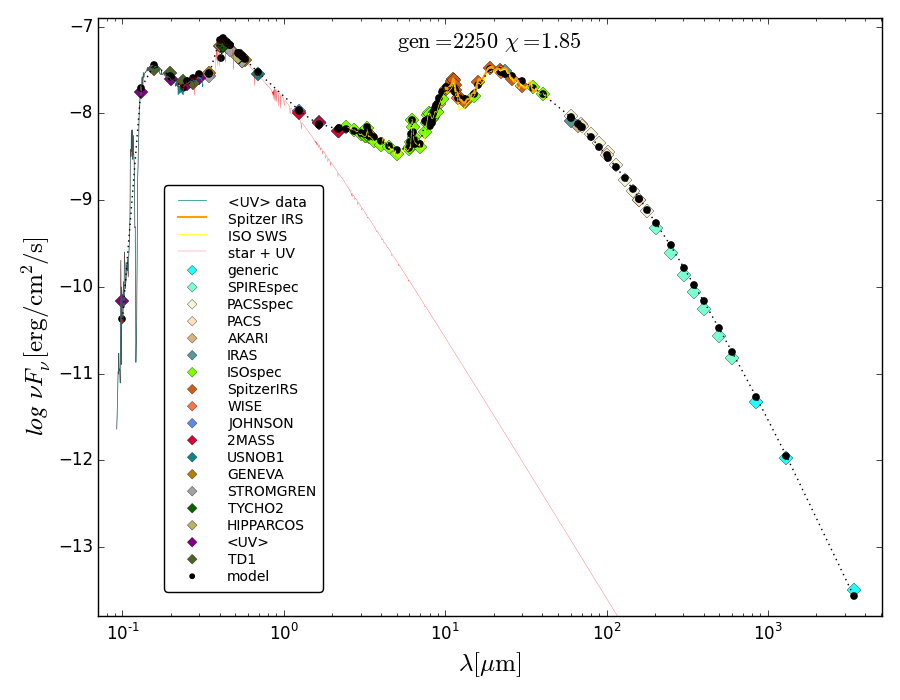} &
  \hspace*{-4mm}\includegraphics[width=87mm,height=74mm]{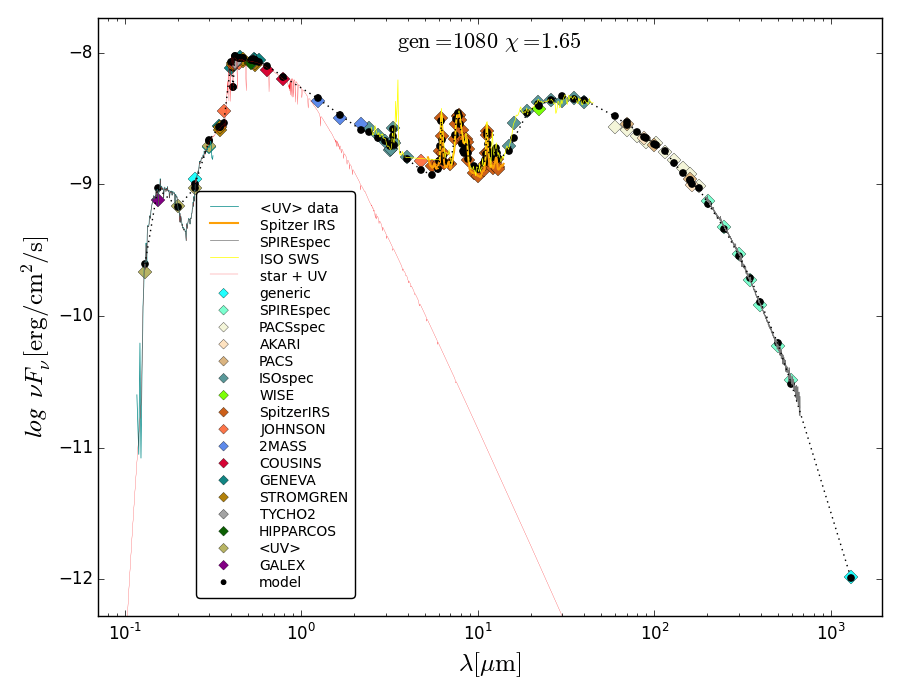}\\[-70mm]
  \sf HD\,100546\hspace*{3mm} & \sf HD\,97048\hspace*{3mm} \\[65mm]
  \hspace*{-5mm}\includegraphics[width=87mm,height=74mm]{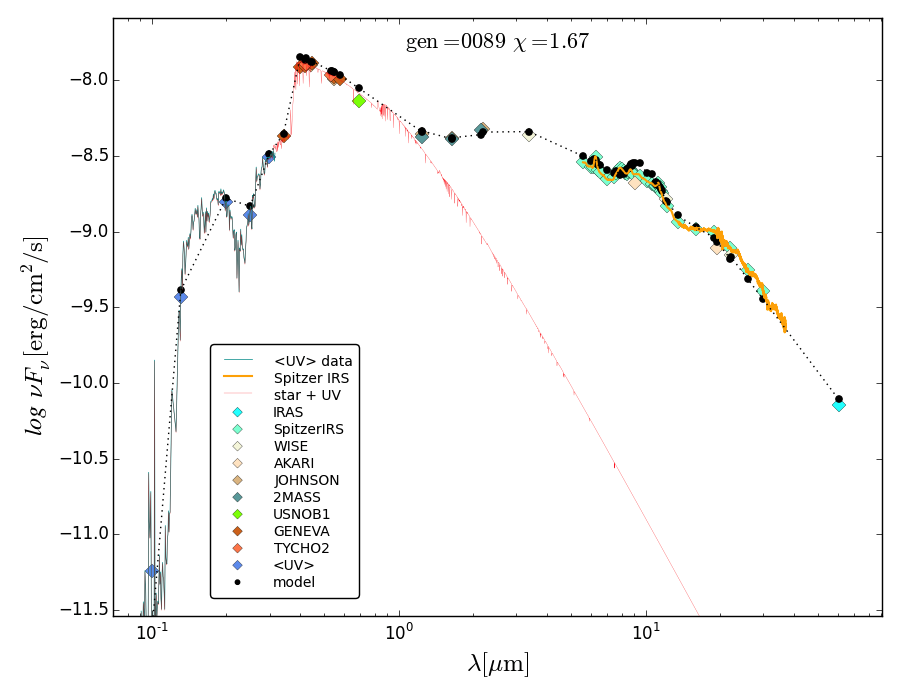} &
  \hspace*{-4mm}\includegraphics[width=87mm,height=74mm]{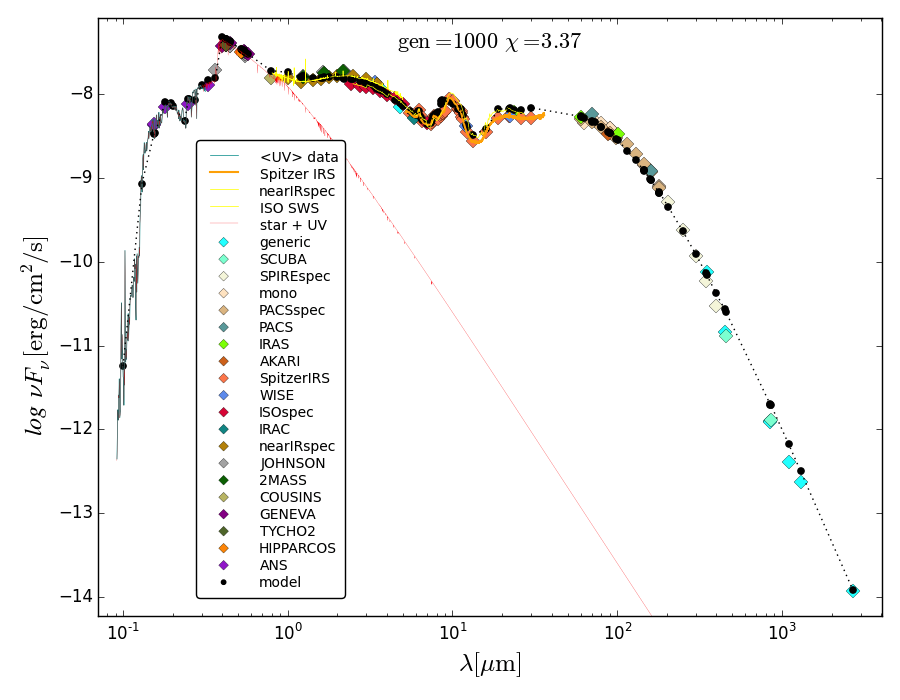} \\[-70mm]
 \sf HD\,95881\hspace*{3mm} & \sf AB\,Aur\hspace*{3mm} \\[65mm]
  \hspace*{-3mm}\includegraphics[width=87mm,height=74mm]{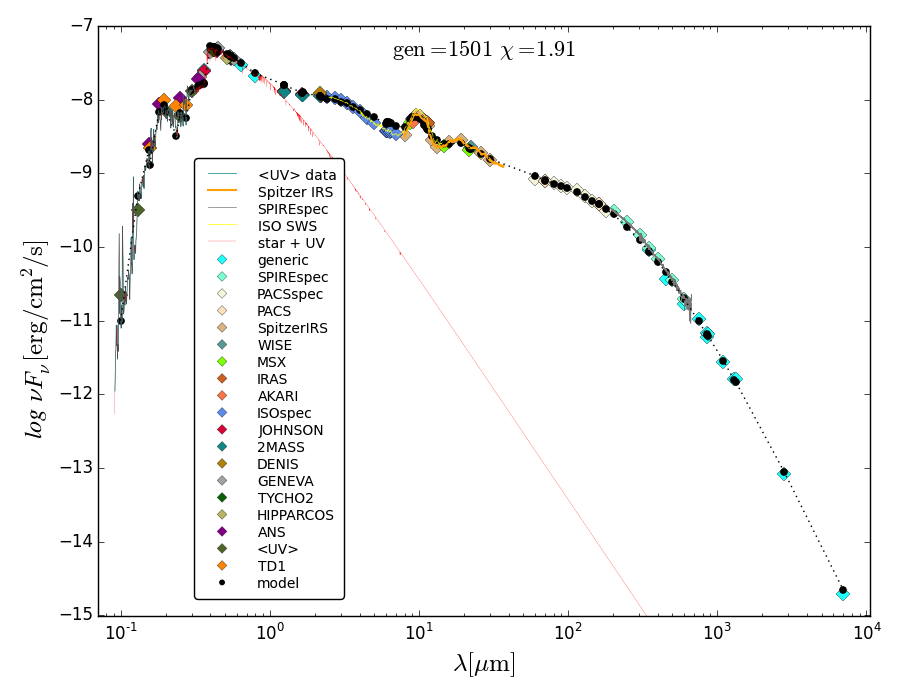} &
  \hspace*{-4mm}\includegraphics[width=87mm,height=74mm]{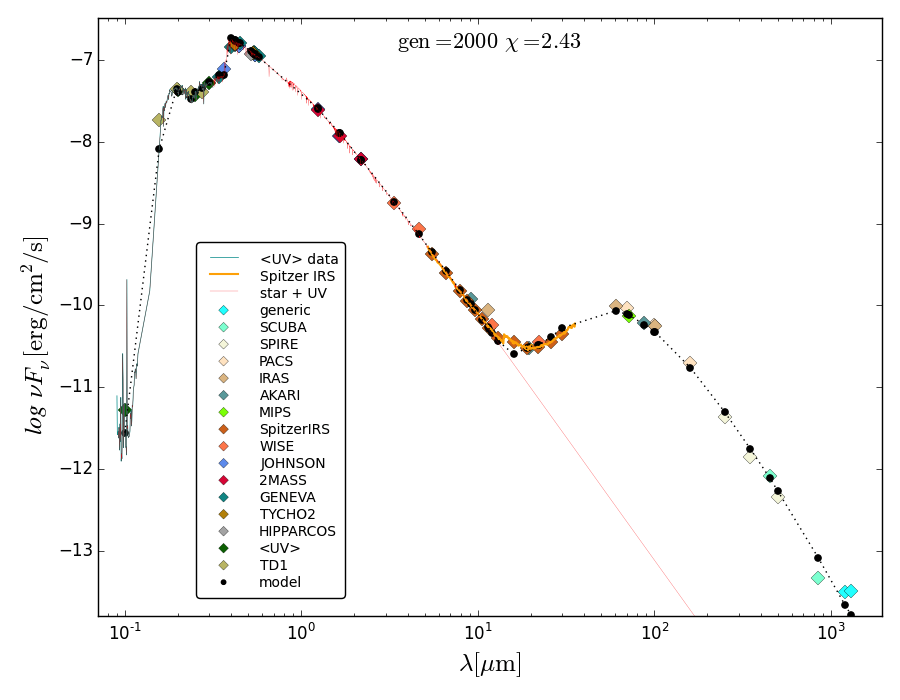}\\[-70mm] 
  \sf HD\,163296\hspace*{4mm} & \sf 49\,Cet\hspace*{3mm}\\[65mm]
  \end{tabular}
  \vspace*{-2mm}
\caption{SED-fitting models after reddening, in comparison to all
  photometric and low-resolution spectroscopic data. All spectroscopic
  data have been converted into a small number of spectral points. The
  red line is the fitted photospheric + UV spectrum of the star. The
  black dots represent the fluxes computed by {\sf MCFOST}, only at
  the wavelength points where we have observations. These model fluxes
  are connected by a black dashed line. The other colored dots and
  lines are the observational data as indicated in the legends. The
  ``generic'' points are individual measurements, usually in the
  mm-region, where a generic filter of type BOL (see
  Eq.\,\ref{filter_flux}) with a relative spectral width 12\% was
  applied.}
  \label{fig:SEDs}
\end{figure}

\setcounter{figure}{0}
\begin{figure}
  \hspace*{-7mm}
  \begin{tabular}{rr}
  \hspace*{-5mm}\includegraphics[width=87mm,height=74mm]{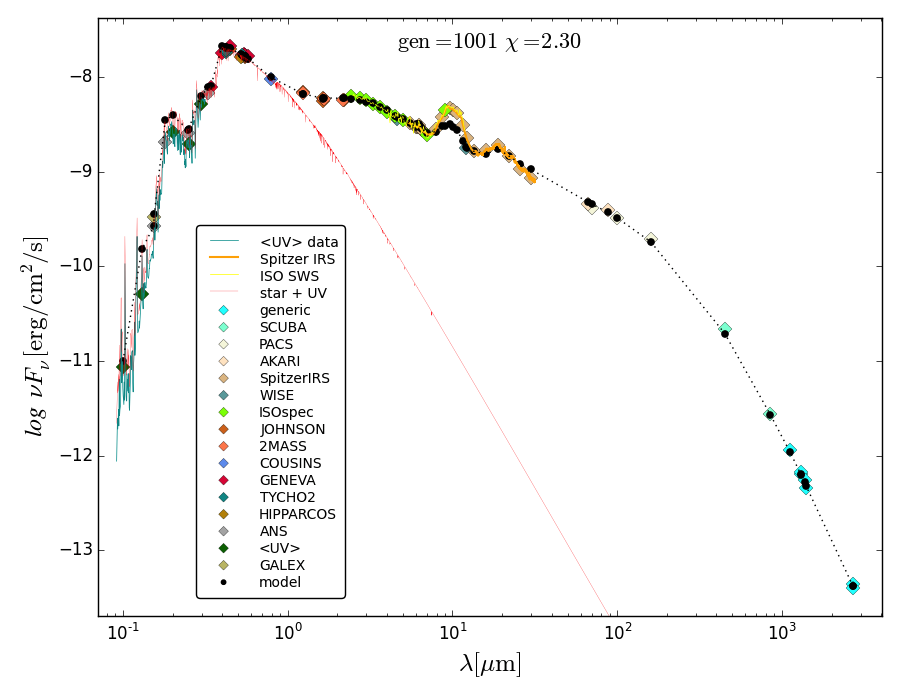} &
  \hspace*{-4mm}\includegraphics[width=87mm,height=74mm]{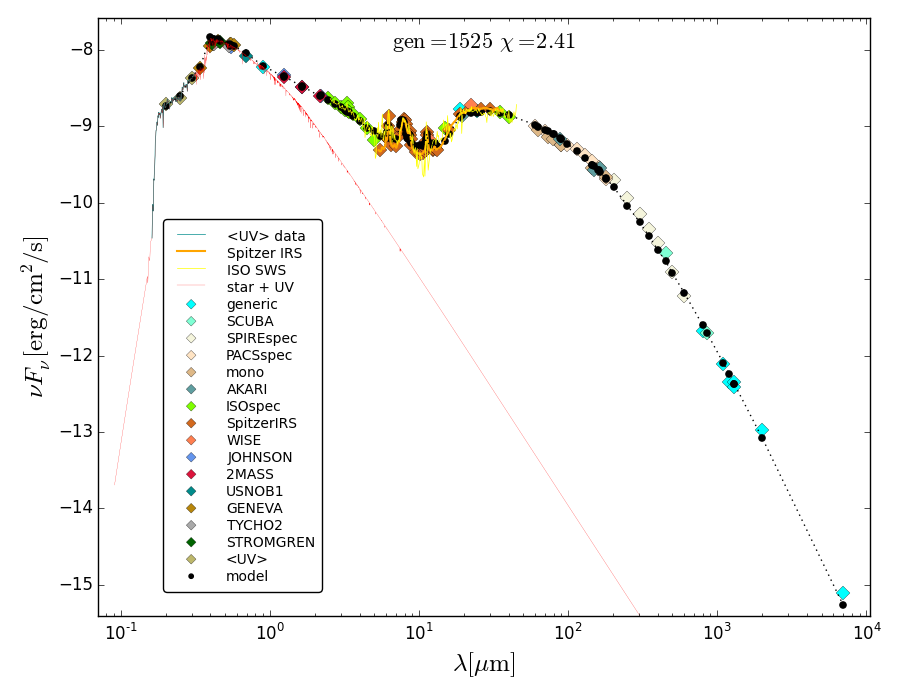} \\[-70mm]
  \sf MWC\,480\hspace*{3mm} & \sf HD\,169142\hspace*{4mm} \\[65mm]
  \hspace*{-5mm}\includegraphics[width=87mm,height=74mm]{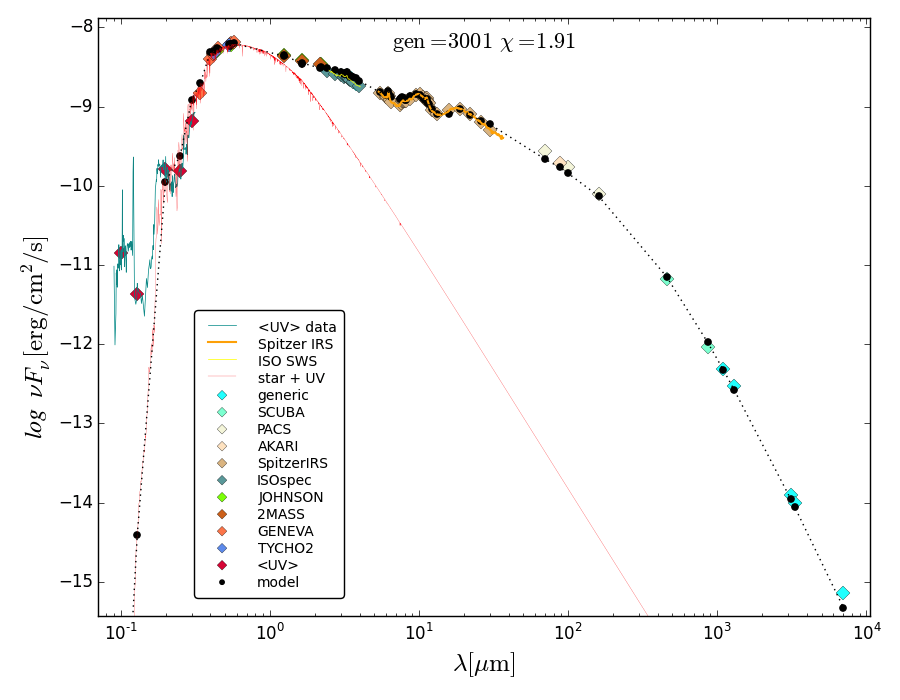} &
  \hspace*{-4mm}\includegraphics[width=87mm,height=74mm]{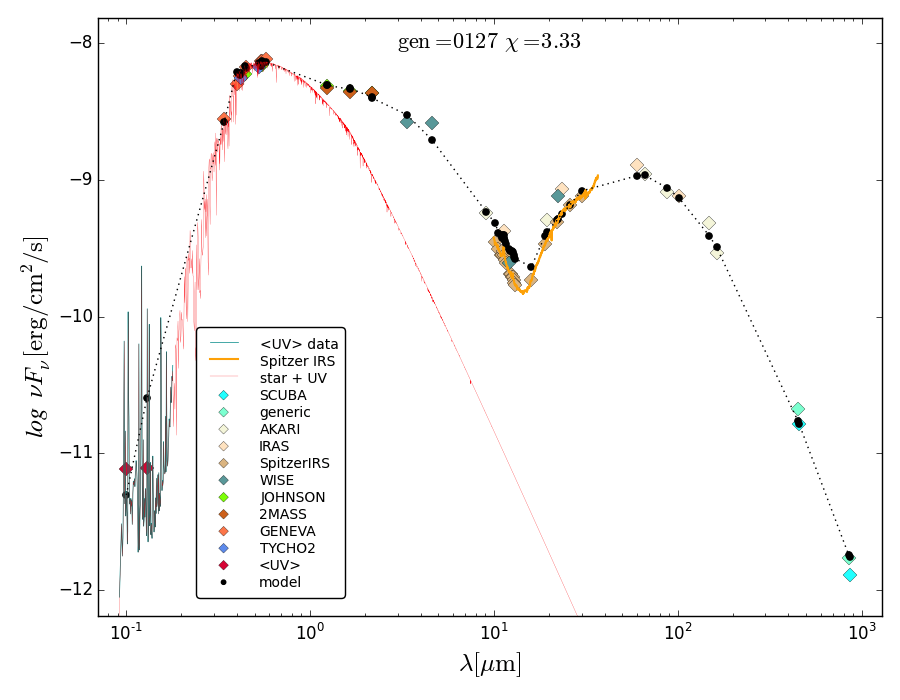} \\[-70mm]
  \sf HD\,142666\hspace*{4mm} & \sf HD\,135344B\hspace*{3mm} \\[65mm]
  \hspace*{-5mm}\includegraphics[width=87mm,height=74mm]{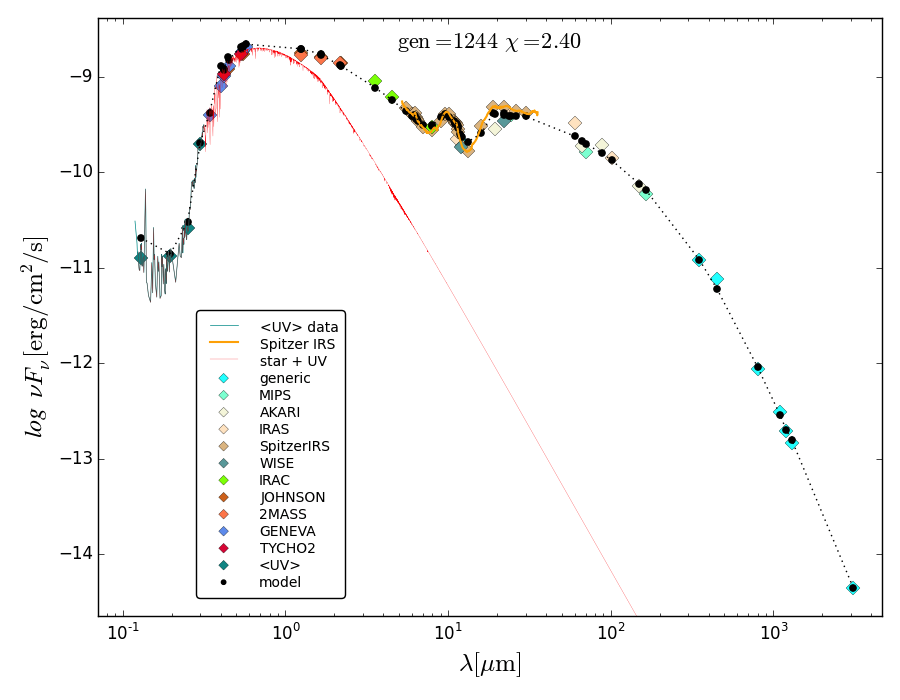} &
  \hspace*{-4mm}\includegraphics[width=87mm,height=74mm]{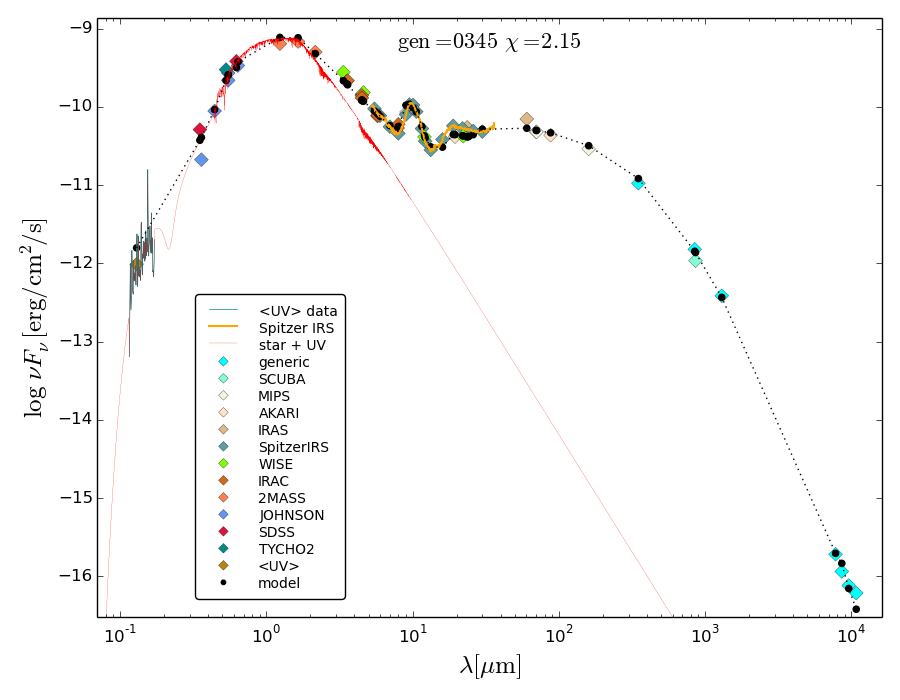}\\[-70mm]
  \sf V\,1149\,Sco\hspace*{3mm} & \sf Lk\,Ca\,15\hspace*{4mm}\\[65mm]
  \end{tabular}
  \vspace*{-2mm}
\caption{(continued)}
\end{figure}

\setcounter{figure}{0}
\begin{figure}
  \hspace*{-7mm}
  \begin{tabular}{rr}
  \hspace*{-5mm}\includegraphics[width=87mm,height=74mm]{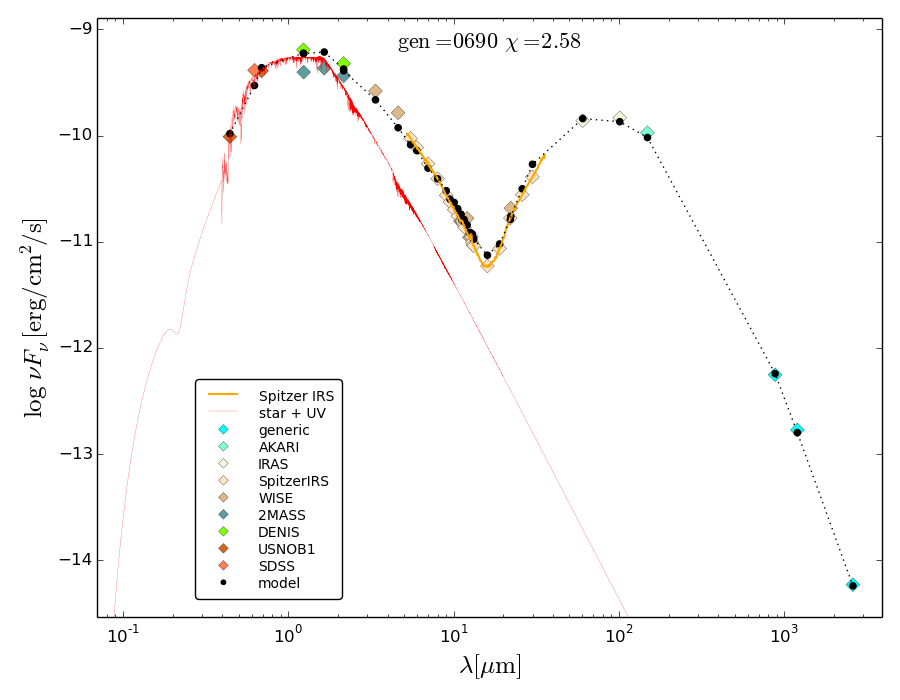} &
  \hspace*{-4mm}\includegraphics[width=87mm,height=74mm]{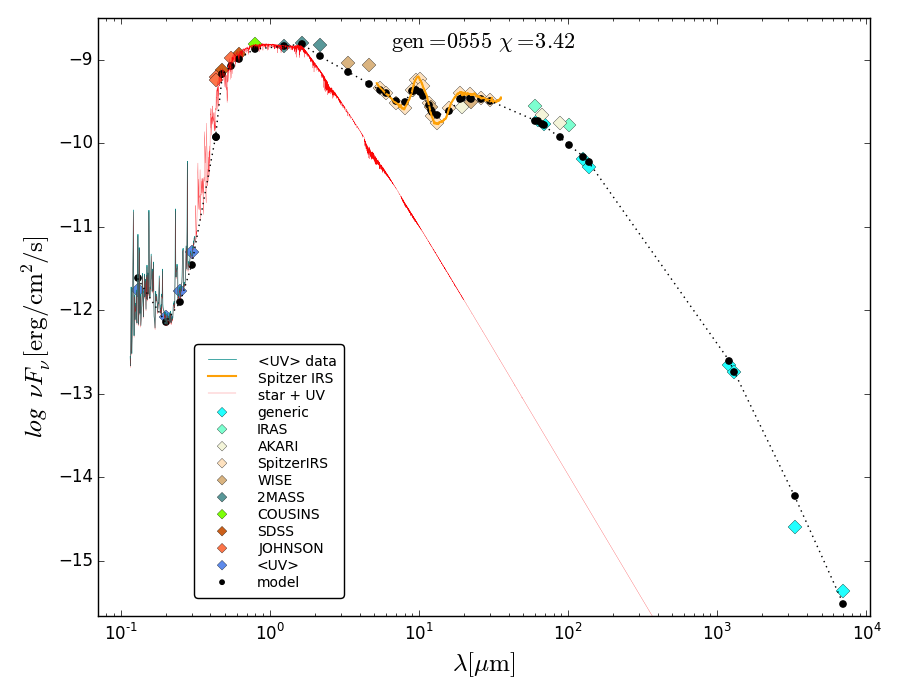} \\[-70mm]
  \sf UScoJ1604-2130\hspace*{4mm} & \sf RY\,Lup\hspace*{5mm} \\[65mm]
  \hspace*{-5mm}\includegraphics[width=87mm,height=74mm]{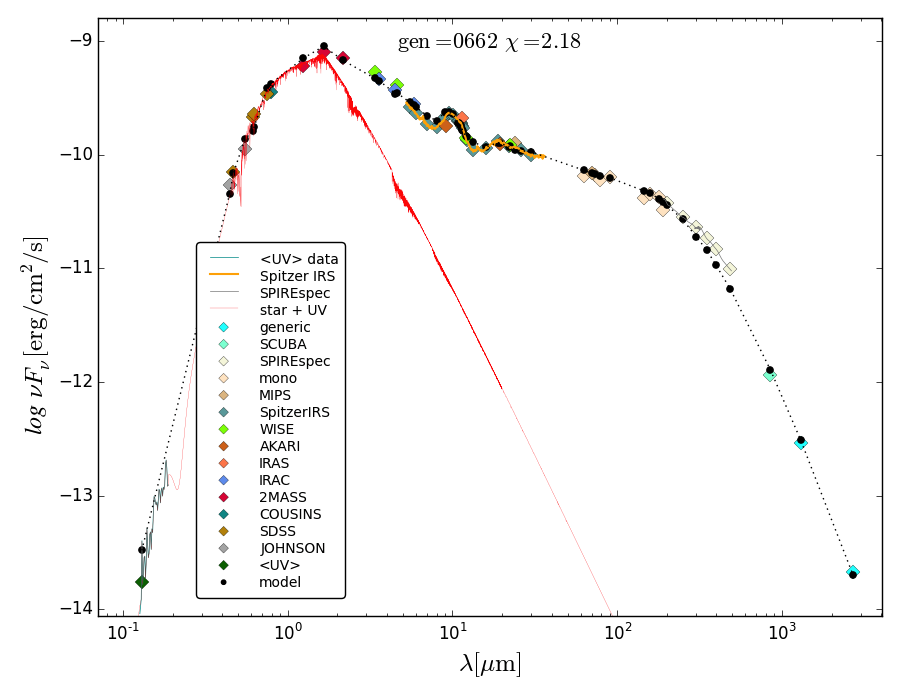} &
  \hspace*{-4mm}\includegraphics[width=87mm,height=74mm]{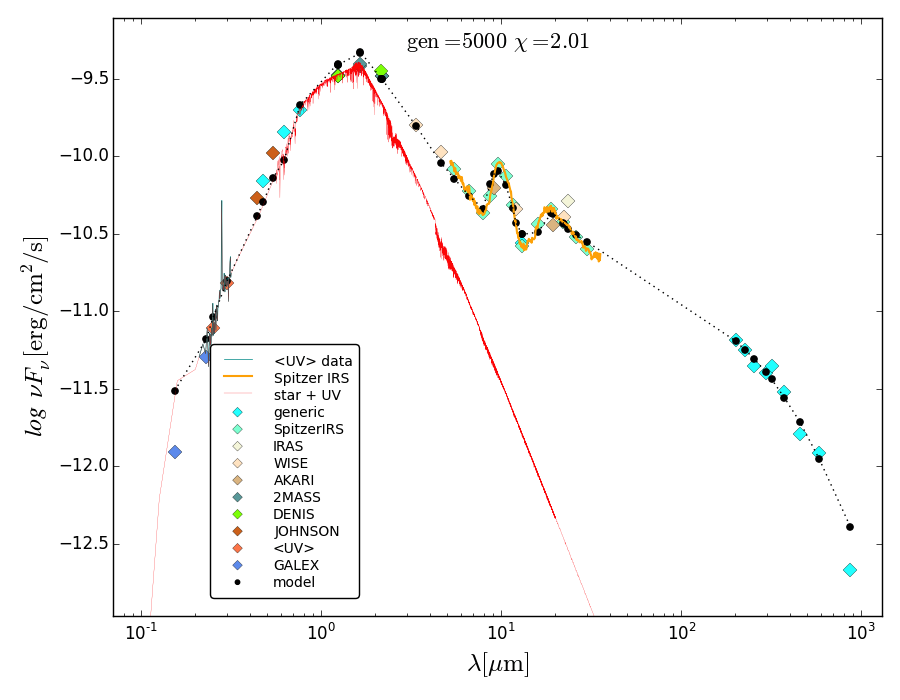} \\[-70mm]
  \sf CI\,Tau\hspace*{3mm} & \sf TW\,Cha\hspace*{3mm} \\[65mm]
  \hspace*{-5mm}\includegraphics[width=87mm,height=74mm]{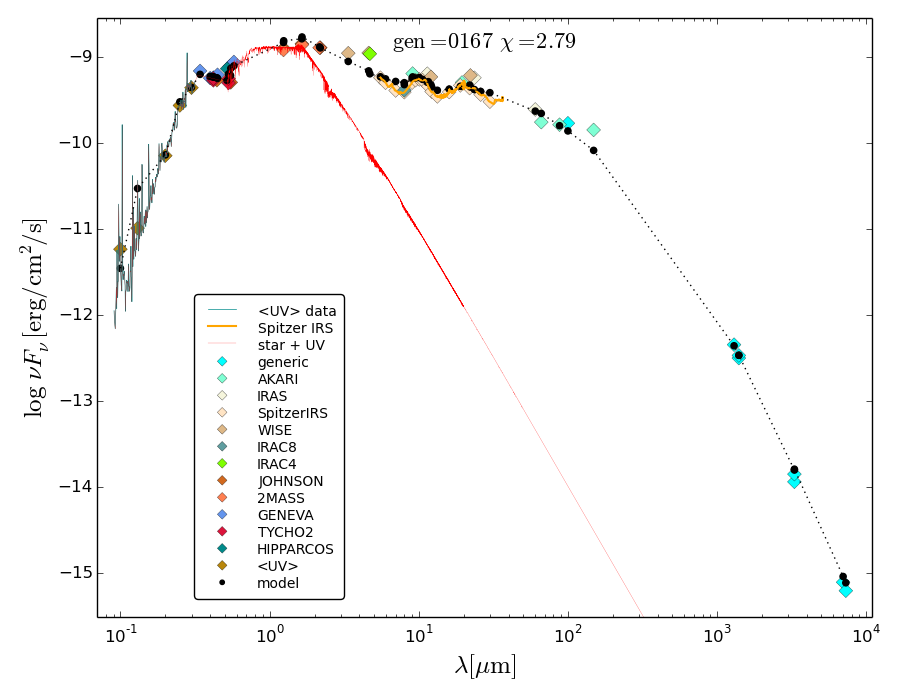} &
  \hspace*{-4mm}\includegraphics[width=87mm,height=74mm]{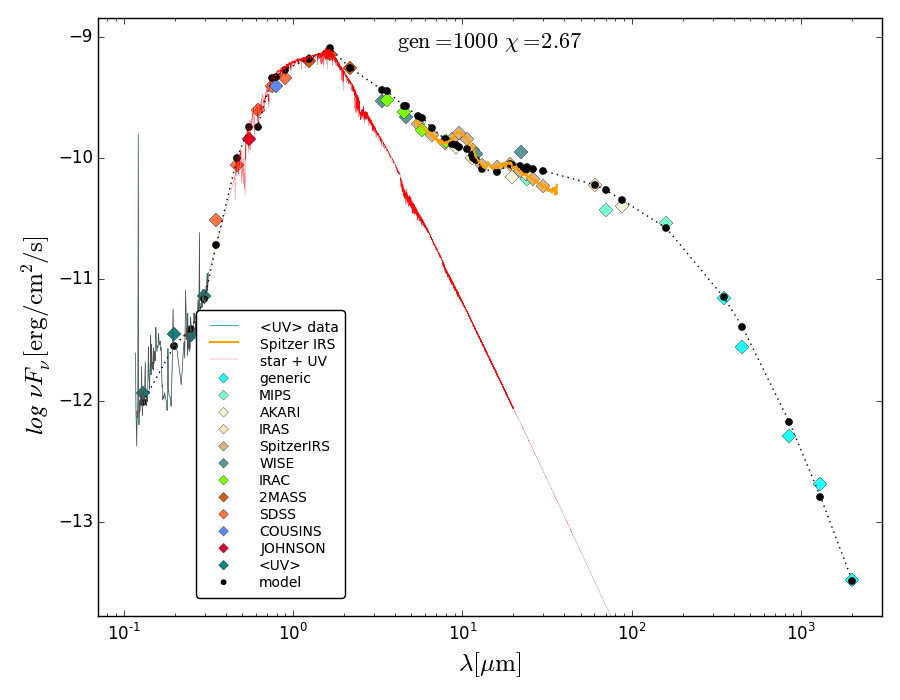} \\[-70mm]
  \sf RU\,Lup\hspace*{4mm} & \sf AA\,Tau\hspace*{3mm} \\[65mm]
  \end{tabular}
  \vspace*{-2mm}
\caption{(continued)}
\end{figure}

\setcounter{figure}{0}
\begin{figure}
  \hspace*{-7mm}
  \begin{tabular}{rr}
  \hspace*{-5mm}\includegraphics[width=87mm,height=74mm]{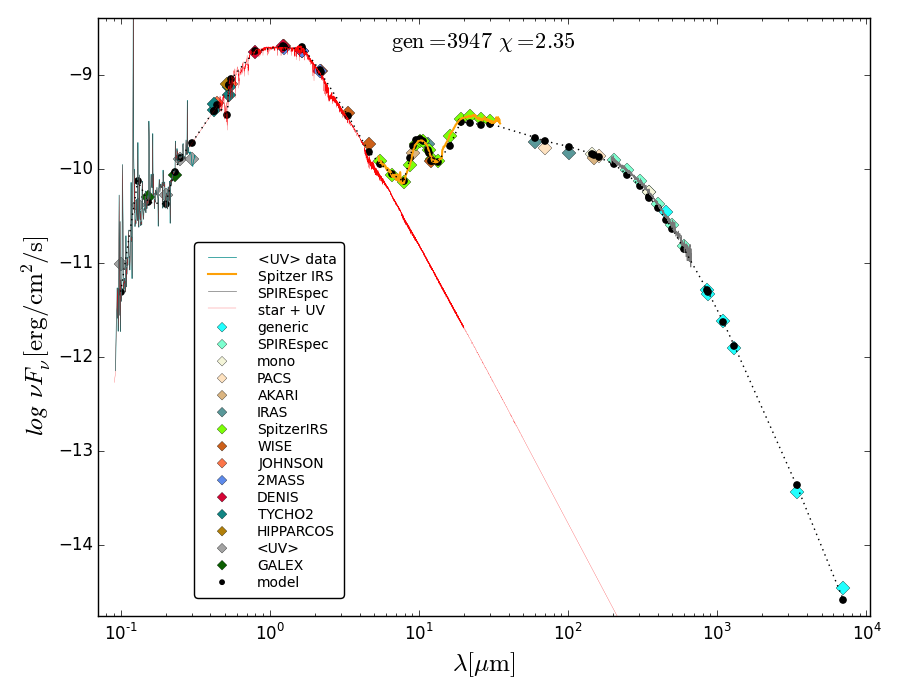} &
  \hspace*{-4mm}\includegraphics[width=87mm,height=74mm]{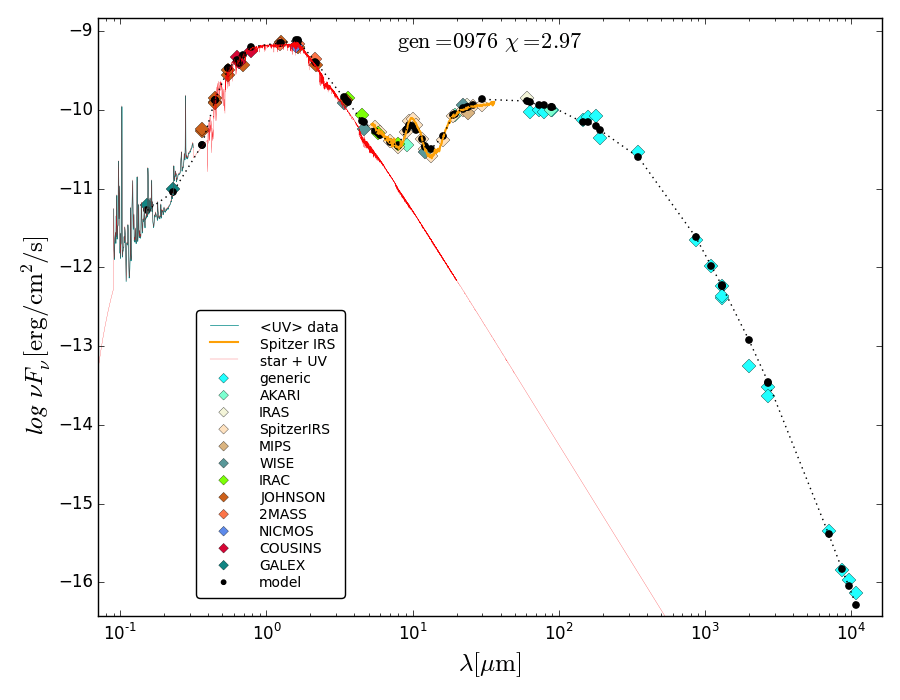}\\[-70mm]
  \sf TW\,Hya\hspace*{4mm} & \sf GM\,Aur\hspace*{3mm}\\[65mm]
  \hspace*{-5mm}\includegraphics[width=87mm,height=74mm]{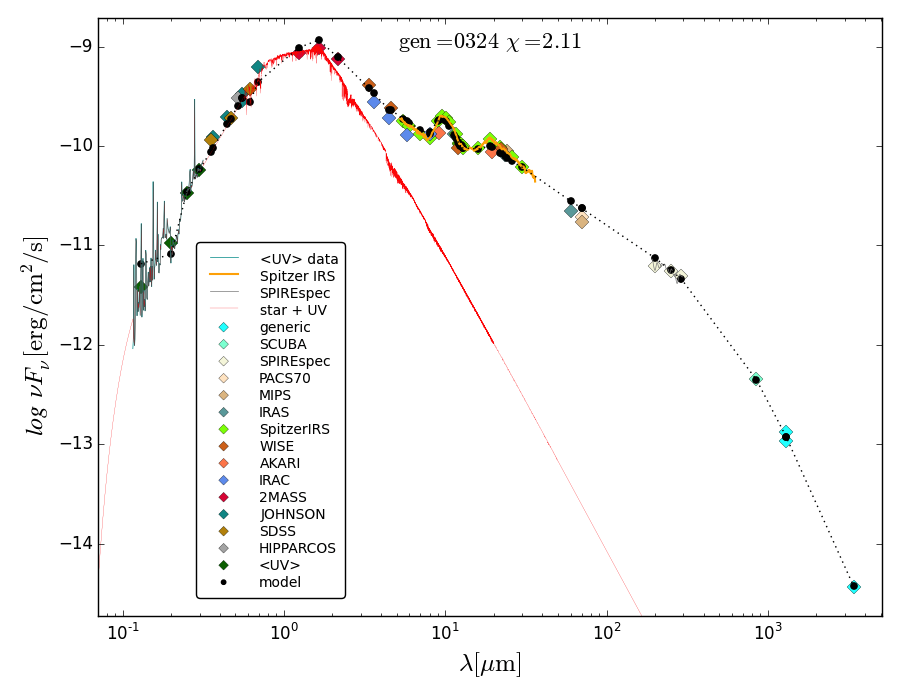} &
  \hspace*{-4mm}\includegraphics[width=87mm,height=74mm]{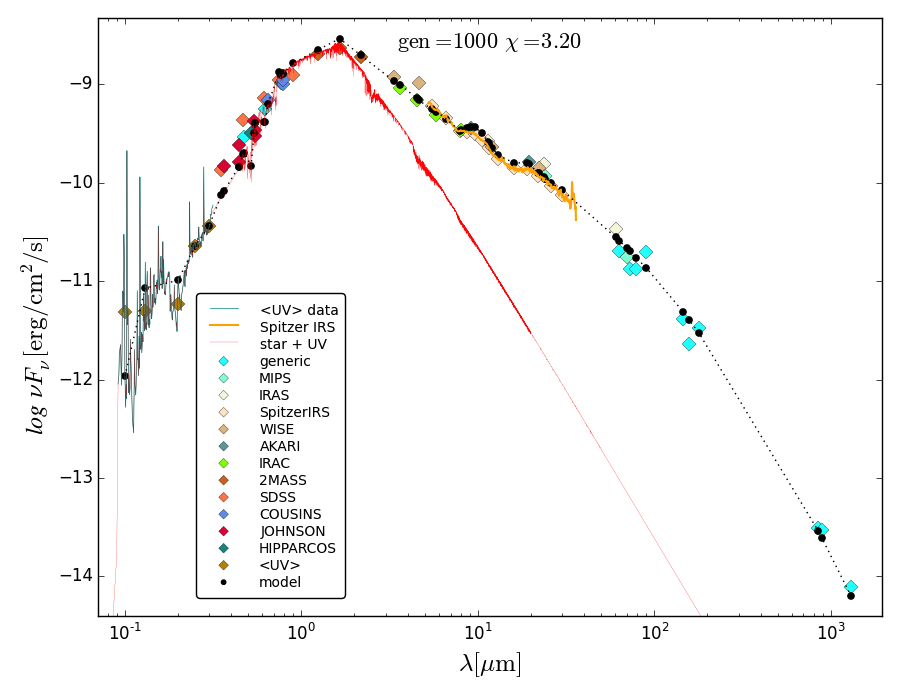} \\[-70mm]
  \sf BP\,Tau\hspace*{3mm} & \sf DF\,Tau\hspace*{3mm} \\[65mm]
  \hspace*{-5mm}\includegraphics[width=87mm,height=74mm]{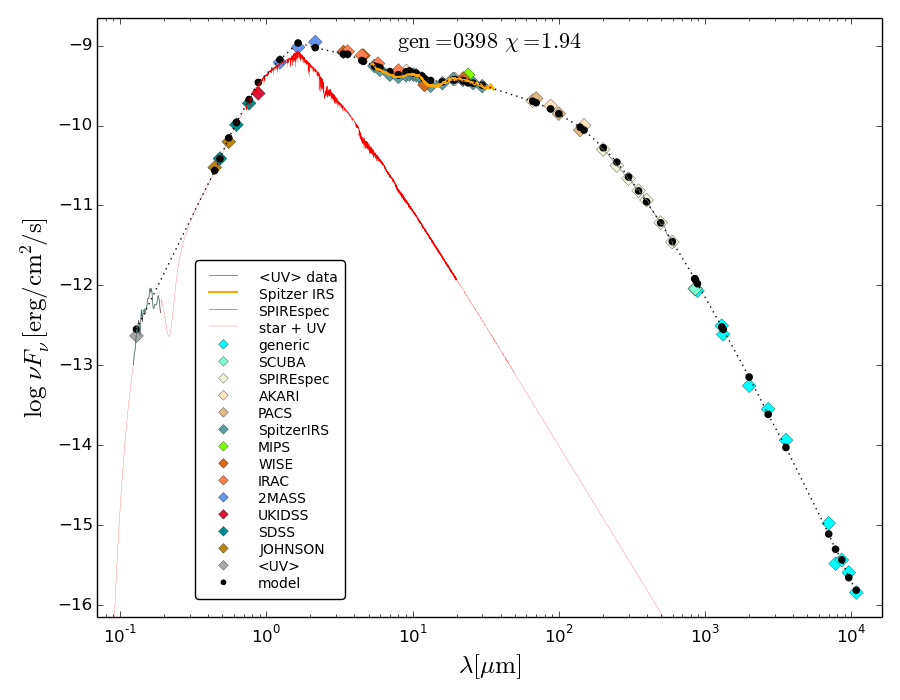} &
  \hspace*{-4mm}\includegraphics[width=87mm,height=74mm]{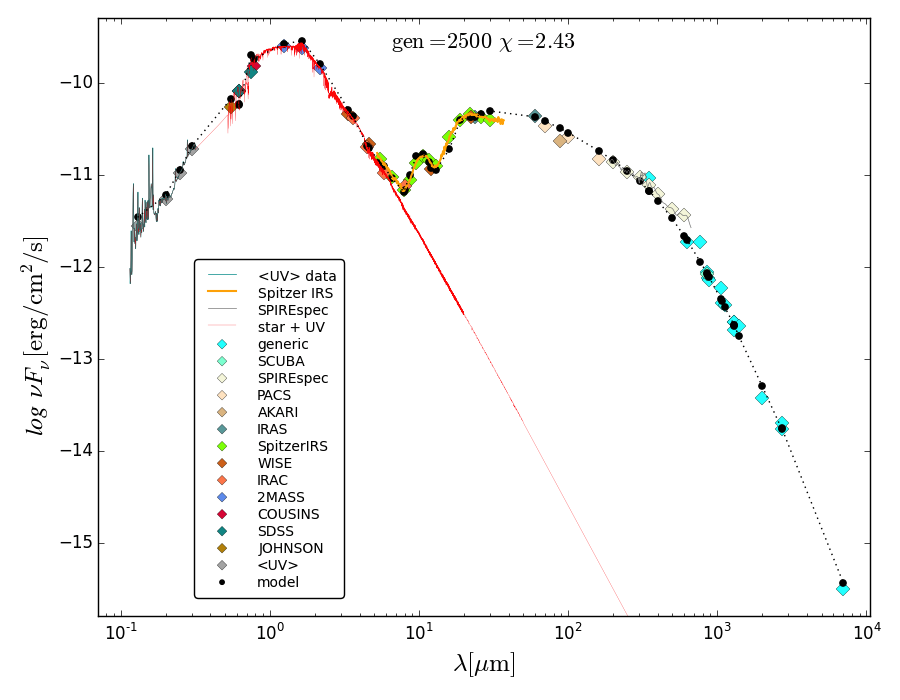} \\[-70mm]
  \sf DO\,Tau\hspace*{4mm} & \sf DM\,Tau\hspace*{4mm} \\[65mm]
  \end{tabular}
  \vspace*{-2mm}
\caption{(continued)}
\end{figure}

\setcounter{figure}{0}
\begin{figure}
  \hspace*{-7mm}
  \begin{tabular}{rr}
  \hspace*{-5mm}\includegraphics[width=87mm,height=74mm]{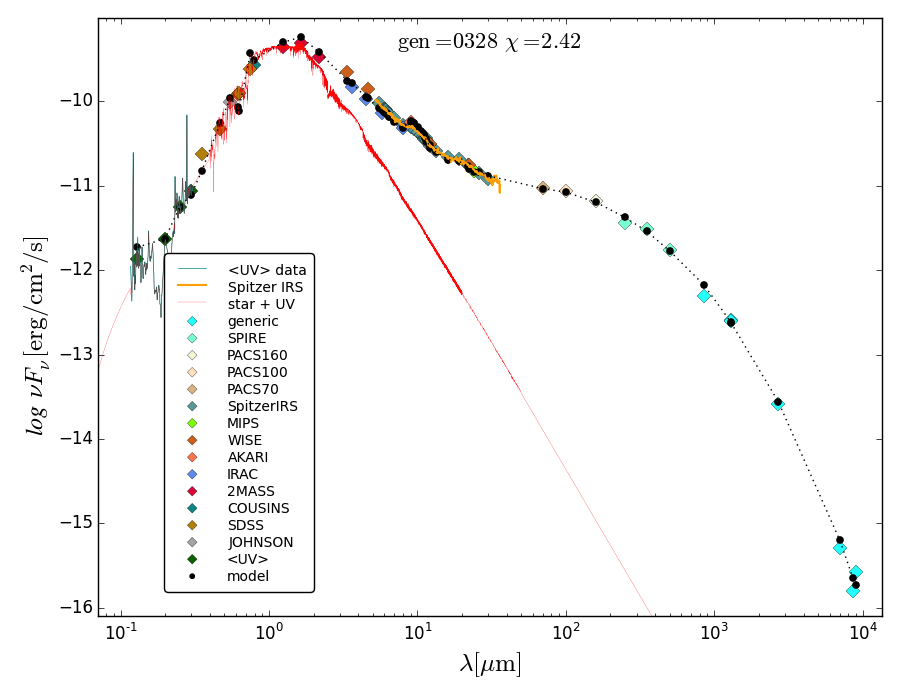} &
  \hspace*{-4mm}\includegraphics[width=87mm,height=74mm]{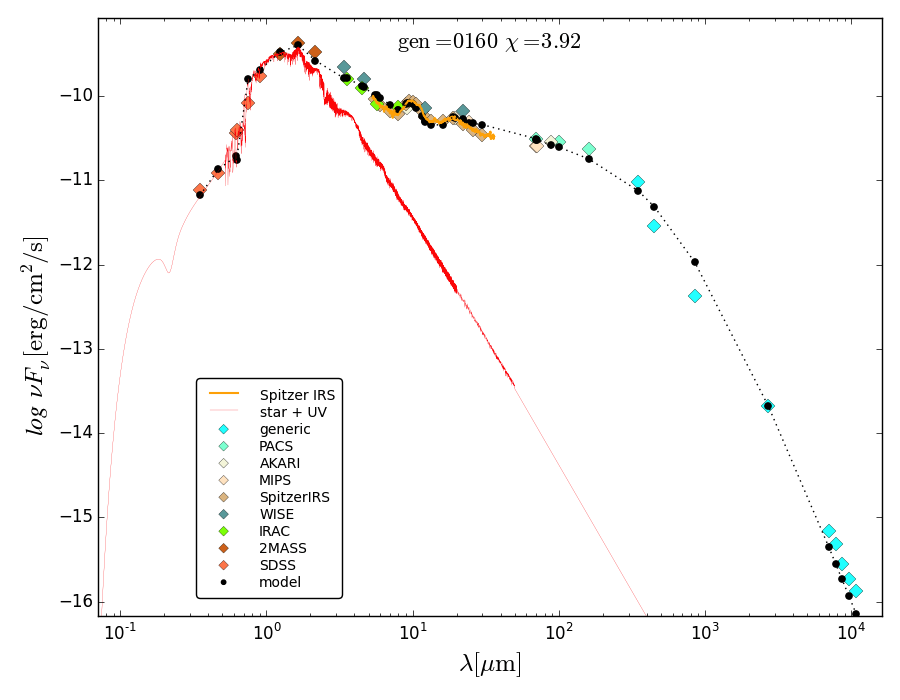} \\[-70mm]
  \sf CY\,Tau\hspace*{3mm} & \sf FT\,Tau\hspace*{3mm}\\[65mm]
  \hspace*{-5mm}\includegraphics[width=87mm,height=74mm]{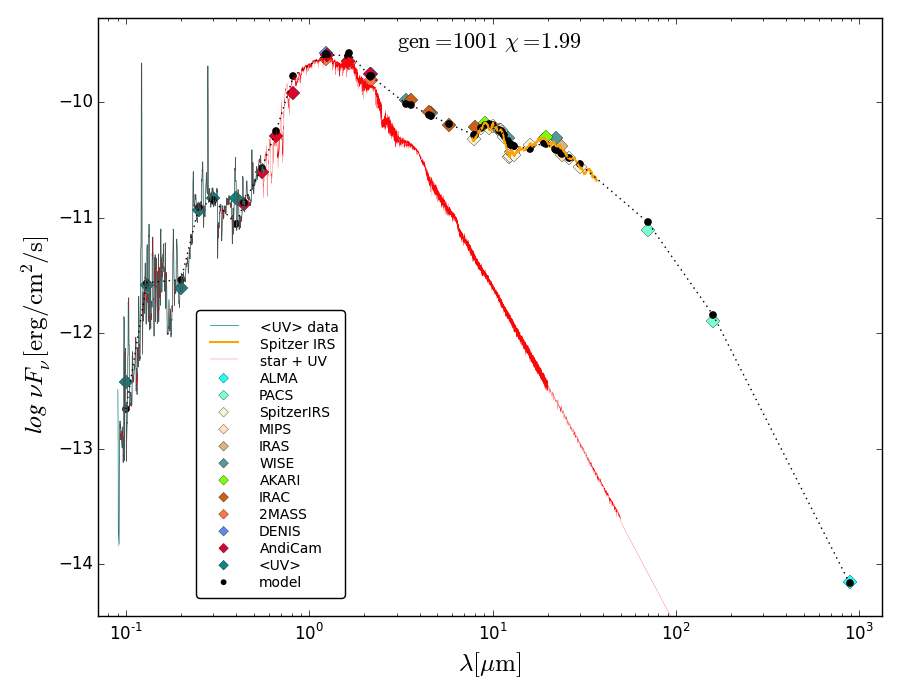}\\[-70mm]
  \sf RECX\,15\hspace*{3mm}\\[65mm]
  \end{tabular}
  \vspace*{-4mm}
\caption{(continued)}
\end{figure}

The SED-fitting models are relatively fast. One radiative transfer
model with {\sf MCFOST} needs about $5-10$ minutes on 6 CPU-cores,
allowing us to complete about $150-1500$ generations ($1800-18000$
models) per target to find a good fit to all photometric and
spectroscopic data, including the Spitzer PAH and silicate emission
features. The same genetic fitting algorithm was used as explained in
Sect.~\ref{sec:step1}. However, a thorough determination of the
errorbars of our results, for example by applying the Markov Chain
Monte Carlo (MCMC) method, is already quite cumbersome in the
SED-fitting stage.  Since we have about 15 free parameters, we would
need to run hundreds of thousands of disk models to sample all
relevant regions of the parameter space, which would correspond to
about $5\times 10^5$ CPU-hours or 20000\,CPU-days. Even if we had 100
processors available to us at all times, we would still need to wait
for more than half a year to finish one of our SED-fitting models with
errorbars. We therefore decided that we do not have the resources to
perform such an analysis.  The uncertainties in our determinations of
disk properties are roughly estimated in App.~\ref{app:uncertainties}.

The obtained SED-fits are visualized in Fig.~\ref{fig:SEDs} with
enlargements of the mid-IR region including the 10 and 20\,$\mu$m
silicate features and the PAH emission bands in Fig.~\ref{fig:midIR}
of App.~\ref{app:midIR}. The SED-fits obtained are very
convincing. The detailed mid-IR spectra, which are plotted on a linear
scale in Fig.\,\ref{fig:midIR}, show some shortcomings of our
global fitting strategy. Of course, one could subtract ``the
continuum'' and use more free parameters for the dust properties (mix
of materials, crystalline/amorphous, dust size and shape distribution
$\to$ opacity fitting), to get a better fit for this limited
wavelength region, but such a model would not be applicable to the
entire SED and would not serve {\sf DIANA}'s purpose of determining
the disk shape and dust properties as preparation for the
thermo-chemical models. Our aim is to fit all available data by a
single model for each object, with a minimum set of free parameters,
here 4 parameters for the kind and size distribution of dust grains,
and 2 parameters for the PAHs.  And in this respect we think that our
results are actually quite remarkable as they broadly capture the
observed wavelength positions and amplitudes of the spectral
variations in many cases.  The observations obtained with different
instruments can also show some ambiguities, with issues due to different
fields of view or variability of the objects. On the chosen
linear scale, one can also start to see the noise in the MC
models. The resulting parameters and physical disk and PAH properties
are continued to be discussed in Sect.~\ref{sec:results}.

\subsection{Modeling step 3:\ \ thermo-chemical disk models 
            (DIANA standard models)}
\label{sec:DIANA}

\noindent Pure SED-fitting is well-known to suffer from various
degeneracies, which can only be resolved by taking into account
additional types of observational data. These degeneracies are often
grounded in certain physical effects, for example:
\begin{itemize}
\item The outer disk radius has very little influence on the SED.  In
  order to determine the radial extension of the dust in a disk,
  continuum images or visibilities at (sub-)mm wavelengths have to be
  taken into account.
\item To determine the radial extension of the gas, we need (sub-)mm
  molecular observations, preferably spatially resolved maps or line
  visibilities. However, already the fluxes and widths of (sub-)mm lines,
  such as low-$J$ rotational CO lines, contain this information.
\item There is a degeneracy between lacking disk flaring and 
  strong dust settling. Both physical mechanisms lead to a flat 
  distribution of dust in the disk, hence to very similar observational 
  consequences for all continuum observations. However, dust settling 
  leaves the vertically extended gas bare and exposed to the stellar 
  UV radiation, leading to higher gas temperatures and stronger gas 
  emission lines in general, hence the opposite effect on the strengths 
  of far IR emission lines \citep{Woitke2016}. By taking into account
  mid or far-IR line flux observations, we can break this degeneracy.
\item More transparent dust in the UV, for example by changing the
  size-distribution parameters of the dust grains, leads to enhanced 
  gas heating and line formation, but has only little influence on the
  appearance of the dust at longer wavelengths. This is an important
  degree of freedom in our models to adjust the emission line fluxes,
  whereas the effects on the continuum appearance are rather subtle 
  and can be compensated for by adjusting other, for example disk
  shape parameters.
\item A tall inner disk zone can efficiently shield the outer disk
  from stellar UV and X-ray photons. Such shielding reduces gas
  heating and emission lines coming from an outer disk.
\end{itemize}
The final step of our data analysis is therefore to run full radiation
thermo-chemical models with an enlarged set of continuum and line
observations.  This is the core of the project, involves running full
{\sf ProDiMo} models, and is by far the computationally most demanding
task. Most published works on fitting gas properties of disks have
fixed the disk dust structure after SED-fitting (multi-stage models,
see Sect.~\ref{sec:intro}), and only adjusted a few remaining gas
parameters (such as the dust-to-gas ratio or the element abundances)
and chemical rate-networks to fit the line observations.  Our
ambitious goal in the {\sf DIANA} project was {\sl not}\, to do that.
From our experience with the dependencies of predicted line
observations as function of dust properties and disk shape, freezing
the spatial distribution and properties of the dust grains may be not
suitable for fitting line observations, because these properties
matter the most for the gas emission lines. The details of our
thermo-chemical models are explained in \citep{Woitke2016,Kamp2017},
which we summarize here as follows
\begin{itemize}
\item usage of detailed UV and X-ray properties of the central stars
  to determine the chemical processes in the disk after detailed
  UV radiative transfer and X-ray extinction in the disk,
\item physical description of dust settling by balancing upward
  turbulent mixing against downward settling, resulting in changes 
  of the dust structure when the gas properties are altered,
\item consistent use of PAH abundance in continuum radiative transfer,
  gas heating and chemistry,
\item the same element abundances in all disks and in all disk zones
  \citep[values from table~5 in][]{Kamp2017},
\item fixed isotope ratios $\rm^{13}C/^{12}C\!=\!0.014$,  
  $\rm^{18}O/^{16}O\!=\!0.0020$ and $\rm^{17}O/^{16}O\!=\!0.00053$ 
  for all disks, no isotope selective photodissociation,
\item fixed dust-to-gas ratios in each zone before dust settling, 
  value can depend on object and on disk zone,
\item small chemical rate network (about 100-200 species) with
  freeze-out, thermal and photodesorption \citep{Kamp2017}, but no
  surface chemistry other than H$_2$ formation on grains
  \citep{Cazaux2004,Cazaux2010},
\item chemical concentrations are taken from the time-independent 
  solution of the chemical rate-network (no time-dependent models).
\item the same standard H$_2$ cosmic ray ionization rate $(1.7\times
  10^{-17}\rm\,s^{-1})$ and the same background interstellar UV field
  strength $(\chi_{\rm ISM}\!=\!1)$ for all objects.
\end{itemize} 
Or fitting approach was to use the SED-fitted models as starting points
in parameter space, but then to {\sl continue varying the dust
  and disk shape parameters}, along with a few additional gas parameters,
{\sl as we fit an enlarged set of line and continuum observations}.
\begin{table}[!t]
\caption{DIANA standard models for 14 disks. See
  Table~\ref{tab:SEDfit} for the (unchanged) model setup concerning 
  number of disk zones and inclusion of PAHs in radiative transfer.}
\label{tab:DIANA}
\vspace*{-7mm}
\begin{center}
\def\z{$\hspace*{-1mm}$} 
\def\zz{$\hspace*{-2mm}$}
\hspace*{-26mm}\resizebox{196mm}{!}{
\begin{tabular}{c|ccc|ccc|cccc|c}
\hline
object & chemistry & visc.heat.? & \#\,free para.\z & 
  $N_{\rm phot}$  & $N_{\rm spec}$ & $N_{\rm image}$ & 
  $N_{\rm lines}$ & $N_{\rm widths}$ & $N_{\rm velo}$ & $N_{\rm maps}$ & 
  final\,$\chi$\\
&&&&&&&&&\\[-2.2ex]
\hline
HD\,97048        & small          & no  & 21 & 41 & 4 & --& 37 & --& --& --& 0.99\\
AB\,Aur          & large          & yes & 22 & 69 & 6 & 4 & 65 & 28& 3 & 1 & 1.87\\
HD\,163296       & small          & yes & 23 & 69 & 4 & 2 & 35 & 5 & 4 & 4 & 1.01\\
MWC\,480         & large          & no  & 10 & 44 & 2 & 2 & 32 & 8 & 4 & 2 & 9.0$^{(3)}$\\
HD\,169142       & small          & no  & 14 & 30 & 4 & --& 2  & 2 & --& --& 3.28\\
HD\,142666       & small          & no  & 13 & 32 & 1 & 1 & 11 & 1 & 1 & --& 1.41\\
Lk\,Ca\,15       & large          & no  & 20 & 48 & 2 & --& 14 & 8 & 8 & --& 2.24\\
\zz USco\,J1604-2130\z & small    & no  & 20 & 18 & 1 & --& 4  & --& --& --& 1.35\\
TW\,Hya          & large$^{(1)}$\zz& yes & 22 & 34 & 2 & 1 & 48 & 12& 3 & 3 & 1.43\\
GM\,Aur          & small          & no  & n.a.$^{(2)}$ 
                                             & 55 & 1 & --& 18 & --& --& --& 3.67\\
BP\,Tau          & small          & yes & 21 & 34 & 2 & --& 6  & 3 & 3 & 1 & 1.39\\
DM\,Tau          & large          & no  & 21 & 32 & 2 & 2 & 13 & 2 & 2 & 2 & 0.92\\
CY\,Tau          & small          & yes & 18 & 30 & 1 & 1 & 7  & 5 & 5 & 3 & 1.94\\
RECX\,15         & small          & yes & 16 & 26 & 1 & --& 10 & 1 & 2 & --& 0.84\\
\hline
\end{tabular}}
\end{center}
\vspace*{-2mm}
\hspace*{0mm}\resizebox{169mm}{!}{\parbox{180mm}{
$N_{\rm phot}\!=$ number of selected photometric data points.\ \  
$N_{\rm spec}\!=$ number of low-resolution spectra, for example Spitzer/IRS,
                Herschel/PACS, Herschel/SPIRE, ISO/SWS or ISO/LWS.\ \ 
$N_{\rm image}\!=$ number of continuum images or visibility data files,
                for example from NICMOS, SUBARU, SMA, ALMA or MIDI.\ \ 
$N_{\rm lines}\!=$ number of observed line fluxes (including upper limits).\ \ 
$N_{\rm velo}\!=$ number of high-resolution line velocity profiles, 
                for example VLT/CRIRES, SMA, ALMA, and two NIRSPEC observations
                probing full series of CO fundamental lines for BP\,Tau and CY\,Tau.\ \ 
$N_{\rm maps}\!=$ number of spatially resolved line datasets,
                converted to line-integrated intensity profiles,
                for example SMA, ALMA.\ \ 
More details about the data, object by object, are given in 
Sect.~\ref{sec:DIANAstand}\\[1mm]
$^{(1)}$: In the TW\,Hya model, D, D$^+$, HD and HD$^+$ have been
         included as additional species to predict the 
         detected HD 1-0 line at 112\,$\mu$m.\\
$^{(2)}$: value not recorded (hand-fitted model)\\
$^{(3)}$: mismatch of NICMOS-image at $\lambda\!=\!1.6\,\mu$m destroys 
         the otherwise fine fit for MWC\,480, possibly a problem related
         to the variability of the object.
}}
\end{table}
All continuum observations used before remain part of the fit quality
$\chi^2$.  The additional observational data include continuum images
and visibilities, line fluxes, line velocity profiles and integrated
line maps, see Table~\ref{tab:DIANA}. For each of these observations
we evaluate a fit quality by calculating additional
$\chi^2_{\rm type}$, which are then added together to form the overall
model $\chi^2$
\begin{equation}
  \chi^2 = w_{\rm phot}\,\chi^2_{\rm phot}
         + w_{\rm spec}\,\chi^2_{\rm spec}
         + w_{\rm image}\,\chi^2_{\rm image}
         + w_{\rm line}\,\chi^2_{\rm line}
  \label{eq:totchi}
\end{equation}
where the weights $w_{\rm type}$ of the different types of
observations are chosen by the modeler and are normalized to $w_{\rm
  phot}\!+\!w_{\rm spec}\!+\!w_{\rm image}\!+\!w_{\rm line}=1$.  The
fit quality of the photometric data $\chi^2_{\rm phot}$ is calculated as
in Eq.\,(\ref{eq:chi}).  $\chi^2_{\rm spec}$ is computed in the same
way by summing up the differences between $\log F_{\lambda_i}^{\rm
  model}$ and $\log F_{\lambda_i}^{\rm obs}$ on all wavelength points
$\lambda_i$ given by the observational spectrum, after reddening and
interpolation in the model spectrum. Concerning the image data, we
have averaged the 2D-intensity data in concentric rings, resulting in
radial intensity profiles.  The model images are treated in the same
way as the observations after rotation and convolution with the
instrument point spread function (PSF). We then apply again
Eq.\,(\ref{eq:chi}) to obtain $\chi^2_{\rm image}$.  The definition of
our $\chi_{\rm line}$ is special and depends on the available data.
We first compute $\chi^2_{\rm flux}$ for one line by comparing the
model and observed line fluxes according to Eq.\,(\ref{eq:chi}). If
the full width half maximum (FWHM) of the line has been measured, we
also compute
\begin{equation}
  \chi^2_{\rm FWHM} = \left(\frac{{\rm FWHM}^{\rm mod}-{\rm FWHM}^{\rm obs}}
                                {\sigma_{\rm FWHM}}\right)^2 \ .
\end{equation}
If velocity profiles are available, we use again Eq.\,(\ref{eq:chi})
to compute $\chi^2_{\rm velo}$, and if a line map is available
(converted to a radial line intensity profile by averaging over
concentric rings), we use Eq.\,(\ref{eq:chi}) to calculate
$\chi^2_{\rm map}$. Finally, these components are added together to
compute
\begin{equation}
   \chi^2_{{\rm line},k} = \left\{\begin{array}{ll}
      \chi^2_{\rm flux}  & \mbox{if only line flux is observed}\\[1mm]
      \frac{1}{2}\,\chi^2_{\rm flux} + \frac{1}{2}\,\chi^2_{\rm FWHM}
                        & \mbox{if line flux and FWHM are observed}\\[1mm]
      \frac{1}{2}\,\chi^2_{\rm flux} + \frac{1}{2}\,\chi^2_{\rm velo}
                        & \mbox{if line flux and line profile are observed}\\[1mm]
      \frac{1}{3}\,\chi^2_{\rm flux} + \frac{1}{3}\,\chi^2_{\rm velo}
                                    + \frac{1}{3}\,\chi^2_{\rm map}
                        & \mbox{if line flux, profile and a map are observed}
   \end{array}\right.  \ ,
\end{equation}
where, for the total $\chi^2_{\rm line}$, we still need to average
over all observed lines $k$ in the dataset. {Our final choices how
  to fit each line for each object are recorded in the individual {\tt
    LINEobs.dat} files, which is contained, for example for DM\,Tau,
  in
  \url{http://www-star.st-and.ac.uk/~pw31/DIANA/DIANAstandard/DMTau_ModelSetup.tgz}}.
Therefore, our $\chi^2$ is not the result of a sound mathematical
procedure.  We have to carefully select and review the data, to see
whether the data quality is sufficient to include them in our fit
quality, and we have to carefully assign some weights to compensate
the different numbers of points associated with each kind of data. For
example, a line flux is one point, a line profile is composed of maybe
10-20 points, but a low resolution spectrum may contain hundreds of
data points.

\bigskip\noindent
In addition to the parameters of our SED-fitting disk models listed in
Sect.~\ref{sec:SED} (the dust-to-gas ratio is listed there already) we
have only the two following additional free model parameters for the
gas in the DIANA standard models:
\begin{enumerate}
\item[1.] the efficiency of exothermal reactions (global parameter) 
  $\gamma_{\rm chem}$, see \citep{Woitke2011} for explanations,
\item[2.] the abundance of PAHs with respect to the gas $f_{\rm PAH}$
  in each zone (only a new parameter when not yet included in the
  SED-fitting model).
\end{enumerate}
All other gas and chemical parameters are fixed throughout the project, in
particular the element abundances.  However, there are two choices
to be made for each object:
\begin{enumerate}
\item[3.] Choice of the size of the chemical rate network, either the
  small or the large {\sf DIANA}-standard chemical setup
  \citep{Kamp2017}. The small network has 12 elements and 100
  molecules and ice species, whereas the large network has 13 elements
  and 235 molecules and ice species.
\item[4.] Choice whether or not viscous heating is taken into
  account as additional heating process, according to a fixed 
  published value of the accretion rate $\dot{M}_{\rm acc}$.
\end{enumerate}
Concerning option~3, the small {\sf DIANA} chemical standard can be
used if line observations are available only for atoms and common
molecules like CO and H$_2$O. If larger and more complicated molecules
are detected such as HCN or HCO$^+$, the large {\sf DIANA} chemical
standard is recommended. Option~4 turned out to be essential to
explain some strong near-IR and mid-IR emission lines detected from
objects with high $\dot{M}_{\rm acc}/M_{\rm disk}$ values. We compute
the total heating rate of an annulus at distance $r$ in
$\rm[erg/cm^2/s]$ according to equation~2 in \citep{Dalessio1998},
\begin{equation}
  F_{\rm vis}(r) = \frac{3\,G M_\star \dot{M}_{\rm acc}}{8\pi r^3}
                  \left(1-\sqrt{\frac{R_\star}{r}}\right) \ ,
\end{equation}
and distribute this amount of heat in the vertical column
$\rm[erg/cm^3/s]$ as
\begin{equation}
  \Gamma_{\rm vis}(r,z) = F_{\rm vis}(r)\,
           \frac{\rho(r,z)^p}{\int_0^\infty \rho(r,z')^p\,dz'} \ ,
\end{equation} 
where $p\!=\!1$ leads to unstoppable heating at high altitudes where
cooling tends to scale as $\propto\!\rho^2$. We avoid this problem 
by setting $p\!=\!1.25$ as global choice. In the passive disk
models discussed in this paper, the dust energy balance is assumed not
to be affected by accretion, only the gas is assumed to receive 
additional heat via the action of viscosity and accretion.

Thermo-chemical disk models which obey the rules and assumptions
listed above and in Sect.~\ref{sec:SED} are henceforth called the {\sl
  DIANA standard models}.  We have not changed these assumptions
throughout the project as we continued to fit the models for more and
more targets. This procedure was highly debated among the team
members, as certain new modeling ideas might have helped to improve
the fits of some objects. However, for the sake of a uniform and
coherent disk modeling, we finally agreed to keep the modeling
assumptions the same for all objects.

One typical radiation thermo-chemical model, including Monte Carlo
radiative transfer, requires about 1 to 3 hours on 6 processors (6-18
CPU hours), depending on the size of the spatial grid, the size of the
chemical rate network, and the quantity and kinds of observational
data to be simulated. We have used the SED-fitted models as starting
points for the DIANA standard models, mostly using the same genetic
fitting algorithm as explained in Sect.~\ref{sec:step1}. However, some
team members preferred to fit just by hand. The number of free
parameters used during this final fitting stage also largely depended
on the judgment of the modeler, see Table~\ref{tab:DIANA}. Some team
members decided to fix as many as possible disk parameters as
determined during the SED-fitting stage, such as the inner disk
radius, the charge and abundances of the PAHs, or the dust masses in
the different disk zones. Other team members decided to leave more
parameters open, for example the dust size distribution, the disk
shape and the dust settling parameters. In such cases, of course, the
continuum radiative transfer must be re-computed.  In particular,
the disk extension and tapering-off parameters can be adjusted to
sub-mm line observations, dust settling and disk flaring can be
disentangled, and the shape of the inner disk, which is usually only
little constrained by the SED, can be fitted to visibility and, for
example, to CO ro-vibrational line data.  The convergence of each fit
was manually monitored, and decisions about data (de-)selection and
fitting weights sometimes needed to be revised on the fly. Again, all
these decisions cannot be automated, they need human expertise.

Computing 300 generations with 12 children per generation requires
3600 DIANA standard models, which can be calculated in about 20000 to
65000 CPU hours per object.  This was at the limit of the
computational resources available to us. Our results are probably
not unique and likely to be influenced by the initial parameter 
values taken from the SED-fitting models. It is probably
fair to state that our computational resources only allowed us to find
a $\chi^2$ minimum in the neighborhood of the SED-fitting model
in parameter space. Running MCMC models to determine errorbars was not
feasible.

\section{Results}
\label{sec:results}

\noindent The full results of our SED-fitting models are available at
\url{http://www-star.st-and.ac.uk/~pw31/DIANA/SEDfit} and the full
results of the {\sf DIANA}-standard models are available at
\url{http://www-star.st-and.ac.uk/~pw31/DIANA/DIANAstandard}. These
files include all continuum and line observations used, the fitted
stellar, disk and dust parameters, the resulting 2D physico-chemical
disk structures, including dust and gas temperatures, chemical
concentrations and dust opacities, and all files required to re-setup
the models and run them again for future purposes. Details about the
content of these files can be found in
Appendix~\ref{app:uploaded-data}.

We offer these results to the community for further analysis, and as
starting points to interpret other or maybe to predict new
observations. It is not our intention in this paper to discuss all
results in a systematic way. We rather want to show a few interesting
properties found for some individual objects, and to highlight a few
trends and results for the overall ensemble of protoplanetary disks
considered. The resulting UV and X-ray properties of the central
stars, derived from modeling step~1, are discussed in
Sect.~\ref{sec:stellar}, the disk dust masses obtained from modeling
step~2 are shown in Sect.~\ref{sec:dustmasses}, and the dust
properties and mm-slopes are discussed in Sect.~\ref{sec:mmslope}, as
well as the resulting PAH properties in Sect.~\ref{sec:PAHprop},
before we turn to the results of the individual {\sf DIANA} standard
models in Sect.~\ref{sec:DIANAstand}.

\subsection{UV and X-ray stellar properties}
\label{sec:stellar}

\noindent The strength and color of the UV and X-ray irradiation have
an important influence on the chemistry, heating and line formation in
the disk. The details about the UV and X-ray data used and methods
applied are explained in \citep{Dionatos2018}.  Considering the total
UV flux between 91.2 to 205\,nm (see Table~\ref{tab:stellar}), the
Herbig~Ae/Be stars are found to be about $10^4$ times brighter, but
this is mostly photospheric, soft UV radiation. If one focuses on the
hard UV from 91.2 to 111\,nm, which can photo-dissociate H$_2$ and CO,
the Herbig~Ae/Be stars are only brighter by a maximum factor of about
100.  Concerning soft X-rays, there is hardly any systematic
difference between the T\,Tauri and Herbig\,Ae stars, and for hard
X-rays between 1\,keV to 10\,keV, the brightest X-ray sources are
actually the T\,Tauri stars RY\,Lup, RU\,Lup, AA\,Tau. We note that
DM\,Tau is also identified as a strong X-ray source with an
unextincted X-ray luminosity of about $3\times 10^{30}\rm\,erg/s$ for
energies $>0.1\,$keV. Earlier Chandra observations \citep{Guedel2007}
only showed $1.8\times 10^{29}\rm\,erg/s$, but then later, XMM
observations \citep{Guedel2010} resulted in a much higher unextincted
X-ray luminosity of $2\times 10^{30}\rm\,erg/s$ for energies
$>0.3\,$keV, which is consistent with $3\times 10^{30}\rm\,erg/s$ for
energies $>0.1\,$keV.  We would like to emphasize again that the UV
and X-ray luminosities listed in Table~\ref{tab:stellar} are not the
observed values, but are the luminosities as seen by the disk,
assuming that the extinction between the emitting source and the disk
is small.

\subsection{Disk dust masses}
\label{sec:dustmasses}

\noindent The dust masses of protoplanetary disks are classically
derived from the observation of (sub-)mm continuum fluxes
\citep[e.g.][]{Andrews2005}, using some well-established values for
the dust absorption opacity and mean dust temperature.  Our results
show that this method can have large uncertainties because of the unknown
mean dust temperature in the disk, the unknown dust opacities, and the
particular behavior of cold disks
\citep[see also Sect.~5.3.3 in][]{Woitke2016}.
If the disk is entirely optically thin at frequency $\nu$, and if all
grains in the disk are warm enough to emit in the Rayleigh-Jeans
limit, the observable spectral flux $F_\nu$ is given by
\begin{equation}
  F_\nu ~=~ \frac{2\nu^2 k}{c^2\,d^2}
                \; \langle T_{\rm dust}\rangle 
                \; M_{\rm dust}
                \; \kappa^{\rm abs}_{850} \ ,
  \label{eq1}
\end{equation}
where $\nu$ is the frequency, $c$ is the speed of light, $d$ the
distance, $k$ the Boltzmann constant, $\langle T_{\rm dust}\rangle$
the mass-averaged dust temperature in the disk, and $\kappa^{\rm
  abs}_{850}$ the absorption coefficient per dust mass [$\rm cm^2/g$]
at 850\,$\mu$m, here assumed to be constant throughout the
disk. Equation\,(\ref{eq1}) is widely used in the literature to
measure total dust masses of protoplanetary disks
\citep[e.g.][]{Andrews2005}, and then drawing conclusions about the
disk gas mass by assuming $M_{\rm disk}\!=\!100\times M_{\rm dust}$.

\begin{table}
\vspace*{-1mm}
\caption{Total dust masses $M_{\rm dust}\,[10^{-4}\rm M_\odot]$ in the
  SED-fitting models compared to the values derived from the spectral flux $F_\nu$
  at $\lambda\!=\!850\,\mu$m. Uncertainties are discussed in
  Appendix~\ref{app:uncertainties}.}
\label{tab:dustmass}
\def\z{$\hspace*{-1mm}$} 
\def\zz{$\hspace*{-2mm}$}
\vspace*{0mm}\hspace*{-10mm}
\resizebox{160mm}{!}{
\begin{tabular}{c|ccc|cc|ccc|r}
\hline
&&&&&&&&&\\[-2.0ex]
object & \z$M_\star\,[M_\odot]$\zz 
       & d\,[pc] 
       & $F_{850}$\,[Jy] 
       & \z$\langle T_{\rm dust}\rangle$\,[K]\zz
       & $\langle\kappa^{\rm abs}_{850}\rangle$
       & $M_{\rm dust}^{\rm class}$\zz 
       & $M_{\rm dust}^{\rm derived}$ \zz
       & $M_{\rm dust}^{\rm model}$
       & \z$M_{\rm disk}/M_\star$\zz \\
&&&&&&&&&\\[-2.0ex]
\hline
&&&&&&&&&\\[-2.0ex]
HD\,100546  & 2.5  & 103  & 1.41   &  56.1  &  4.69  & 2.67   & 0.710  & 0.708& 0.28\,\%\\
HD\,97048   & 2.5  & 171  & 2.02   &  34.9  &  2.42  & 10.6   & 8.75   & 13.20& 5.3\,\% \\
HD\,95881   & 2.5  & 171  & 0.0346 &  131   &  7.93  & 0.181  & 0.0122 & 0.394& 0.16\,\%\\
AB\,Aur     & 2.5  & 144  & 0.585  &  29.7  &  2.85  & 2.17   & 1.79   & 2.20 & 0.88\,\%\\
HD\,163296  & 2.47 & 118  & 1.93   &  30.1  &  6.00  & 4.86   & 1.88   & 5.29 & 2.1\,\% \\
49\,Cet     & 2.0  & 59.4 & 0.0173 &  58.7  &  4.78  & 0.011  & 0.0027 & 0.0028 & 0.0014\,\%\\
MWC\,480    & 1.97 & 137  & 0.748  &  19.7  &  5.96  & 2.51   & 1.49   & 2.18 & 1.1\,\% \\
HD\,169142  & 1.8  & 145  & 0.607  &  38.3  &  9.36  & 2.28   & 0.445  & 0.581& 0.32\,\%\\
HD\,142666  & 1.6  & 116  & 0.307  &  26.8  &  3.78  & 0.740  & 0.512  & 0.840& 0.53\,\%\\
\zz HD\,135344B\z & 1.65  & 140 & 0.552 & 34.3 & 7.74 & 1.94  & 0.511  & 0.602& 0.37\,\%\\
V\,1149\,Sco& 1.3  & 145  & 0.209  &  27.6  &  4.46  & 0.785  & 0.447  & 0.761& 0.59\,\%\\
Lk\,Ca\,15  & 1.0  & 140  & 0.425  &  11.6  &  5.70  & 1.49   & 1.57   & 3.53 & 3.53\,\%\\
USco\,J1604-2130 & 1.2& 145& 0.190 &  26.4  &  8.02  & 0.716  & 0.237  & 0.376& 0.31\,\%\\
RY\,Lup     & 1.38 & 185  & 0.210  &  29.1  &  4.93  & 1.29   & 0.627  & 2.05 & 1.5\,\%\\
CI\,Tau     & 0.9  & 140  & 0.371  &  22.9  &  12.2  & 1.30   & 0.326  & 1.19 & 1.32\,\%\\
TW\,Cha     & 1.0  & 160  & 0.125  &  14.2  &  8.70  & 0.570  & 0.323  & 1.02 & 1.0\,\%\\
RU\,Lup     & 1.15 & 150  & 0.479  &  41.8  &  4.94  & 1.93   & 0.654  & 3.09 & 2.7\,\%\\
AA\,Tau     & 0.85 & 140  & 0.176  &  28.6  &  7.23  & 0.617  & 0.209  & 1.31 & 1.5\,\%\\
TW\,Hya     & 0.75 &  51  & 1.52   &  23.1  &  11.1  & 0.705  & 0.193  & 0.889& 1.2\,\%\\
GM\,Aur     & 0.7  & 140  & 0.681  &  17.9  &  3.78  & 2.44   & 2.47   & 11.1 & 16\,\%\\
BP\,Tau     & 0.65 & 140  & 0.131  &  16.7  &  10.5  & 0.460  & 0.184  & 0.701& 1.1\,\%\\
DF\,Tau     & 1.17 & 140  & 0.0077 &  32.4  &  3.78  & 0.027  & 0.015  & 0.020& 0.020\,\%\\
DO\,Tau     & 0.52 & 140  & 0.346  &  37.3  &  7.92  & 1.21   & 0.287  & 0.721& 1.4\,\%\\
DM\,Tau     & 0.53 & 140  & 0.251  &  10.2  &  9.42  & 0.878  & 0.642  & 1.63 & 3.1\,\%\\
CY\,Tau     & 0.43 & 140  & 0.193  &   9.3  &  6.69  & 0.676  & 0.762  & 10.0 & 23\,\% \\
FT\,Tau     & 0.30 & 140  & 0.310  &  19.7  &  9.16  & 1.09   & 0.421  & 3.03 & 10\,\% \\
RECX\,15    & 0.28 & 94.3 & 0.0023 &  64.8  &  11.8  & 0.0036 & 0.0033 & 0.0038 & 0.0014\,\%\\[1mm]
\hline
&&&&&&&&&\\[-2.0ex]
mean value$^\star$  &      &       &        &  27.7   &  6.32 &  &  &  & 0.72\,\%\\[1mm]
\hline
\end{tabular}}\\[1mm]
\hspace*{8mm}\resizebox{146mm}{!}{\parbox{155mm}{ $M_{\rm dust}^{\rm
      class}$, $M_{\rm dust}^{\rm derived}$ and $M_{\rm dust}^{\rm
      model}$ are listed in units $[10^{-4}\rm M_\odot]$. $F_{850}$ is
    the spectral flux at 850\,$\mu$m taken from the SED-fitting
    models. The dust mass $M_{\rm dust}^{\rm class}$ is calculated
    from $F_{850}$, using the classical Eq.\,(\ref{eq1}) with standard
    values for opacity $\kappa^{\rm abs}_{850} = 3.5\rm\,cm^2/g(dust)$
    and mean dust temperature $\langle T_{\rm dust}\rangle = 20\,$K
    \citep{Andrews2005}.  $M_{\rm dust}^{\rm derived}$ is also derived
    from Eq.\,(\ref{eq1}), but using the proper opacities and
    mass-averaged dust temperatures $\langle T_{\rm dust}\rangle$ as
    found in the SED-fitting models. $M_{\rm dust}^{\rm model}$ is the
    actual dust mass in the model. The mass-mean dust absorption
    opacities $\langle\kappa^{\rm abs}_{850}\rangle$ are listed in units
    [cm$^2$/g(dust)]. $M_{\rm disk} = M_{\rm dust}^{\rm model}\times
    100$ is assumed here. $^\star$: logarithmic mean value
    $=\exp(\langle\log.\rangle)$.}}
\end{table}

\begin{figure}
\centering 
\vspace*{-2mm}
\includegraphics[width=120mm]{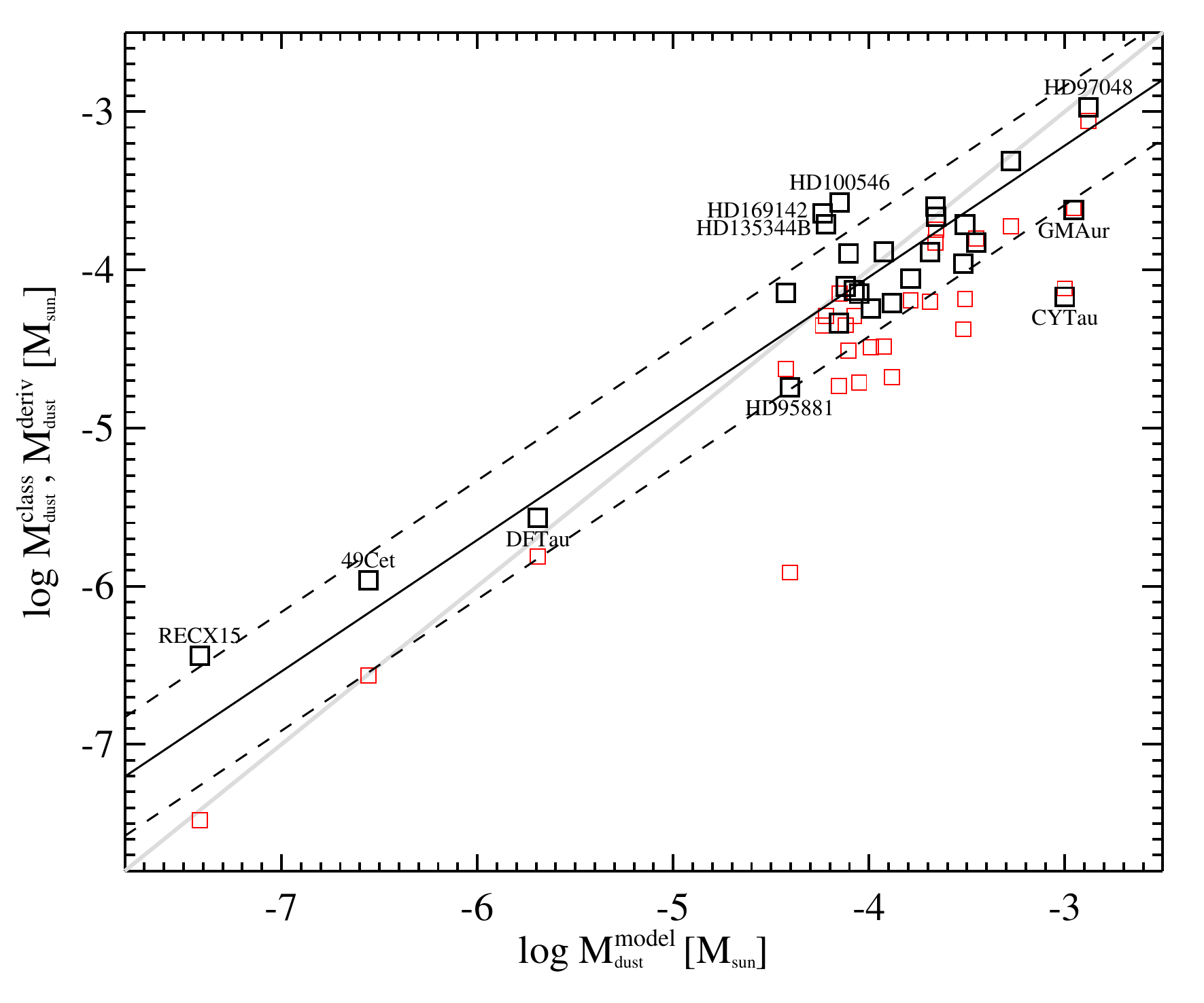}\\[-4mm]
\caption{Black squares show the dust masses $M_{\rm dust}^{\rm class}$
  computed from the 850\,$\mu$m flux using the classical method
  (standard opacity of 3.5~cm$^{2}$/g(dust), dust temperature of
  20\,K) versus the actual dust masses $M_{\rm dust}^{\rm model}$ used
  in the SED-fitting models to produce those fluxes. The black full
  and dashed lines show a linear fit to these data
  (Eq.\,\ref{eq:Mfit}). The red squares show the dust masses $M_{\rm
    dust}^{\rm derived}$ derived in the classical way, but using the
  proper mean dust opacities and mean dust temperatures present in the
  models.}
\label{fig:dustmasses}
\end{figure}

Table~\ref{tab:dustmass} shows a comparison between the actual dust
masses in the SED-fitting models $(M_{\rm dust}^{\rm model})$, and the
results that would be obtained if Eq.\,(\ref{eq1}) was applied to the
850\,$\mu$m flux $(M_{\rm dust}^{\rm class})$. In many cases, the two
dust mass results are fairly similar, however, there are deviations of
up to a factor of 15 for individual objects. In fact,
Table~\ref{tab:dustmass} shows that Eq.\,(\ref{eq1}) {\it is not
  valid}, because if we use the actual, proper mass averaged dust
temperatures and opacities used in the models, $\langle T_{\rm
  dust}\rangle$ and $\langle\kappa^{\rm abs}_{850}\rangle$, the
agreement is worse and shows a systematic offset with respect to
$M_{\rm dust}^{\rm model}$ (Fig.~\ref{fig:dustmasses}). There are
three effects we need to understand in order to explain this
behavior.  1) Our mean dust opacities are usually larger than
$3.5\rm\,cm^2/g$, which generally leads to lower dust masses at fixed
850\,$\mu$m flux. 2) Many disks are not optically thin at 850\,$\mu$m,
which leads to higher dust masses at fixed flux. 3) Small or tenuous 
disks can be much warmer than 20\,K, but massive extended disks can also be much
cooler than 20\,K. Those cool disks have very cold midplane regions,
which emit less than in the Rayleigh-Jeans limit at 850\,$\mu$m. 
Consequently, such cold disks need to be more massive in order
to produce a certain given 850\,$\mu$m flux. To explain the deviations
between $M_{\rm dust}^{\rm class}$ and $M_{\rm dust}^{\rm model}$,
mechanism (3) is actually the most significant one. All three effects
superimposed lead to roughly $M_{\rm dust}^{\rm class}\approx M_{\rm
  dust}^{\rm model}$, but cause an uncertainty of about 0.5\,dex, see
Fig.~\ref{fig:dustmasses}. A linear fit results in
\begin{equation}
  \log_{10} M_{\rm dust}[M_\odot] = \big(0.87\pm0.45\big) 
       + \big(1.20\pm0.11\big)\log_{10} M_{\rm dust}^{\rm class}[M_\odot]\ .
  \label{eq:Mfit}
\end{equation}  
This leads us to the following important conclusion.  For warm disks,
such as the Herbig\,Ae/Be in general, and HD\,100546 and HD\,169142 in
particular, the classical method tends to overestimate the dust
masses. In contrast, cold T\,Tauri disks, such as CY\,Tau and DM\,Tau, 
might be much more massive than previously thought. Using the classical dust
mass determination method, observations tentatively indicate a linear correlation
between disk dust mass with stellar mass
\citep[e.g.][]{Andrews2013,Mohanty2013,Pascucci2016}, but also report
on significant scatter with deviations of up to a factor of 100.  In
our rather small sample such a relation is not obvious. Our sample
also includes a debris disk (49\,Cet) and a strongly truncated disk
(RECX\,15). However, in summary, our derived disk mass - stellar mass ratios are
well within the observed ranges and are consistent with a mean mass
ratio of about 1\,\%.

\subsection{Dust properties, and the mm-slope}
\label{sec:mmslope}

\begin{table}[!t]
\vspace*{-1mm}
\caption{Dust opacity and SED-slopes in the millimeter and centimeter
  regimes as affected by dust input parameters and disk mean
  temperatures. We list results with up to 3 digits precision in order
  to discuss the differences $\Delta$.}
\label{tab:mm-slope}
\def\z{$\hspace*{-1mm}$} 
\def\zz{$\hspace*{-2mm}$}
\vspace*{0mm}\hspace*{-19mm}
\resizebox{174mm}{!}{
\begin{tabular}{c|cccr|c|ccc|ccc}
\hline
&&&&&&&&&&&\\[-2.0ex]
object & \z$a_{\rm min}\rm[\mu m]$\z
       & \zz$a_{\rm max}\rm[\mu m]$\zz
       & \z$a_{\rm pow}$\z
       & amC{\ }
       & \z$\langle T_{\rm dust}\rangle$[K]\z
       & $\beta_{\rm abs}^{\rm mm}$
       & $\alpha_{\rm SED}^{\rm mm}$ 
       & $\Delta^{\rm mm}$
       & $\beta_{\rm abs}^{\rm cm}$
       & $\alpha_{\rm SED}^{\rm cm}$ 
       & $\Delta^{\rm cm}$\\
&&&&&&&&&&&\\[-2.0ex]
\hline
&&&&&&&&&&&\\[-2.0ex]
HD\,100546     & 0.042 &  2980  & 3.34 & 16.8\% & 56.1 & 0.90 & 2.69 & -0.20 &  1.33 &  3.30 & -0.03\\
HD\,97048      & 0.054 &    96  & 3.49 &  8.2\% & 34.9 & 1.93 & 3.48 & -0.45 &  1.71 &  3.69 & -0.02\\
HD\,95881      & 0.046 &  3030  & 3.67 & 18.8\% & 131  & 1.06 & 2.07 & -0.99 &    -- &    -- & --   \\
AB\,Aur        & 0.026 &  4560  & 3.92 &  8.2\% & 29.7 & 1.55 & 3.31 & -0.24 &  1.71 &  3.46 & -0.25\\
HD\,163296     & 0.052 & 10000  & 3.80 & 17.4\% & 30.1 & 1.16 & 2.41 & -0.75 &  1.44 &  3.06 & -0.38\\
49\,Cet        & 0.037 &  1260  & 3.20 & 15.6\& & 58.7 & 0.93 & 2.76 & -0.17 &    -- &    -- & --   \\
MWC\,480       & 0.020 &  4280  & 3.62 & 18.9\% & 19.7 & 1.05 & 2.56 & -0.49 &    -- &    -- & --   \\
HD\,169142     & 0.046 &  6510  & 3.76 & 22.4\% & 38.3 & 1.01 & 2.63 & -0.39 &  1.21 &  3.19 & -0.02\\
HD\,142666     & 0.082 & 10000  & 3.53 & 17.6\% & 26.8 & 0.93 & 2.39 & -0.54 &  1.19 &  2.95 & -0.24\\
HD\,135344B    & 0.05  &  5440  & 3.73 & 19.7\% & 34.3 & 1.05 & 2.81 & -0.24 &  1.35 &  3.30 & -0.05\\ 
V\,1149\,Sco   & 0.003 &  4950  & 3.78 & 12.5\% & 27.6 & 1.31 & 2.69 & -0.62 &  1.57 &  3.55 & -0.02\\
Lk\,Ca\,15     & 0.005 &  2570  & 3.60 & 14.5\% & 11.6 & 1.20 & 2.30 & -0.90 &  1.83 &  3.38 & -0.45\\
USco\,J1604-21 & 0.014 &    44  & 3.02 & 21.4\% & 26.4 & 1.59 & 3.17 & -0.42 &  1.54 &  3.52 & -0.02\\
RY\,Lup        & 0.004 &  3880  & 3.62 & 13.4\% & 29.1 & 1.15 & 2.36 & -0.79 &  1.54 &  3.10 & -0.43\\
CI\,Tau        & 0.084 &  1990  & 4.24 & 23.4\% & 22.9 & 1.35 & 2.40 & -0.95 &  1.49 &  3.36 & -0.13\\
TW\,Cha        & 0.075 &  5650  & 3.55 & 25.0\% & 14.2 & 0.76 & 1.90 & -0.86 &    -- &    -- & --   \\
RU\,Lup        & 0.051 &  4620  & 3.52 & 15.0\% & 41.8 & 0.98 & 2.29 & -0.80 &  1.35 &  2.79 & -0.56\\
AA\,Tau        & 0.42  &  1120  & 3.67 & 13.9\% & 28.6 & 1.29 & 2.36 & -0.93 &    -- &    -- & --   \\
TW\,Hya        & 0.11  &  6650  & 3.99 & 23.0\% & 23.1 & 1.22 & 2.23 & -0.99 &  1.32 &  2.98 & -0.34\\ 
GM\,Aur        & 0.006 &  1070  & 3.99 & 10.7\% & 17.9 & 1.64 & 2.43 & -1.21 &  1.95 &  3.88 & -0.07\\
BP\,Tau        & 0.078 &  4290  & 3.97 & 22.1\% & 16.7 & 1.23 & 2.09 & -1.14 &  1.44 &  3.18 & -0.27\\
DF\,Tau        & 0.055 &  3000  & 3.25 & 13.7\% & 32.4 & 0.94 & 2.62 & -0.32 &    -- &    -- & --   \\
DO\,Tau        & 0.037 &  7880  & 3.55 & 24.9\% & 37.3 & 0.74 & 2.26 & -0.48 &  0.94 &  2.74 & -0.21\\
DM\,Tau        & 0.015 &  3990  & 3.70 & 21.8\% & 10.2 & 1.01 & 2.13 & -0.88 &  1.38 &  3.24 & -0.14\\
CY\,Tau        & 0.076 &  1540  & 3.72 & 14.3\% &  9.3 & 1.29 & 1.48 & -1.80 &  2.26 &  3.74 & -0.52\\
FT\,Tau        & 0.052 &  8630  & 4.97 & 23.8\% & 19.7 & 1.43 & 2.13 & -1.30 &  1.49 &  3.23 & -0.26\\
RECX\,15       & 0.021 &  3000  & 3.47 & 27.4\% & 64.8 & 0.65 & 2.42 & -0.23 &    -- &    -- & --\\[1mm]
\hline
&&&&&&&&&&&\\[-2.0ex]
mean value & 0.034$^\star$ & 2820$^\star$ & 3.69 & 18\%  &&&&&&&\\[1mm]
\hline
\end{tabular}}\\[2mm]
\hspace*{7mm}\resizebox{14.8cm}{!}{\parbox{15.5cm}{$a_{\rm min}$ and
    $a_{\rm max}$ are the minimum and maximum dust radii assumed in
    the (outer disk of the) model, and $a_{\rm pow}$ is the dust size
    distribution powerlaw index. amC is the volume fraction of
    amorphous carbon in the dust material, and $\langle T_{\rm
      dust}\rangle$ the mass averaged dust temperature in the
    disk. $\beta_{\rm abs}$ is the dust opacity slope according to
    Eq.\,(\ref{eq:opac-slope}) and $\alpha_{\rm SED}$ the observable
    log-log flux gradient in the SED, at millimeter (0.85\,mm to
    1.3\,mm) and centimeter wavelengths (5\,mm to 10\,mm),
    respectively. The deviation between expected and actual
    SED-slope, according to Eq.\,(\ref{eq:alphaSED}), is
    $\Delta=\alpha_{\rm SED}-(\beta_{\rm abs}+2)$. HD\,95881, 49\,Cet,
    MWC\,480, TW\,Cha, AA\,Tau, DF\,Tau and RECX\,15 have no cm-data.
    \ \ $^\star$: logarithmic mean value
    $=\exp(\langle\log.\rangle)$.}}
\end{table}

\noindent Table~\ref{tab:mm-slope} shows the results from our
SED-fitting work concerning the mm-slope, generally considered as
important indicators for grain size, dust growth and disk evolution,
see e.g.\,\citet{Natta2007} and \citet{Testi2014}. However, the
material composition of the grains, in particular the inclusion of
conducting materials like, in our case, amorphous carbon, can have an
important influence on the opacity mm-slope, too. \citet{Min2016} have
shown that if conducting materials are included, dust aggregate
opacity computations, using the discrete dipole approximation, result
in shallow opacity slopes in the millimeter regime even for small
grains, known as the 'antenna effect'. This behavior can be
reproduced by using Mie computations with effective medium theory and
a distribution of hollow spheres \citep{Min2016}. The dust size
distributions in our SED-fitted models typically ranges from a few
0.01\,$\mu$m to a few millimeters, with power-law exponents between about 3.2
and 4.0, with two outliers CI\,Tau and FT\,Tau and a mean value of
about 3.7.  The amorphous carbon volume fraction is found to be about
$10\%-25\%$.

The observable SED-slope $\alpha_{\rm SED}$ and the dust opacity slope
$\beta_{\rm abs}$ are given by
\begin{eqnarray}
  \hspace*{8mm}
  \alpha_{\rm SED} &=& -\frac{d\log F_\nu}{d\log \lambda}
  \label{eq:mm-slope}\\
  \beta_{\rm abs}  &=& -\frac{d\log \kappa^{\rm abs}}{d\log \lambda}
  \label{eq:opac-slope} \ .
\end{eqnarray}
For optically thin disks, which are warm enough to emit in the
Rayleigh-Jeans limit (where Eq.\,\ref{eq1} holds), the following
equation is valid
\begin{equation}
  \alpha_{\rm SED} = \beta_{\rm abs} + 2 \ .
  \label{eq:alphaSED}
\end{equation}
Table~\ref{tab:mm-slope} shows results from our SED-fitting models
where we know $\beta_{\rm abs}$ and can measure in how far
Eq.\,(\ref{eq:alphaSED}) is valid. We consider the mm-slope by using
the wavelength interval 0.85\,mm to 1.3\,mm, and the cm-slope by using
wavelengths 5\,mm to 10\,mm. Steep SEDs can be explained either by the
complete absence of millimeter-sized dust grains (small $a_{\rm max}$
such as 0.1~mm for HD\,97048) or by a steep dust size distribution
function (large size distribution powerlaw index $a_{\rm pow}\!>\!3.9$
as for AB\,Aur and GM\,Aur). However, both explanations require in
addition that the amorphous carbon volume fraction is small
(amC$\,<\!10$\%). To break this degeneracy, additional continuum and
line observations have to be taken into account. For example, amC is
crucial for the dust albedo in the near IR, and thereby controls the
primary heating of the dust in the disk surface by stellar photons,
with large implications on the shape of the SED in the near and mid IR
regions.

\begin{figure}
\centering 
\vspace*{-2mm}
\includegraphics[width=115mm]{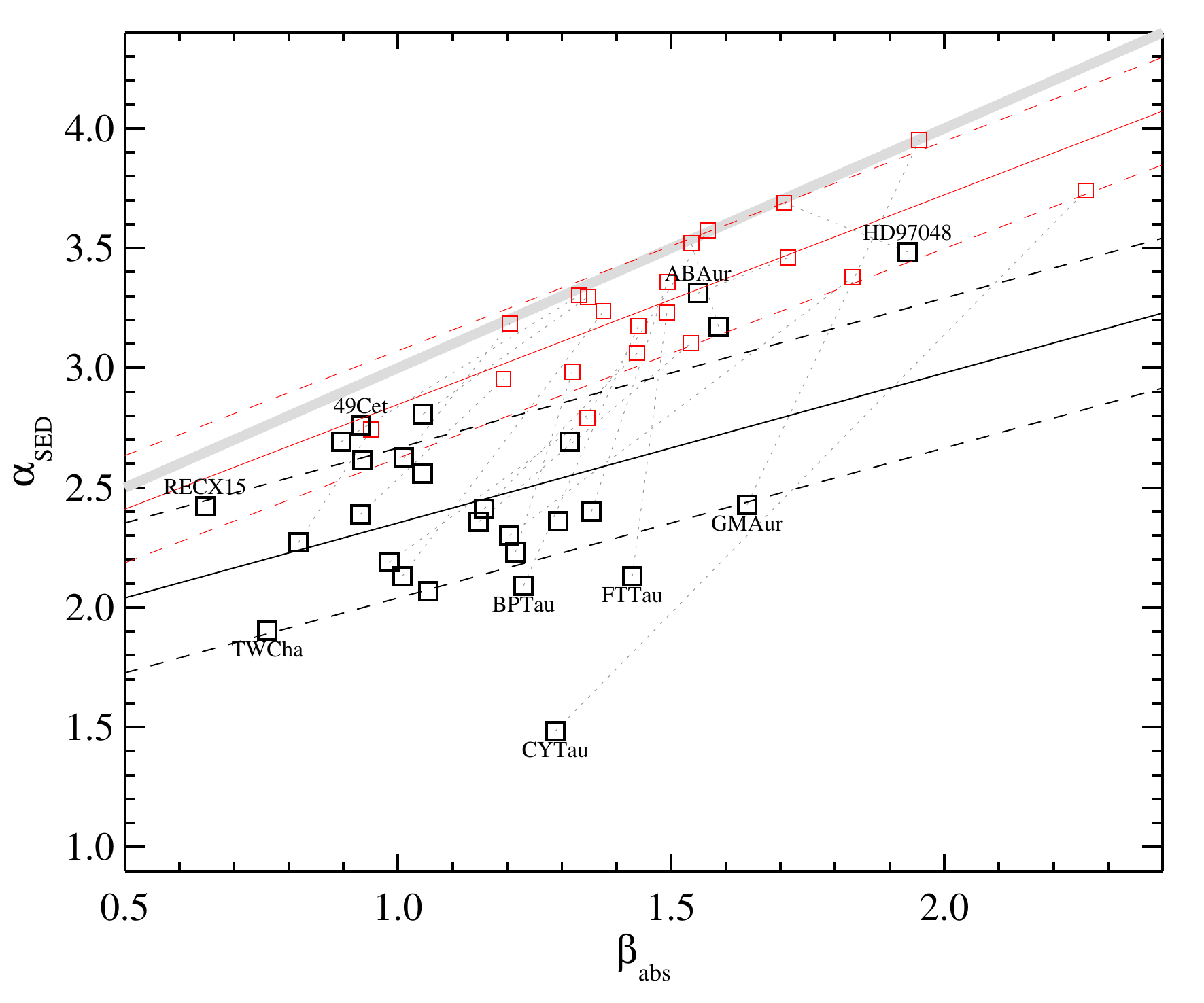}\\[-4mm]
\caption{Black squares show the millimeter SED-slopes $\alpha_{\rm
    SED}^{\rm mm}$ as function of the dust opacity slopes $\beta_{\rm
    abs}^{\rm mm}$ that we used in the models to produce those
  SED-slopes. The black full and dashed lines show a linear fit to
  these results (Eq.\,\ref{eq:slopefit}). The red squares show the
  same for centimeter wavelength.  The expectation $\alpha_{\rm SED} =
  \beta_{\rm abs} + 2$ for a warm, optically thin disk is shown by a
  thick gray line.}
\label{fig:slopes}
\end{figure}

Interestingly, the deviation from the optically thin warm case
(Eq.\,\ref{eq:alphaSED}), $\Delta\!=\!\alpha_{\rm SED} - (\beta_{\rm
  abs} +2)$, can be quite substantial. In the millimeter-region
(0.85\,mm to 1.3\,mm), we find the largest deviation to be $-1.8$ for
CY\,Tau (because it is a very cold disk in the model), followed by
$-1.3$ for FT\,Tau (because it is a cold and optically thick disk in
the model). Deviations between $-0.5$ and $-1.0$ are quite typical
among our models. Only a few warm and optically thin disks
(HD\,100546, AB\,Aur, HD\,135344B, RECX\,15) have deviations as small
as $-0.2$ to $-0.3$. For very cold disks like CY\,Tau, the
SED-slope flattens, despite an opacity slope that is
relatively steep. These trends continue into the centimeter region,
albeit less pronounced. Linear fits to these results lead to
\begin{eqnarray}
  \alpha_{\rm SED}^{\rm mm} &=& \big(1.72\pm0.31\big) 
  + \big(0.63\pm0.26\big)\beta^{\rm mm}_{\rm abs} \\
  \alpha_{\rm SED}^{\rm cm} &=& \big(1.97\pm0.22\big) 
  + \big(0.87\pm0.15\big)\beta^{\rm cm}_{\rm abs}  \ .
  \label{eq:slopefit}
\end{eqnarray}
All millimeter and centimeter modeling results are visualized in
Fig.~\ref{fig:slopes}, together with these linear fits.  The coolest
disks, showing the largest deviations from the expectations
(Eq.~\ref{eq:alphaSED}), gather at the bottom of this plot. The gray
line, which visualizes those expectations, is an upper limit to these
dependencies.  It always overestimates $\alpha_{\rm SED}$ or
underestimates $\beta_{\rm abs}$, respectively, depending on what
quantity is considered to be given.  The deviations are much less
pronounced at centimeter wavelengths, but still relevant. We conclude
that the determination of dust sizes from the observation of
SED-slopes $\alpha_{\rm SED}$ has large principle uncertainties.  This
analysis may already fail in its first step, namely the determination
of $\beta_{\rm abs}$, because disks may be partly optically thick and
too cold to allow for the application of the Rayleigh-Jeans
approximation, so the problem becomes non-linear. Other unknowns, such
as the impact of amorphous carbon or other conducting materials on the
opacity slope, can affect the second analysis step as well, namely the
determination of dust sizes from $\beta_{\rm abs}$.

\begin{table}
\centering
\caption{Properties of polycyclic aromatic hydrocarbon molecules
  (PAHs) in 14 SED-fitting models$^{(1)}$. All models use the spectral
  properties of circumcoronene with 54 carbon atoms and 18 hydrogen
  atoms, with a constant mixture of charged and neutral opacities.}
\def\z{$\hspace*{-1mm}$}
\label{tab:PAHprop}
\def\zz{$\hspace*{-2mm}$}
\vspace*{0mm}
\begin{tabular}{c|cc|c}
\hline
&&\\[-2.0ex]
object & $f_{\rm PAH}^{(5)}$
       & $M_{\rm PAH}/M_{\rm dust}$
       & $\langle$PAH-charge$\rangle^{(2)}$\\
&&\\[-2.0ex]
\hline
&&\\[-2.0ex]
HD\,100546  & 0.0028            & 3.7(-5) &  0.89\\
HD\,97048   & 0.42              & 5.5(-3) &  0.63\\
HD\,95881   & 0.36              & 4.8(-3) &  0.61\\
AB\,Aur     & 0.40              & 5.3(-3) &  0.90\\
HD\,163296  & 0.050             & 6.6(-4) &  0.80\\
MWC\,480    & 0.1$^{(3)}$        & 1.3(-3)$^{(3)}$\zz\z & 1$^{(3)}$\\
HD\,169142  & 0.77              & 1.0(-2) &  0.98\\
HD\,142666  & 0.050             & 6.6(-4) &  0.70\\
HD\,135344B & 0.2$^{(3)}$        & 2.7(-3)$^{(3)}$\zz\z & 1$^{(3)}$\\
V\,1149\,Sco& 0.5               & 6.6(-3) &  1\\
CI\,Tau     & 0.1$^{(3)}$       & 1.3(-3) &  0$^{(3)}$\\
BP\,Tau     & 0.3$^{(3)}$       & 4.0(-3) &  0$^{(3)}$\\
CY\,Tau     & 0.007$^{(3)}$     & 8.9(-5) &  0$^{(3)}$\\
RECX\,15    & 0.0005$^{(3),(4)}$\zz & 3.4(-4)$^{(3),(4)}$\zz\zz & 0$^{(3)}$\\[1mm]
\hline
\end{tabular}\\[2mm]
\hspace*{0mm}\resizebox{145mm}{!}{\parbox{160mm}{
 $^{(1)}$: Only objects with detected PAH features are listed. There
    are two well-known ``PAH-sources'' (HD\,97048, HD\,169142) which
    show a number of very prominent PAH features, see
    Fig.\,\ref{fig:midIR}.  However, many other objects do have at
    least one PAH band detected, which allows us to determine some PAH
    properties as well.\\[1mm]
 $^{(2)}$: The mean PAH charge is the mixing ratio between charged and
    (neutral + charged) PAHs, 0 means all PAHs neutral, 1 means all
    PAHs charged.\\[1mm]
 $^{(3)}$: Only one PAH band was detected, or the spectroscopic data
    are of too poor quality to perform a detailed analysis including
    the PAH-charge. The PAH-charge is assumed in those cases.\\[1mm]
 $^{(4)}$: The model for RECX\,15 has a very peculiar gas/dust mass
    ratio of 5200. The PAH/dust mass ratio is comparable to the other
    objects listed.\\[1mm]
 $^{(5)}$: $f_{\rm PAH}$ is the PAH concentration with respect to
    hydrogen nuclei, normalized to standard conditions in the
    interstellar medium (ISM), i.\,e.\, $f_{\rm PAH}\!=\!1$ means that
    PAHs are as abundant in the disk as in the standard ISM
    \citep[$10^{-6.52}$ PAH molecules/H-nucleus][]{Tielens2008}.
    $f_{\rm PAH} \propto M_{\rm PAH}/M_{\rm dust} \cdot \delta$, with
    $\delta$ being the dust/gas mass ratio.}}
\end{table}

\subsection{PAH properties}
\label{sec:PAHprop}

\noindent The results of our simultaneous fits of the the dust and PAH
properties in the disks to the SED and mid-IR spectral data are
summarized in Table~\ref{tab:PAHprop}. Small PAHs with 54 carbon atoms
('circumcoronene') are sufficient to explain the observations, with
abundances $f_{\rm PAH} \approx 0.005 - 0.8$, i.e.\ somewhat
underabundant with respect to standard interstellar medium (ISM)
conditions ($10^{-6.52}$ PAH molecules/H-nucleus), which agrees with
the results of \citet{Geers2007,Geers2009}.  We note, however, that
our analysis is based on full radiative transfer disk models.  In all
cases where a couple of different PAH bands are detected (HD\,100546,
HD\,97048, HD\,95881, AB\,Aur, HD\,169142, HD\,142666) we find a mean
PAH charge of about $0.6-0.98$, but see no clear trend of PAH charge
with disk gaps as reported by \citet{Maaskant2014}.

\newpage
\subsection{DIANA Standard Models}
\label{sec:DIANAstand}

\noindent We computed full {\sf DIANA}-standard disk models for 14
objects, see Table~\ref{tab:discpara}. Full information about the
fitted stellar, disk shape, dust and gas parameters, the full 2D
modeling results and predicted observations are available at
\url{http://www-star.st-and.ac.uk/~pw31/DIANA/DIANAstandard}.  The
corresponding data products are described in
Appendix~\ref{app:uploaded-data}. The collected observational data can
also be downloaded from \url{http://www.univie.ac.at/diana}.

In the following pages, we highlight a few results for each object.
The figures show the fitted disk density structures, the surface
density profiles of dust and gas, and a few selected graphical
comparisons between observed and predicted line properties, followed
by the complete list of observed and modeled line properties. The
SED-plots are not repeated here, as they are very similar to those
shown in Fig.\,\ref{fig:SEDs}, although, admittedly, $\chi^2_{\rm
  phot}$ and $\chi^2_{\rm spec}$ usually increased slightly as we
started fitting the images and lines, too. This is a natural
consequence of including more data into the fit quality (see
Eq.\,\ref{eq:totchi}).  In cases where the genetic fitting algorithm
started to actually loose the SED fit, the program was stopped and
re-run with larger weights $w_{\rm phot}$ and $w_{\rm spec}$. Full
information about the final choice of fitting weights is contained in
the downloadable \,{\tt ModelSetup.tgz}\, files, see \,{\tt
  Parameter.in}.

\begin{table}[!b]
\def\z{$\hspace*{-1mm}$}
\def\zz{$\hspace*{-2mm}$}
\def\zzz{$\hspace*{-3mm}$}
\centering
\vspace*{-2mm}
\caption{Disk shape, dust opacity and PAH parameters of the {\sf DIANA}
standard models.$^{(1)}$}
\label{tab:discpara}
\hspace*{-15mm}\resizebox{130mm}{!}{\begin{tabular}{rl|ccccccc}
\hline
&&&\\[-1.8ex]
&
& \sffamily\bfseries \z ABAur\z
& \sffamily\bfseries \z HD163296\z
& \sffamily\bfseries \z MWC480\z
& \sffamily\bfseries \z HD169142\z
& \sffamily\bfseries \z HD142666\z
& \sffamily\bfseries \z GMAur\z
& \sffamily\bfseries \z TWHya\z\\
\hline
&&&\\[-1.8ex]
$M_\star$\zzz & $[\rm M_\odot]$
& 2.50 & 2.47 & 1.97 & 1.80 & 1.60
& 0.70 & 0.75 \\
$T_{\rm eff}$\zzz & $\rm[K]$
& 9550 & 9000 & 8250 & 7800 & 7050 & 4000
& 4000 \\
$L_\star$\zzz & $[\rm L_\odot]$
& 42.1 & 34.7 & 13.7 & 9.8 & 6.3 & 0.60
& 0.24 \\
\multicolumn{2}{r|}{$\dot{M}_{\rm acc}$$^{(5)}$ $[\rm M_\odot/yr]$}
& 1.4(-7) & 4.5(-7) & -- & -- & -- & --
& 1.5(-9) \\[1mm]
\hline
\\*[-3mm]
\multicolumn{9}{l}{\sffamily\bfseries dust and PAH parameters}\\
\hline
&&&\\[-1.8ex]
$a_{\rm min}$\zzz & $\rm[\mu m]$
& 0.047 & 0.020 & 0.020 & 0.046 & 0.057 & 0.050
& 0.0011 \\
$a_{\rm max}$$^{(2)}$\zzz & [mm]\z
& 7.5 & 8.2$^{(2)}$ & 4.3 & 6.5 & 2.8 & 3.0
& 5.7 \\
$a_{\rm pow}$\zzz &
& 4.00 & 3.71 & 3.62 & 3.76 & 3.33 & 3.84
& 3.99 \\
$\rm amC$\zzz & [\%]
& 5.0\% & 6.0\% & 18.9\% & 22.4\% & 25.0\% & 15.0\%
& 24.9\% \\
$\alpha_{\rm settle}$\zzz &
& 1.0(-3) & 6.6(-3) & 7.3(-5) & 3.8(-3) & 2.8(-3) & 1.0(-1)
& 5.2(-3) \\
$f_{\rm PAH}$$^{(2)}$\zzz &
& 0.061 & 0.076 & 0.1 & 0.77$^{(2)}$ & 0.05 & 0.01
& 0.081$^{(2)}$ \\
\zz$\rm PAH_{charge}$$^{(3)}$\zzz & [\%]
& 73\% & 1.1\% & 100\% & 97.6\% & 0\% & --
& -- \\[1mm]
\hline
\\*[-3mm]
\multicolumn{9}{l}{\sffamily\bfseries inner disk zone}\\
\hline
&&&\\[-1.8ex]
\zzz $R_{\rm in}$\zzz & [AU]
& 0.46 & 0.41 & -- & 0.10 & 0.10 & 0.19
& 0.078 \\
\zzz $R_{\rm taper}$\zzz & [AU]
& -- & -- & -- & -- & -- & -- & -- \\
\zzz $R_{\rm out}$\zzz & [AU]
& 88 & 3.7 & -- & 5.0 & 10.3 & 5.0
& 4.6 \\
\zzz gas mass\zzz & $\rm[M_\odot]$
& 4.2(-4) & 1.3(-4) & -- & 7.9(-7) & 6.8(-6) & 1.0(-7)
& 1.1(-6) \\
\zzz dust mass\zzz & $\rm[M_\odot]$
& 6.8(-7) & 1.5(-9) & -- & 7.9(-9) & 6.8(-8) & 1.0(-9)
& 1.3(-9) \\
\zzz gas/dust\zzz &
& 618 & 86667 & -- & 100 & 100 & 100 &
847 \\
\zzz col.\,dens.\,$\epsilon$\zzz &
& 1.28 & 1.11 & -- & 1.38 & 0.12 & 0.90
& -0.78 \\
\zzz tapering $\gamma$\zzz &
& -- & -- & -- & -- & -- & -- & --\\
\zzz $H$\,@\,1\,AU\zzz & [AU]
& 0.092 & 0.077 & -- & 0.035 & 0.20 & 0.036
& 0.028 \\
\zzz flaring $\beta$\zzz &
& 0.99 & 1.00 & -- & 0.80 & 1.15 & 1.22
& 1.21 \\[1mm]
\hline
\\*[-3mm]
\multicolumn{9}{l}{\sffamily\bfseries outer disk zone}\\
\hline
&&&\\[-1.8ex]
\zzz $R_{\rm in}$\zzz & [AU]
& 88 & 3.7 & 0.28 & 22 & 10.3 & 20.0
& 4.6 \\
\zzz $R_{\rm taper}$\zzz & [AU]
& 174 & 133 & 100 & 140 & 53 & 100
& 48 \\
\zzz $R_{\rm out}$$^{(4)}$\zzz & [AU]
& 680 & 488 & 474 & 457 & 344 & 521
& 192 \\
\zzz gas mass\zzz & $\rm[M_\odot]$
& 1.9(-2) & 5.8(-1) & 2.2(-2) & 5.8(-3) & 3.3(-3) & 3.3(-2)
& 4.5(-2) \\
\zzz dust mass\zzz & $\rm[M_\odot]$
& 1.8(-4) & 1.7(-3) & 2.2(-4) & 5.8(-5) & 3.3(-5) & 3.3(-4)
& 1.0(-4) \\
\zzz gas/dust\zzz &
& 106 & 341 & 100 & 100 & 100 & 100 &
450 \\
\zzz col.\,dens.\,$\epsilon$\zzz &
& 0.72 & 0.95 & 0.66 & 1.0 & 1.0 & 1.3
& 1.52 \\
\zzz tapering $\gamma$\zzz &
& 0.54 & 0.2 & 0.66 & 0.50 & 1.0 & 0.8
& 0.45 \\
\zz\zzz $H$\,@\,100\,AU\zzz & [AU]
& 16.8 & 6.5 & 12.3 & 9.6 & 10.0 & 8.0
& 6.3 \\
\zzz flaring $\beta$\zzz &
& 0.95 & 1.11 & 1.10 & 1.07 & 1.15 & 1.21 
& 1.21 \\[0mm]
\hline
\end{tabular}}
\noindent\resizebox{13.5cm}{!}{\parbox{14.5cm}{
\begin{itemize}
\item[$^{(1)}$:] amC is the volume fraction of amorphous carbon in the
dust material, $\epsilon$ is the column density powerlaw index,
$\gamma$ is the powerlaw index for outer disk tapering, $H$ is the
scale height, $\beta$ the flaring index. For further explanations of
the parameter symbols see \citet{Woitke2016}. Numbers written
$A(-B)$ mean $A\times 10^{-B}$.
\item[$^{(2)}$:] In the outer disk zone. Parameters not
marked with $^{(2)}$ are assumed to be unique throughout the disk.
\item[$^{(3)}$:] '--' indicates that PAH emission features are not
detected, hence PAHs are not included in the radiative transfer, but
they still have an influence on the model because of the photoelectric
heating by PAHs.
\item[$^{(4)}$:] Where the hydrogen nuclei particle density reaches
$10^{20}\rm\,cm^{-2}$.
\item[$^{(5)}$:] The mass accretion rate $\dot{M}_{\rm acc}$ enters
the model via an additional heating rate for the gas. Entries '--'
mean that this heating process was not included (passive disk model).
\end{itemize}}}
\vspace*{-4mm}
\end{table}

\addtocounter{table}{-1}
\begin{table}
\def\z{$\hspace*{-1mm}$}
\def\zz{$\hspace*{-2mm}$}
\def\zzz{$\hspace*{-3mm}$}
\centering
\caption{continued}
\hspace*{-15mm}\resizebox{130mm}{!}{\begin{tabular}{rl|ccccccc}
\hline
&&&\\[-1.8ex]
&&\sffamily\bfseries \z BPTau\z
& \sffamily\bfseries \z DMTau\z
& \sffamily\bfseries \z CYTau\z
& \sffamily\bfseries \z RECX15\z
& \sffamily\bfseries \z LkCa15\z
& \sffamily\bfseries \z UScoJ1604-2130\z
& \sffamily\bfseries \z HD97048\zz\\
\hline
&&&\\[-1.8ex]
$M_\star$\zzz & $[\rm M_\odot]$
& 0.65 & 0.53 & 0.43 & 0.28 & 1.00 & 1.20 & 2.50\\
$T_{\rm eff}$\zzz & $\rm[K]$
& 3950 & 3780 & 3640 & 3400 & 4730 & 4550 & 10000\\
$L_\star$\zzz & $[\rm L_\odot]$
& 0.89 & 0.23 & 0.36 & 0.091 & 1.20 & 0.76 & 39.4\\
\multicolumn{2}{r|}{$\dot{M}_{\rm acc}$$^{(5)}$ $[\rm M_\odot/yr]$}
& 2.8(-8) & -- & 7.5(-9) & 1.0(-9) & -- & -- & -- \\[1mm]
\hline
\\*[-3mm]
\multicolumn{9}{l}{\sffamily\bfseries dust and PAH parameters}\\
\hline
&&&\\[-1.8ex]
$a_{\rm min}$\zzz & $\rm[\mu m]$
& 0.049 & 0.019 & 0.050 & 0.019 & 0.0067 & 0.015 & 0.024 \\
$a_{\rm max}$$^{(2)}$\zzz & [mm]\z
& 3.1 & 4.5 & 3.0 & 1.6 & 2.2$^{(2)}$ & 0.040 & 0.037\\
$a_{\rm pow}$\zzz &
& 3.97 & 3.73 & 3.68 & 3.46 & 3.64 & 2.90 & 3.42\\
$\rm amC$\zzz & [\%]
& 17.2\% & 19.0\% & 12.0\% & 22.2\% & 14.8\% & 21.4\% & 17.4\%\\
$\alpha_{\rm settle}$\zzz &
& 6.0(-5) & 2.1(-3) & 6.4(-5) & 1.0(-4) & 2.2(-4) & 2.3(-3) & 5.9(-2)\\
$f_{\rm PAH}$$^{(2)}$\zzz &
& 0.12 & 0.1 & 0.009 & 0.0009 & 0.014 & 0.0074$^{(2)}$ & 0.42$^{(2)}$\\
\zz$\rm PAH_{charge}$$^{(3)}$\zzz & [\%]
& 100\% & 0\% & 0\% & 0\% & 0\% & 100\% & 63\%\\[1mm]
\hline
\\*[-3mm]
\multicolumn{9}{l}{\sffamily\bfseries inner disk zone}\\
\hline
&&&\\[-1.8ex]
\zzz $R_{\rm in}$\zzz & [AU]
& 0.060 & 0.98 & 0.035 & 0.028 & 0.1 & 0.044 & 0.33 \\
\zzz $R_{\rm taper}$\zzz & [AU]
& -- & -- & -- & 1.8 & -- & -- & -- \\
\zzz $R_{\rm out}$\zzz & [AU]
& 1.3 & 12.5 & 0.72 & 7.5 & 10.0 & 0.050 & 7.18 \\
\zzz gas mass\zzz & $\rm[M_\odot]$
& 7.1(-7) & 1.2(-6) & 6.4(-7) & 1.3(-4) & 1.9(-8) & 1.8(-9) & 1.5(-6)\\
\zzz dust mass\zzz & $\rm[M_\odot]$
& 6.9(-9) & 1.2(-8) & 6.3(-9) & 3.5(-8) & 1.9(-10) & 1.8(-11) & 9.8(-9)\\
\zzz gas/dust\zzz &
& 103 & 100 & 102 & 3715 & 100 & 100 & 153 \\
\zzz col.\,dens.\,$\epsilon$\zzz &
& 0.52 & -0.81 & 0.38 & 0.53 & 1.34 & 0.42 & 1.29\\
\zzz tapering $\gamma$\zzz &
& -- & -- & -- & 0.31 & -- & -- & -- \\
\zzz $H$\,@\,1\,AU\zzz & [AU]
& 0.12 & 0.42 & 0.098 & 0.14 & 0.11 & 0.10 & 0.02\\
\zzz flaring $\beta$\zzz &
& 1.22 & 0.60 & 1.01 & 1.24 & 1.50 & 1.02 & 1.16\\[1mm]
\hline
\\*[-3mm]
\multicolumn{9}{l}{\sffamily\bfseries outer disk zone}\\
\hline
&&&\\[-1.8ex]
\zzz $R_{\rm in}$\zzz & [AU]
& 1.3 & 12.5 & 4.2 & -- & 40.1 & 46.4 & 62.6 \\
\zzz $R_{\rm taper}$\zzz & [AU]
& 32 & 149 & 58 & -- & 205 & 56 & 76 \\
\zzz $R_{\rm out}$$^{(4)}$\zzz & [AU]
& 166 & 570 & 220 & -- & 400 & 165 & 650 \\
\zzz gas mass\zzz & $\rm[M_\odot]$
& 6.4(-3) & 1.6(-2) & 1.2(-1) & -- & 2.2(-2) & 2.1(-3) & 9.9(-2) \\
\zzz dust mass\zzz & $\rm[M_\odot]$
& 1.4(-4) & 2.7(-4) & 1.2(-3) & -- & 2.8(-4) & 4.0(-5) & 5.3(-4) \\
\zzz gas/dust\zzz &
& 46 & 59 & 100 & -- & 80 & 53 & 187 \\
\zzz col.\,dens.\,$\epsilon$\zzz &
& 0.65 & 0.50 & 0.11 & -- & 0.95 & 0.73 & 0.87 \\
\zzz tapering $\gamma$\zzz &
& 0.67 & 0.50 & -0.34 & -- & -0.93 & 0.036 & 0.94 \\
\zz\zzz $H$\,@\,100\,AU\zzz & [AU]
& 7.1 & 6.0 & 7.7 & -- & 4.0 & 19.9 & 5.9\\
\zzz flaring $\beta$\zzz &
& 1.12 & 1.17 & 1.15 & -- & 1.12 & 1.21 & 1.19 \\[0mm]
\hline
\end{tabular}}
\noindent\resizebox{13.5cm}{!}{\parbox{14.5cm}{
\begin{itemize}
\item[$^{(1)}$:] amC is the volume fraction of amorphous carbon in the
dust material, $\epsilon$ is the column density powerlaw index,
$\gamma$ is the powerlaw index for outer disk tapering, $H$ is the
scale height, $\beta$ the flaring index. For further explanations of
the parameter symbols see \citet{Woitke2016}. Numbers written
$A(-B)$ mean $A\times 10^{-B}$.
\item[$^{(2)}$:] In outer disk zone. Parameters not
marked with $^{(2)}$ are assumed to be unique throughout the disk.
\item[$^{(3)}$:] '--' indicates that PAH emission features are not
detected, hence PAHs are not included in the radiative transfer, but
they still have an influence on the model because of the photoelectric
heating by PAHs.
\item[$^{(4)}$:] Where the hydrogen nuclei particle density reaches
$10^{20}\rm\,cm^{-2}$.
\item[$^{(5)}$:] The mass accretion rate $\dot{M}_{\rm acc}$ enters
the model via an additional heating rate for the gas. Entries '--'
mean that this heating process was not included (passive disk model).
\end{itemize}}}
\end{table}

\clearpage
\newpage

\begin{figure}
\centering
\vspace*{2mm}
{\large\sffamily\bfseries\underline{AB Aur}}\\[2mm]
$M_{\rm gas} = 1.85\times 10^{-2}\,M_\odot$, 
$M_{\rm dust} = 1.75\times 10^{-4}\,M_\odot$, 
dust settling $\alpha_{\rm settle} = 1.0\times10^{-3}$\\
(full set of model parameters are given in Table~\ref{tab:discpara})\\
\hspace*{-3mm}
\begin{tabular}{cc}
\begin{minipage}{75mm}
\includegraphics[width=78mm]{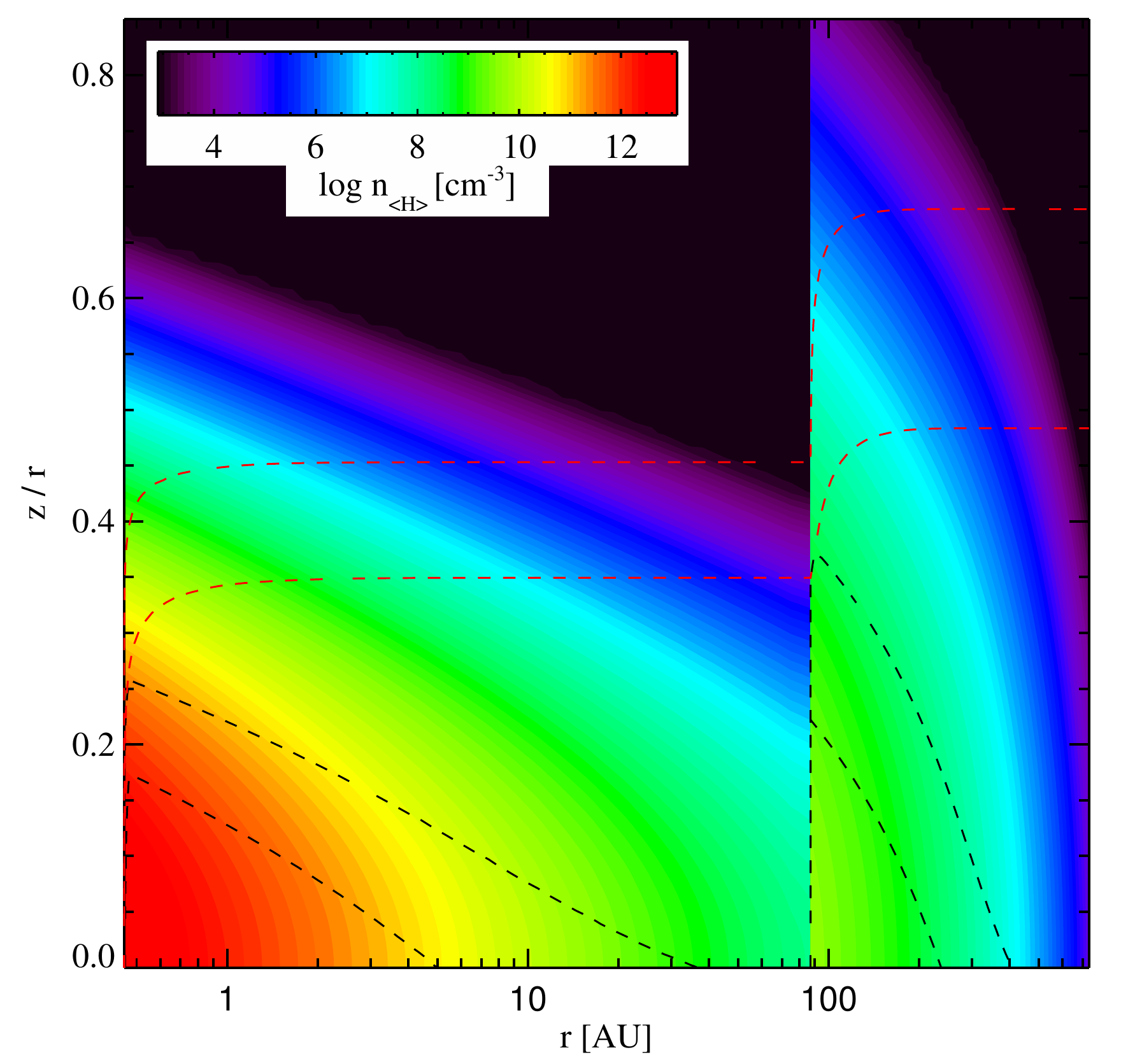}
\end{minipage}
&
\begin{minipage}{87mm}
\vspace*{1mm}
\includegraphics[width=87mm]{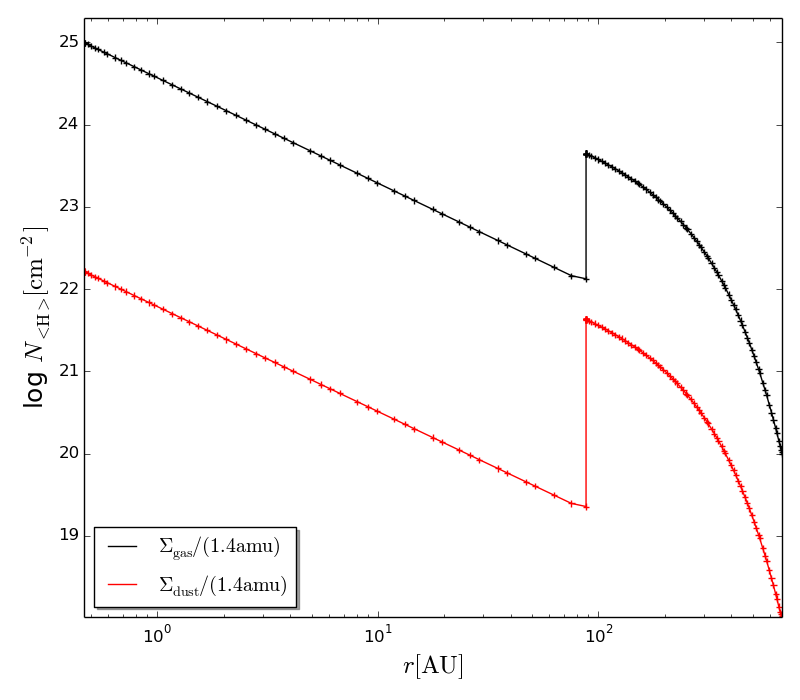}
\end{minipage}
\end{tabular}\\[0mm]
\caption{The left contour plot shows the gas density structure of
  AB\,Aur in the model with overplotted optical depth contours
  corresponding to radial $A_V\!=\!0.01$ and $A_V\!=\!1$ (dashed red),
  and vertical $A_V\!=\!1$ and $A_V\!=\!10$ (dashed black). The right
  plot shows the dust (red) and gas (black) column densities. The plus
  signs illustrate the distribution of radial grid points in the disk model.}
\label{fig:ABAur-1}
\end{figure}

\begin{figure}
\vspace*{-7mm}\hspace*{-9mm}
\begin{tabular}{cc}
\begin{minipage}{87mm}
\includegraphics[width=90mm]{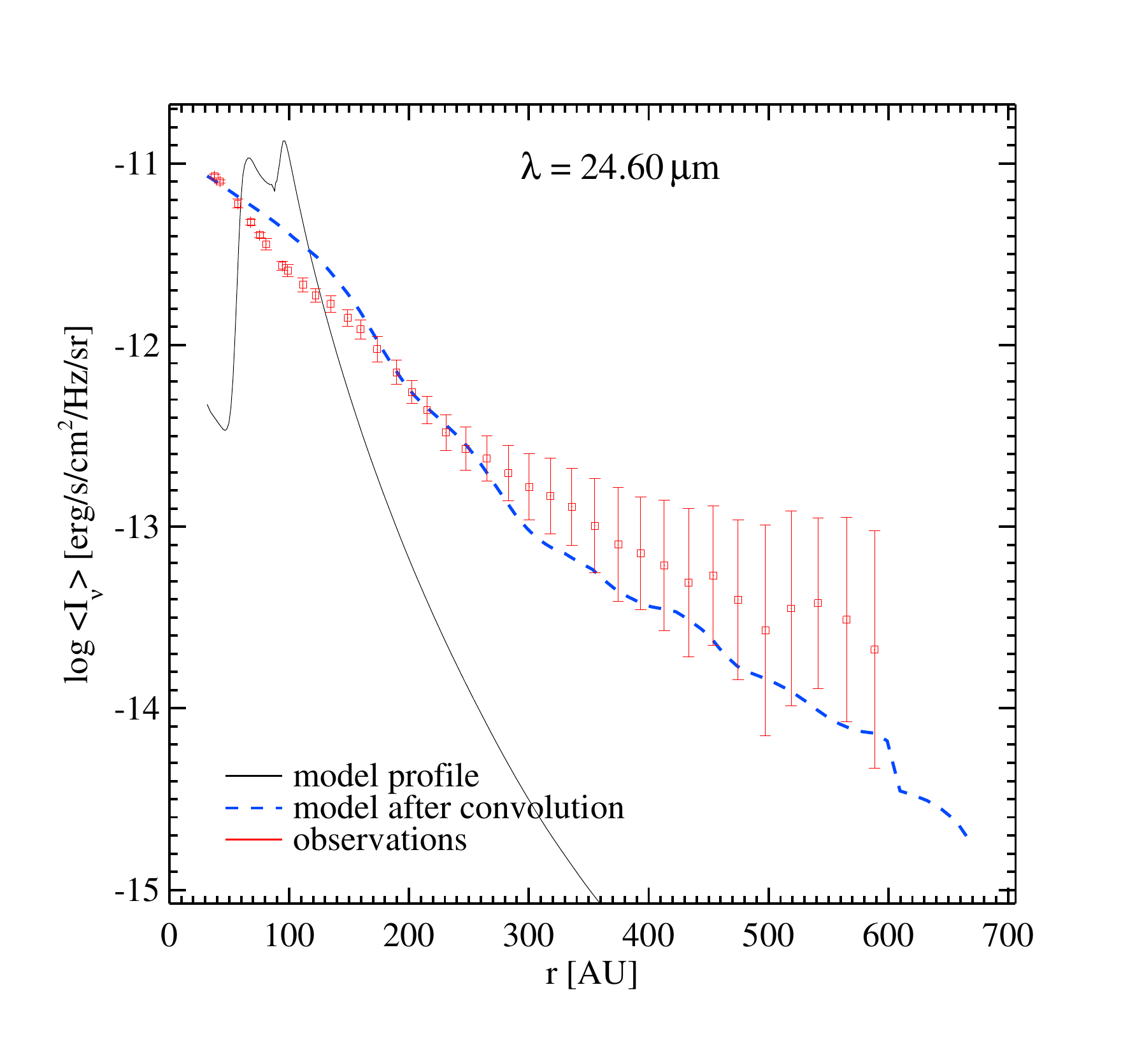}
\end{minipage} &
\hspace*{-6mm}
\begin{minipage}{87mm}
\includegraphics[width=87mm]{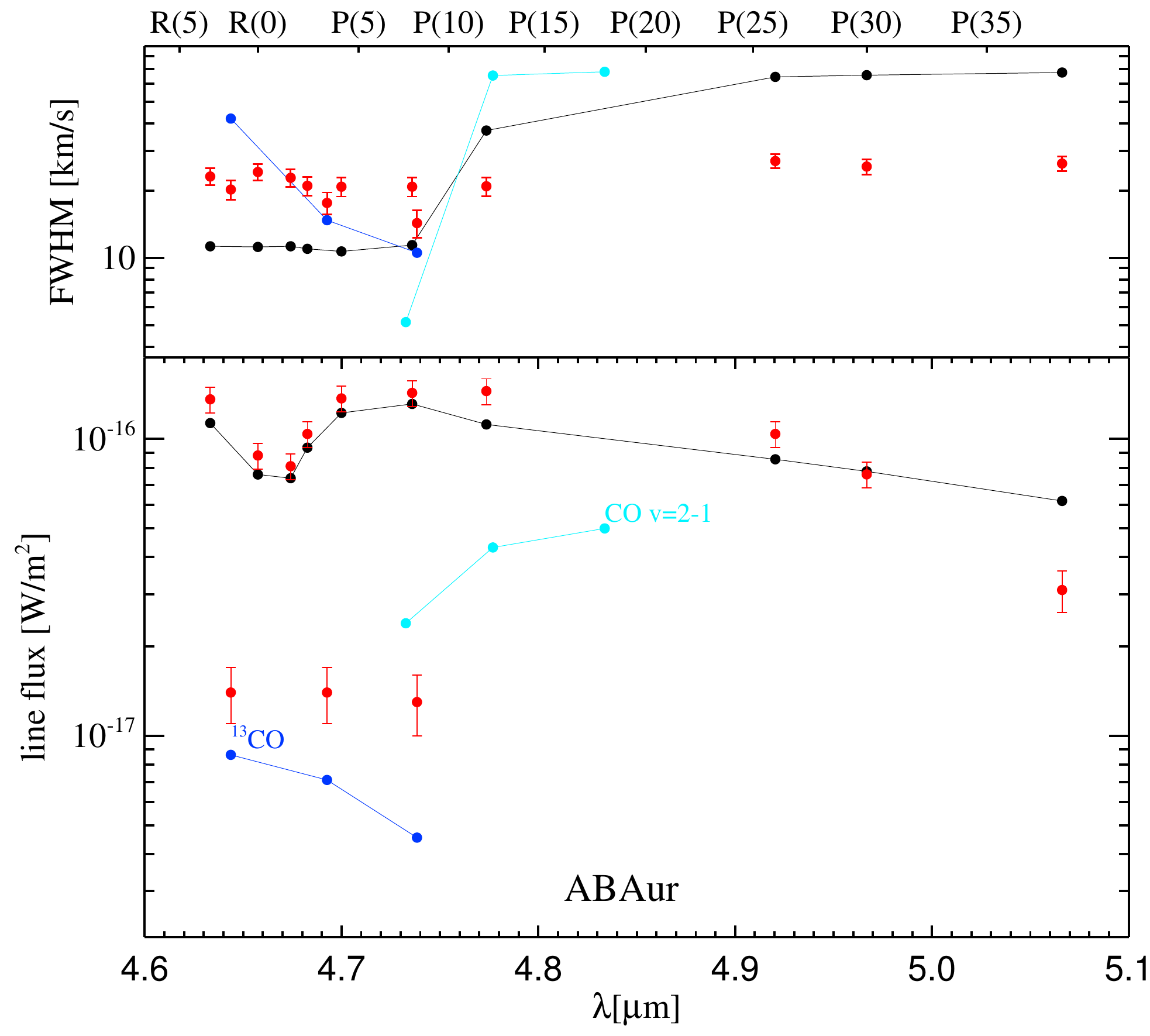}
\end{minipage}
\end{tabular}\\*[-5mm]
\caption{{\bf Left:} Fit of mid-IR radial intensity profile of AB\,Aur
  at 25\,$\mu$m. The model (black full line) is rotated and convolved
  with the instrument PSF (blue dashed line) before comparing it to
  the SUBARU data (red errorbars). {\bf Right:} Results for the CO
  ro-vibrational lines, with the FWHM plotted on top and line fluxes
  below. Dots and lines show the model results ($\rm
  CO\,v\!=\!1\!-\!0$ in black, $\rm CO\,v\!=\!2\!-\!1$ in cyan and
  $\rm^{13}CO\,v\!=\!1\!-\!0$ in blue color), and red dots and errorbars are
  the measurements.}
\label{fig:ABAur-2}
\end{figure}

\noindent{\sffamily\bfseries\underline{1.\ \ AB\,Aur}}\ \ is one of
the best studied Herbig~Ae/Be disks. Low-resolution spectral data are
available (see Fig.\,\ref{fig:SEDs}) from about 1\,$\mu$m (SpeX/IRTF),
over ISO/SWS, Spitzer/IRS, ISO/LWS and Herschel/PACS, all the way up
to about 400\,$\mu$m (Herschel/SPIRE). Our data collection comprises four
continuum images: NICMOS 1.1\,$\mu$m \citep{Perrin2009}, SUBARU/CIAO
H-band \citep{Fukagawa2004}, SUBARU/COMICS 25\,$\mu$m
\citep{Honda2010}, and archival SMA 850\,$\mu$m data.  In addition, we have
69 line fluxes, three line velocity profiles, and one line image (CO 3-2
from SMA). However, both continuum and line data are partly confused
by the massive envelope around AB\,Aur, seen as ``nebulosity'' in the
optical. Although observers have tried to carefully disentangle
disk emission and envelope emission/absorption, the
observational data is often puzzling. For example, concerning the
$^{12}$CO~2-1 and $^{13}$CO~2-1 lines as published by \citet[][their
  Fig.\,2b]{Fuente2010} and \citet[][their Table~3]{Guilloteau2013},
respectively, the $^{12}$CO 2-1 line seems weaker than
the $^{13}$CO 2-1 line, which no disk model can explain. \label{page:ABAur}

Given these observational uncertainties, our 2D disk model for AB\,Aur
(Fig.~\ref{fig:ABAur-1}) provides a reasonable fit to a surprisingly
large number of continuum and line observations by a simple two-zone
disk model with a discontinuity around 80\,AU; this model does not
include any envelope component. The disk model manages to explain the
huge near-IR excess (about $9\,L_\odot$) by scattering and thermal
re-emission from a tall inner disk. The equally impressive mid and
far-IR excess (together about $10\,L_\odot$) are caused by an even
taller outer disk starting outside of about 80\,AU. Together, these
two disk zones result in a relative height $z/r\!\approx\!0.5$ where
the radial optical depth in the visual approaches one, i.e.\ the disk
starts to obscure the star already at inclination angles
$i\ga\!63^{\rm o}$. Note however, that the specific disk structure
resulting from the fitting procedure could be partially biased by the
fact that we did not include an envelope component.

\begin{table}
\caption{Derived physical properties of the disk model and comparison
  of computed spectral line properties with observations from {\sf
    ABAur.properties}. The observational data in the second table is
  given in the form $\rm value\pm\sigma$ for detections, and in the
  form $<\!3\sigma$ in case of non-detections. The size in the last
  column is the radius in the image plane that contains 95\% of the
  flux according to the model.}  \footnotesize
\centering\begin{tabular}{rr}
\hline
DIANA standard fit model properties &  \\
\hline
minimum dust temperature [K] & 21 \\
maximum dust temperature [K] & 1830 \\
mass-mean dust temperature [K] & 36 \\
minimum  gas temperature [K] & 26 \\
maximum  gas temperature [K] & 23349 \\
mass-mean  gas temperature [K] & 54 \\
mm-opacity-slope (0.85-1.3)mm & 1.7 \\
cm-opacity-slope (5-10)mm & 1.8 \\
mm-SED-slope (0.85-1.3)mm & 3.5 \\
cm-SED-slope (5-10)mm & 3.6 \\
10$\mathrm{\mu m}$ silicate emission amplitude & 1.6 \\
naked star luminosity [$L_\odot$] & 41.3 \\
bolometric luminosity [$L_\odot$] & 70.4 \\
near IR excess ($\lambda=$2.01-6.82$\mathrm{\mu m}$)
 [$L_\odot$] & 8.56 \\
mid IR excess ($\lambda=$6.82-29.3$\mathrm{\mu m}$)
 [$L_\odot$] & 5.12 \\
far IR excess ($\lambda=$29.3-999.$\mathrm{\mu m}$)
 [$L_\odot$] & 4.74 \\
\hline
\end{tabular}

\vspace{0.2cm}\centering\begin{tabular}{lrccccr}
\hline & & \multicolumn{2}{c}{line flux $[\mathrm{W\,m^{-2}}$]} & \multicolumn{2}{c}{FWHM $[\mathrm{km\,s^{-1}}]$} &  \\
species & $\lambda\,$[$\mathrm{\mu m}]$ & observed & model & observed & model & size[AU] \\
\hline
$\mathrm{OI}$ & 63.183 & 8.5$\pm$0.2(-16) & 2.6(-15) &  & 4.2 & 386.2 \\
$\mathrm{OI}$ & 145.525 & 4.5$\pm$1.5(-17) & 1.8(-16) &  & 3.9 & 319.6 \\
$\mathrm{OI}$ & 0.630 & 9.7$\pm$3.0(-16) & 1.7(-16) & 20.0$\pm$2.0 & 5.7 & 98.9 \\
$\mathrm{CII}$ & 157.740 & 5.1$\pm$0.8(-17) & 8.1(-17) &  & 2.7 & 615.6 \\
$\mathrm{C}$ & 370.415 & 5.6$\pm$2.8(-18) & 1.5(-17) &  & 2.8 & 536.6 \\
$\mathrm{HCO^+}$ & 3361.334 & 1.4$\pm$0.2(-21) & 1.1(-21) & 4.0$\pm$0.5 & 3.3 & 338.9 \\
$\mathrm{HCO^+}$ & 1120.478 & 4.8$\pm$1.0(-20) & 9.2(-20) & 2.5$\pm$0.5 & 3.4 & 384.5 \\
$\mathrm{CO}$ & 866.963 & 1.0$\pm$0.2(-18) & 2.7(-18) & 3.0$\pm$0.5 & 3.1 & 577.2 \\
$\mathrm{CO}$ & 1300.403 & 1.2$\pm$0.2(-19) & 7.5(-19) & 3.5$\pm$0.5 & 3.1 & 484.2 \\
$\mathrm{^{13}CO}$ & 1360.227 & 2.7$\pm$0.2(-19) & 2.1(-19) & 2.5$\pm$0.5 & 3.1 & 422.1 \\
$\mathrm{C^{18}O}$ & 1365.430 & 3.9$\pm$0.8(-20) & 8.5(-20) & 2.5$\pm$0.5 & 3.2 & 387.7 \\
$\mathrm{C^{17}O}$ & 1334.098 & 2.0$\pm$0.2(-20) & 3.6(-20) & 2.5$\pm$0.5 & 3.3 & 371.8 \\
$\mathrm{CN}$ & 1321.390 & 3.2$\pm$0.3(-21) & 8.5(-20) & 2.5$\pm$0.5 & 2.5 & 563.2 \\
$\mathrm{o\!-\!H_2CO}$ & 1419.394 & 2.6$\pm$0.3(-21) & 2.9(-21) & 2.5$\pm$0.5 & 3.1 & 415.1 \\
$\mathrm{o\!-\!H_2CO}$ & 1328.291 & 9.0$\pm$0.9(-21) & 3.6(-21) & 2.5$\pm$0.5 & 3.1 & 411.3 \\
$\mathrm{HCN}$ & 1127.520 & 1.1$\pm$0.2(-20) & 1.3(-19) & 2.5$\pm$0.5 & 2.6 & 486.5 \\
$\mathrm{CS}$ & 2039.834 & 1.8$\pm$0.4(-21) & 1.8(-21) & 2.5$\pm$0.5 & 2.7 & 515.9 \\
$\mathrm{SO}$ & 1454.060 & 1.4$\pm$0.3(-21) & 1.4(-21) & 2.5$\pm$0.5 & 3.6 & 293.6 \\
$\mathrm{SO}$ & 1363.006 & 1.9$\pm$0.4(-21) & 2.9(-21) & 2.5$\pm$0.5 & 3.6 & 300.3 \\
$\mathrm{CO}$ & 371.650 & 3.5$\pm$0.3(-17) & 3.0(-17) &  & 3.1 & 483.9 \\
$\mathrm{CO}$ & 325.225 & 4.3$\pm$0.4(-17) & 4.0(-17) &  & 3.2 & 472.6 \\
$\mathrm{CO}$ & 289.120 & 4.3$\pm$0.4(-17) & 5.2(-17) &  & 3.2 & 458.5 \\
$\mathrm{CO}$ & 260.239 & 4.1$\pm$0.4(-17) & 6.3(-17) &  & 3.2 & 441.5 \\
$\mathrm{CO}$ & 236.613 & 4.9$\pm$0.5(-17) & 7.2(-17) &  & 3.3 & 419.9 \\
$\mathrm{CO}$ & 216.927 & 3.8$\pm$0.4(-17) & 7.9(-17) &  & 3.5 & 389.2 \\
$\mathrm{CO}$ & 200.272 & 4.4$\pm$0.5(-17) & 8.2(-17) &  & 3.7 & 341.0 \\
$\mathrm{^{13}CO}$ & 388.743 & 3.8$\pm$2.5(-18) & 9.5(-18) &  & 3.2 & 416.9 \\
$\mathrm{^{13}CO}$ & 302.414 & 7.1$\pm$4.0(-18) & 1.4(-17) &  & 3.4 & 384.5 \\
$\mathrm{CO}$ & 72.842 & $<$3.8(-17) & 9.9(-19) &  & 5.7 & 97.5 \\
$\mathrm{CO}$ & 79.359 & $<$3.4(-17) & 2.9(-18) &  & 7.3 & 95.1 \\
$\mathrm{CO}$ & 90.162 & 3.1$\pm$0.4(-17) & 1.8(-17) &  & 7.5 & 78.4 \\
$\mathrm{CO}$ & 108.762 & $<$4.5(-17) & 7.5(-17) &  & 6.9 & 80.0 \\
$\mathrm{CO}$ & 113.457 & 5.9$\pm$1.2(-17) & 8.2(-17) &  & 6.8 & 81.1 \\
$\mathrm{CO}$ & 118.580 & $<$4.6(-17) & 8.5(-17) &  & 6.8 & 82.9 \\
$\mathrm{CO}$ & 124.193 & $<$3.5(-17) & 8.5(-17) &  & 6.8 & 86.3 \\
$\mathrm{CO}$ & 130.368 & 3.9$\pm$0.9(-17) & 8.3(-17) &  & 6.8 & 106.2 \\
$\mathrm{CO}$ & 137.196 & $<$2.9(-17) & 8.1(-17) &  & 6.7 & 129.0 \\
$\mathrm{CO}$ & 144.784 & 2.7$\pm$0.9(-17) & 8.1(-17) &  & 6.4 & 150.4 \\
$\mathrm{CO}$ & 153.266 & 4.1$\pm$1.4(-17) & 8.1(-17) &  & 5.8 & 173.9 \\
$\mathrm{CO}$ & 162.811 & $<$2.7(-17) & 8.2(-17) &  & 4.9 & 200.4 \\
$\mathrm{CO}$ & 173.631 & 6.4$\pm$1.6(-17) & 8.2(-17) &  & 4.4 & 234.1 \\
$\mathrm{CO}$ & 185.999 & $<$3.0(-17) & 8.2(-17) &  & 4.0 & 280.8 \\
\hline
\end{tabular}
\end{table}

\addtocounter{table}{-1}
\begin{table}
\caption{(continued)}
\footnotesize
\vspace*{-1mm}
\centering\begin{tabular}{lrccccr}
\hline\\[-2mm]
  & & \multicolumn{2}{c}{line flux $[\mathrm{W\,m^{-2}}$]} & \multicolumn{2}{c}{FWHM[$\mathrm{km\,s^{-1}}$]} &  \\
species & $\lambda\,$[$\mathrm{\mu m}]$ & observed & model & observed & model & size[AU] \\
\hline
$\mathrm{CO}$ & 4.633 & 1.4$\pm$0.2(-16) & 1.1(-16) & 23.1$\pm$2.0 & 11.3 & 92.1 \\
$\mathrm{CO}$ & 4.657 & 8.8$\pm$0.5(-17) & 7.6(-17) & 24.2$\pm$2.0 & 11.2 & 92.3 \\
$\mathrm{CO}$ & 4.674 & 8.1$\pm$0.5(-17) & 7.4(-17) & 22.8$\pm$2.0 & 11.3 & 92.3 \\
$\mathrm{CO}$ & 4.682 & 1.0$\pm$0.2(-16) & 9.4(-17) & 21.0$\pm$2.0 & 11.0 & 92.2 \\
$\mathrm{CO}$ & 4.699 & 1.4$\pm$0.2(-16) & 1.2(-16) & 20.8$\pm$2.0 & 10.7 & 92.2 \\
$\mathrm{CO}$ & 4.735 & 1.4$\pm$0.2(-16) & 1.3(-16) & 20.8$\pm$2.0 & 11.4 & 92.3 \\
$\mathrm{CO}$ & 4.773 & 1.4$\pm$0.2(-16) & 1.1(-16) & 20.9$\pm$2.0 & 37.1 & 92.1 \\
$\mathrm{CO}$ & 4.920 & 1.0$\pm$0.2(-16) & 8.5(-17) & 27.1$\pm$2.0 & 64.5 & 1.3 \\
$\mathrm{CO}$ & 4.966 & 7.6$\pm$0.5(-17) & 7.8(-17) & 25.6$\pm$2.0 & 65.6 & 1.1 \\
$\mathrm{CO}$ & 5.066 & 3.1$\pm$0.5(-17) & 6.2(-17) & 26.4$\pm$2.0 & 67.4 & 0.8 \\
$\mathrm{^{13}CO}$ & 4.643 & 1.4$\pm$0.3(-17) & 8.6(-18) & 20.2$\pm$2.0 & 42.0 & 0.5 \\
$\mathrm{^{13}CO}$ & 4.692 & 1.4$\pm$0.3(-17) & 7.1(-18) & 17.6$\pm$2.0 & 14.7 & 95.6 \\
$\mathrm{^{13}CO}$ & 4.738 & 1.3$\pm$0.3(-17) & 4.5(-18) & 14.3$\pm$2.0 & 10.5 & 101.8 \\
$\mathrm{OH}$ & 65.131 & 7.0$\pm$2.0(-17) & 4.0(-17) &  & 9.4 & 94.6 \\
$\mathrm{OH}$ & 65.278 & 1.2$\pm$0.2(-16) & 6.0(-17) &  & 7.2 & 96.2 \\
$\mathrm{OH}$ & 71.170 & 4.5$\pm$0.6(-17) & 2.2(-17) &  & 12.1 & 93.7 \\
$\mathrm{OH}$ & 71.215 & 4.5$\pm$0.6(-17) & 2.6(-17) &  & 10.3 & 95.2 \\
$\mathrm{OH}$ & 79.115 & 2.5$\pm$0.8(-17) & 2.4(-17) &  & 5.9 & 101.1 \\
$\mathrm{OH}$ & 79.179 & $<$2.4(-17) & 2.4(-17) &  & 6.0 & 101.2 \\
$\mathrm{OH}$ & 84.420 & 1.0$\pm$0.1(-16) & 5.4(-17) &  & 7.7 & 96.2 \\
$\mathrm{OH}$ & 84.596 & 1.0$\pm$0.1(-16) & 7.2(-17) &  & 7.2 & 97.7 \\
$\mathrm{OH}$ & 119.234 & 2.8$\pm$1.4(-17) & 3.7(-17) &  & 7.2 & 100.5 \\
$\mathrm{OH}$ & 119.441 & 3.1$\pm$1.4(-17) & 4.5(-17) &  & 7.0 & 103.8 \\
$\mathrm{o\!-\!H_2}$ & 17.033 & 6.9$\pm$2.6(-18) & 1.5(-16) & 7.8$\pm$1.2 & 8.5 & 139.6 \\
$\mathrm{p\!-\!H_2}$ & 12.277 & 5.6$\pm$2.0(-18) & 3.7(-17) & 8.3$\pm$1.2 & 7.8 & 113.5 \\
$\mathrm{p\!-\!H_2}$ & 8.025 & 1.5$\pm$0.3(-17) & 4.2(-18) & 10.4$\pm$2.0 & 6.0 & 111.0 \\
$\mathrm{o\!-\!H_2}$ & 4.694 & 9.0$\pm$2.0(-18) & 4.1(-19) & 15.9$\pm$2.0 & 11.1 & 102.3 \\
\hline
\end{tabular}

\label{tab:ABAur}
\end{table}

The gas/dust ratio is found to be close to 100 in this model with a
total disk mass of about $0.019\,M_\odot$ (close to the value
$0.022\,M_\odot$ obtained from the pure SED-fit, see
Table~\ref{tab:dustmass}). Fig.\,\ref{fig:ABAur-2} shows that the fit
of the CO ro-vibrational line flux data is excellent, although the
measurements of the FWHM show that two disk zones with sharp inner
edges cannot fully explain the CO ro-vibrational observations. In our
model, the CO emission region switches from the outer disk ($\rm
FWHM\!\approx\!10\,km/s$) to the inner disk ($\rm
FWHM\!\approx\!50\,km/s$) around $P$(10).

Table~\ref{tab:ABAur} shows that the [O\,{\sc i}]\,63\,$\mu$m line is
too strong in the model by a factor $\sim$3, and some of the far-IR
emission lines (e.g.\ high-$J$ CO) are somewhat too strong as well,
whereas the OH lines are fine. The results for the sub-mm and mm-lines
are a bit diverse, probably due to the aforementioned observational
issues with cloud absorption. $^{12}$CO~3-2 (866.963~$\mu$m), HCO$^+$
(3361.334 and 1120.478~$\mu$m), $^{13}$CO~2-1 (1360.227~$\mu$m),
o-H$_2$CO (1419.394 and 1328.291~$\mu$m) and CS (2039.834~$\mu$m)
lines are in good agreement with the observations (better than a
factor $2.5$), but the $^{12}$CO~2-1 line (1300.403~$\mu$m) is a
factor $6.3$ too strong, probably due to cloud issues, see
page~\pageref{page:ABAur}.

\clearpage
\newpage

\begin{figure}
\centering
\vspace*{2mm}
{\large\sffamily\bfseries\underline{HD\,97048}}\\[2mm]
$M_{\rm gas} = 9.9\times 10^{-2}\,M_\odot$,
$M_{\rm dust} = 5.3\times 10^{-4}\,M_\odot$, 
dust settling $\alpha_{\rm settle} = 5.9\times 10^{-2}$\\
\hspace*{-4mm}
\begin{tabular}{cc}
\begin{minipage}{73mm}
\includegraphics[width=78mm]{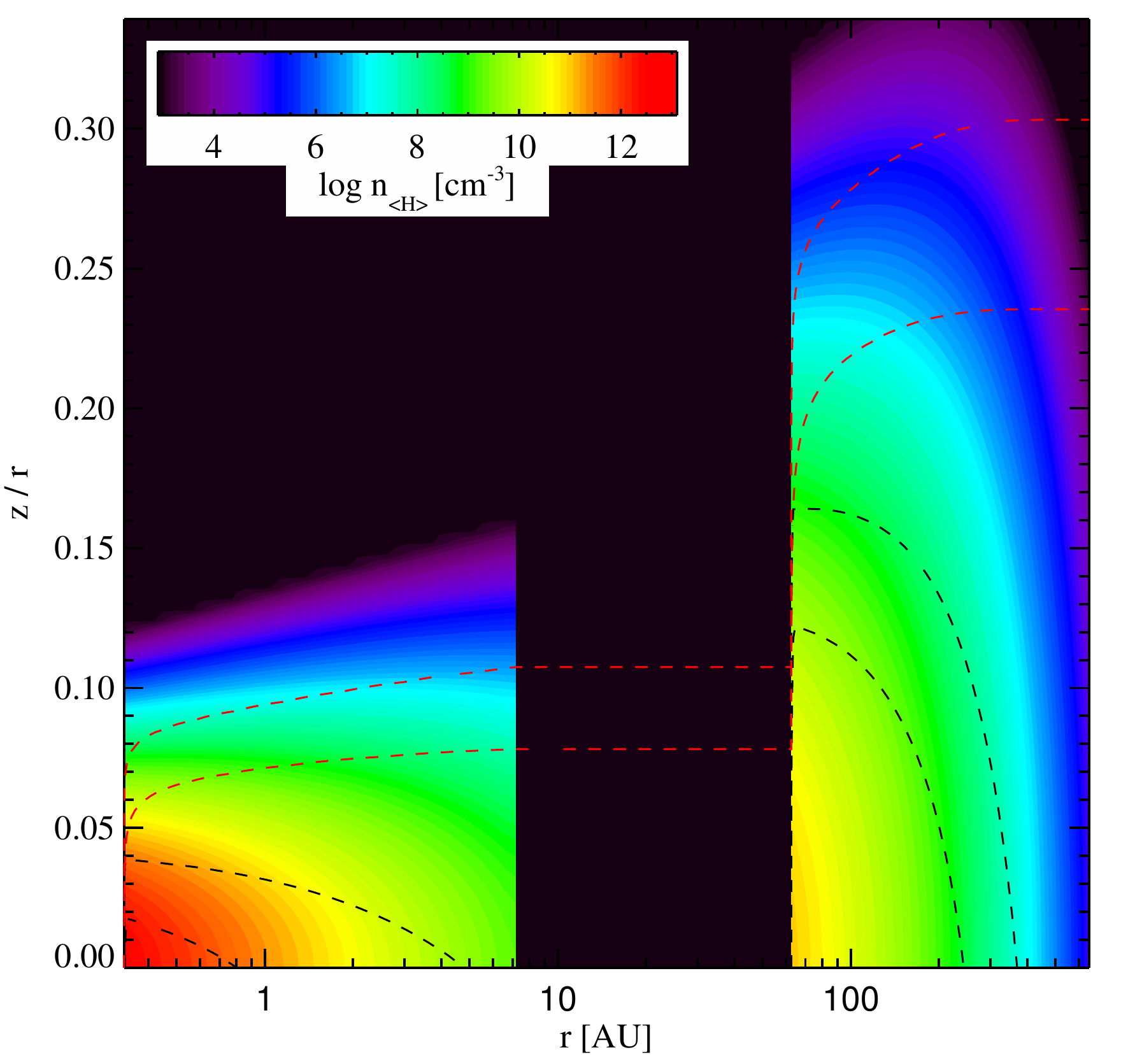}
\end{minipage}
&
\begin{minipage}{87mm}
\vspace*{1mm}
\includegraphics[width=87mm]{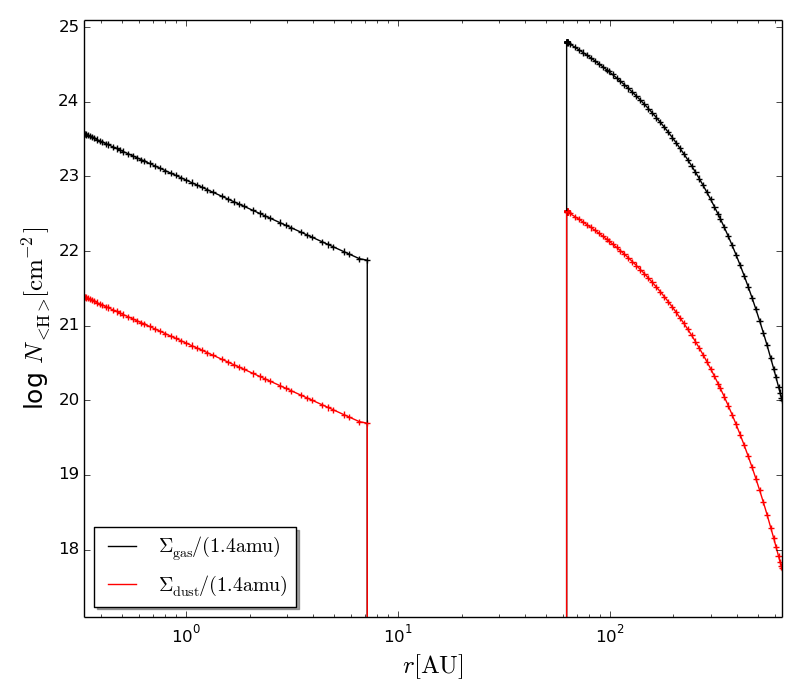}
\end{minipage}
\end{tabular}\\[-1mm]
\caption{HD\,97048 density and surface density plots. For details see
  Fig.\,\ref{fig:ABAur-1}.}
\label{fig:HD97048-1}
\end{figure}

\begin{figure}
\centering
\vspace*{-8mm}
\includegraphics[width=95mm]{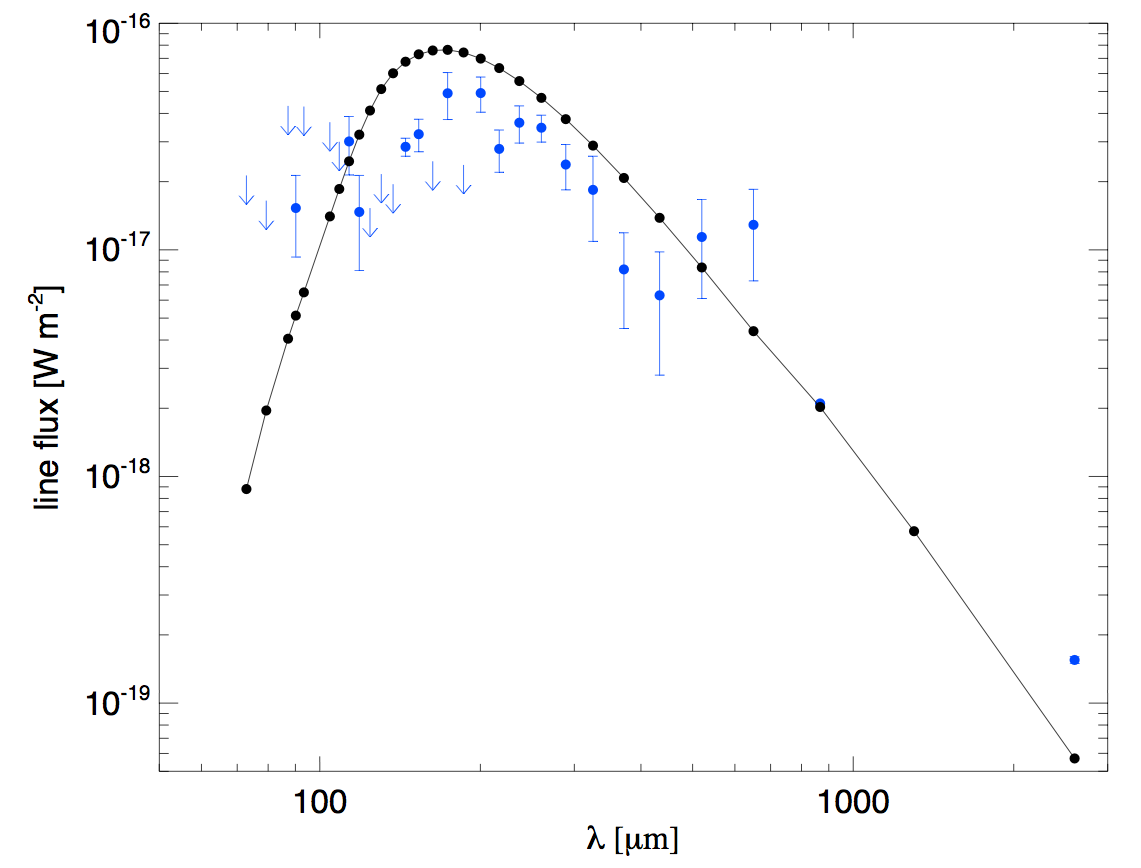} 
\vspace*{-2mm}
\caption{CO SLED in comparison to observations for HD\,97048. {The
  measurement points have $1\,\sigma$ errorbars attached.} The blue
arrows indicate upper limits, drawn from $3\,\sigma$ down to $1\,\sigma$.}
\label{fig:HD97048-2}
\end{figure}

\begin{table}
\caption{Model properties and comparison of computed spectral line
  properties with observations from {\sf HD97048.properties}. }
\footnotesize
\centering\begin{tabular}{rr}
\hline
DIANA standard fit model properties &  \\
\hline
minimum dust temperature [K] & 13 \\
maximum dust temperature [K] & 1777 \\
mass-mean dust temperature [K] & 36 \\
minimum  gas temperature [K] & 16 \\
maximum  gas temperature [K] & 18292 \\
mass-mean  gas temperature [K] & 38 \\
mm-opacity-slope (0.85-1.3)mm & 1.6 \\
cm-opacity-slope (5-10)mm & 1.6 \\
mm-SED-slope (0.85-1.3)mm & 3.1 \\
cm-SED-slope (5-10)mm & 3.6 \\
10$\mathrm{\mu m}$ silicate emission amplitude & 1.8 \\
naked star luminosity [$L_\odot$] & 40.0 \\
bolometric luminosity [$L_\odot$] & 53.9 \\
near IR excess ($\lambda=$2.02-6.72$\mathrm{\mu m}$)
 [$L_\odot$] & 2.09 \\
mid IR excess ($\lambda=$6.72-29.5$\mathrm{\mu m}$)
 [$L_\odot$] & 3.91 \\
far IR excess ($\lambda=$29.5-991.$\mathrm{\mu m}$)
 [$L_\odot$] & 5.21 \\
\hline
\end{tabular}

\vspace{0.5cm}\centering\begin{tabular}{lrccccr}
\hline & & \multicolumn{2}{c}{line flux $[\mathrm{W\,m^{-2}}$]} & \multicolumn{2}{c}{FWHM $[\mathrm{km\,s^{-1}}]$} &  \\
species & $\lambda\,$[$\mathrm{\mu m}]$ & observed & model & observed & model & size[AU] \\
\hline
$\mathrm{OI}$ & 63.183 & 1.6$\pm$0.1(-15) & 1.6(-15) &  & 5.1 & 349.0 \\
$\mathrm{OI}$ & 145.525 & 6.6$\pm$0.3(-17) & 8.5(-17) &  & 4.9 & 321.3 \\
$\mathrm{CII}$ & 157.740 & 1.1$\pm$0.3(-16) & 2.9(-17) &  & 3.5 & 553.6 \\
$\mathrm{CH^+}$ & 60.245 & 2.9$\pm$1.5(-17) & 7.1(-20) &  & 7.9 & 67.6 \\
$\mathrm{CH^+}$ & 72.137 & $<$1.9(-17) & 4.0(-19) &  & 8.1 & 72.3 \\
$\mathrm{CH^+}$ & 90.010 & 1.7$\pm$0.5(-17) & 5.7(-19) &  & 8.0 & 78.1 \\
$\mathrm{CH^+}$ & 179.593 & $<$1.8(-17) & 1.8(-19) &  & 7.5 & 169.8 \\
$\mathrm{OH}$ & 79.115 & $<$1.7(-17) & 9.6(-18) &  & 7.9 & 85.0 \\
$\mathrm{OH}$ & 79.179 & $<$1.9(-17) & 9.8(-18) &  & 7.9 & 85.6 \\
$\mathrm{o\!-\!H_2O}$ & 63.323 & $<$1.6(-17) & 2.2(-17) &  & 7.9 & 63.0 \\
$\mathrm{o\!-\!H_2O}$ & 71.946 & $<$1.7(-17) & 1.8(-17) &  & 7.8 & 63.2 \\
$\mathrm{o\!-\!H_2O}$ & 78.742 & $<$1.8(-17) & 1.2(-17) &  & 7.8 & 64.0 \\
$\mathrm{p\!-\!H_2O}$ & 89.988 & 1.7$\pm$0.5(-17) & 4.0(-18) &  & 7.9 & 63.2 \\
$\mathrm{o\!-\!H_2O}$ & 179.526 & $<$1.8(-17) & 1.9(-18) &  & 5.8 & 310.9 \\
$\mathrm{o\!-\!H_2O}$ & 180.488 & $<$1.4(-17) & 5.7(-19) &  & 7.9 & 114.2 \\
$\mathrm{CO}$ & 72.842 & $<$2.1(-17) & 8.8(-19) &  & 7.8 & 66.8 \\
$\mathrm{CO}$ & 79.359 & $<$1.6(-17) & 2.0(-18) &  & 7.8 & 68.8 \\
$\mathrm{CO}$ & 87.190 & $<$4.3(-17) & 4.1(-18) &  & 7.8 & 90.9 \\
$\mathrm{CO}$ & 90.162 & 1.5$\pm$0.6(-17) & 5.1(-18) &  & 7.7 & 100.4 \\
$\mathrm{CO}$ & 93.349 & $<$4.3(-17) & 6.5(-18) &  & 7.7 & 109.5 \\
$\mathrm{CO}$ & 104.444 & $<$3.7(-17) & 1.4(-17) &  & 7.1 & 135.5 \\
$\mathrm{CO}$ & 108.762 & $<$3.0(-17) & 1.9(-17) &  & 6.8 & 144.1 \\
$\mathrm{CO}$ & 113.457 & 3.0$\pm$0.9(-17) & 2.5(-17) &  & 6.6 & 152.6 \\
$\mathrm{CO}$ & 118.580 & 1.5$\pm$0.7(-17) & 3.2(-17) &  & 6.4 & 161.1 \\
$\mathrm{CO}$ & 124.193 & $<$1.5(-17) & 4.1(-17) &  & 6.3 & 169.9 \\
$\mathrm{CO}$ & 130.368 & $<$2.2(-17) & 5.1(-17) &  & 6.1 & 180.4 \\
$\mathrm{CO}$ & 137.196 & $<$1.9(-17) & 6.0(-17) &  & 5.9 & 190.8 \\
$\mathrm{CO}$ & 144.784 & 2.9$\pm$0.3(-17) & 6.8(-17) &  & 5.7 & 203.3 \\
$\mathrm{CO}$ & 153.266 & 3.2$\pm$0.5(-17) & 7.3(-17) &  & 5.5 & 218.1 \\
$\mathrm{CO}$ & 162.811 & $<$2.5(-17) & 7.6(-17) &  & 5.3 & 235.6 \\
$\mathrm{CO}$ & 173.631 & 4.9$\pm$1.1(-17) & 7.6(-17) &  & 5.1 & 255.4 \\
$\mathrm{CO}$ & 185.999 & $<$2.4(-17) & 7.4(-17) &  & 4.9 & 276.9 \\
$\mathrm{CO}$ & 200.272 & 4.9$\pm$0.9(-17) & 7.0(-17) &  & 4.7 & 299.2 \\
$\mathrm{CO}$ & 216.927 & 2.8$\pm$0.6(-17) & 6.3(-17) &  & 4.6 & 321.8 \\
$\mathrm{CO}$ & 236.613 & 3.6$\pm$0.7(-17) & 5.6(-17) &  & 4.5 & 344.8 \\
$\mathrm{CO}$ & 260.239 & 3.5$\pm$0.5(-17) & 4.7(-17) &  & 4.4 & 368.0 \\
$\mathrm{CO}$ & 289.120 & 2.4$\pm$0.5(-17) & 3.8(-17) &  & 4.3 & 389.1 \\
$\mathrm{CO}$ & 325.225 & 1.8$\pm$0.8(-17) & 2.9(-17) &  & 4.3 & 407.4 \\
$\mathrm{CO}$ & 371.650 & 8.2$\pm$3.7(-18) & 2.1(-17) &  & 4.2 & 422.9 \\
$\mathrm{CO}$ & 433.556 & 6.3$\pm$3.5(-18) & 1.4(-17) &  & 4.2 & 435.9 \\
$\mathrm{CO}$ & 520.231 & 1.1$\pm$0.5(-17) & 8.4(-18) &  & 4.1 & 446.1 \\
$\mathrm{CO}$ & 650.251 & 1.3$\pm$0.6(-17) & 4.4(-18) &  & 4.1 & 453.0 \\
\hline
\end{tabular}

\label{tab:HD97048}
\end{table}

\noindent{\sffamily\bfseries\underline{2.\ HD\,97048}}\ \ is a
well-studied Herbig group\,{\sc i} disk. It has an inner cavity of
$34\pm4$~AU derived from Q-band imaging \citep[T-ReCS on the Gemini
  South telescope,][]{Maaskant2013}. In addition, a small optically
thick inner disk (0.3-2.5\,AU) is required to fit the strong near-IR
excess in the SED. Our best fit model has a similar geometry, albeit
slightly different radial zones (inner disk from 0.3 to 7.18~AU, outer
disk starting at 62.6\,AU (Fig.~\ref{fig:HD97048-1}). The differences
can arise from the choice of dust opacities (e.g.\ different $a_{\rm
  min}$, $a_{\rm max}$ and composition). Recent spatially resolved
ALMA data at 302 and 346~GHz are consistent with an inner cavity of
$\sim\!50$~AU \citep{Walsh2016}.

HD\,97048 has been observed with the VISIR instrument in the PAH band
at $8.6~\mu$m \citep{Lagage2006}. Our flaring angle of
$\beta\!=\!1.19$ is slightly smaller than the one inferred from the
VISIR image ($\beta\!=\!1.26\pm0.05$). The surface scale height found
from the VISIR PAH image is 51.3\,AU at 135\,AU distance from the
star. Our model has a gas scale height of 8.43\,AU at 135\,AU, and
typical factors between the gas scale height and the surface scale
height are of the order of three to five. So our best disk model
agrees with this within a factor $\sim\!2$.

Both [O\,{\sc i}] fine structure lines at 63 and 145~$\mu$m were
detected with Herschel/PACS \citep{Meeus2012} and our model reproduces
those fluxes within 30\%
(Table~\ref{tab:HD97048}). Figure~\ref{fig:HD97048-2} shows that the
modeled CO ladder (high J lines) agrees quite well with the observed
one \citep{Meeus2012,Fedele2013,vanderWiel2014}. The CH$^+$ emission
from our model is a factor 30 fainter than the detected Herschel/PACS
flux \citep{Fedele2013}. \citet{Thi2011} showed that CH$^+$ emission
in HD\,100546 originates from the surface of the inner wall of the
outer disk and this holds also for our disk model of HD\,97048.

Since the construction of this DIANA model, CO ro-vibrational line
fluxes were measured by \citet{vanderPlas2015}. For the $v\!=\!1-0$
band, they range from $\sim\!10^{-16}$~W/m$^2$ (low $J$) to
$\sim\!10^{-17}$~W/m$^2$ (high $J$). Our model has values a factor
$\sim\!10$ smaller. The ro-vibrational emission originates largely in
the inner disk ($\sim\!75$\%), but partially also in the inner wall of
the outer disk.

Also, new sub-mm data became available from APEX and ALMA. Foreground
extinction causes the total flux of CO $J\!=\!1-0$ and $J\!=\!3-2$ as
observed with ALMA ($74.28\pm 0.14$~Jy km/s and $8.22\pm 0.28$~Jy
km/s) to be a lower limit \citep{vanderPlas2017} and makes the line
profiles difficult to fit; the $J\!=\!3-2$ flux is close to that of
\citep{Hales2014} who reported 1.4~K~km/s (resulting in $2.13\times
10^{-19}$~W/m$^2$ using a beam efficiency of 0.6 and the APEX beam
size of 18"). Our model underpredicts the ALMA HCO$^+$ flux by a
factor $\sim\!7$. However, our model used the small chemical network
presented in \citet{Kamp2017} which does not fully capture the HCO$^+$
chemistry. With the large chemical network, HCO$^+$ line fluxes
increase typically by a factor $3-4$ \citep{Kamp2017}, which would
bring our HD\,97048 model closer to the observed value.

The ALMA CO $J\!=\!3-2$ data of HD\,97048 \citep{vanderPlas2017}
suggests an even larger gas disk ($R_{\rm out}\!\sim\!820$\,AU) than
used in our model. They also show that CO and HCO$^+$ gas extends
inside the dust cavity of this disk, which could have an effect on the
CH$^+$ and CO ro-vibrational line fluxes (see above) and bring them
closer to the observed values. Based on the strong tapering outer edge
of our current disk model, we predict large differences in emitting
size (85\% of radial flux) between the three isotopologue $J\!=\!2-1$
lines of $^{12}$CO, $^{13}$CO and C$^{18}$O of 430, 370, and 290\,AU
and flux ratios of 6 and 3.02 for $^{12}$CO/$^{13}$CO and
$^{13}$CO/C$^{18}$O. This can be tested with high spatial resolution
ALMA data. To summarize, the disk model should be refined in the
future using the wealth of existing and upcoming ALMA data.

\clearpage
\newpage

\begin{figure}[!t]
\centering
\vspace*{2mm}
{\large\sffamily\bfseries\underline{HD 163296}}\\[2mm]
$M_{\rm gas} = 0.58\,M_\odot$,
$M_{\rm dust} = 1.7\times 10^{-3}\,M_\odot$, 
dust settling $\alpha_{\rm settle} = 6.6\times 10^{-3}$\\
\hspace*{-3mm}
\begin{tabular}{cc}
\begin{minipage}{76mm}
\includegraphics[width=79mm]{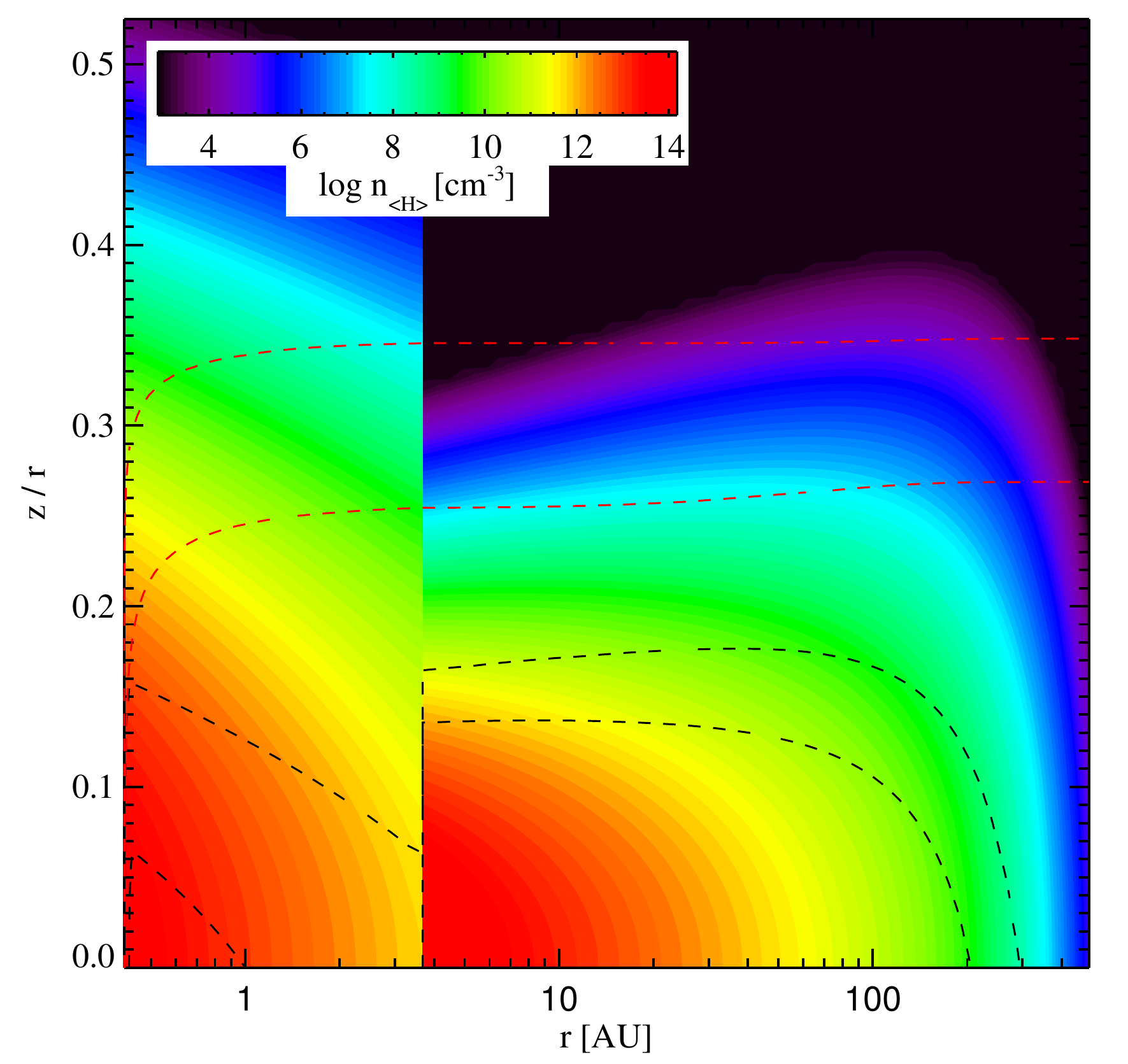}
\end{minipage}
&
\begin{minipage}{87mm}
\vspace*{1mm}
\includegraphics[width=87mm]{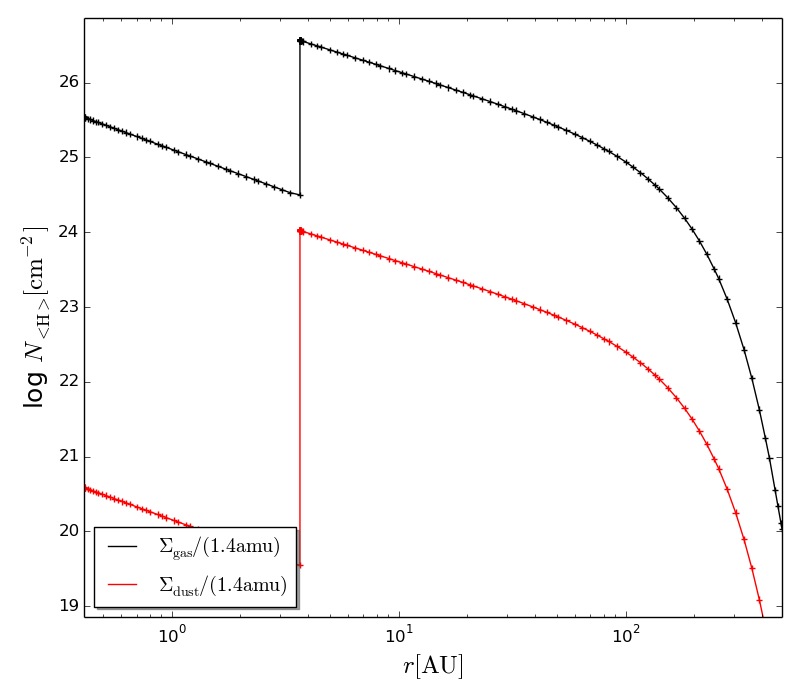}
\end{minipage}
\end{tabular}\\[-2mm]
\caption{HD\,163296 density and surface density plots. For details see
  Fig.\,\ref{fig:ABAur-1}}
\label{fig:HD163296-1}
\end{figure}

\begin{table}
\caption{Model properties and comparison of computed spectral
  line properties with observations from {\sf HD163296.properties}. }
\footnotesize
\centering\begin{tabular}{rr}
\hline
DIANA standard fit model properties &  \\
\hline
minimum dust temperature [K] & 10 \\
maximum dust temperature [K] & 1946 \\
mass-mean dust temperature [K] & 27 \\
minimum  gas temperature [K] & 12 \\
maximum  gas temperature [K] & 15203 \\
mass-mean  gas temperature [K] & 29 \\
mm-opacity-slope (0.85-1.3)mm & 1.4 \\
cm-opacity-slope (5-10)mm & 1.6 \\
mm-SED-slope (0.85-1.3)mm & 2.5 \\
cm-SED-slope (5-10)mm & 3.1 \\
10$\mathrm{\mu m}$ silicate emission amplitude & 2.0 \\
naked star luminosity [$L_\odot$] & 34.4 \\
bolometric luminosity [$L_\odot$] & 48.7 \\
near IR excess ($\lambda=$2.05-6.90$\mathrm{\mu m}$)
 [$L_\odot$] & 4.10 \\
mid IR excess ($\lambda=$6.90-29.3$\mathrm{\mu m}$)
 [$L_\odot$] & 2.18 \\
far IR excess ($\lambda=$29.3-970.$\mathrm{\mu m}$)
 [$L_\odot$] & 0.70 \\
\hline
\end{tabular}

\vspace{0.2cm}\centering\begin{tabular}{lrccccr}
\hline & & \multicolumn{2}{c}{line flux $[\mathrm{W\,m^{-2}}$]} & \multicolumn{2}{c}{FWHM $[\mathrm{km\,s^{-1}}]$} &  \\
species & $\lambda\,$[$\mathrm{\mu m}]$ & observed & model & observed & model & size[AU] \\
\hline
$\mathrm{CO}$ & 72.842 & $<$1.1(-17) & 3.2(-18) &  & 51.9 & 2.9 \\
$\mathrm{CO}$ & 79.359 & $<$1.6(-17) & 3.6(-18) &  & 49.2 & 3.0 \\
$\mathrm{CO}$ & 90.162 & $<$7.4(-18) & 3.4(-18) &  & 48.1 & 3.1 \\
$\mathrm{CO}$ & 144.784 & 6.7$\pm$1.5(-18) & 2.8(-18) &  & 20.1 & 55.2 \\
$\mathrm{CO}$ & 866.963 & 1.1$\pm$0.0(-18) & 1.3(-18) & 4.5$\pm$0.1 & 4.4 & 388.8 \\
$\mathrm{CO}$ & 1300.404 & 4.0$\pm$0.1(-19) & 3.7(-19) & 4.5$\pm$0.2 & 4.5 & 380.9 \\
$\mathrm{^{13}CO}$ & 1360.227 & 1.4$\pm$0.2(-19) & 1.2(-19) & 4.9$\pm$0.3 & 4.9 & 313.2 \\
$\mathrm{C^{18}O}$ & 1365.421 & 4.5$\pm$0.2(-20) & 5.4(-20) & 5.5$\pm$0.3 & 5.3 & 284.2 \\
$\mathrm{^{13}CO}$ & 2720.406 & 1.2$\pm$0.7(-20) & 1.1(-20) & 4.5$\pm$0.5 & 5.2 & 279.9 \\
$\mathrm{OI}$ & 63.183 & 1.7$\pm$0.1(-16) & 5.3(-16) &  & 4.2 & 447.1 \\
$\mathrm{OI}$ & 145.525 & $<$6.0(-18) & 4.7(-17) &  & 4.1 & 407.0 \\
$\mathrm{o\!-\!H_2O}$ & 29.836 & 4.3$\pm$0.4(-17) & 4.8(-17) &  & 49.5 & 3.1 \\
$\mathrm{o\!-\!H_2O}$ & 63.323 & 1.6$\pm$0.4(-17) & 1.1(-17) &  & 50.3 & 3.2 \\
$\mathrm{p\!-\!H_2O}$ & 63.458 & 1.1$\pm$0.3(-17) & 1.0(-17) &  & 50.3 & 3.1 \\
$\mathrm{o\!-\!H_2O}$ & 71.946 & 1.4$\pm$0.5(-17) & 8.1(-18) &  & 50.2 & 3.3 \\
$\mathrm{o\!-\!H_2O}$ & 78.742 & 1.1$\pm$0.3(-17) & 6.6(-18) &  & 50.1 & 6.9 \\
$\mathrm{p\!-\!H_2O}$ & 78.928 & $<$1.4(-17) & 5.8(-18) &  & 50.2 & 3.2 \\
$\mathrm{p\!-\!H_2O}$ & 89.988 & $<$5.1(-18) & 5.3(-18) &  & 4.4 & 264.9 \\
$\mathrm{p\!-\!H_2O}$ & 144.517 & $<$1.2(-17) & 1.1(-18) &  & 49.9 & 3.4 \\
$\mathrm{p\!-\!H_2O}$ & 158.311 & $<$1.4(-17) & 6.7(-19) &  & 49.6 & 3.1 \\
$\mathrm{o\!-\!H_2O}$ & 179.526 & $<$1.7(-17) & 1.1(-17) &  & 4.9 & 336.6 \\
$\mathrm{o\!-\!H_2O}$ & 180.488 & $<$1.7(-17) & 2.5(-18) &  & 4.6 & 318.8 \\
$\mathrm{OH}$ & 79.115 & 1.2$\pm$0.3(-17) & 2.8(-18) &  & 47.3 & 93.6 \\
$\mathrm{OH}$ & 79.179 & $<$9.0(-18) & 2.8(-18) &  & 46.3 & 120.0 \\
$\mathrm{CO}$ & 650.251 & $<$2.1(-17) & 2.9(-18) &  & 4.4 & 386.2 \\
$\mathrm{CO}$ & 520.231 & 1.0$\pm$0.4(-17) & 5.5(-18) &  & 4.4 & 388.6 \\
$\mathrm{CO}$ & 433.556 & 7.4$\pm$2.9(-18) & 8.7(-18) &  & 4.4 & 389.1 \\
$\mathrm{CO}$ & 371.650 & 9.0$\pm$3.0(-18) & 1.3(-17) &  & 4.4 & 387.9 \\
$\mathrm{CO}$ & 325.225 & 1.2$\pm$0.6(-17) & 1.7(-17) &  & 4.3 & 385.3 \\
$\mathrm{CO}$ & 289.120 & 9.1$\pm$3.8(-18) & 2.0(-17) &  & 4.3 & 381.2 \\
$\mathrm{CO}$ & 260.239 & 1.2$\pm$0.3(-17) & 2.2(-17) &  & 4.4 & 375.0 \\
$\mathrm{CO}$ & 236.613 & 1.1$\pm$0.3(-17) & 2.2(-17) &  & 4.4 & 367.8 \\
$\mathrm{CO}$ & 216.927 & 1.2$\pm$0.3(-17) & 1.7(-17) &  & 4.5 & 361.8 \\
$\mathrm{CO}$ & 200.272 & 1.5$\pm$0.4(-17) & 1.2(-17) &  & 6.0 & 352.7 \\
$\mathrm{C}$ & 370.415 & 5.0$\pm$2.0(-18) & 4.9(-18) &  & 4.2 & 418.4 \\
$\mathrm{N^+}$ & 205.240 & 1.7$\pm$0.2(-17) & 4.5(-22) &  & 5.8 & 478.4 \\
\hline
\end{tabular}

\label{tab:HD163296}
\end{table}

\begin{figure}[!ht]
\vspace*{-7mm}\hspace*{-9mm}
\begin{tabular}{cc}
\begin{minipage}{89mm}
\includegraphics[width=92mm]{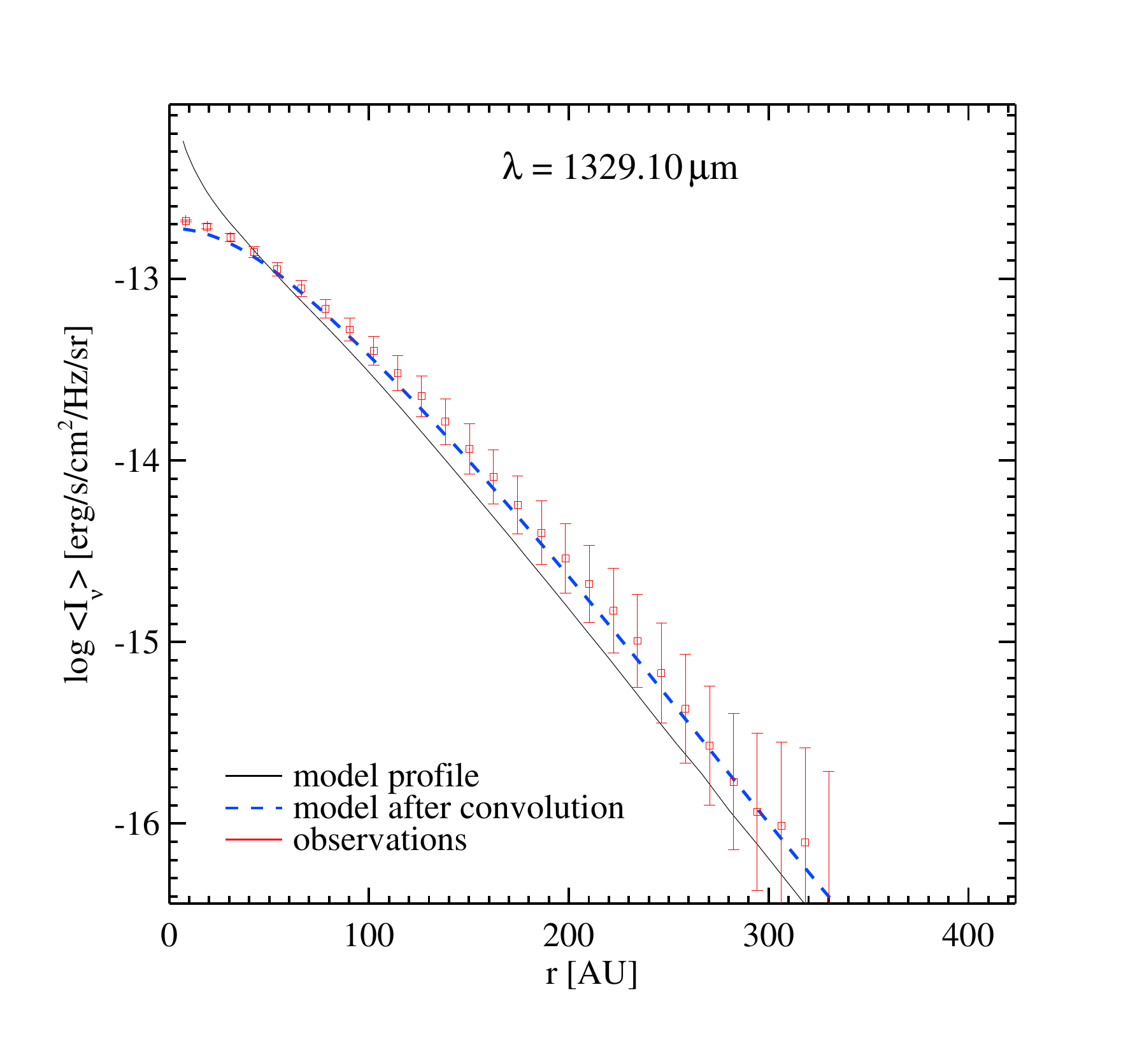} 
\end{minipage} &
\hspace*{-6mm}
\begin{minipage}{85mm}
\includegraphics[width=85mm]{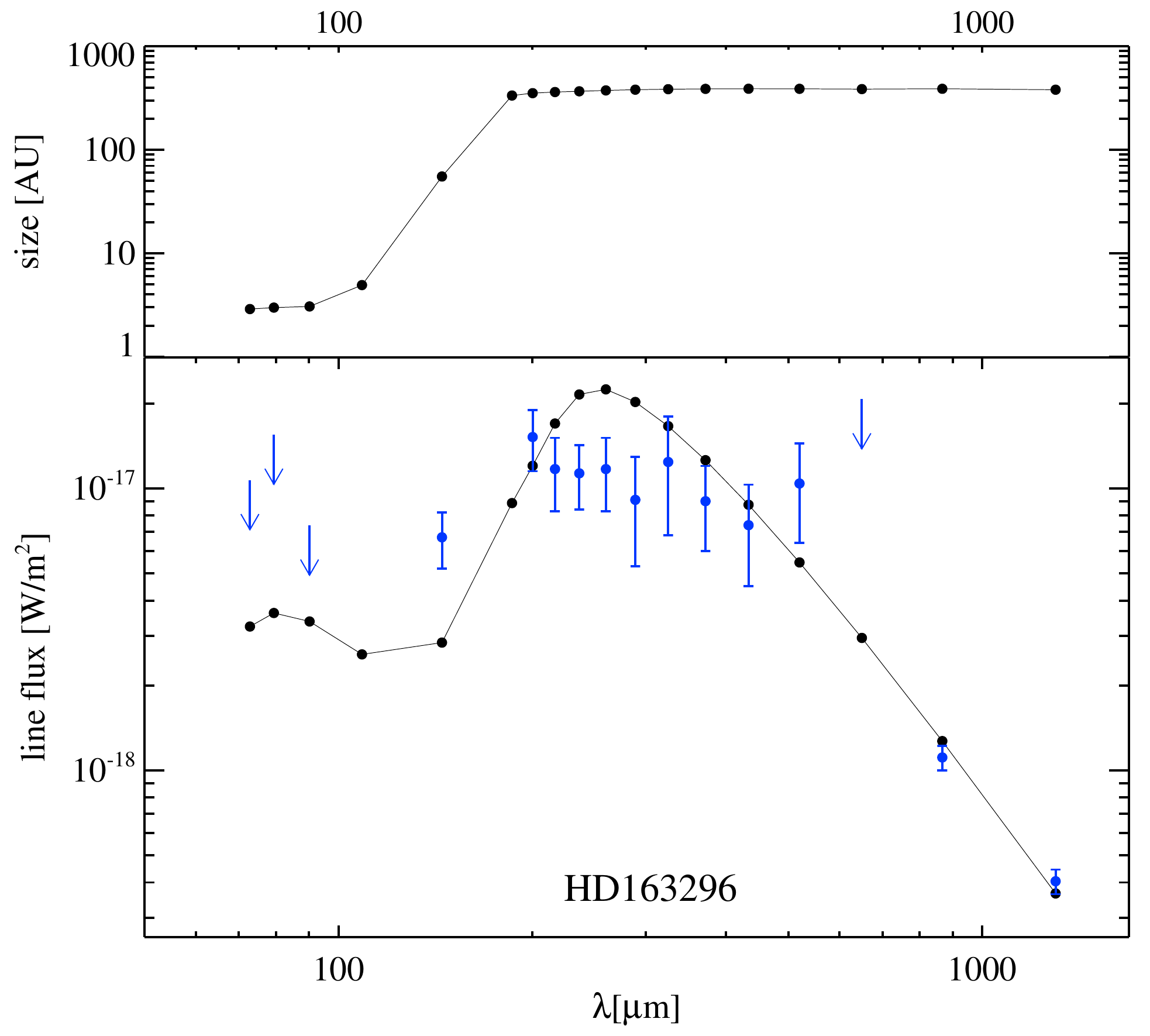}
\end{minipage}
\end{tabular}\\*[-7mm]
\caption{{\bf Left:} Fit of radial intensity profile of HD\,163296 at 1.3\,mm
  (band~6 ALMA science verification data). {\bf Right:} CO SLED model of
  HD\,163296 (black lines and dots) in comparison to multi-instrument
  observational data (blue errorbars, arrows are $3\,\sigma$ non-detections).}
\label{fig:HD163296-2}
\end{figure}

\begin{figure}[!ht]
\centering
\hspace*{-6mm}\begin{tabular}{cc}
\hspace*{-6mm}\includegraphics[width=60mm]{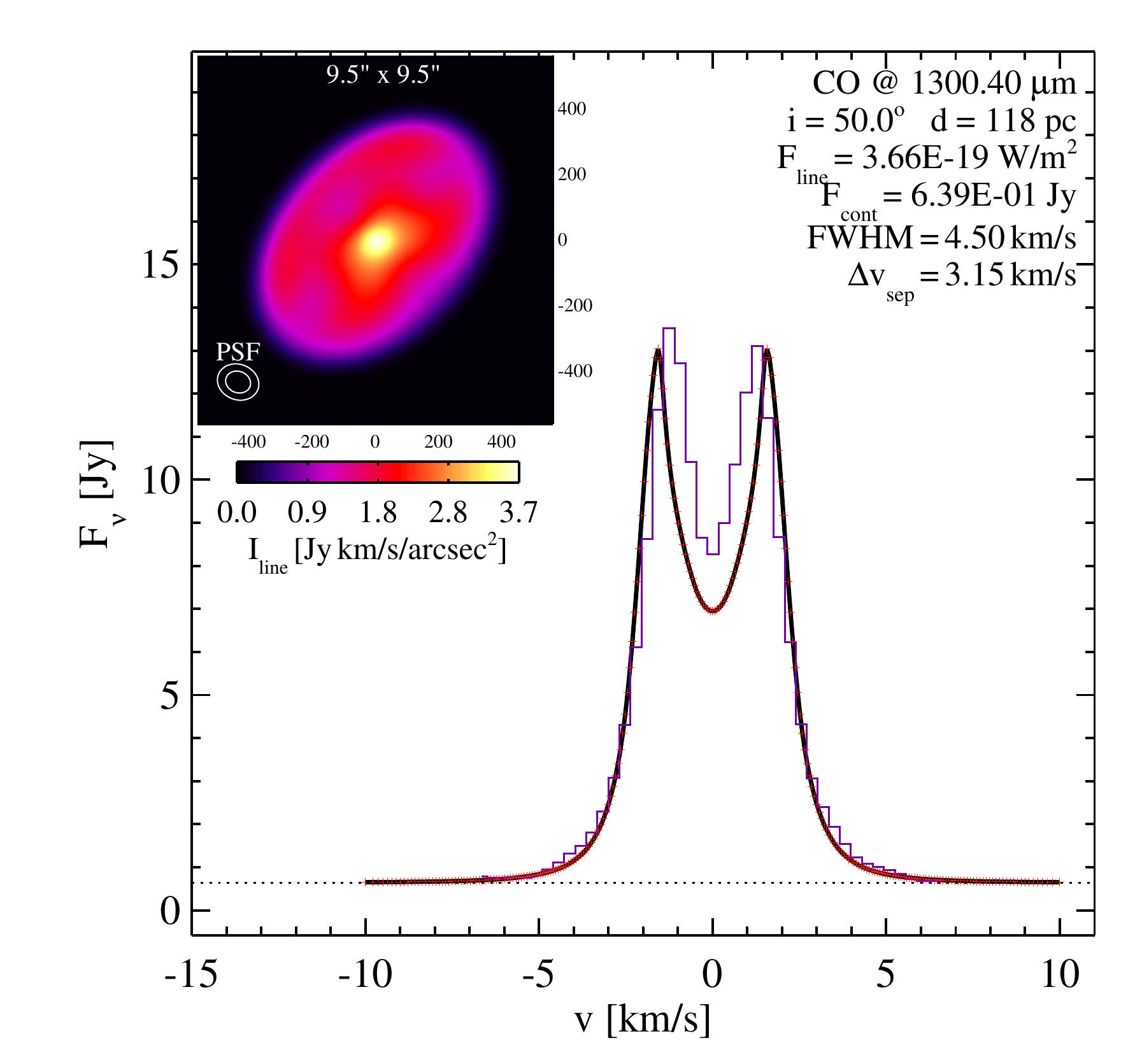} &
\hspace*{-6mm}\includegraphics[width=60mm]{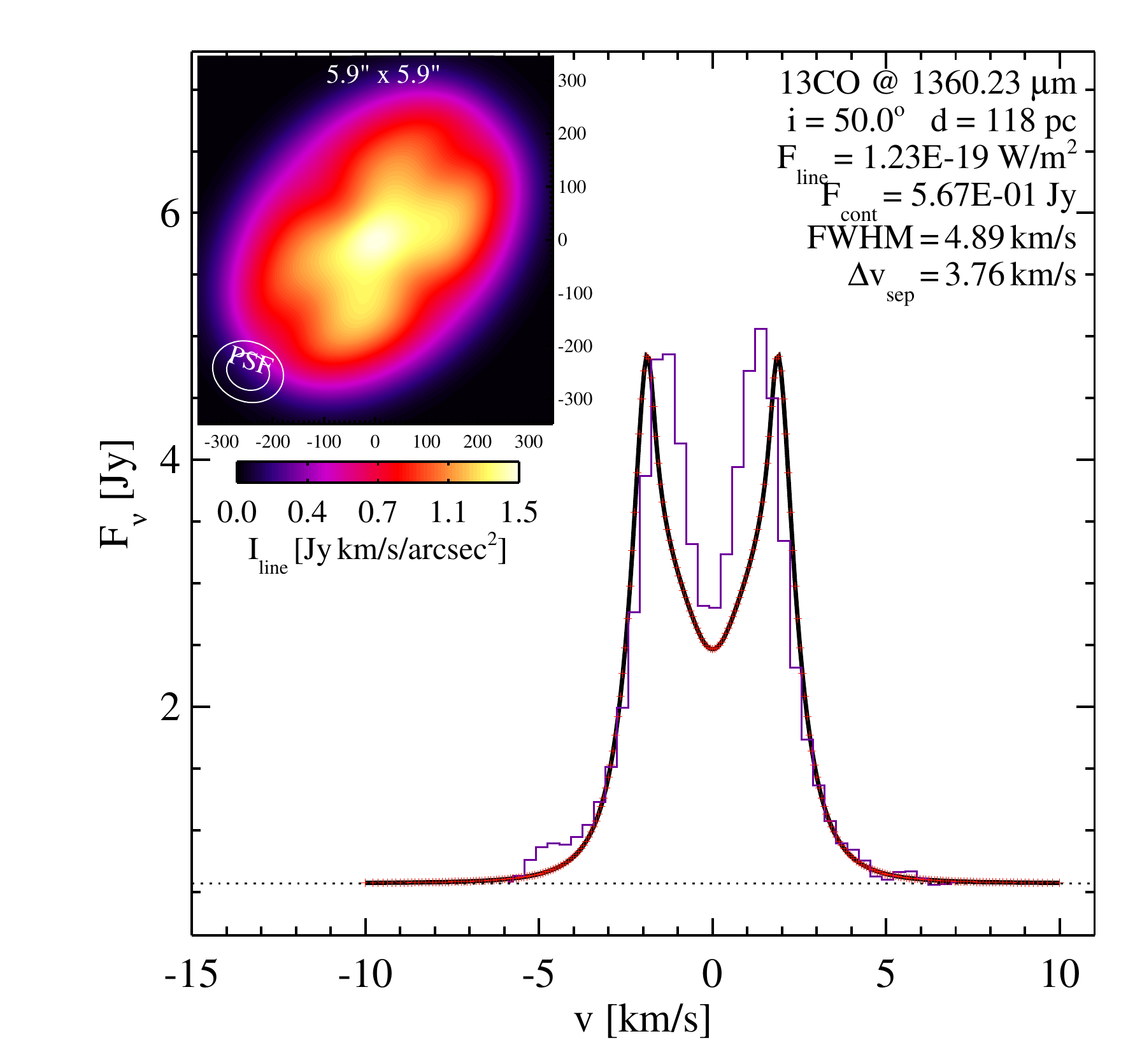} \\
\hspace*{-6mm}\includegraphics[width=60mm]{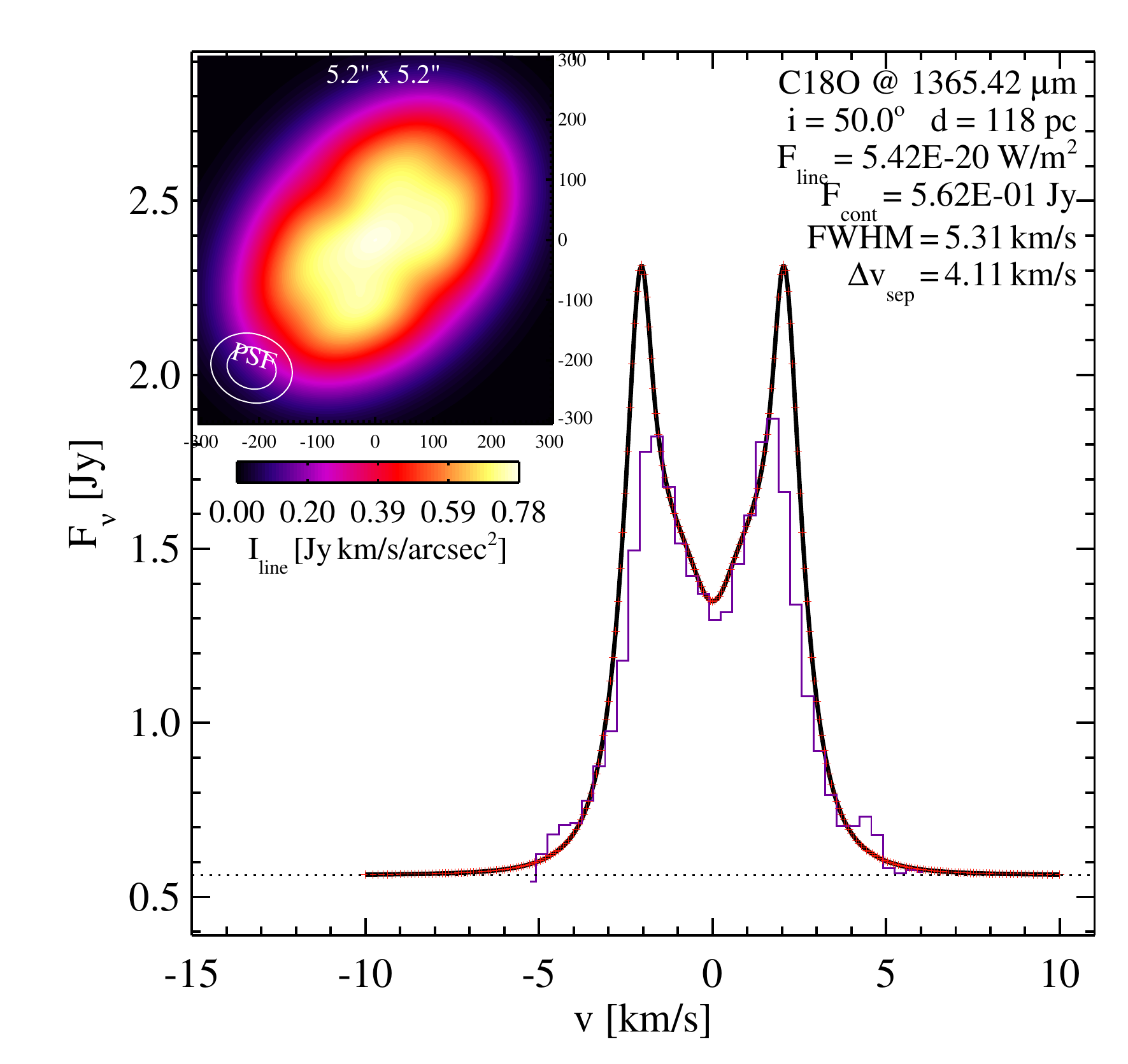} &
\hspace*{4mm}\includegraphics[width=40mm]{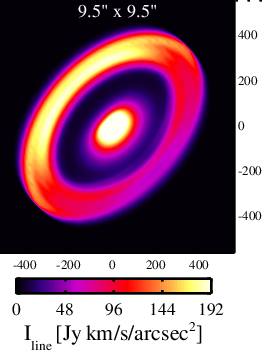}
\end{tabular}\\[-1mm]
\caption{CO isotopologue lines around 1.3\,mm for HD\,163296. The
  lower right panel shows the [O\,{\sc i}]\,63\,$\mu$m image from the
  disk model of HD\,163296.}
\label{fig:HD163296-3}
\end{figure}

\noindent{\sffamily\bfseries\underline{3.\ \ HD\,163296}}\ \ is a
bright isolated Herbig\,Ae star with a large and apparently almost
perfectly symmetric Keplerian disk. The object has ALMA science
verification continuum data partly shown in Fig.~\ref{fig:HD163296-2},
and line data partly shown in the lower row of plots in
Fig.\,\ref{fig:HD163296-3}. Our data collection includes a number of
low resolution spectra (ISO/SWS, Spitzer/IRS, Herschel/PACS and
Herschel/SPIRE), two (sub-)mm ALMA continuum images (band 6 and 7), 36
line observations and four ALMA line maps with derived intensity
profiles.  HD\,163296 also has an almost complete coverage of observed
CO line observations, from $J\!=$2-1 at 1.3mm up to $J\!=$36-35 at
72.8\,$\mu$m, the so-called CO spectral line energy distribution
(SLED), Fig.\,\ref{fig:HD163296-2}.

The model follows a pattern we have already seen before. A tall inner
disk casts a shadow onto a massive, cold and mildly flaring outer disk
with tapered outer edge (Fig.\,\ref{fig:HD163296-1}). The model also
suggests a large depletion of dust in the inner disk. The model
provides an excellent SED-fit, and can explain most of the line
observations within a factor two, from CO isotopologue lines over
high-$J$ CO and high-excitation water lines to the neutral carbon line
at 370\,$\mu$m. The [O\,{\sc i}]\,63\,$\mu$m line is overpredicted by
a factor of three, though, and the [N\,{\sc ii}] line at 205\,$\mu$m
is several orders of magnitude too weak.

The radial continuum and line intensity profiles show that the
tapered-edged disk model can, to some extent, naturally explain the
somewhat larger extension of the disk in $^{12}$CO (sub-)mm lines
($\sim\,$400\,AU) as compared to the (sub-)mm continuum
($\sim\,$200\,AU), because the CO-lines stay optically thick even at
larger radii where the optically thin continuum already vanishes in
the background noise. There is no obvious need in this model to
introduce a gas/dust ratio that changes with radius, as our
model with a constant gas/dust ratio of about 350 in the outer disk
zone fits both, the disk extension in mm-continuum and CO
lines. However, dust radial migration is expected to 
result in a changing gas/dust ratio \citep[e.g.][]{Facchini2017},
and closer inspection reveals that the model actually
arrived at some kind of compromise. The disk extension of the model is
slightly too small in CO lines, and slightly too large in the
continuum. An abrupt disappearance of the mm-continuum signal around
the outer edge has indeed been reported by \citep{deGregorio2013}, which
can be considered as true evidence for inward radial drift of mm-sized dust 
particles in HD\,163296.

This {\sf DIANA}-standard model has an unprecedented level of physical
consistency and agreement with a large suite of multi-wavelength line
and continuum data (Table~\ref{tab:HD163296}). The outer disk is found
to be very massive in this model (0.58\,M$_\odot$), the heaviest disk
among all {\sf DIANA}-standard models, with a gas/dust mass ratio of
$\sim\,$350. This is different from the results of the pure SED-fit,
where the disk mass was estimated to be only 0.053\,M$_\odot$, see
Table~\ref{tab:dustmass}. The inner disk is also found to be even more
gas-enriched (gas/dust $\sim\,$85000), which gives a boost to all
emission lines at shorter wavelength that originate in the inner
disk, similar to RECX\,15.

The shadow casted by the inner disk hits the outer disk around 200\,AU
in this model, exciting far-IR emission lines in the surface of the
outer disk only outside of that radius. This particular geometry leads
to the prediction of a ring-like appearance of far-IR emission lines
like [O\,{\sc i}]\,63\,$\mu$m (lower right panel of
Fig~\ref{fig:HD163296-3}) and high-$J$ CO lines, which, unfortunately,
cannot be resolved by any current instrumentation.  These results
demonstrate how important the stellar UV irradiation is for the
heating and line formation in the model, as both the gas and the dust
are assumed to be entirely smooth and continuous around
200\,AU in the model. Latest VLT/SPHERE and ALMA continuum data, however, show
ring-like substructures at 80, 124 and 200\,AU at millimeter
wavelengths, but only the innermost of these in scattered light
\citep{Muro-Arena2018}. These new observations have not been included
in our data collection for HD\,163296.

\clearpage
\newpage

\begin{figure}
\centering
\vspace*{2mm}
{\large\sffamily\bfseries\underline{MWC 480 (HD 31648)}}\\[2mm]
$M_{\rm gas} = 2.2\times 10^{-2}\,M_\odot$,
$M_{\rm dust} = 2.2\times 10^{-4}\,M_\odot$, 
dust settling $\alpha_{\rm settle} = 7.3\times 10^{-5}$\\
\includegraphics[width=80mm]{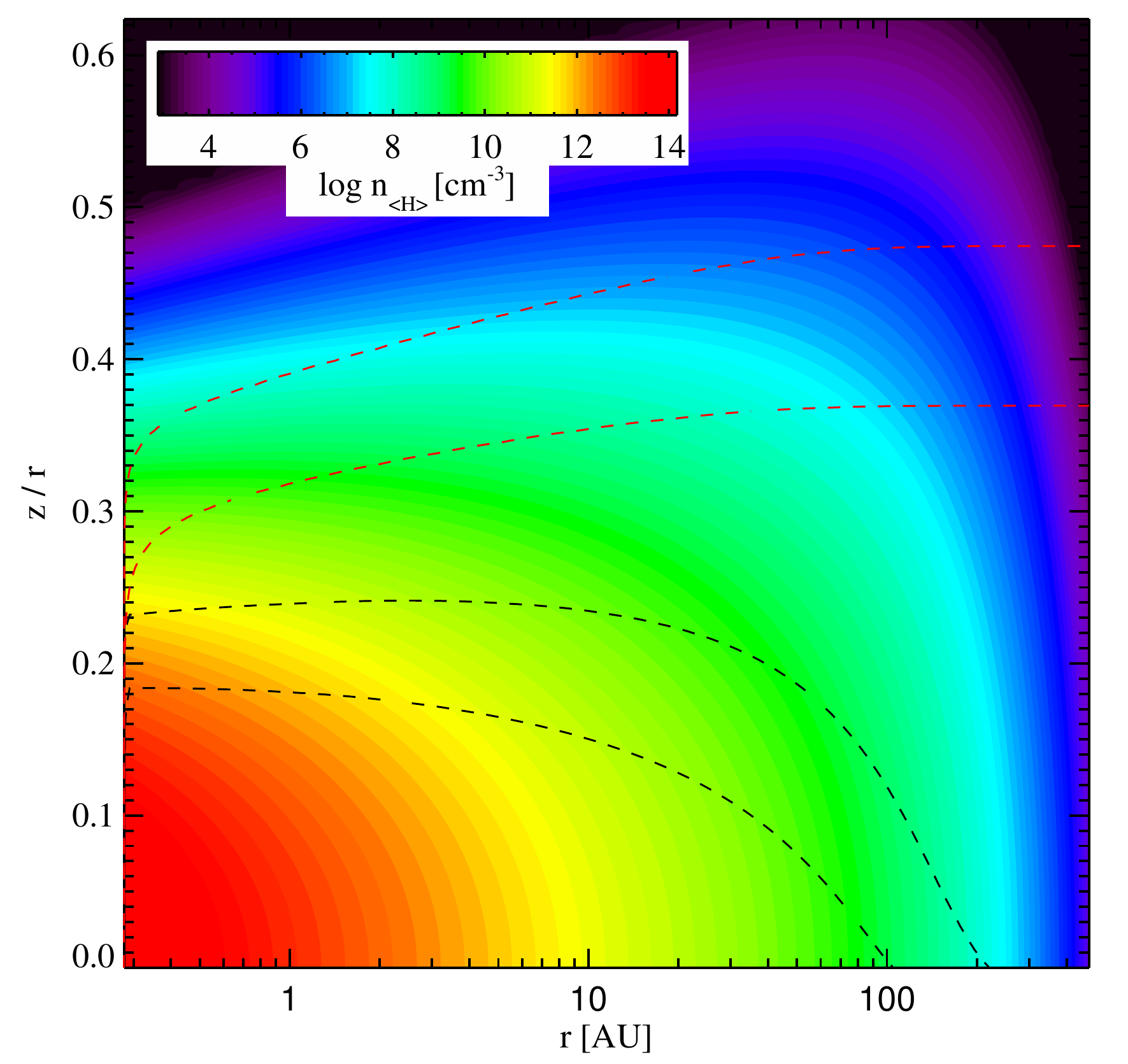}
\includegraphics[width=87mm]{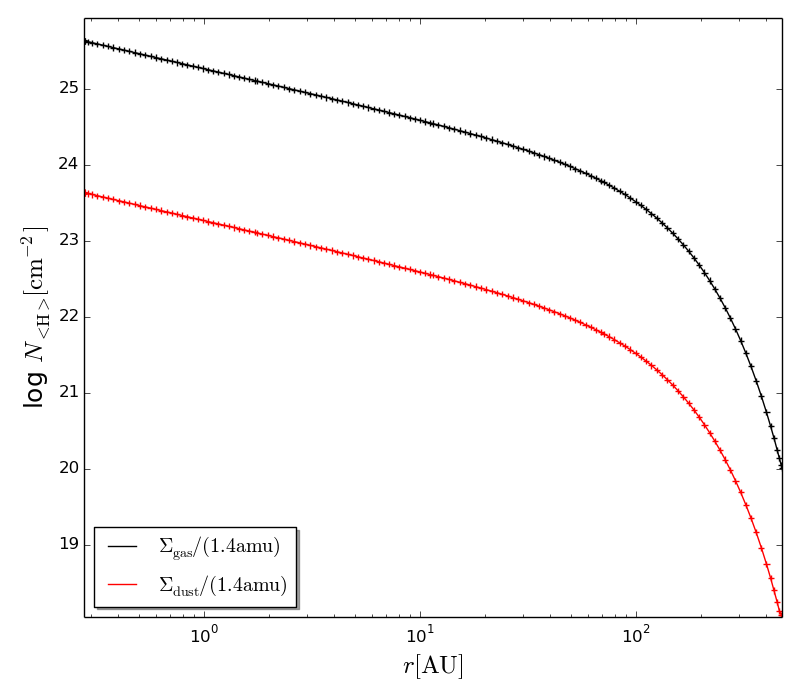}
\vspace*{-2mm}
\caption{MWC\,480 density and surface density plots. For details see Fig.\,\ref{fig:ABAur-1}.}
\label{fig:MWC480-1}
\end{figure}

\begin{figure}
\centering
\vspace*{-7mm}
\includegraphics[width=82mm]{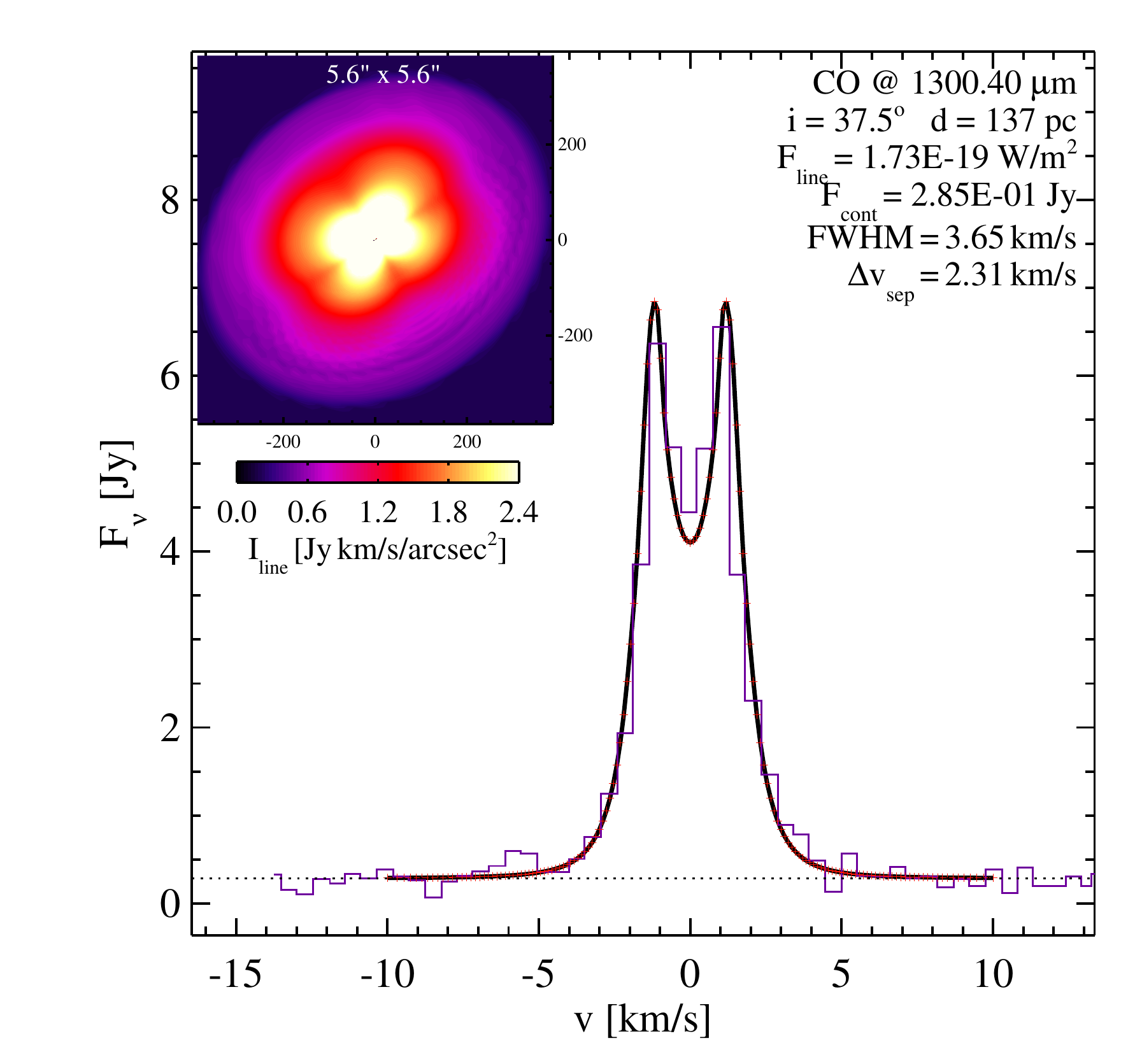} 
\includegraphics[width=82mm]{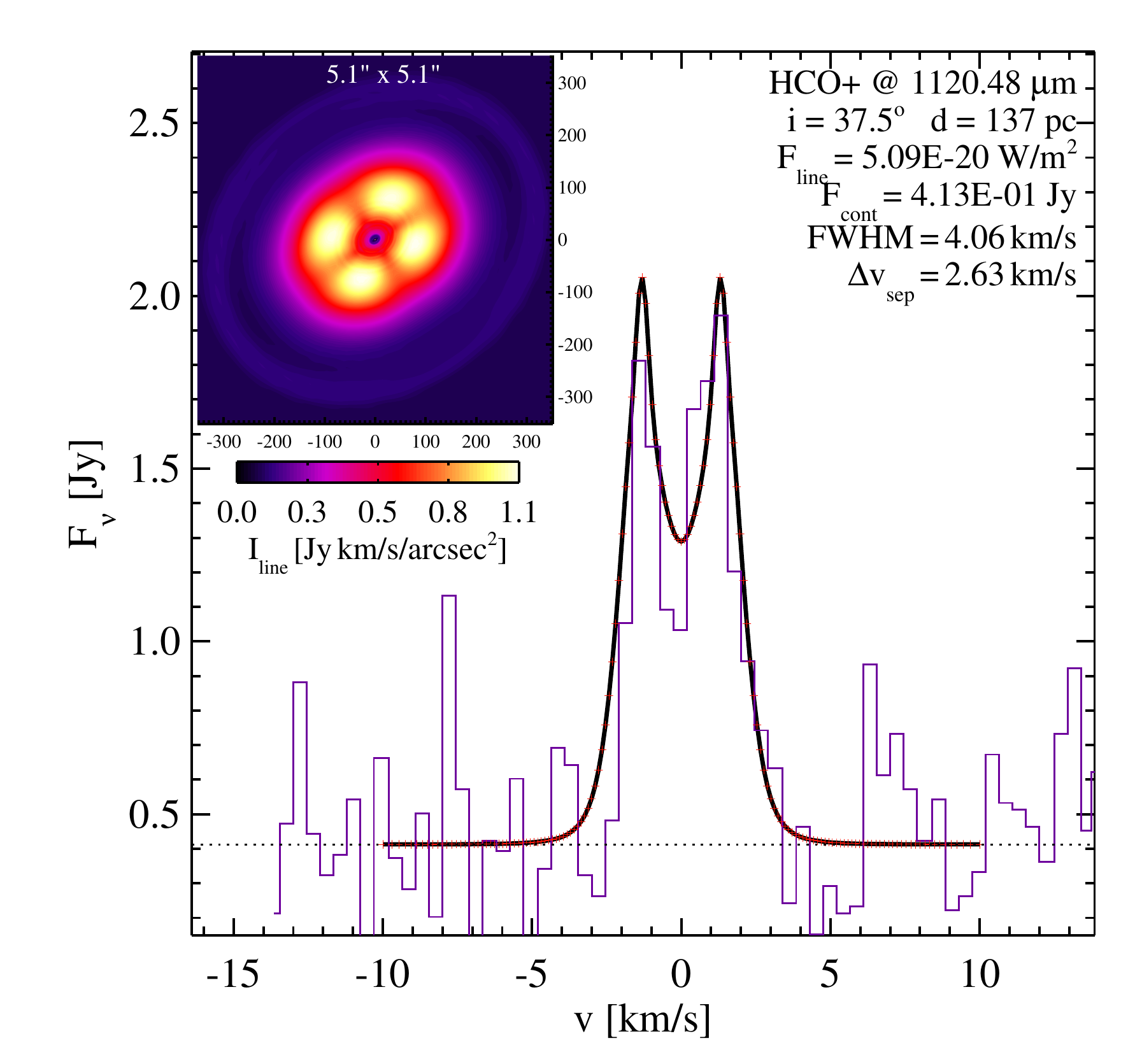} 
\vspace*{-3mm}
\caption{CO $J\!=$2-1 and HCO$^+$ $J\!=$3-2 lines in comparison to observations for MWC\,480.}
\label{fig:MWC480-2}
\end{figure}

\begin{table}
\caption{Model properties and comparison of computed spectral line
  properties with observations from {\sf MWC480.properties}. }
\footnotesize
\centering\begin{tabular}{rr}
\hline
DIANA standard fit model properties &  \\
\hline
minimum dust temperature [K] & 11 \\
maximum dust temperature [K] & 1599 \\
mass-mean dust temperature [K] & 24 \\
minimum  gas temperature [K] & 11 \\
maximum  gas temperature [K] & 23865 \\
mass-mean  gas temperature [K] & 36 \\
mm-opacity-slope (0.85-1.3)mm & 1.0 \\
cm-opacity-slope (5-10)mm & 1.3 \\
mm-SED-slope (0.85-1.3)mm & 2.4 \\
cm-SED-slope (5-10)mm & n.a. \\
10$\mathrm{\mu m}$ silicate emission amplitude & 1.6 \\
naked star luminosity [$L_\odot$] & 13.3 \\
bolometric luminosity [$L_\odot$] & 20.3 \\
near IR excess ($\lambda=$2.05-6.82$\mathrm{\mu m}$)
 [$L_\odot$] & 2.88 \\
mid IR excess ($\lambda=$6.82-29.1$\mathrm{\mu m}$)
 [$L_\odot$] & 1.68 \\
far IR excess ($\lambda=$29.1-981.$\mathrm{\mu m}$)
 [$L_\odot$] & 0.58 \\
\hline
\end{tabular}

\vspace{0.5cm}\centering\begin{tabular}{lrccccr}
\hline & & \multicolumn{2}{c}{line flux $[\mathrm{W\,m^{-2}}$]} & \multicolumn{2}{c}{FWHM $[\mathrm{km\,s^{-1}}]$} &  \\
species & $\lambda\,$[$\mathrm{\mu m}]$ & observed & model & observed & model & size[AU] \\
\hline
$\mathrm{OI}$ & 63.183 & 9.5$\pm$0.3(-17) & 1.1(-16) &  & 8.3 & 151.6 \\
$\mathrm{OI}$ & 145.525 & $<$7.8(-18) & 7.9(-18) &  & 7.5 & 117.4 \\
$\mathrm{CII}$ & 157.740 & $<$9.6(-18) & 6.8(-18) &  & 3.2 & 410.4 \\
$\mathrm{CO}$ & 4.633 & 3.1$\pm$0.6(-17) & 4.0(-17) & 47.5$\pm$10.0 & 94.0 & 6.1 \\
$\mathrm{CO}$ & 4.641 & 3.2$\pm$0.6(-17) & 3.6(-17) & 53.7$\pm$10.0 & 93.7 & 8.2 \\
$\mathrm{CO}$ & 4.682 & 3.7$\pm$0.7(-17) & 2.9(-17) & 69.2$\pm$14.0 & 92.9 & 15.3 \\
$\mathrm{CO}$ & 4.726 & 3.6$\pm$0.7(-17) & 4.7(-17) & 48.1$\pm$10.0 & 93.2 & 7.8 \\
$\mathrm{CO}$ & 4.735 & 3.6$\pm$0.7(-17) & 4.8(-17) & 47.6$\pm$10.0 & 93.5 & 6.2 \\
$\mathrm{CO}$ & 4.920 & 3.4$\pm$0.7(-17) & 4.3(-17) & 69.9$\pm$13.4 & 95.5 & 0.3 \\
$\mathrm{CO}$ & 4.966 & 2.7$\pm$0.5(-17) & 4.1(-17) & 71.0$\pm$14.2 & 95.5 & 0.3 \\
$\mathrm{CO}$ & 4.990 & 3.2$\pm$0.6(-17) & 4.0(-17) & 80.0$\pm$16.0 & 95.5 & 0.3 \\
$\mathrm{CO}$ & 72.842 & $<$9.6(-18) & 2.7(-20) &  & 37.4 & 7.6 \\
$\mathrm{CO}$ & 79.359 & $<$1.6(-17) & 4.4(-20) &  & 33.9 & 9.0 \\
$\mathrm{CO}$ & 90.162 & $<$9.6(-18) & 1.4(-19) &  & 26.8 & 12.1 \\
$\mathrm{CO}$ & 144.784 & 6.9$\pm$3.2(-18) & 3.2(-18) &  & 10.8 & 50.3 \\
$\mathrm{CO}$ & 866.963 & 6.1$\pm$0.1(-19) & 5.6(-19) &  & 3.6 & 350.6 \\
$\mathrm{^{13}CO}$ & 906.846 & 1.1$\pm$0.2(-19) & 2.5(-19) &  & 4.1 & 275.9 \\
$\mathrm{CO}$ & 1300.404 & 1.7$\pm$0.1(-19) & 1.7(-19) &  & 3.6 & 346.5 \\
$\mathrm{o\!-\!H_2O}$ & 63.323 & $<$6.6(-18) & 9.1(-19) &  & 34.1 & 7.1 \\
$\mathrm{o\!-\!H_2O}$ & 71.946 & $<$1.2(-18) & 9.1(-19) &  & 30.4 & 9.3 \\
$\mathrm{o\!-\!H_2O}$ & 78.742 & $<$1.8(-17) & 1.7(-18) &  & 17.0 & 35.5 \\
$\mathrm{OH}$ & 79.115 & $<$9.9(-18) & 2.0(-18) &  & 8.5 & 106.6 \\
$\mathrm{OH}$ & 79.179 & $<$1.0(-17) & 2.1(-18) &  & 8.3 & 110.2 \\
$\mathrm{CN}$ & 881.097 & 6.7$\pm$0.7(-20) & 1.6(-20) &  & 3.1 & 394.9 \\
$\mathrm{CN}$ & 1321.390 & 2.5$\pm$0.1(-20) & 7.4(-21) &  & 3.0 & 388.9 \\
$\mathrm{HCN}$ & 845.663 & 3.6$\pm$0.4(-20) & 9.1(-20) &  & 4.0 & 294.8 \\
$\mathrm{HCN}$ & 1127.520 & 2.1$\pm$0.2(-20) & 4.5(-20) &  & 3.8 & 325.4 \\
$\mathrm{HCO^+}$ & 840.380 & 7.5$\pm$1.5(-20) & 1.0(-19) &  & 4.3 & 312.9 \\
$\mathrm{HCO^+}$ & 1120.478 & 4.2$\pm$0.3(-20) & 5.1(-20) &  & 4.1 & 334.2 \\
$\mathrm{N_2H^+}$ & 804.439 & $<$6.7(-21) & 2.9(-21) &  & 3.7 & 221.4 \\
$\mathrm{N_2H^+}$ & 1072.557 & $<$6.9(-21) & 1.8(-21) &  & 3.7 & 221.9 \\
$\mathrm{CH^+}$ & 72.137 & $<$2.2(-17) & 1.1(-20) &  & 21.0 & 18.5 \\
\hline
\end{tabular}

\label{tab:MWC480}
\end{table}

\noindent{\sffamily\bfseries\underline{4.\ \ MWC\,480}}\ \ is one of
only two DIANA sources where the final fit uses just a single-zone
disk model (Fig.~\ref{fig:MWC480-1}). The SED in Fig.\,\ref{fig:SEDs}
can be conveniently fitted with a mildly flared, strongly settled
disk, where the near-IR excess of about $3\,L_\mathrm{\odot}$ is a
natural by-product. The data collection of MWC 480 includes two
continuum images (at $850\,\mathrm{\mu m}$ and a NICMOS scattered
light image at $1.6\,\mathrm{\mu m}$), 32 line observations with three
velocity-profiles and two line intensity maps. MWC~480 is particularly
well-observed in (sub-)mm lines including CO, $^{13}$CO, HCO$^+$, CN
and HCN, and the model manages to reproduce all these observations,
though less convincing for CN and HCN (see
Table~\ref{tab:MWC480}). Figure~\ref{fig:MWC480-2} shows the excellent
line flux and profile agreement for the $^{12}$CO and HCO$^+$ sub-mm
lines. The CO ro-vibrational line fluxes also fit astonishingly well
-- for a single-zone model -- but the lines are too broad. The high-J
CO lines seem a bit too weak, indicating that the disk is not warm
enough in the inner regions of the model. Similar conclusions can be
drawn from some of the mm-line intensity profiles, where the line
signals from the inner disk regions are somewhat too weak. A
vertically more extended, less dense inner disk (as for HD 163296, HD
142666, CY Tau) might also improve the fit of some high energy
emission lines in the case of MWC 480.

The gas/dust mass ratio was not varied during fitting this model, and
the total disk gas mass of $0.022\,M_\mathrm{\odot}$ is in accordance
with the SED-fit (Table~\ref{tab:dustmass}). The derived mass value
agrees within a factor of three with the values reported in
\citet{Mannings1997b} and \citet{Meeus2012} but are not consistent
with the much higher values
($M_\mathrm{disk}\gtrsim0.2\,M_\mathrm{\odot}$) derived from pure dust
modeling by \citet{Guilloteau2011} or \citet{Sitko2008a}.

A remarkable feature of the MWC\,480 disk is its observed variability
in the infrared, including the silicate feature. We only focused on
one epoch of observational data but we did run several models where we
changed the scale height of the disk (a possible origin of the
variability, \citealt{Sitko2008a,Grady2010b}) and found that such
changes do not have a significant impact on the spectral line
emission. Recently \citet{Fernandes2018} proposed dusty outflows/winds
as an origin of the infrared variability but also azimuthally
asymmetric features in the inner disk (clumps) might play a role
\citep{Jamialahmadi2018}. A more detailed study of CO ro-vibrational
lines and possible future observation with CRIRES+ on the VLT would
provide important constraints for the proposed variability
scenarios. Our consistent dust and gas model is an excellent starting
point for such investigations.

MWC~480 is also an excellent topic to study disk chemistry
\citep[e.g.][]{Pietu2007,Henning2010n,Chapillon2012}. For example the
detection of CH$_3$CN and HC$_3$N
\citep{Chapillon2012a,Oeberg2015,Huang2017,Bergner2018} together with
the spatially resolved observations of CN and HCN and its
isotopologues \citep{Guzman2015} make it a perfect target to study
cyanide chemistry. Our model  provides a detailed physical structure
for detailed chemical studies and to test, for example, the importance
of excited molecular hydrogen for CN/HCN chemistry as recently
proposed by \citet{Cazzoletti2018} which might improve our fit for the
CN and HCN lines.

\clearpage
\newpage

\begin{figure}
\centering
{\large\sffamily\bfseries\underline{HD 169142}}\\[2mm]
$M_{\rm gas} = 5.8\times 10^{-3}\,M_\odot$, 
$M_{\rm dust} = 5.8\times 10^{-5}\,M_\odot$, 
dust settling $\alpha_{\rm settle} = 3.7\times 10^{-3}$\\
\includegraphics[width=80mm]{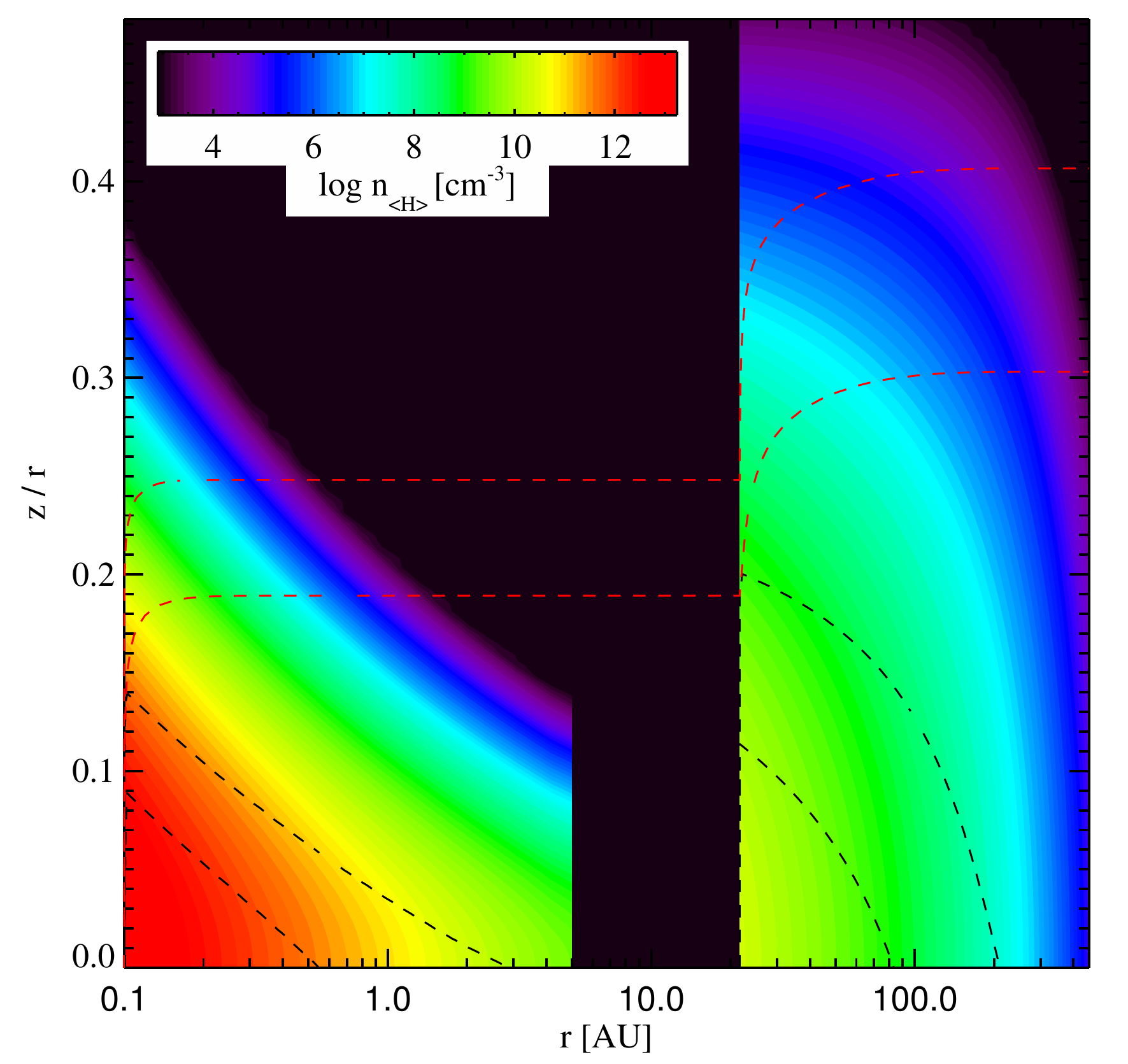}
\includegraphics[width=87mm]{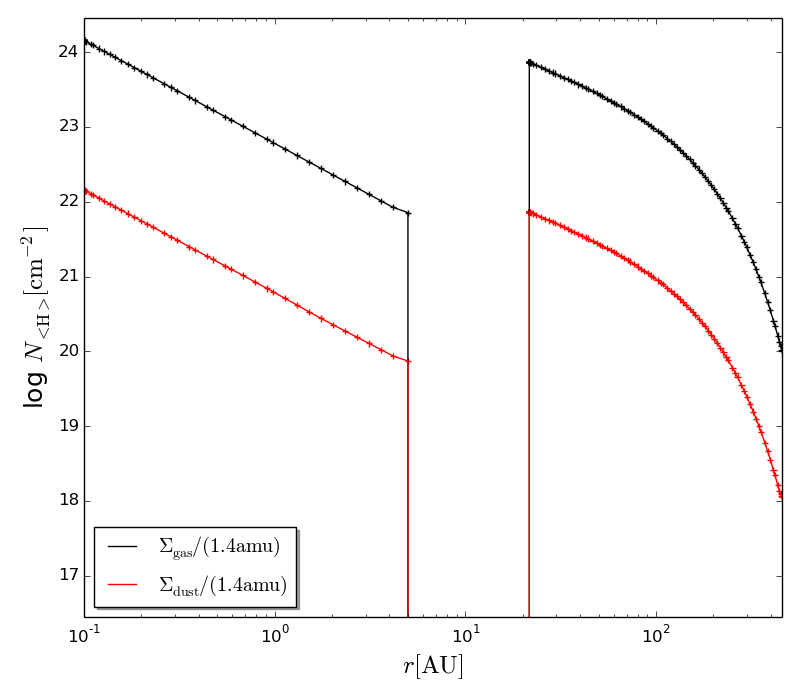}
 \vspace*{-2mm}
\caption{HD\,169142 density and surface density plots. For details see
  Fig.\,\ref{fig:ABAur-1}.}
\label{fig:HD169142}
\end{figure}

\noindent{\sffamily\bfseries\underline{5.\ \ HD\,169142}}\ \ This is a
simple model on top of the SED-fit with slightly modified disk
extension (Fig.~\ref{fig:HD169142}).  The model has the disk dust
cavity at 20\,AU as seen in near-IR \citep{Quanz2013} and sub-mm
continuum observations \citep[e.g.][]{Fedele2017}, an inner disk
extending from 0.1 to 5\,AU, thus a disk gap between 5 and 20\,AU.
The disk masses and gas/dust ratio are not altered with respect to the
SED-fitting results.  The model fits the [O\,{\sc i}]\,63\,$\mu$m line
and CO 2-1 isotopologue lines in the (sub-)mm reasonably well
(Table~\ref{tab:HD169142}). However, the CO fundamental ro-vibrational
lines are too strong and too narrow, leading to similar issues as for
GM Aur, see page~\pageref{page:GMAur}. In our model the $^{12}$CO 4.7
$\mu$m emission is dominated by the inner wall of the outer disk. To
improve the fit of these lines, one could consider a vertically
extended but transparent inner disk, which would shield the inner wall
of the outer disk from the stellar UV field.  CRIRES observations
(Carmona in prep.) show, however, that the $^{12}$CO ro-vibrational
emission region extends inside to at least 1\,AU, i.e.\ well through
the cavity and into the inner disk. In contrast, the observations show
that the narrower $^{13}$CO and C$^{18}$O ro-vibrational lines (not
shown in Table~\ref{tab:HD169142}) are emitted from the outer disk
$>\!20\,$AU.  Since our model has no gas between 5\,AU and 20\,AU, it
fails to predict these properties of the CO isotopologue
ro-vibrational lines.

\begin{table}
 \vspace*{-5mm}
\caption{Model properties and comparison of computed spectral line
  properties with observations from {\sf HD169142.properties}. }
\noindent\resizebox{68mm}{!}{
\hspace*{-7mm}\begin{tabular}{rr}
\hline
DIANA standard fit model properties &  \\
\hline
minimum dust temperature [K] & 12 \\
maximum dust temperature [K] & 2325 \\
mass-mean dust temperature [K] & 28 \\
minimum  gas temperature [K] & 14 \\
maximum  gas temperature [K] & 40000 \\
mass-mean  gas temperature [K] & 36 \\
mm-opacity-slope (0.85-1.3)mm & 1.0 \\
cm-opacity-slope (5-10)mm & 1.2 \\
mm-SED-slope (0.85-1.3)mm & 2.8 \\
cm-SED-slope (5-10)mm & 3.2 \\
10$\mathrm{\mu m}$ silicate emission amplitude & 1.6 \\
naked star luminosity [$L_\odot$] & 10.0 \\
bolometric luminosity [$L_\odot$] & 13.7 \\
near IR excess ($\lambda=$2.02-6.72$\mathrm{\mu m}$)
 [$L_\odot$] & 0.88 \\
mid IR excess ($\lambda=$6.72-29.5$\mathrm{\mu m}$)
 [$L_\odot$] & 0.81 \\
far IR excess ($\lambda=$29.5-991.$\mathrm{\mu m}$)
 [$L_\odot$] & 0.99 \\
\hline
\end{tabular}
}
\hspace*{-10mm}\resizebox{113mm}{!}{
\begin{tabular}{lrccccr}
\hline & & \multicolumn{2}{c}{line flux $[\mathrm{W\,m^{-2}}$]} & \multicolumn{2}{c}{FWHM $[\mathrm{km\,s^{-1}}]$} &  \\
species & $\lambda\,$[$\mathrm{\mu m}]$ & observed & model & observed & model & size[AU] \\
\hline
$\mathrm{OI}$ & 63.183 & 7.2$\pm$0.4(-17) & 4.5(-17) &  & 3.2 & 121.8 \\
$\mathrm{OI}$ & 145.525 & $<$1.1(-17) & 1.6(-18) &  & 3.7 & 72.8 \\
$\mathrm{CII}$ & 157.740 & $<$6.6(-18) & 2.8(-18) &  & 1.2 & 436.4 \\
$\mathrm{CO}$ & 1300.404 & 9.3$\pm$0.4(-20) & 1.7(-19) & 2.1$\pm$0.1 & 1.4 & 415.8 \\
$\mathrm{^{13}CO}$ & 1360.227 & 4.8$\pm$0.4(-20) & 4.2(-20) & 2.1$\pm$0.1 & 1.6 & 323.6 \\
$\mathrm{C^{18}O}$ & 1365.421 & 2.0$\pm$0.4(-20) & 1.0(-20) &  & 1.7 & 299.1 \\
$\mathrm{o\!-\!H_2O}$ & 179.526 & $<$8.7(-18) & 1.5(-17) &  & 2.3 & 338.7 \\
$\mathrm{o\!-\!H_2O}$ & 78.742 & $<$1.1(-17) & 1.9(-17) &  & 3.2 & 125.1 \\
$\mathrm{CO}$ & 72.842 & $<$1.6(-17) & 2.6(-20) &  & 4.1 & 23.6 \\
$\mathrm{CO}$ & 90.162 & $<$1.1(-17) & 5.4(-19) &  & 4.1 & 24.0 \\
$\mathrm{CO}$ & 4.754 & 2.9$\pm$0.2(-18) & 2.0(-17) & 7.3$\pm$0.5 & 3.8 & 22.1 \\
$\mathrm{CO}$ & 4.773 & 2.6$\pm$0.3(-18) & 1.4(-17) & 7.0$\pm$0.6 & 3.8 & 22.1 \\
\hline
\end{tabular}
}

\label{tab:HD169142}
\end{table}
  
\clearpage
\newpage

\begin{figure}[!t]
\centering
\vspace*{2mm}
{\large\sffamily\bfseries\underline{HD 142666}}\\[2mm]
$M_{\rm gas} = 3.3\times 10^{-3}\,M_\odot$, 
$M_{\rm dust} = 3.3\times 10^{-5}\,M_\odot$, 
dust settling $\alpha_{\rm settle} = 0.1$\\
\hspace*{-3mm}
\begin{tabular}{cc}
\begin{minipage}{75mm}
\includegraphics[width=79mm]{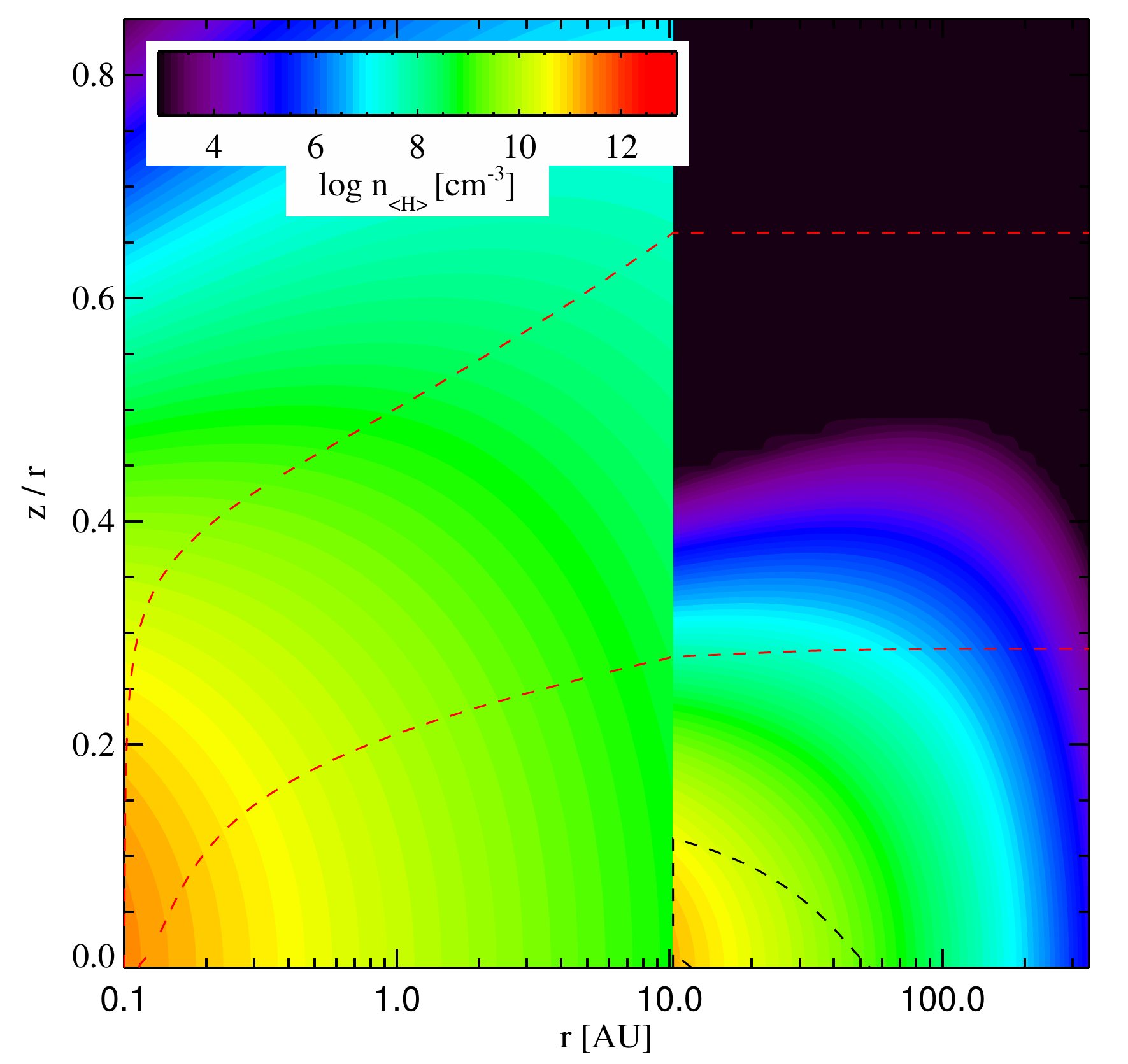}
\end{minipage}
&
\begin{minipage}{87mm}
\vspace*{1mm}
\includegraphics[width=87mm]{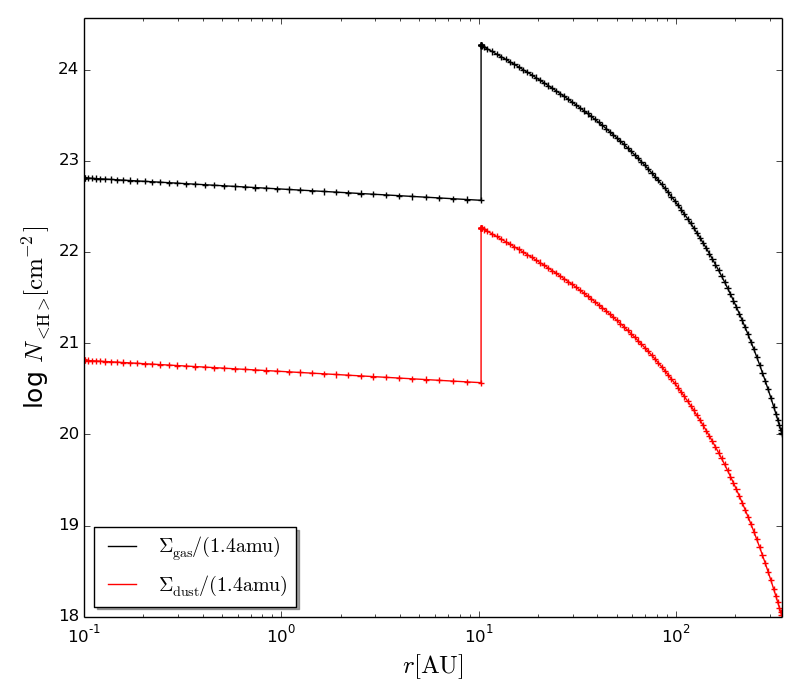}
\end{minipage}
\end{tabular}\\[-2mm]
\caption{HD\,142666 density and surface density plots. For details see Fig.\,\ref{fig:ABAur-1}.}
\label{fig:HD142666-1}
\vspace*{-2mm}
\end{figure}

\begin{figure}[!h]
\vspace*{-2mm}\hspace*{-8mm}
\begin{tabular}{cc}
\begin{minipage}{84mm}
\vspace*{2.5mm}
\includegraphics[width=84mm]{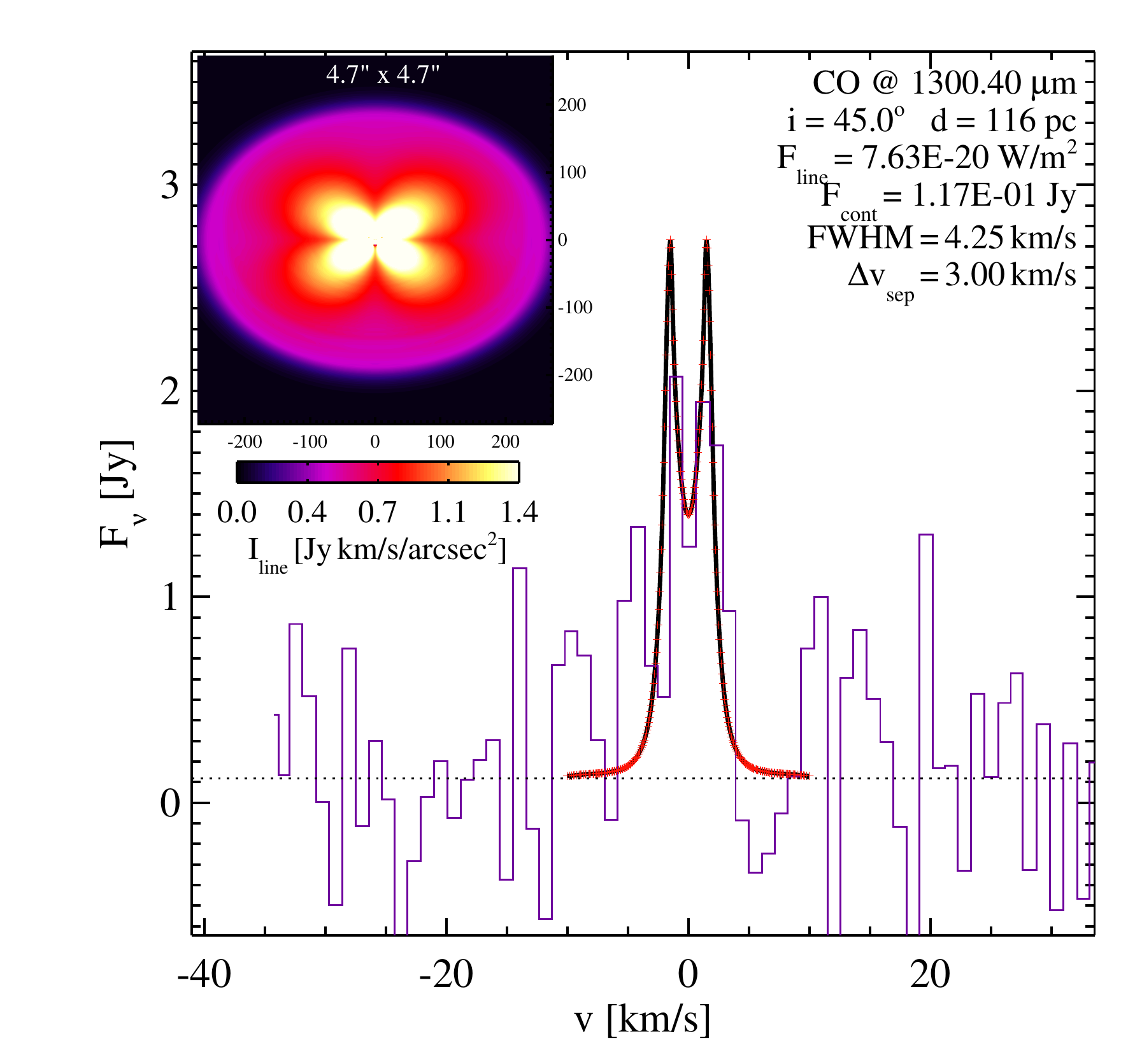}
\end{minipage} &
\hspace*{-6mm}
\begin{minipage}{87mm}
\includegraphics[width=86mm]{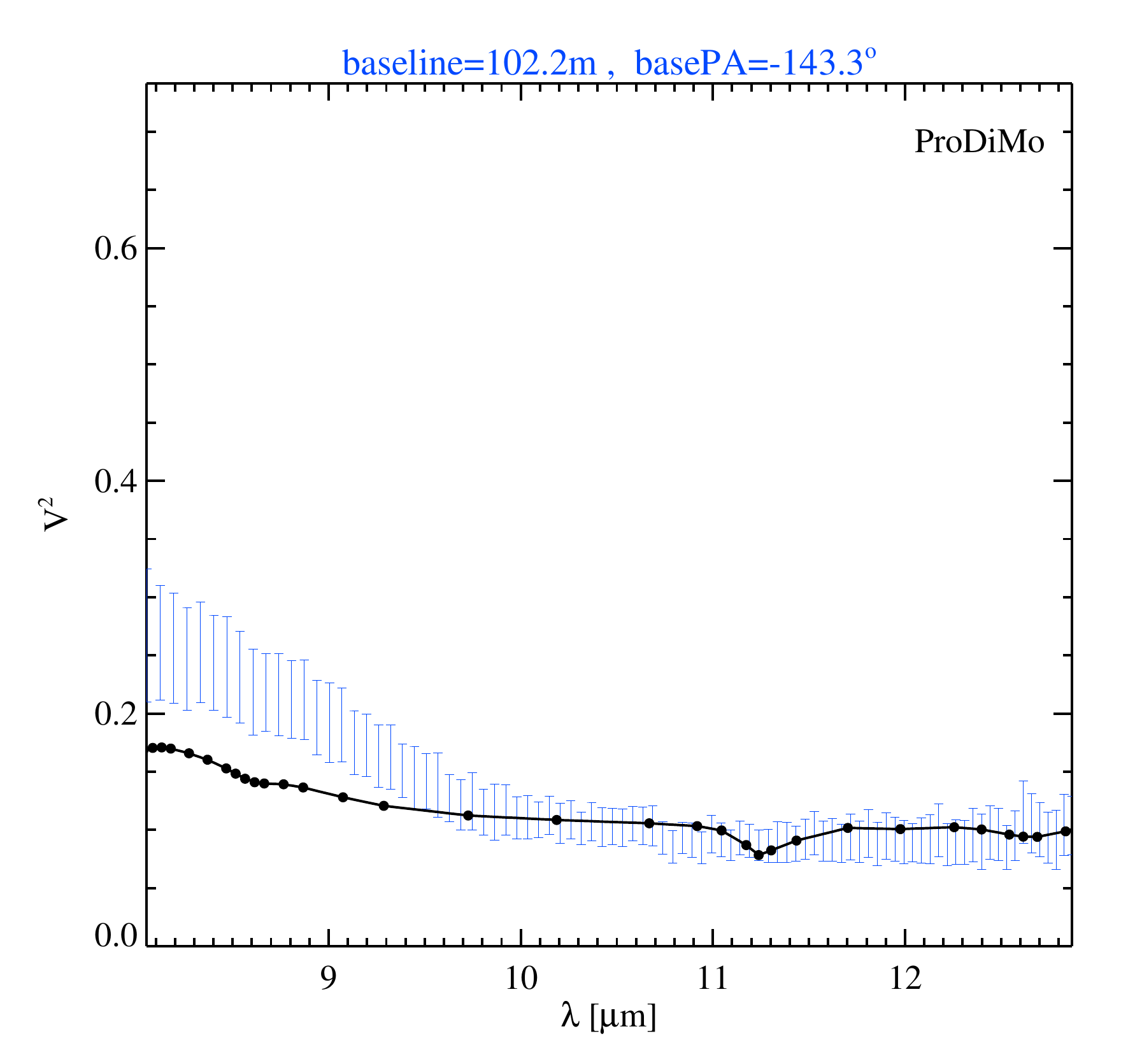}
\end{minipage}
\end{tabular}\\*[-3mm]
\caption{{\bf Left:} Fit of noisy CO 2-1 archival SMA data of
  HD\,142666. {\bf Right:} Mid-IR visibilities computed by ProDiMo
  (black) in comparison to archival MIDI data (J.~Menu, priv.comm.,
  blue) for HD\,142666.}
\label{fig:HD142666-2}
\end{figure}

\begin{table}
\caption{Model properties and comparison of computed spectral line
  properties with observations from {\sf HD142666.properties}. }
\centering\begin{tabular}{rr}
\hline
DIANA standard fit model properties &  \\
\hline
minimum dust temperature [K] & 7 \\
maximum dust temperature [K] & 1890 \\
mass-mean dust temperature [K] & 25 \\
minimum  gas temperature [K] & 10 \\
maximum  gas temperature [K] & 24910 \\
mass-mean  gas temperature [K] & 34 \\
mm-opacity-slope (0.85-1.3)mm & 0.6 \\
cm-opacity-slope (5-10)mm & 0.9 \\
mm-SED-slope (0.85-1.3)mm & 2.2 \\
cm-SED-slope (5-10)mm & 2.8 \\
10$\mathrm{\mu m}$ silicate emission amplitude & 1.8 \\
naked star luminosity [$L_\odot$] & 6.3 \\
bolometric luminosity [$L_\odot$] & 9.1 \\
near IR excess ($\lambda=$2.02-6.72$\mathrm{\mu m}$)
 [$L_\odot$] & 0.81 \\
mid IR excess ($\lambda=$6.72-29.5$\mathrm{\mu m}$)
 [$L_\odot$] & 0.60 \\
far IR excess ($\lambda=$29.5-991.$\mathrm{\mu m}$)
 [$L_\odot$] & 0.22 \\
\hline
\end{tabular}

\vspace{0.5cm}\centering\begin{tabular}{lrccccr}
\hline & & \multicolumn{2}{c}{line flux $[\mathrm{W\,m^{-2}}$]} & \multicolumn{2}{c}{FWHM $[\mathrm{km\,s^{-1}}]$} &  \\
species & $\lambda\,$[$\mathrm{\mu m}]$ & observed & model & observed & model & size[AU] \\
\hline
$\mathrm{CO}$ & 1300.404 & 1.1$\pm$0.1(-19) & 7.6(-20) & 6.4$\pm$1.3 & 4.2 & 245.2 \\
$\mathrm{CO}$ & 2600.758 & 4.4$\pm$0.4(-20) & 8.9(-21) & 10.6$\pm$1.3 & 4.4 & 231.6 \\
$\mathrm{CO}$ & 72.842 & $<$3.0(-17) & 5.0(-20) &  & 122.4 & 0.6 \\
$\mathrm{CO}$ & 130.368 & $<$1.2(-17) & 1.0(-18) &  & 27.7 & 9.3 \\
$\mathrm{OI}$ & 63.183 & 1.9$\pm$0.3(-17) & 1.7(-17) &  & 21.2 & 46.7 \\
$\mathrm{OI}$ & 145.525 & $<$4.7(-18) & 1.9(-18) &  & 15.4 & 39.9 \\
$\mathrm{CII}$ & 157.740 & $<$9.0(-18) & 1.2(-18) &  & 3.6 & 304.8 \\
$\mathrm{CO}$ & 72.842 & $<$1.6(-17) & 5.0(-20) &  & 122.4 & 0.6 \\
$\mathrm{CO}$ & 79.359 & $<$1.9(-17) & 8.5(-20) &  & 104.1 & 0.8 \\
$\mathrm{CO}$ & 90.162 & $<$1.3(-17) & 1.8(-19) &  & 79.5 & 1.8 \\
$\mathrm{CO}$ & 144.784 & $<$6.2(-18) & 1.2(-18) &  & 23.0 & 11.5 \\
$\mathrm{CO}$ & 4.652 & $<$5.6(-18) & 9.3(-18) &  & 146.1 & 1.8 \\
$\mathrm{CO}$ & 4.990 & $<$3.1(-18) & 4.4(-18) &  & 146.2 & 0.2 \\
\hline
\end{tabular}
\label{tab:HD142666}
\end{table}

\noindent{\sffamily\bfseries\underline{6.\ \ HD\,142666:}}\ \ was
classified as a group II disk by \citep{Meeus2001} based on
its SED (Fig.\,\ref{fig:SEDs}), which shows
a smooth curvature, starting out with a strong near-IR excess
($0.8\rm\,L_\odot$), low-amplitude 10\,$\mu$m and 20\,$\mu$m silicate
emission features and clearly detected PAH features
(Fig.\,\ref{fig:midIR}), followed by a smooth and steady decline into
the millimeter region.  The strength of the PAH bands \citep{Acke2010} as
well as the location in the N-band size-color diagram \citep{Menu2015}
suggest that the geometry of HD\,142666 may have some similarity
with the transitional, gaped group I sources. Indeed,
\citet{Rubinstein2018} find evidence from ALMA data for a large
cavity of mm-side grains in the disk of HD142666, and
\citet{Garufi2017} detect the disk in scattered light. This
illustrates that there is overlap in disk geometry between group I and
II sources. The inner disk structure is complex as evidenced by recent
near-IR interferometry \citep{Davies2018}. HD\,142666 has no
clear line detections other than [O\,{\sc i}]\,63\,$\mu$m. The CO 3-2
and 2-1 (sub-)mm lines are detected, but the line data is noisy on a
relatively bright continuum level, see Fig.~\ref{fig:HD142666-2}, and
the two measured FWHMs contradict each other (they should be very
similar if emitted from a disk). 

Our DIANA standard model fits all continuum observations by a
tall and marginally transparent inner disk zone casting a shadow on
the main outer disk (Fig.~\ref{fig:HD142666-1}). The fit to the MIDI
visibilities shows that the size of the
10\,$\mu$m-continuum emission region in the model is about correct
(Fig.~\ref{fig:HD142666-2}). The fit to the [O\,{\sc i}]\,63\,$\mu$m
data is good, but the CO fundamental ro-vibrational lines are somewhat
too strong and very broad (Table~\ref{tab:HD142666}).  The line fits
worsen considerably if the stellar UV-excess is taken into account,
which would lead to stronger heating and brighter emission
lines. HD\,142666 is known as a highly variable source
\citep[e.g.][]{Zwintz2009}.

\clearpage
\newpage

\begin{figure}
\centering
\vspace*{2mm}
{\large\sffamily\bfseries\underline{Lk\,Ca\,15}}\\[2mm]
$M_{\rm gas} = 2.2\times 10^{-2}\,M_\odot$,
$M_{\rm dust} = 2.8\times 10^{-4}\,M_\odot$, 
dust settling $\alpha_{\rm settle} = 2.2\times 10^{-4}$\\
\hspace*{-4mm}
\begin{tabular}{cc}
\begin{minipage}{75mm}
\includegraphics[width=78mm]{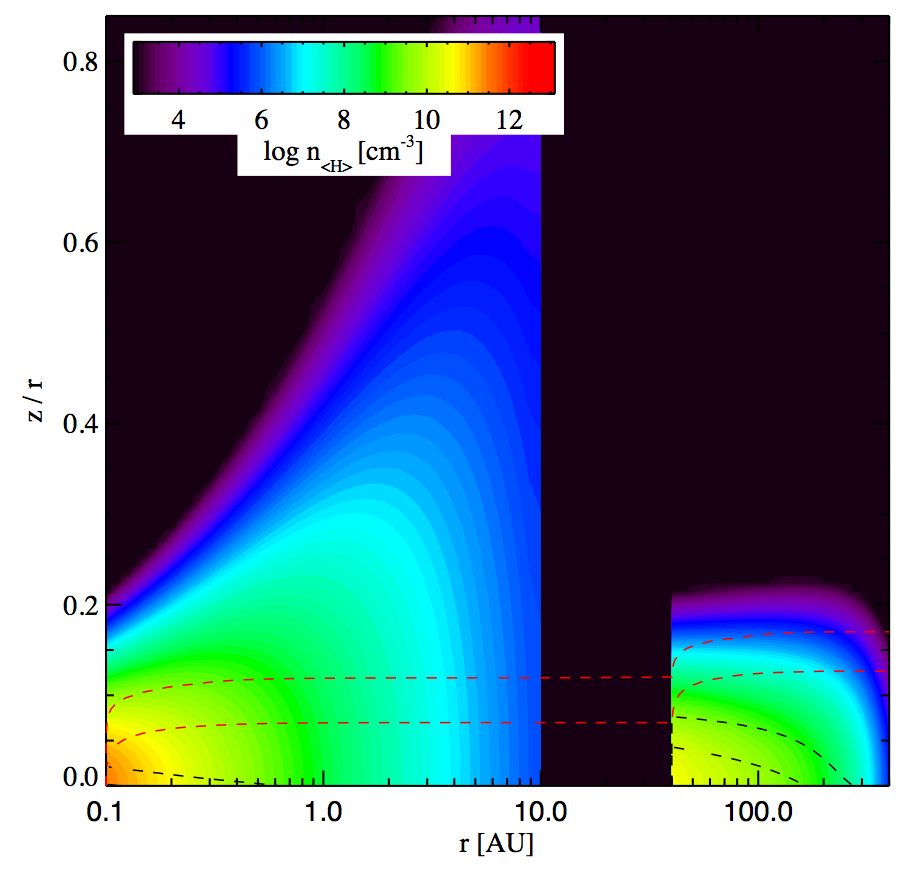}
\end{minipage}
&
\begin{minipage}{87mm}
\vspace*{1mm}
\includegraphics[width=87mm]{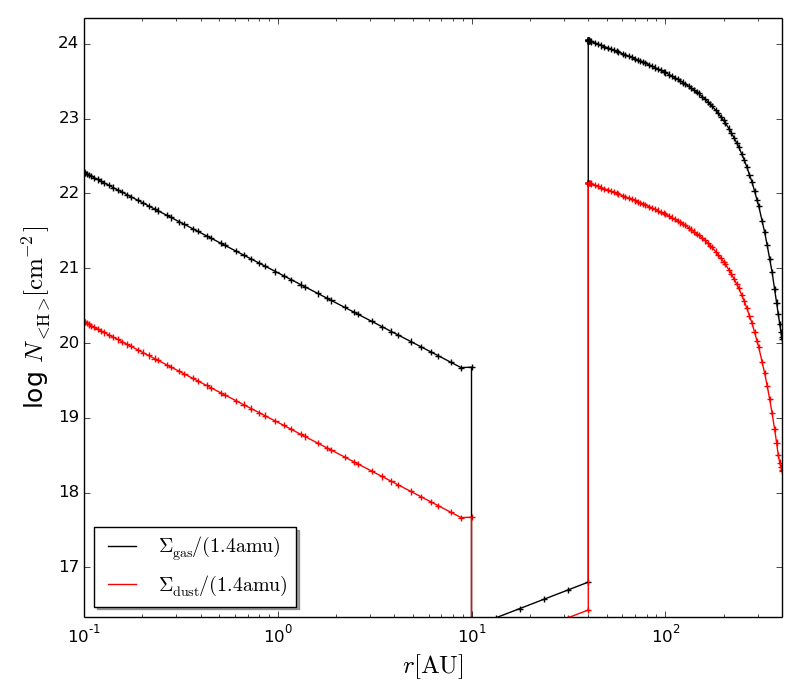}
\end{minipage}
\end{tabular}\\[-1mm]
\caption{Lk\,Ca\,15 density and surface density plots. For details see
  Fig.\,\ref{fig:ABAur-1}.}
\label{fig:LkCa15-1}
\end{figure}

\begin{figure}[!h]
\vspace*{0mm}\hspace*{-9mm}
\begin{tabular}{cc}
\begin{minipage}{86mm}
\includegraphics[width=86mm,trim=0 0 0 19, clip]{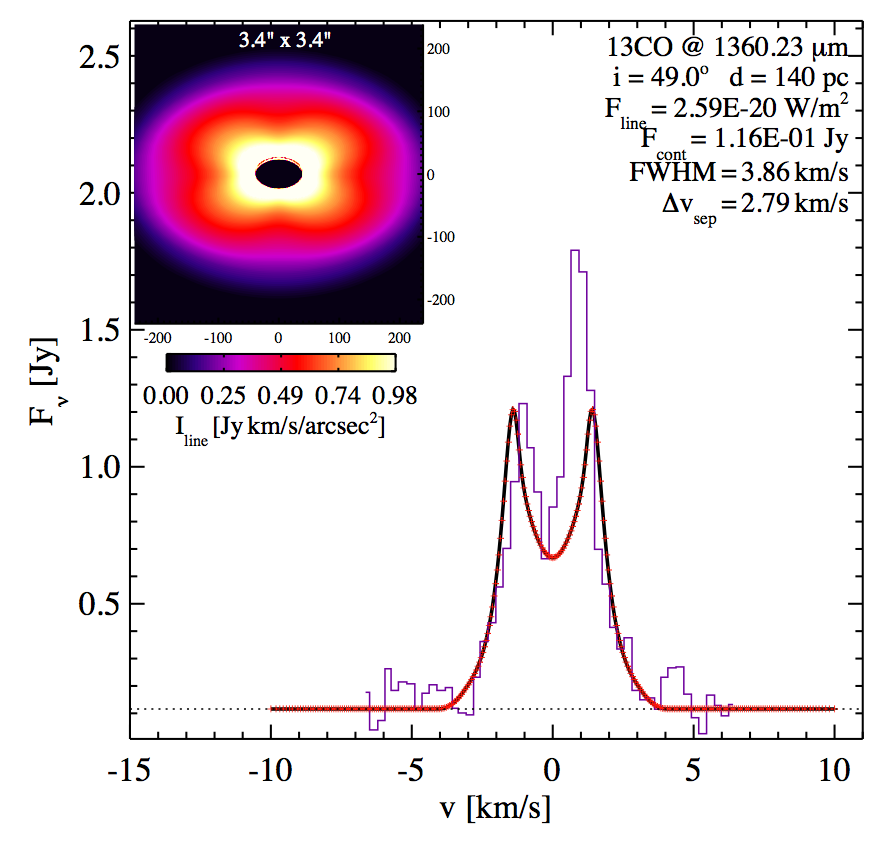}
\end{minipage} &
\hspace*{-5mm}
\begin{minipage}{85mm}
\includegraphics[width=85mm,trim=0 0 0 19, clip]{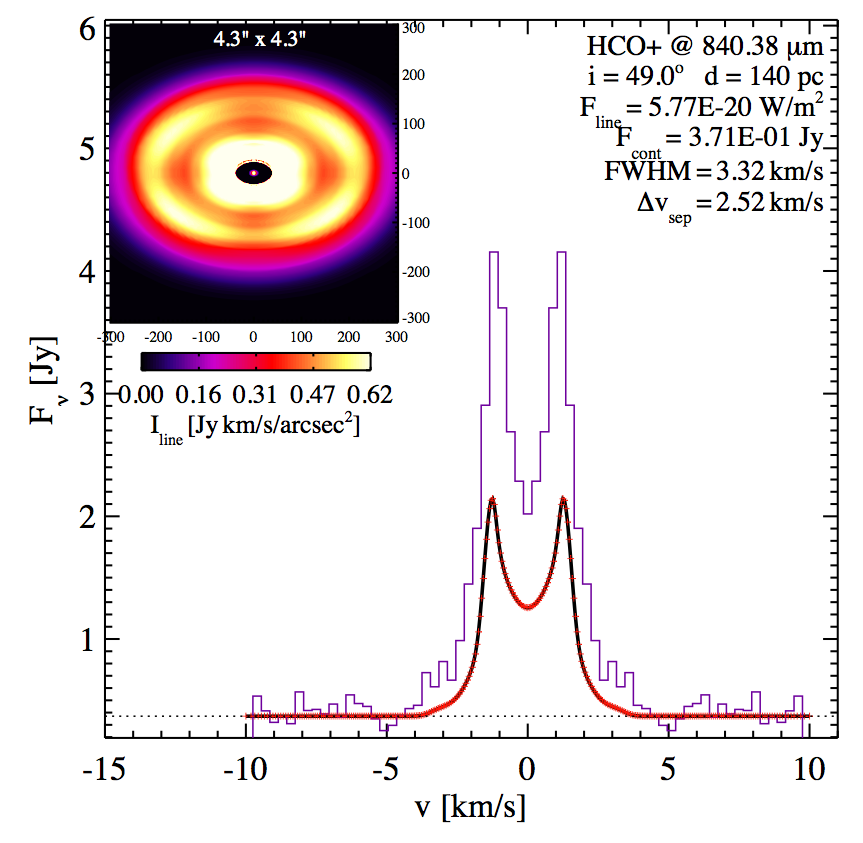}
\end{minipage}
\end{tabular}
\caption{$^{13}$CO $J\!=$2-1 and HCO$^+$ $J\!=$3-2 lines in comparison
  to observations for Lk\,Ca\,15.}
\label{fig:LkCa15-2}
\end{figure}

\begin{table}
\caption{Model properties and comparison of computed spectral line
  properties with observations from {\sf LkCa15.properties}. }
\centering\begin{tabular}{rr}
\hline
DIANA standard fit model properties &  \\
\hline
minimum dust temperature [K] & 6 \\
maximum dust temperature [K] & 1538 \\
mass-mean dust temperature [K] & 12 \\
minimum  gas temperature [K] & 6 \\
maximum  gas temperature [K] & 27008 \\
mass-mean  gas temperature [K] & 18 \\
mm-opacity-slope (0.85-1.3)mm & 1.2 \\
cm-opacity-slope (5-10)mm & 1.8 \\
mm-SED-slope (0.85-1.3)mm & 2.4 \\
cm-SED-slope (5-10)mm & 3.5 \\
10$\mathrm{\mu m}$ silicate emission amplitude & 2.2 \\
naked star luminosity [$L_\odot$] & 1.4 \\
bolometric luminosity [$L_\odot$] & 1.6 \\
near IR excess ($\lambda=$2.02-6.72$\mathrm{\mu m}$)
 [$L_\odot$] & 0.06 \\
mid IR excess ($\lambda=$6.72-29.5$\mathrm{\mu m}$)
 [$L_\odot$] & 0.05 \\
far IR excess ($\lambda=$29.5-991.$\mathrm{\mu m}$)
 [$L_\odot$] & 0.07 \\
\hline
\end{tabular}

\vspace{0.5cm}\centering\begin{tabular}{lrccccr}
\hline & & \multicolumn{2}{c}{line flux $[\mathrm{W\,m^{-2}}$]} & \multicolumn{2}{c}{FWHM $[\mathrm{km\,s^{-1}}]$} &  \\
species & $\lambda\,$[$\mathrm{\mu m}]$ & observed & model & observed & model & size[AU] \\
\hline
$\mathrm{OI}$ & 63.183 & 1.0$\pm$0.2(-17) & 8.4(-17) &  & 6.7 & 161.2 \\
$\mathrm{OI}$ & 145.525 & $<$9.0(-18) & 2.6(-18) &  & 7.0 & 102.7 \\
$\mathrm{CII}$ & 157.740 & $<$1.1(-17) & 3.8(-18) &  & 3.5 & 312.5 \\
$\mathrm{CO}$ & 1300.404 & 8.7$\pm$0.2(-20) & 7.6(-20) & 2.8$\pm$0.1 & 3.6 & 269.5 \\
$\mathrm{^{13}CO}$ & 1360.227 & 3.2$\pm$0.2(-20) & 2.6(-20) & 2.9$\pm$0.3 & 3.9 & 220.2 \\
$\mathrm{C^{18}O}$ & 1365.421 & 5.3$\pm$0.4(-21) & 9.3(-21) &  & 4.3 & 201.2 \\
$\mathrm{HCO^+}$ & 1120.478 & 4.7$\pm$0.2(-20) & 2.9(-20) & 3.3$\pm$0.2 & 3.2 & 289.4 \\
$\mathrm{HCN}$ & 1127.520 & 5.1$\pm$0.3(-20) & 4.1(-20) & 4.0$\pm$0.3 & 3.2 & 300.4 \\
$\mathrm{CS}$ & 1223.964 & 1.2$\pm$0.2(-20) & 6.1(-21) & 3.4$\pm$0.6 & 3.2 & 265.6 \\
$\mathrm{o\!-\!H_2CO}$ & 1419.394 & 9.4$\pm$0.5(-21) & 1.2(-20) &  & 2.9 & 337.3 \\
$\mathrm{p\!-\!H_2CO}$ & 1373.794 & 5.4$\pm$0.5(-21) & 1.4(-20) &  & 2.9 & 339.0 \\
$\mathrm{o\!-\!H_2CO}$ & 1328.291 & 8.0$\pm$0.4(-21) & 1.4(-20) & 2.7$\pm$0.4 & 2.9 & 335.1 \\
$\mathrm{CO}$ & 433.556 & 6.2$\pm$0.3(-19) & 1.3(-18) & 5.4$\pm$0.5 & 3.8 & 243.2 \\
$\mathrm{HCO^+}$ & 840.380 & 1.4$\pm$0.1(-19) & 5.8(-20) & 3.4$\pm$0.2 & 3.3 & 273.9 \\
\hline
\end{tabular}
\label{tab:LkCa15}
\end{table}

\noindent{\sffamily\bfseries\underline{7.\ Lk\,Ca\,15}}\ \ is an
important source known to have bright and rich mm-emission lines
including bio-molecules. The SED requires a very flat outer disk,
partly in the shadow of a high inner disk to reproduce the 10~$\mu$m
silicate feature (Fig.~\ref{fig:LkCa15-1}). The scattered light image
at 1~$\mu$m \citep{Thalmann2010} and the thermal emission image at
850~$\mu$m \citep{Andrews2011} are consistent with a gap of
$\sim\!35-50$\,AU, very much in line with our best fit model.

\citet{Drabek2016} show with thermo-chemical disk models that their
new HCO$^+$ data requires the presence of gas inside 50\,AU and a large
scale height for that inner gas disk. Based on their work, we refined
the model further within the DIANA modeling framework. We still
require an inner gas disk with a large scale height ($H\!=0\!.11$\,AU
at 1\,AU, Fig.~\ref{fig:LkCa15-1}). The slightly revised DIANA disk
model fits a number of (sub-)mm lines within a factor three including
$^{12}$CO, $^{13}$CO, HCO$^+$, HCN, CS, and H$_2$CO
(Table~\ref{tab:LkCa15}). Some remaining discrepancies are seen in the
wings of the HCO$^+$~$J\!=\!3-2$, $4-3$ (Fig.~\ref{fig:LkCa15-2},
right panel) and HCN~$J\!=\!3-2$ lines. Some profiles are slightly
asymmetric like $^{13}$CO~$J\!=\!2-1$ (Fig.~\ref{fig:LkCa15-2}, left
panel) and CS~$J\!=\!5-4$. This is a feature not captured in our disk
model. The H$_2$CO lines are overall a factor three too strong in our
model.

The CO~$J\!=\!6-5$ line and also the [O\,{\sc i}]~63\,$\mu$m line are
a factor two and eight too strong, respectively, in our disk model,
possibly suggesting that the outer disk is in fact colder than our
model shows. The inner disk, though vertically extended, does neither
completely shield the strong and hard X-rays from the central source,
nor the strong UV excess.

\clearpage
\newpage

\begin{figure}
\centering
\vspace*{2mm}
{\large\sffamily\bfseries\underline{USco\,J1604-2130}}\\[2mm]
$M_{\rm gas} = 2.1\times 10^{-3}\,M_\odot$,
$M_{\rm dust} = 4.0\times 10^{-5}\,M_\odot$, 
dust settling $\alpha_{\rm settle} = 2.3\times 10^{-3}$\\
\hspace*{-4mm}
\begin{tabular}{cc}
\begin{minipage}{75mm}
\includegraphics[width=78mm]{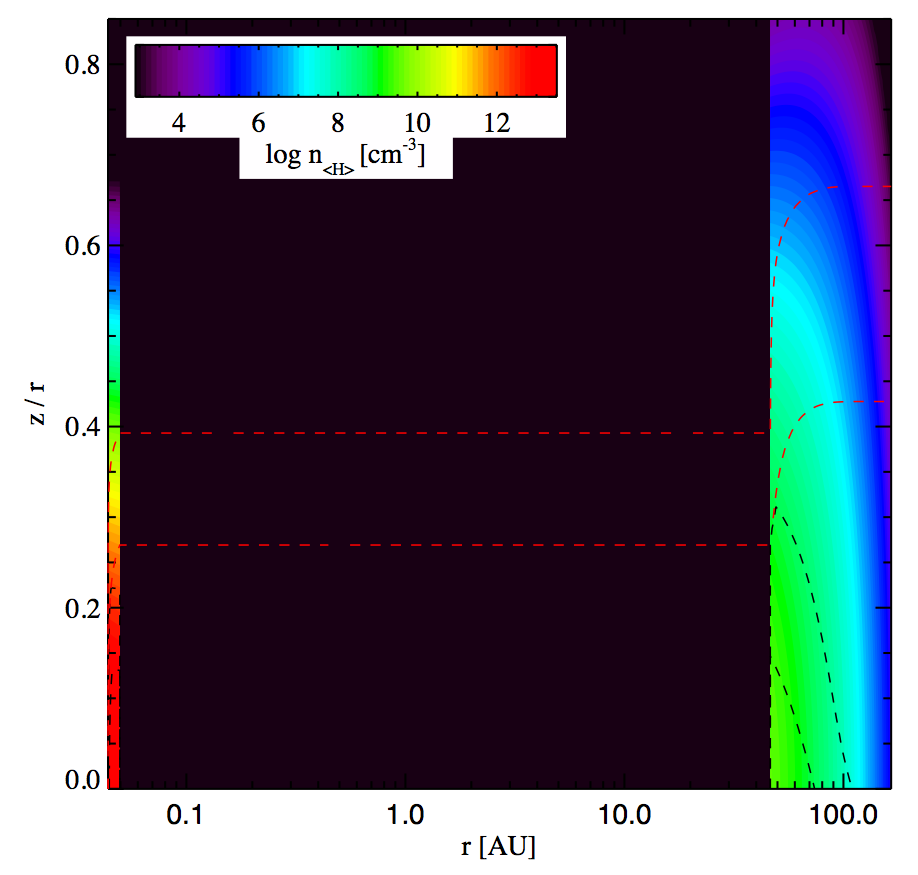}
\end{minipage}
&
\begin{minipage}{87mm}
\vspace*{1mm}
\includegraphics[width=87mm]{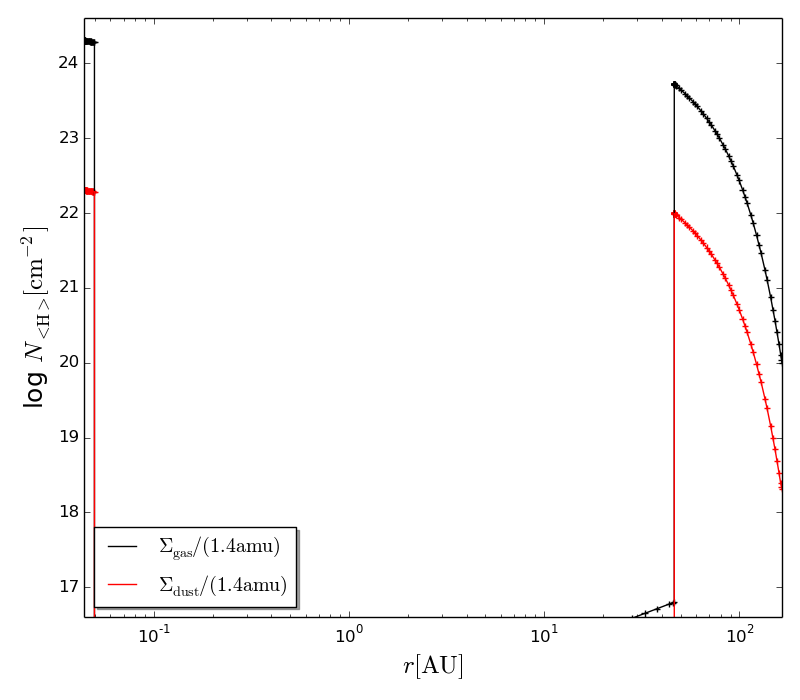}
\end{minipage}
\end{tabular}\\[-1mm]
\caption{USco\,J1604-2130 density and surface density plots. For
  details see Fig.\,\ref{fig:ABAur-1}.}
\label{fig:UScoJ1604-2130-1}
\end{figure}

\begin{table}
\caption{Model properties and comparison of computed spectral line
  properties with observations from {\sf UScoJ1604-2130.properties}. }
\noindent\resizebox{68mm}{!}{
\hspace*{-7mm}\begin{tabular}{rr}
\hline
DIANA standard fit model properties &  \\
\hline
minimum dust temperature [K] & 13 \\
maximum dust temperature [K] & 1869 \\
mass-mean dust temperature [K] & 24 \\
minimum  gas temperature [K] & 14 \\
maximum  gas temperature [K] & 27686 \\
mass-mean  gas temperature [K] & 31 \\
mm-opacity-slope (0.85-1.3)mm & 1.6 \\
cm-opacity-slope (5-10)mm & 1.5 \\
mm-SED-slope (0.85-1.3)mm & 3.2 \\
cm-SED-slope (5-10)mm & 3.5 \\
10$\mathrm{\mu m}$ silicate emission amplitude & 1.1 \\
naked star luminosity [$L_\odot$] & 0.8 \\
bolometric luminosity [$L_\odot$] & 1.1 \\
near IR excess ($\lambda=$2.02-6.72$\mathrm{\mu m}$)
 [$L_\odot$] & 0.07 \\
mid IR excess ($\lambda=$6.72-29.5$\mathrm{\mu m}$)
 [$L_\odot$] & 0.02 \\
far IR excess ($\lambda=$29.5-991.$\mathrm{\mu m}$)
 [$L_\odot$] & 0.16 \\
\hline
\end{tabular}
}
\hspace*{-10mm}\resizebox{113mm}{!}{
\begin{tabular}{lrccccr}
\hline & & \multicolumn{2}{c}{line flux $[\mathrm{W\,m^{-2}}$]} & \multicolumn{2}{c}{FWHM $[\mathrm{km\,s^{-1}}]$} &  \\
species & $\lambda\,$[$\mathrm{\mu m}]$ & observed & model & observed & model & size[AU] \\
\hline
$\mathrm{OI}$ & 63.183 & 2.9$\pm$0.2(-17) & 3.1(-17) &  & 1.6 & 85.8 \\
$\mathrm{CII}$ & 157.740 & 5.9$\pm$1.7(-18) & 1.6(-18) &  & 1.3 & 146.3 \\
$\mathrm{CO}$ & 866.963 & 6.5$\pm$0.1(-20) & 8.8(-20) &  & 1.4 & 139.5 \\
$\mathrm{CO}$ & 2600.758 & 2.5$\pm$0.9(-21) & 3.0(-21) &  & 1.4 & 135.5 \\
\hline
\end{tabular}
}

\label{tab:UScoJ1604-2130}
\end{table}

\noindent{\sffamily\bfseries\underline{8.\ USco\,J1604-2130}}\ \ is a
transition disks with a large gap between 0.05 and 46\,AU. The H and
K-band infrared excess requires dust at very small distances from the
star (inner disk), while SMA imaging at 880~$\mu$m
\citep{mathews2012b} reveals the clear presence of a large gap. The
small inner disk is optically thick and vertically extended, thus
shielding the outer disk partially from heating/dissociating UV
radiation.

The far-IR and sub-mm line emission arises entirely in the outer disk
and the model line fluxes match the observed ones within 30\% except
for the [C\,{\sc ii}] line (Table~\ref{tab:UScoJ1604-2130}). The
latter shows some extended emission on the Herschel/PACS footprint
\citep{Mathews2013}. Since PACS did not resolve the disks, we cannot
entirely exclude extended emission to contribute to the measured
[C\,{\sc ii}] flux from the central spaxel.

Mid- and near-IR CO lines (wavelength below $\sim\!100~\mu$m) have an
ever increasing contribution also from the small inner disk:
$\sim\!40$\% for the low $J$ CO $v\!=\!1-0$ lines increasing to
$100$\% for $J>20$ and $\sim\!60$\% for the low $J$ CO $v\!=\!2-1$
lines increasing rapidly with $J$ to 100\%. However, line fluxes for
ro-vibrational CO lines from our model are below
$10^{-18}$~W/m$^2$. Due to the low inclination of $10^{\rm o}$, the CO
sub-mm lines are narrow with typical FWHM of 1.4~km/s
(Table~\ref{tab:UScoJ1604-2130}).

\clearpage
\newpage

\begin{figure}[!t]
\centering
\vspace*{2mm}
{\large\sffamily\bfseries\underline{TW Hya}}\\[2mm]
$M_{\rm gas} = 4.5\times 10^{-2}\,M_\odot$,
$M_{\rm dust} = 1.0\times 10^{-4}\,M_\odot$, 
dust settling $\alpha_{\rm settle} = 5.2\times 10^{-3}$\\
\hspace*{-3mm}
\begin{tabular}{cc}
\begin{minipage}{75mm}
\includegraphics[width=79mm]{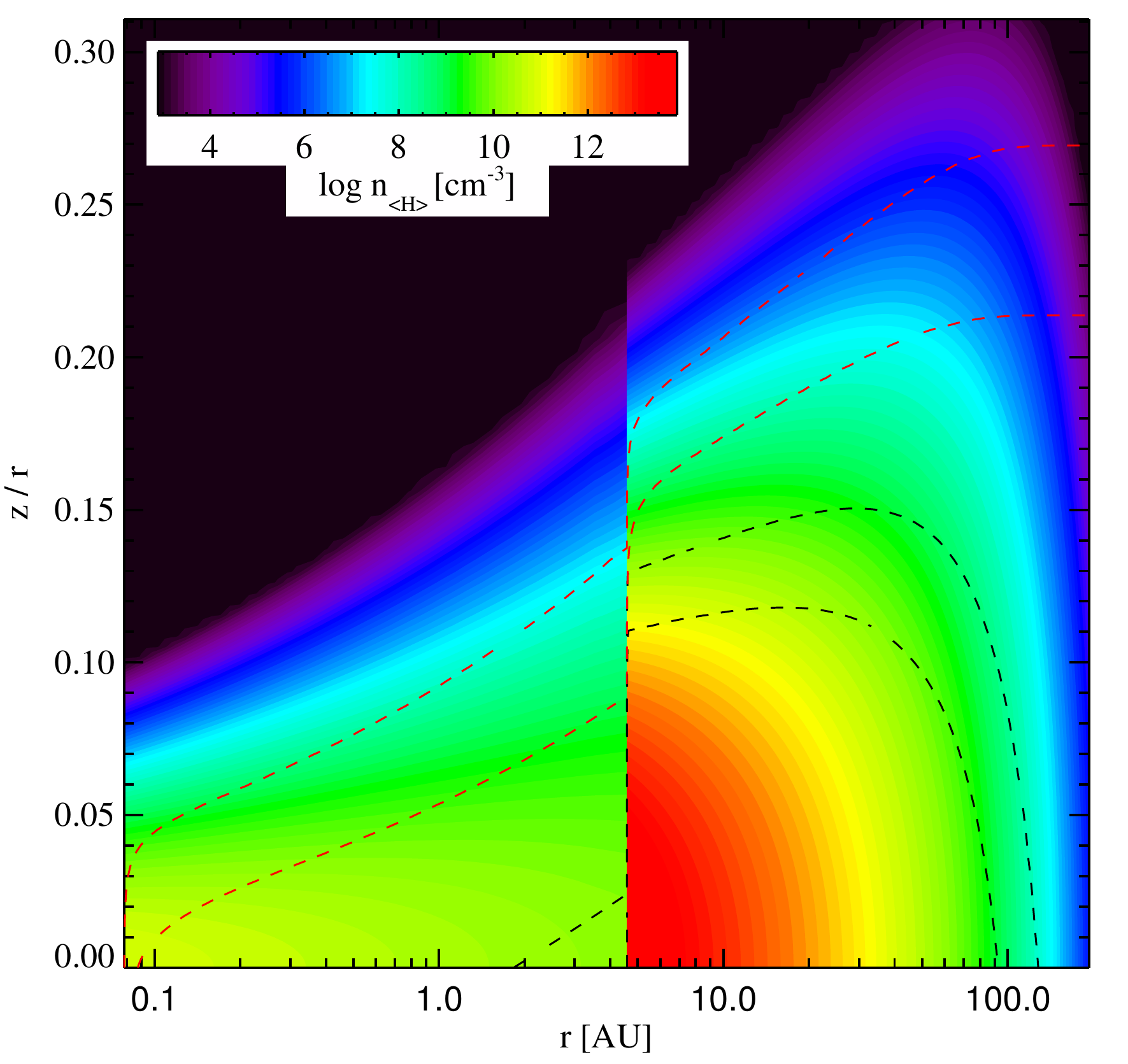}
\end{minipage}
&
\begin{minipage}{87mm}
\vspace*{1mm}
\includegraphics[width=87mm]{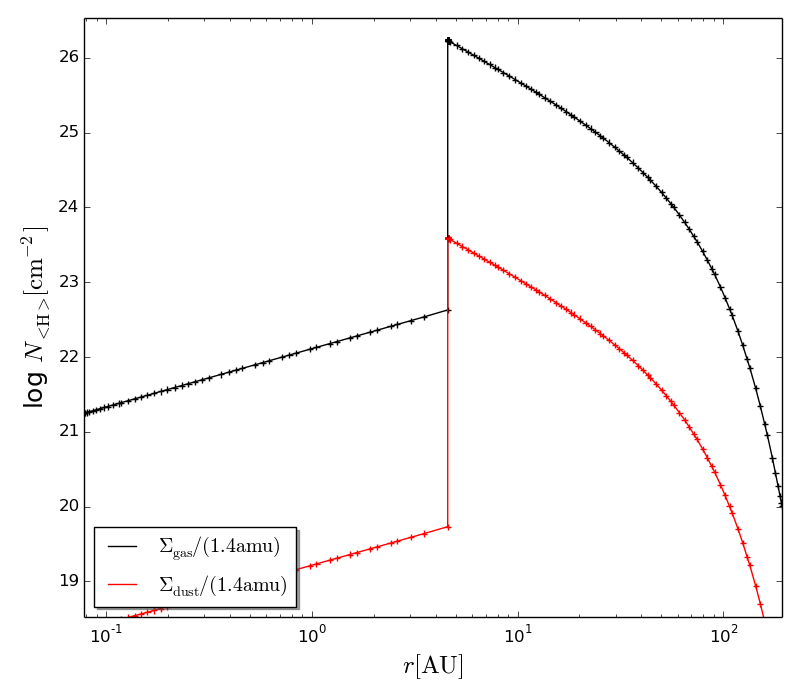}
\end{minipage}
\end{tabular}\\[-1mm]
\caption{TW\,Hya density and surface density plots. For details see Fig.\,\ref{fig:ABAur-1}.}
\label{fig:TWHya-1}
\end{figure}

\begin{table}
\caption{Model properties and comparison of computed spectral line
  properties with observations from {\sf TWHya.properties}. }
\footnotesize
\centering\begin{tabular}{rr}
\hline
DIANA standard fit model properties &  \\
\hline
minimum dust temperature [K] & 7 \\
maximum dust temperature [K] & 1112 \\
mass-mean dust temperature [K] & 21 \\
minimum  gas temperature [K] & 7 \\
maximum  gas temperature [K] & 40000 \\
mass-mean  gas temperature [K] & 21 \\
mm-opacity-slope (0.85-1.3)mm & 1.2 \\
cm-opacity-slope (5-10)mm & 1.4 \\
mm-SED-slope (0.85-1.3)mm & 2.2 \\
cm-SED-slope (5-10)mm & 3.0 \\
10$\mathrm{\mu m}$ silicate emission amplitude & 2.2 \\
naked star luminosity [$L_\odot$] & 0.3 \\
bolometric luminosity [$L_\odot$] & 0.3 \\
near IR excess ($\lambda=$2.02-6.99$\mathrm{\mu m}$)
 [$L_\odot$] & 0.01 \\
mid IR excess ($\lambda=$6.99-29.4$\mathrm{\mu m}$)
 [$L_\odot$] & 0.03 \\
far IR excess ($\lambda=$29.4-972.$\mathrm{\mu m}$)
 [$L_\odot$] & 0.04 \\
\hline
\end{tabular}

\vspace{2mm}\centering\begin{tabular}{lrccccr}
\hline & & \multicolumn{2}{c}{line flux $[\mathrm{W\,m^{-2}}$]} & \multicolumn{2}{c}{FWHM $[\mathrm{km\,s^{-1}}]$} &  \\
species & $\lambda\,$[$\mathrm{\mu m}]$ & observed & model & observed & model & size[AU] \\
\hline
$\mathrm{OI}$ & 63.183 & 3.7$\pm$0.1(-17) & 1.4(-16) &  & 1.3 & 97.6 \\
$\mathrm{OI}$ & 145.525 & $<$3.6(-18) & 2.9(-18) &  & 1.5 & 99.5 \\
$\mathrm{CII}$ & 157.740 & $<$3.6(-18) & 2.4(-18) &  & 1.0 & 159.8 \\
$\mathrm{o\!-\!H_2O}$ & 20.341 & 5.2$\pm$2.0(-18) & 1.8(-17) &  & 11.3 & 0.8 \\
$\mathrm{o\!-\!H_2O}$ & 23.859 & 1.1$\pm$0.2(-17) & 2.6(-17) &  & 10.2 & 2.1 \\
$\mathrm{o\!-\!H_2O}$ & 25.365 & 6.6$\pm$2.0(-18) & 2.7(-17) &  & 9.8 & 3.5 \\
$\mathrm{o\!-\!H_2O}$ & 30.525 & 2.4$\pm$0.2(-17) & 2.5(-17) &  & 9.0 & 4.7 \\
$\mathrm{o\!-\!H_2O}$ & 30.870 & 1.6$\pm$0.2(-17) & 2.1(-17) &  & 9.2 & 4.6 \\
$\mathrm{OH\_H}$ & 27.393 & 1.5$\pm$0.2(-17) & 2.6(-18) &  & 12.3 & 0.8 \\
$\mathrm{OH\_H}$ & 30.277 & 1.6$\pm$0.2(-17) & 2.4(-18) &  & 12.0 & 0.8 \\
$\mathrm{OH\_H}$ & 33.875 & 2.9$\pm$0.2(-17) & 2.0(-18) &  & 11.8 & 0.9 \\
$\mathrm{o\!-\!H_2O}$ & 63.323 & $<$7.2(-18) & 5.3(-18) &  & 6.7 & 15.3 \\
$\mathrm{o\!-\!H_2O}$ & 71.946 & $<$6.3(-18) & 4.6(-18) &  & 3.4 & 26.7 \\
$\mathrm{o\!-\!H_2O}$ & 78.742 & $<$6.0(-18) & 5.4(-18) &  & 2.1 & 43.3 \\
$\mathrm{p\!-\!H_2O}$ & 89.988 & 3.1$\pm$0.9(-18) & 3.4(-18) &  & 2.1 & 46.7 \\
$\mathrm{o\!-\!H_2O}$ & 179.526 & $<$7.8(-18) & 3.0(-18) &  & 1.5 & 65.3 \\
$\mathrm{o\!-\!H_2O}$ & 180.488 & $<$8.4(-18) & 6.1(-19) &  & 2.0 & 44.1 \\
$\mathrm{p\!-\!H_2O}$ & 269.272 & 6.1$\pm$0.4(-19) & 1.2(-18) &  & 1.3 & 95.3 \\
$\mathrm{o\!-\!H_2O}$ & 538.288 & 1.7$\pm$0.4(-19) & 2.5(-19) &  & 1.1 & 105.1 \\
$\mathrm{CO}$ & 1300.403 & 1.1$\pm$0.1(-19) & 1.7(-19) &  & 0.8 & 162.4 \\
$\mathrm{CO}$ & 866.963 & 3.9$\pm$0.2(-19) & 5.7(-19) & 0.7$\pm$0.1 & 0.8 & 173.1 \\
$\mathrm{CO}$ & 650.251 & 5.9$\pm$0.6(-19) & 1.3(-18) &  & 0.8 & 164.2 \\
$\mathrm{CO}$ & 433.556 & 6.1$\pm$0.6(-19) & 3.0(-18) &  & 0.9 & 150.2 \\
$\mathrm{CO}$ & 260.239 & 2.1$\pm$0.2(-18) & 3.7(-18) &  & 1.2 & 80.4 \\
$\mathrm{CO}$ & 200.272 & 1.1$\pm$0.2(-17) & 1.6(-18) &  & 1.8 & 51.0 \\
$\mathrm{CO}$ & 144.784 & 3.5$\pm$1.2(-18) & 8.7(-19) &  & 4.4 & 7.6 \\
$\mathrm{CO}$ & 113.457 & 4.4$\pm$1.2(-18) & 6.0(-19) &  & 6.6 & 4.5 \\
$\mathrm{^{13}CO}$ & 1360.227 & 2.0$\pm$0.1(-20) & 2.5(-20) &  & 1.0 & 123.8 \\
$\mathrm{^{13}CO}$ & 906.846 & 4.4$\pm$2.7(-20) & 9.5(-20) &  & 1.0 & 119.7 \\
$\mathrm{^{13}CO}$ & 453.497 & 1.8$\pm$0.3(-19) & 4.0(-19) & 1.4$\pm$0.1 & 1.1 & 98.2 \\
$\mathrm{^{13}CO}$ & 272.204 & 3.6$\pm$0.4(-19) & 1.9(-19) &  & 1.5 & 66.1 \\
$\mathrm{C^{18}O}$ & 1365.430 & 5.0$\pm$1.4(-21) & 6.0(-21) &  & 1.1 & 109.7 \\
$\mathrm{HCO^+}$ & 3361.334 & 2.5$\pm$0.4(-21) & 1.2(-21) &  & 0.7 & 175.9 \\
$\mathrm{HCO^+}$ & 1120.478 & 1.1$\pm$0.2(-19) & 6.6(-20) &  & 0.8 & 177.5 \\
$\mathrm{HCO^+}$ & 840.380 & 2.7$\pm$0.1(-19) & 1.2(-19) & 0.7$\pm$0.1 & 0.8 & 177.8 \\
$\mathrm{HCN}$ & 1127.520 & 7.5$\pm$1.5(-20) & 8.4(-20) &  & 0.8 & 170.5 \\
$\mathrm{HCN}$ & 845.663 & 1.1$\pm$0.3(-19) & 1.5(-19) &  & 0.8 & 161.7 \\
$\mathrm{N_2H^+}$ & 1072.557 & 2.0$\pm$0.4(-20) & 6.6(-21) &  & 0.8 & 125.0 \\
$\mathrm{N_2H^+}$ & 804.439 & 5.7$\pm$0.9(-20) & 1.1(-20) &  & 0.8 & 119.2 \\
$\mathrm{HD}$ & 112.053 & 6.3$\pm$0.7(-18) & 4.8(-19) &  & 1.9 & 63.9 \\
$\mathrm{HD}$ & 56.223 & $<$8.1(-18) & 8.2(-20) &  & 3.3 & 75.8 \\
$\mathrm{OH}$ & 55.890 & 2.2$\pm$0.1(-17) & 5.0(-18) &  & 8.2 & 4.5 \\
$\mathrm{OH}$ & 55.950 & 2.7$\pm$0.1(-17) & 5.8(-18) &  & 7.8 & 4.6 \\
$\mathrm{CO}$ & 4.609 & 6.4$\pm$0.2(-18) & 5.6(-18) & 7.2$\pm$0.2 & 12.3 & 4.6 \\
$\mathrm{CO}$ & 4.657 & 3.7$\pm$0.2(-18) & 2.8(-18) & 7.2$\pm$0.3 & 4.9 & 4.8 \\
$\mathrm{CO}$ & 4.682 & 4.1$\pm$0.2(-18) & 3.4(-18) & 7.7$\pm$0.3 & 4.9 & 5.0 \\
$\mathrm{CO}$ & 4.754 & 6.2$\pm$0.1(-18) & 6.5(-18) & 7.1$\pm$0.2 & 11.7 & 4.6 \\
$\mathrm{CO}$ & 4.793 & 7.2$\pm$0.3(-18) & 6.7(-18) & 7.7$\pm$0.3 & 12.6 & 4.6 \\
$\mathrm{CO}$ & 4.966 & 3.9$\pm$1.0(-18) & 3.2(-18) & 18.5$\pm$1.0 & 16.4 & 0.5 \\
$\mathrm{CO}$ & 4.990 & 2.3$\pm$0.7(-18) & 2.8(-18) & 11.8$\pm$1.0 & 16.5 & 0.5 \\
$\mathrm{OI}$ & 0.630 & 1.2$\pm$0.1(-16) & 2.2(-16) & 10.0$\pm$1.0 & 12.2 & 6.2 \\
$\mathrm{OI}$ & 0.557 & 1.7$\pm$0.5(-17) & 4.3(-17) & 10.0$\pm$1.0 & 19.2 & 0.8 \\
$\mathrm{S^+}$ & 0.406 & 4.4$\pm$0.5(-17) & 1.7(-19) &  & 8.7 & 9.3 \\
$\mathrm{Ne^+}$ & 12.814 & 7.0$\pm$0.2(-17) & 3.4(-17) &  & 4.5 & 45.8 \\
$\mathrm{Ne^{++}}$ & 15.554 & 5.0$\pm$2.0(-18) & 1.2(-17) &  & 4.3 & 88.5 \\
$\mathrm{o\!-\!H_2}$ & 17.033 & 1.2$\pm$0.2(-17) & 5.9(-19) &  & 4.7 & 24.9 \\
$\mathrm{p\!-\!H_2}$ & 12.277 & 7.0$\pm$2.0(-18) & 2.4(-19) &  & 5.6 & 40.9 \\
\hline
\end{tabular}

\label{tab:TWHya}
\end{table}

\begin{figure}[!h]
\vspace*{-7mm}\hspace*{-9mm}
\begin{tabular}{cc}
\begin{minipage}{86mm}
\includegraphics[width=94mm]{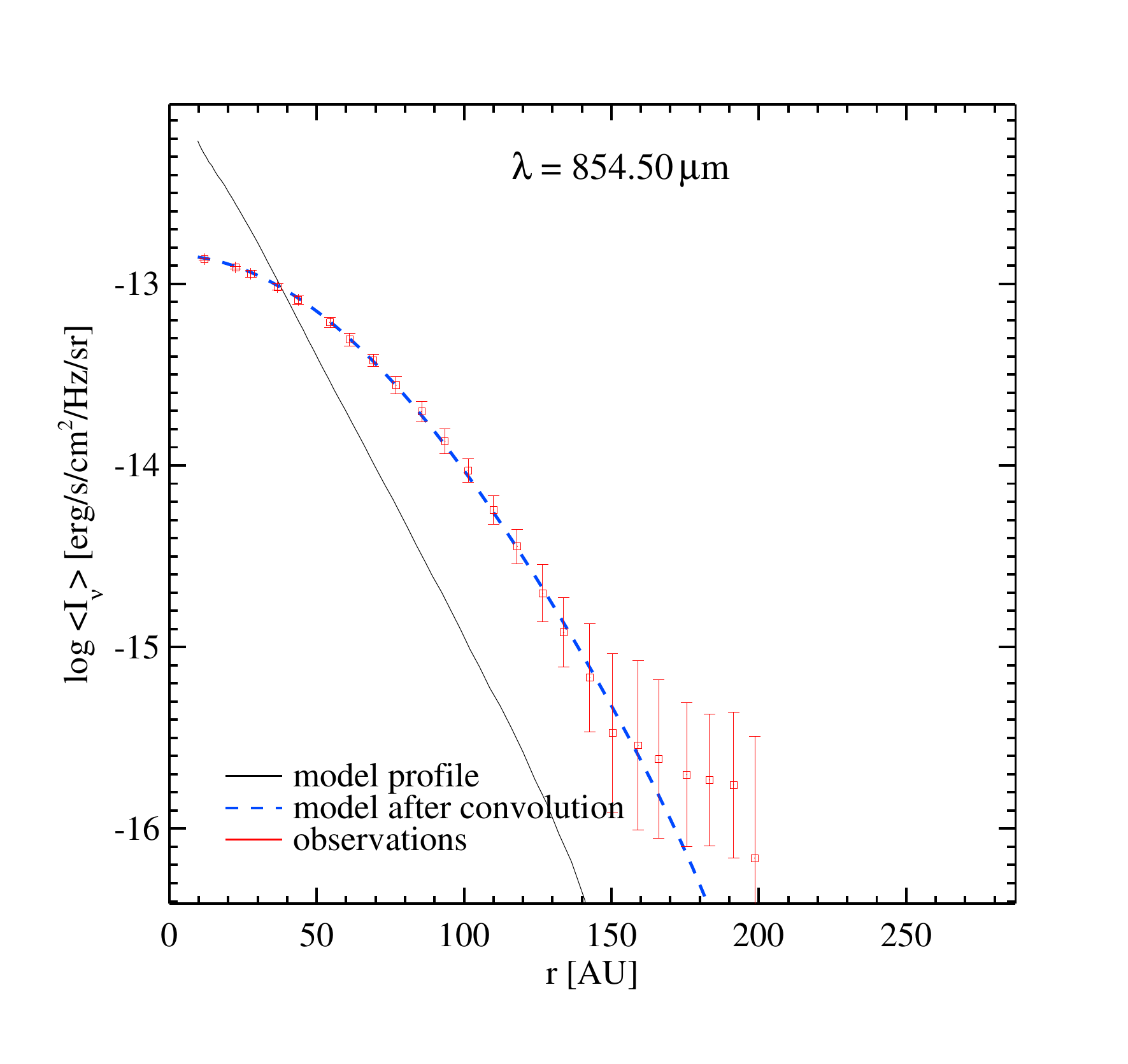} 
\end{minipage} &
\hspace*{-6mm}
\begin{minipage}{85mm}
\includegraphics[width=85mm]{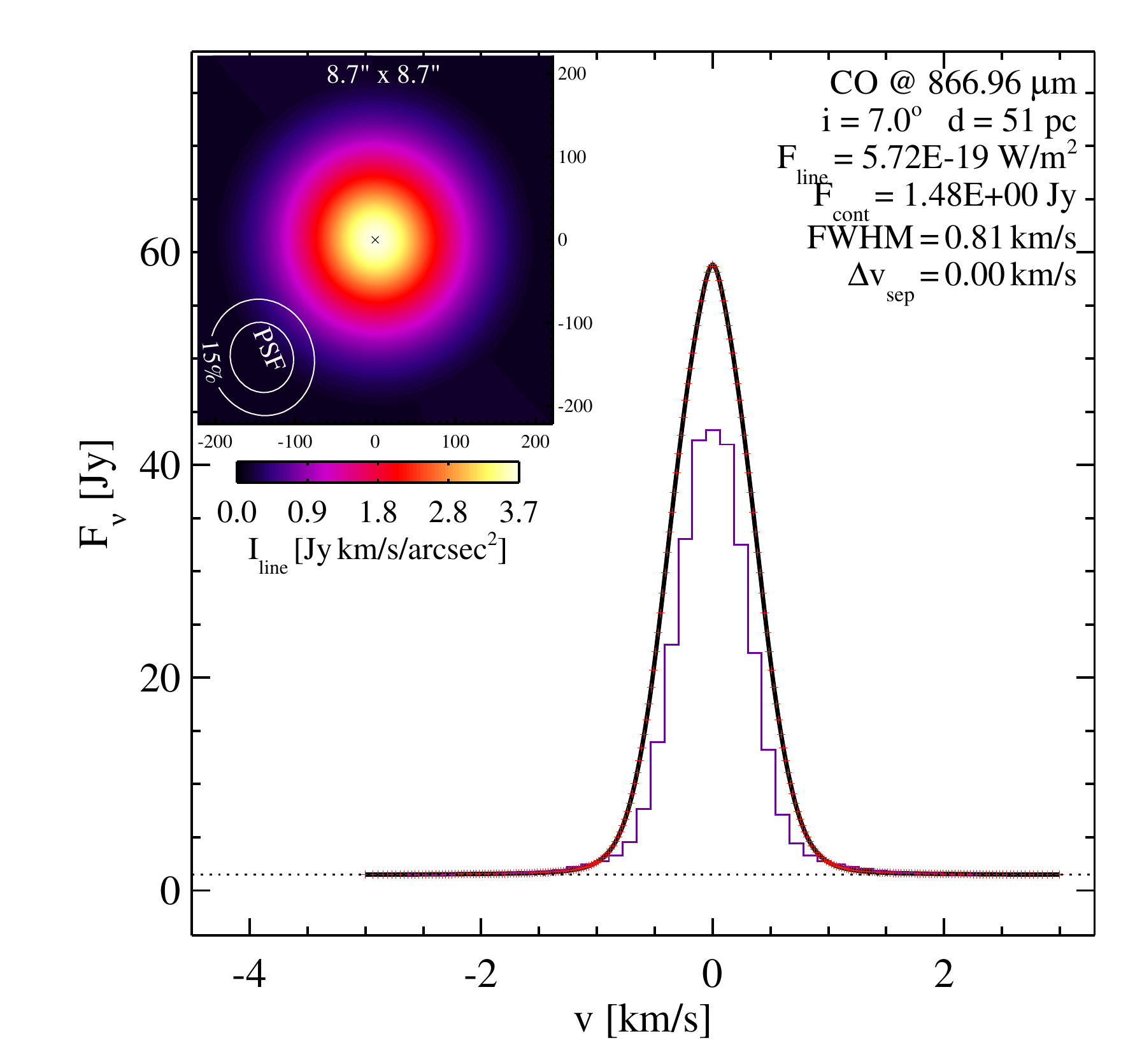}
\end{minipage}
\end{tabular}\\*[-7mm]
\caption{{\bf Left:} Fit of radial intensity profile to archival ALMA
  850\,$\mu$m continuum data. {\bf Right:} CO $J\!=$3-2 line from
  TW\,Hya disk model in comparison to ALMA data.}
\label{fig:TWHya-2}
\vspace*{-4mm}
\end{figure}

\noindent{\sffamily\bfseries\underline{9.\ \ TW\,Hya}}\ \ probably is
the best studied protoplanetary disk around a T\,Tauri
star. Practically every suitable instrument developed in the past 30
years was pointed at this object, producing datasets of varying
quality and scientific usefulness, yet the shape of the disk around
TW\,Hya is still debated \citep[see e.g.][]{Menu2014,Andrews2016}.
Our data collection of TW\,Hya provides exquisite spectral (UV, X-ray,
Spitzer/IRS, SPIRE) and photometric data (see Fig.\,\ref{fig:SEDs}),
one continuum image with derived intensity profile, and 57 lines,
among them three with velocity and intensity profiles.

Figure~\ref{fig:TWHya-1} illustrates the geometry of the disk and the
surface densities of dust and gas assumed.  The model results in good
fits of the SED and most line observations, from CO fundamental
ro-vibrational lines over high-excitation water lines in the
Spitzer-spectrum, [Ne\,{\sc ii}]\,12.81\,$\mu$m, far-IR high-$J$ CO
and $^{13}$CO lines, to fundamental water lines (HIFI
instrument) and a number of (sub-)mm lines, such as CO isotopologue
lines, HCO$^+$, N$_2$H$^+$ and HCN (Table~\ref{tab:TWHya}). Due to the
wealth of line data, the model aims at providing appropriate line
excitation conditions in very different parts of the disk, therefore
it is not surprising that the fits of individually selected lines
(such as CO $J\!=$3-2 shown in Fig.\,\ref{fig:TWHya-2}) are not
perfect.

Unfortunately, two of the key lines observed with Herschel do not fit
very well. The [O\,{\sc i}]\,63\,$\mu$m line is too strong by a factor
4, and the HD\,112\,$\mu$m line is too weak by a factor of 13. Given
our standard disk modeling approach, it is impossible to adjust the
disk and dust opacity parameters to fit both lines. Matching the {
  partially optically thick HD line} requires a more massive disk
where the gas is substantially warmer than the dust in deep layers;
the [O\,{\sc i}]\,63\,$\mu$m line requires just the opposite, a
cooler, less massive disk.

Similar to HD\,163296, we find a gas-rich disk with
gas/dust\,$\sim$\,450, with an even more gas-enriched inner disk
(gas/dust\,$\sim$\,800), which could be the result of radial migration
during disk evolution (in the outer disk) and a planet located at the
transition between inner and outer disk which traps the dust, keeping the
larger grains in the outer disk. Similar to HD\,163296, it seems
essential for TW\,Hya to assume a gas-rich inner disk with little
dust, to boost all gas lines at shorter wavelengths, see also
\citet{Thi2010} and \citet{Kamp2013}.

Despite some remaining deviations between model and observations, we
consider this TW\,Hya disk model as our ``flagship'', because of the
unprecedented degree of physical and chemical consistency in the
model, its simplicity, and the agreement of the results with a large suite of
multi-wavelength line and continuum data. Recently, \citet{Du2015} and
\citet{Kama2016} published TW\,Hya models aiming at a consistent
fitting of the HD and CO lines and the CO and [C\,{\sc i}] lines
respectively. Both models require a strong depletion of the elements
oxygen and carbon in the surface layers of the disk around
TW\,Hya. While the element depletion remains an issue of debate we
raise here some points that might explain the
differences. \citet{Du2015} use a three component model where one of
the components has a tapered outer edge at 50~AU; the detailed
structure and tapering of the outer edge of the disk has been shown to
have a profound effect on the CO line fluxes, especially also the line
ratios of the isotopologues \citep{Woitke2016}. Also, dust settling is
shown to affect the rarer isotopologues; the settling parameters in
our model $\alpha_{\rm set}\!=\!5\times 10^{-3}$ indicates only
moderate turbulence, which profoundly changes the gas-to-dust mass
ratio in the line forming regions of the far-IR lines, especially in
the outer disk where gas densities are low. The model presented here
fits the CO isotopologue lines in the sub-mm and the water lines
within a factor two, without any additional assumptions about peculiar
element abundances. So, while enlarging the disk mass and
simultaneously decreasing the carbon or oxygen abundances may be a
tempting option to improve some line fits, it might worsen the fits in
other spectral regions, for example the mid-IR and sub-mm and water 
lines \citep{Kamp2013}. \citet{Kama2016b} showed in a more detailed
parameter study using the DALI code \citep{Bruderer2009} that the
flaring angle and tapering radius both affect the carbon fine
structure and CO sub-mm lines in the same way as the carbon elemental
abundance. In addition, the differences in quoted observed line fluxes
from various papers amount to a factor 2-3 as well. Clearly more work
is needed to reconcile the remaining model discrepancies.

\clearpage
\newpage

\begin{figure}[!t]
\centering
\vspace*{2mm}
{\large\sffamily\bfseries\underline{GM Aur}}\\[2mm]
$M_{\rm gas} = 3.3\times 10^{-2}\,M_\odot$, 
$M_{\rm dust} = 3.3\times 10^{-4}\,M_\odot$, 
dust settling $\alpha_{\rm settle} = 0.1$\\
\hspace*{-3mm}
\begin{tabular}{cc}
\begin{minipage}{75mm}
\includegraphics[width=78mm]{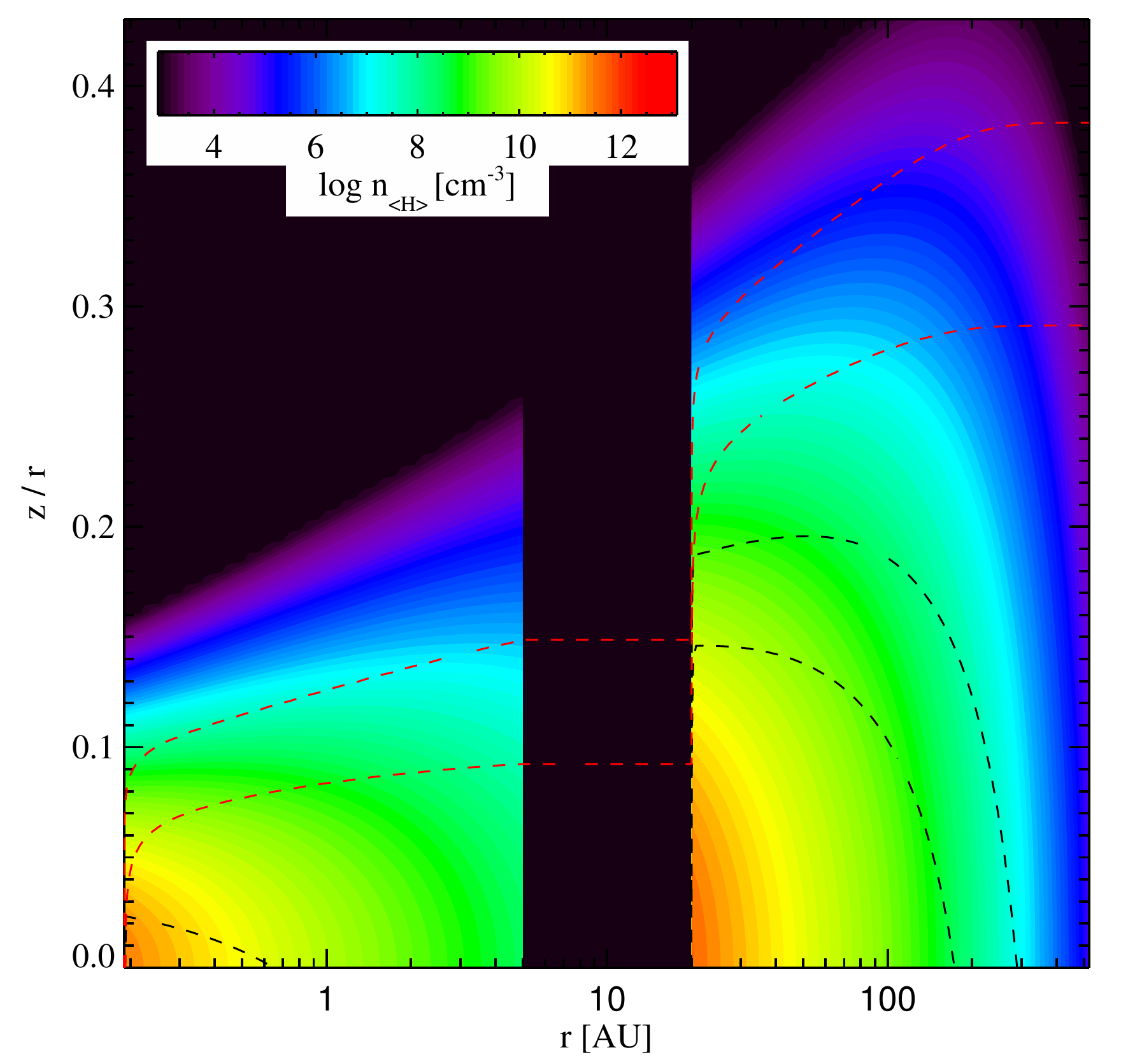}
\end{minipage}
&
\begin{minipage}{87mm}
\vspace*{1mm}
\includegraphics[width=87mm]{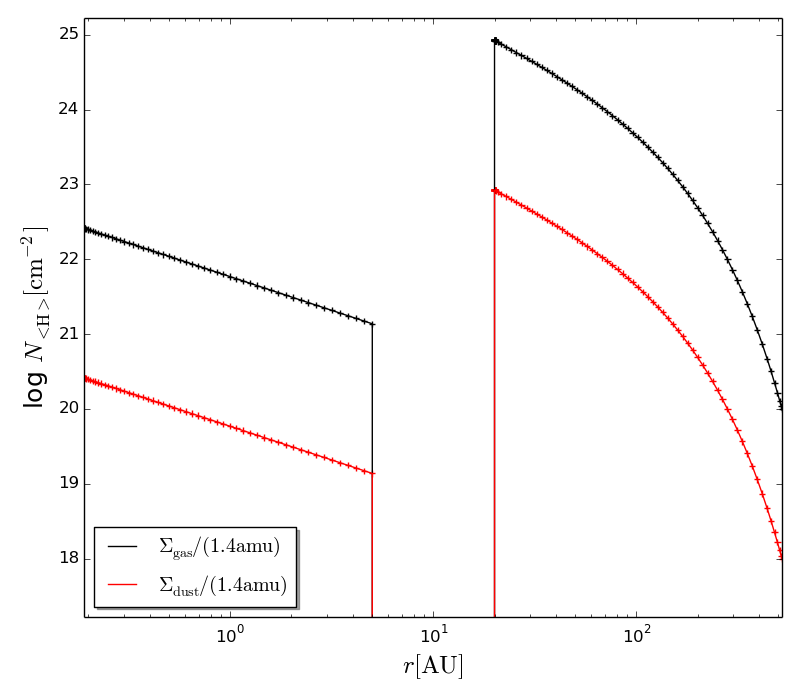}
\end{minipage}
\end{tabular}\\[-2mm]
\caption{GM\,Aur density and surface density plots. For details see
  Fig.\,\ref{fig:ABAur-1}.}
\label{fig:GMAur-1}
\end{figure}

\begin{figure}[!h]
\vspace*{-3mm}\hspace*{-9mm}
\begin{tabular}{cc}
\begin{minipage}{82mm}
\includegraphics[width=85mm]{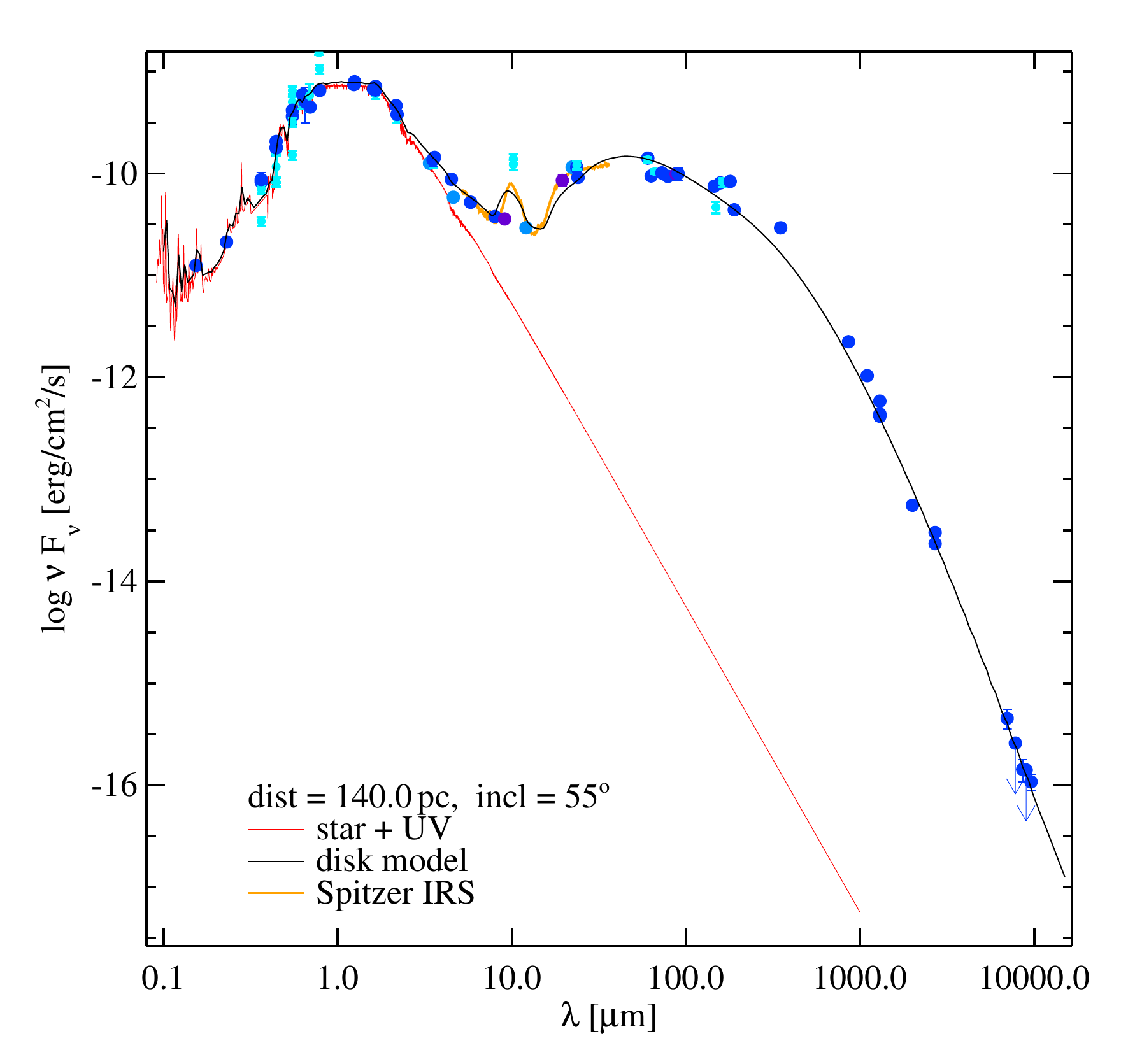} 
\end{minipage} &
\hspace*{-6mm}
\begin{minipage}{85mm}
\vspace*{1mm}
\includegraphics[width=86mm]{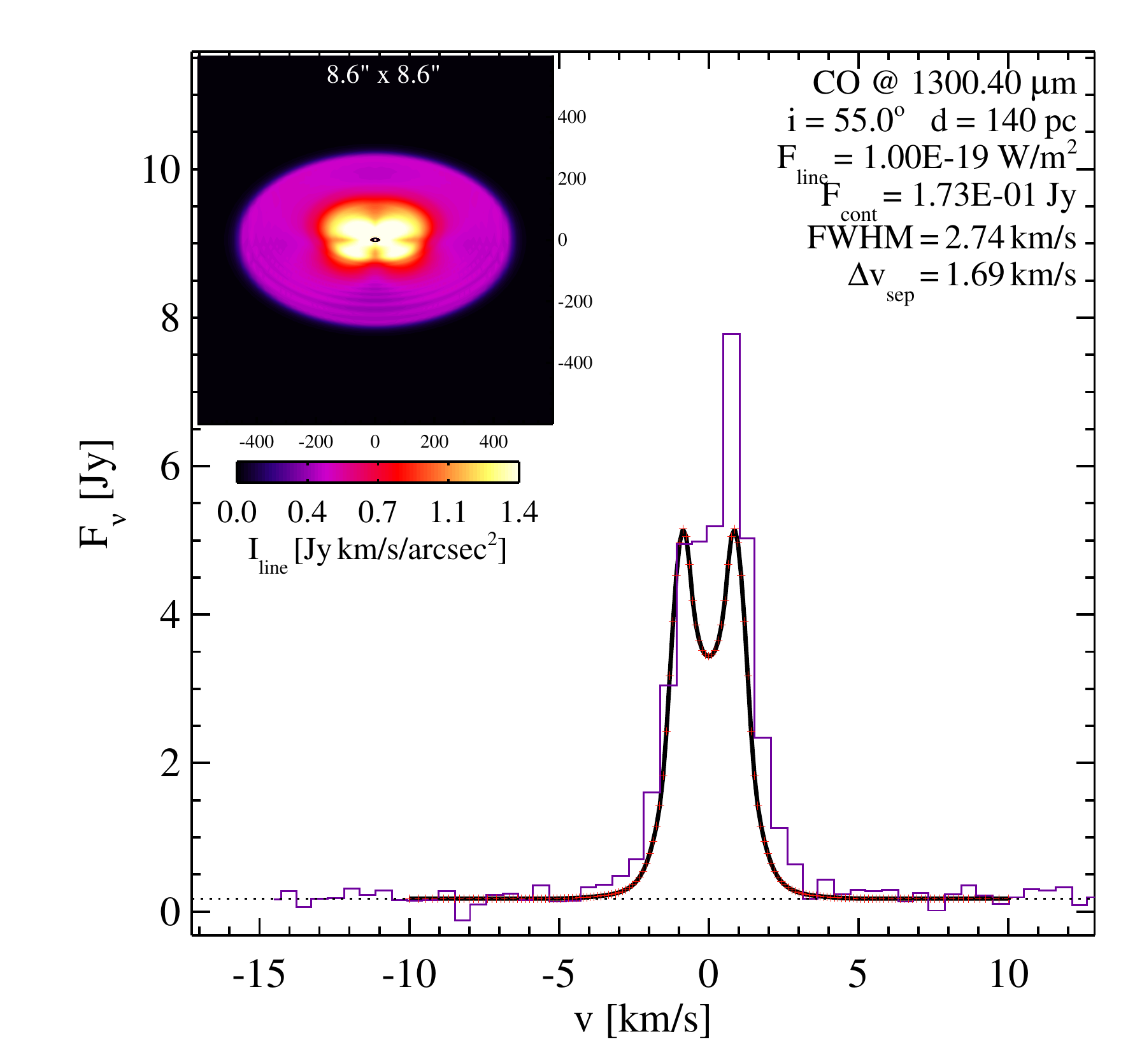}
\end{minipage}
\end{tabular}\\*[-2mm]
\caption{{\bf Left:} SED-fit of GM\,Aur. {\bf Right:} CO $J\!=$2-1
  line in comparison to the GM\,Aur data.}
\label{fig:GMAur-2}
\end{figure}

\begin{table}
\caption{Model properties and comparison of computed spectral line
  properties with observations from {\sf GMAur.properties}. }
\centering\begin{tabular}{rr}
\hline
DIANA standard fit model properties &  \\
\hline
minimum dust temperature [K] & 6 \\
maximum dust temperature [K] & 886 \\
mass-mean dust temperature [K] & 18 \\
minimum  gas temperature [K] & 7 \\
maximum  gas temperature [K] & 26157 \\
mass-mean  gas temperature [K] & 19 \\
mm-opacity-slope (0.85-1.3)mm & 1.3 \\
cm-opacity-slope (5-10)mm & 1.7 \\
mm-SED-slope (0.85-1.3)mm & 2.4 \\
cm-SED-slope (5-10)mm & 3.5 \\
10$\mathrm{\mu m}$ silicate emission amplitude & 1.8 \\
naked star luminosity [$L_\odot$] & 0.7 \\
bolometric luminosity [$L_\odot$] & 0.9 \\
near IR excess ($\lambda=$2.07-6.84$\mathrm{\mu m}$)
 [$L_\odot$] & 0.03 \\
mid IR excess ($\lambda=$6.84-29.9$\mathrm{\mu m}$)
 [$L_\odot$] & 0.05 \\
far IR excess ($\lambda=$29.9-998.$\mathrm{\mu m}$)
 [$L_\odot$] & 0.14 \\
\hline
\end{tabular}

\vspace{0.5cm}\centering\begin{tabular}{lrccccr}
\hline & & \multicolumn{2}{c}{line flux $[\mathrm{W\,m^{-2}}$]} & \multicolumn{2}{c}{FWHM $[\mathrm{km\,s^{-1}}]$} &  \\
species & $\lambda\,$[$\mathrm{\mu m}]$ & observed & model & observed & model & size[AU] \\
\hline
$\mathrm{OI}$ & 0.630 & 3.9$\pm$1.0(-17) & 1.2(-17) & 42.0$\pm$0.0 & 10.4 & 27.5 \\
$\mathrm{CO}$ & 4.649 & 1.3$\pm$0.9(-18) & 1.2(-16) & 22.8$\pm$0.0 & 9.0 & 19.4 \\
$\mathrm{CO}$ & 4.657 & 9.0$\pm$8.0(-19) & 8.3(-17) & 15.6$\pm$0.0 & 9.1 & 19.4 \\
$\mathrm{CO}$ & 4.699 & 1.0$\pm$0.8(-18) & 2.0(-16) & 19.6$\pm$0.0 & 9.0 & 19.4 \\
$\mathrm{CO}$ & 4.708 & 1.1$\pm$0.8(-18) & 1.9(-16) & 18.0$\pm$0.0 & 9.1 & 19.4 \\
$\mathrm{CO}$ & 4.717 & 1.1$\pm$0.8(-18) & 1.7(-16) & 18.0$\pm$0.0 & 9.1 & 19.4 \\
$\mathrm{CO}$ & 4.726 & 1.5$\pm$0.9(-18) & 1.3(-16) & 26.9$\pm$0.0 & 9.1 & 19.5 \\
$\mathrm{CO}$ & 4.735 & 1.6$\pm$0.9(-18) & 9.7(-17) & 25.4$\pm$0.0 & 9.1 & 19.5 \\
$\mathrm{CO}$ & 4.745 & 2.2$\pm$1.1(-18) & 6.7(-17) & 37.2$\pm$0.0 & 9.1 & 19.5 \\
$\mathrm{CO}$ & 4.754 & 2.7$\pm$1.2(-18) & 4.5(-17) & 44.6$\pm$0.0 & 9.1 & 19.5 \\
$\mathrm{CO}$ & 4.763 & 2.3$\pm$1.1(-18) & 3.0(-17) & 35.6$\pm$0.0 & 9.1 & 19.5 \\
$\mathrm{CO}$ & 4.773 & 2.9$\pm$1.2(-18) & 2.0(-17) & 42.9$\pm$0.0 & 9.1 & 19.6 \\
$\mathrm{CO}$ & 4.793 & 2.8$\pm$1.2(-18) & 9.1(-18) & 46.3$\pm$0.0 & 9.1 & 19.7 \\
$\mathrm{Ne^+}$ & 12.814 & 1.2$\pm$0.3(-17) & 6.8(-18) &  & 9.5 & 53.3 \\
$\mathrm{OI}$ & 63.183 & 3.8$\pm$0.5(-17) & 7.2(-17) &  & 5.4 & 230.2 \\
$\mathrm{o\!-\!H_2O}$ & 63.323 & $<$7.3(-18) & 4.3(-18) &  & 9.1 & 20.2 \\
$\mathrm{CO}$ & 1300.404 & 1.5$\pm$0.3(-19) & 1.0(-19) &  & 2.7 & 402.8 \\
$\mathrm{HCO^+}$ & 1120.478 & 3.9$\pm$0.1(-20) & 1.2(-20) &  & 3.5 & 346.6 \\
\hline
\end{tabular}

\label{tab:GMAur}
\end{table}

\noindent{\sffamily\bfseries\underline{10.\ \ GM\,Aur}}\ \ has been
fitted by hand, independent of the SED-fit shown in
Fig.\,\ref{fig:SEDs}. We therefore show the obtained SED-fit by this
model in Fig.~\ref{fig:GMAur-2} as well. The density and column
density structure are shown in Fig.~\ref{fig:GMAur-1}. Not all model
parameters have been varied, so some of the dust size and material
parameters, and the gas/dust ratio, still have their default values
for T\,Tauri stars as recommended in \citep{Woitke2016}. These
circumstances allow us to assess the uncertainties in mass
determination. The {\sf DIANA}-standard model shown here is about
a factor of 3 less massive as the SED-fit model which resulted in a
total disk mass of $0.11\,M_\odot$. \label{page:GMAur}

A good fit of the SED, [O\,{\sc i}]\,63\,$\mu$m and CO 2-1 and HCO$^+$
3-2 lines has been obtained. However, the CO fundamental
ro-vibrational lines are too strong in the model, and too narrow
(Table~\ref{tab:GMAur}). In the model, these lines originate from the
inner wall of the outer disk at 20\,AU, and this wall emission is
about a factor of 100 too bright, and also too narrow by a factor 2-3.
A tall but tenuous inner disk might improve the fit of the CO
fundamental lines, to shield the stellar UV field, similar to CY\,Tau
and BP\,Tau.

\citet{McClure2016} derives a total disk mass of
$M_\mathrm{disk}=0.18\,M_\mathrm{\odot}$ from SED modeling (gas to
dust ratio of 100) and from HD$\,J=1-0$ line observations they
estimated $2.5-20.4\times10^{-2}\,M_\mathrm{\odot}$. They claim a CO
gas phase depletion of up to two orders of magnitude using a rough
estimate for the CO disk mass. However, due to the large uncertainties
in disk masses derived via CO and HD also much lower depletion factors
are possible. In the DIANA standard model we get a disk gas mass of
$M_\mathrm{disk}=3.3\times10^{-2}\,M_\mathrm{sun}$ which is consistent
with the HD derived disk mass but about an order of magnitude higher
than the CO disk mass estimate of \citet{McClure2016}. This implies
that for our model no additional CO gas depletion is
necessary. However, only the $^{12}$CO$\,J=2-1$ is included for the
modeling. Further observations of CO isotopologues (e.g. with ALMA)
and higher quality HD observations (e.g. possibly with SOFIA) are
required to better constrain the disk masses and possibly the
depletion of CO in GM\,Aur.

\clearpage
\newpage

\begin{figure}[!t]
\vspace*{2mm}
\centering
{\large\sffamily\bfseries\underline{BP Tau}}\\[2mm]
$M_{\rm gas} = 6.4\times 10^{-3}\,M_\odot$, 
$M_{\rm dust} = 1.42\times 10^{-4}\,M_\odot$, 
dust settling $\alpha_{\rm settle} = 6.0\times10^{-5}$\\
\hspace*{-5mm}
\begin{tabular}{cc}
\begin{minipage}{75mm}
\includegraphics[width=79mm]{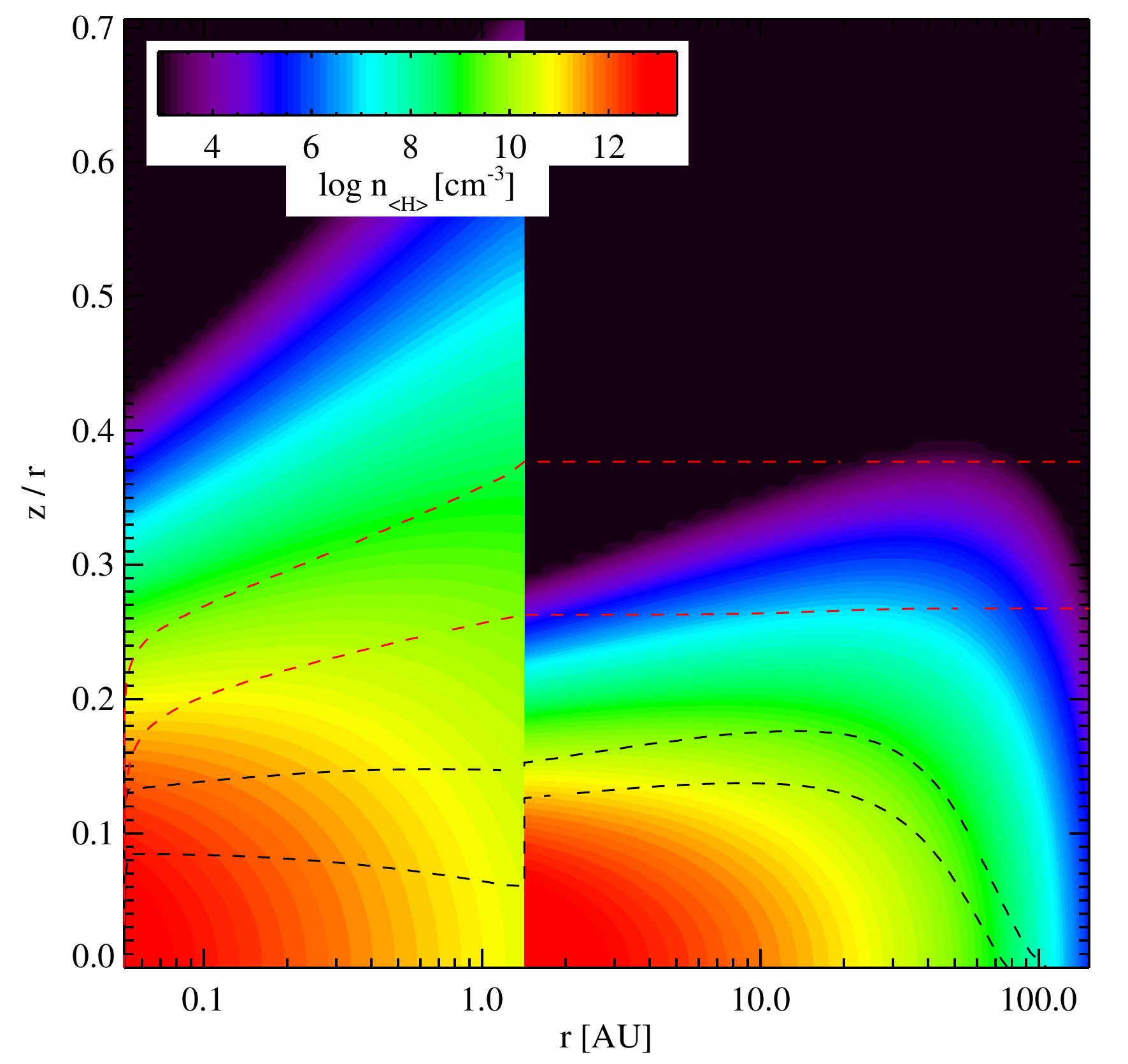}
\end{minipage}
&
\begin{minipage}{87mm}
\vspace*{1mm}
\includegraphics[width=87mm]{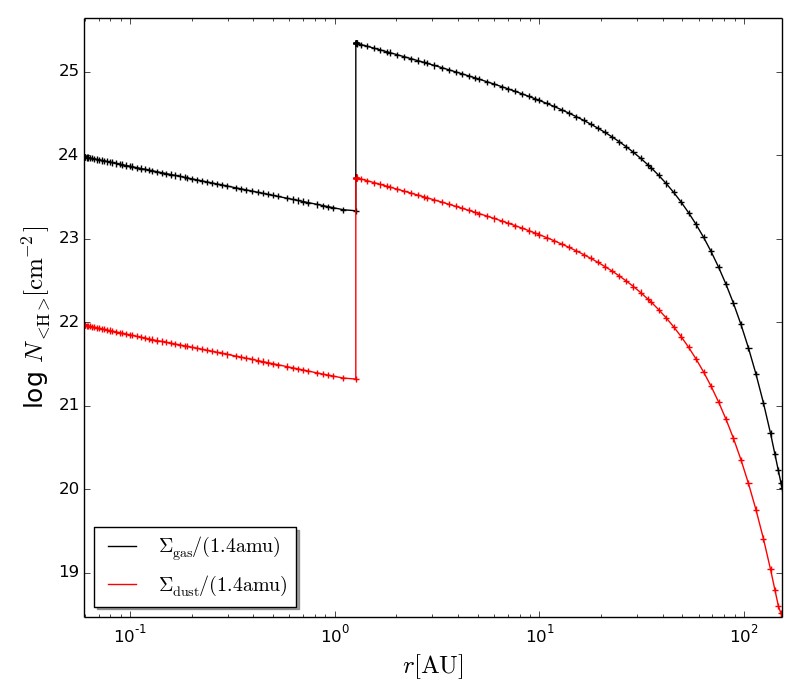}
\end{minipage}
\end{tabular}\\[-1mm]
\caption{BP\,Tau density and surface density plots. For details see Fig.\,\ref{fig:ABAur-1}.}
\label{fig:BPTau-1}
\end{figure}

\begin{figure}[!h]
\centering
\includegraphics[width=150mm]{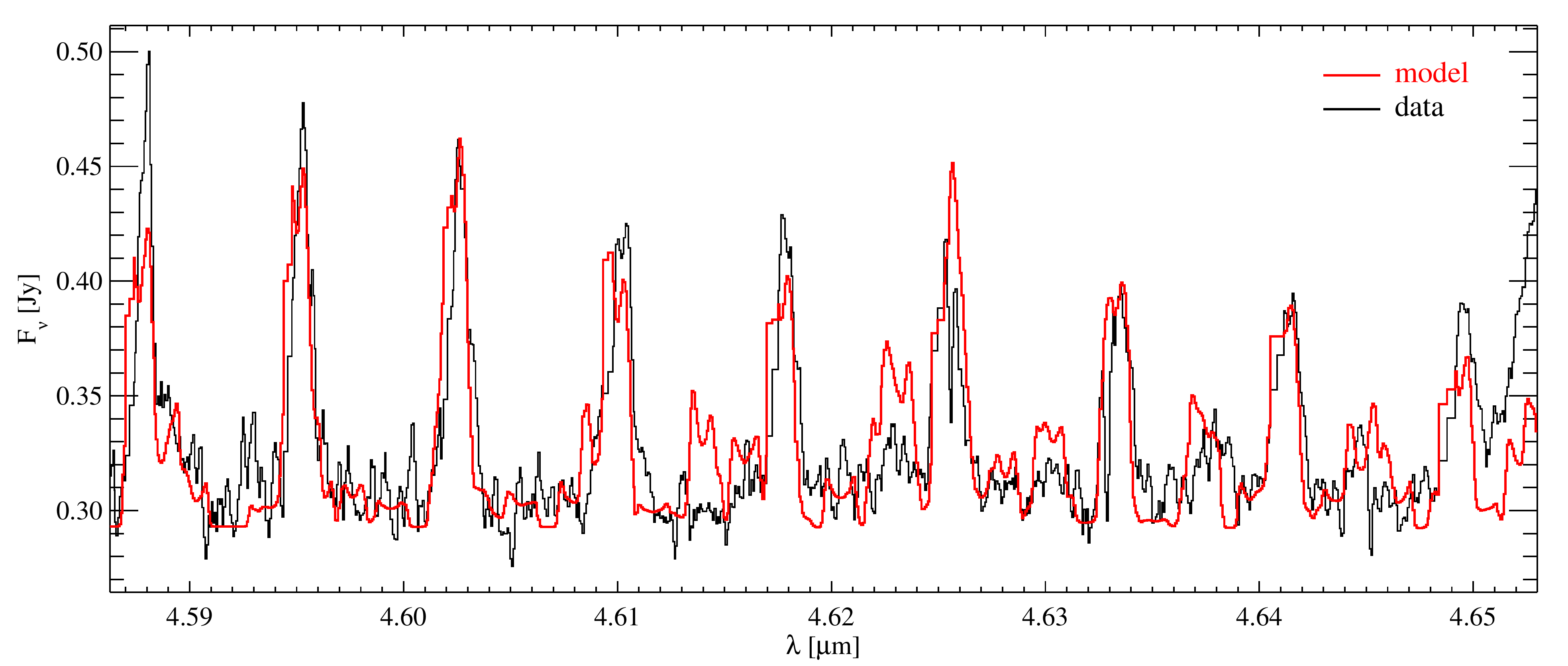}
\vspace*{-2mm}
\caption{Model prediction and observed high-resolution Keck/NIRSPEC spectrum of the $R$-branch of fundamental 
      $\rm CO\ v\!=\!1\!-\!0$ around 4.6\,$\mu$m in BP\,Tau.}
\label{fig:BPTau-2}
\end{figure}

\begin{table}
\caption{Model properties and comparison of computed spectral
  line properties with observations from {\sf BPTau.properties}. }
\centering\begin{tabular}{rr}
\hline
DIANA standard fit model properties &  \\
\hline
minimum dust temperature [K] & 4 \\
maximum dust temperature [K] & 1723 \\
mass-mean dust temperature [K] & 15 \\
minimum  gas temperature [K] & 4 \\
maximum  gas temperature [K] & 40000 \\
mass-mean  gas temperature [K] & 18 \\
mm-opacity-slope (0.85-1.3)mm & 1.4 \\
cm-opacity-slope (5-10)mm & 1.6 \\
mm-SED-slope (0.85-1.3)mm & 1.9 \\
cm-SED-slope (5-10)mm & 3.3 \\
10$\mathrm{\mu m}$ silicate emission amplitude & 1.7 \\
naked star luminosity [$L_\odot$] & 1.0 \\
bolometric luminosity [$L_\odot$] & 1.3 \\
near IR excess ($\lambda=$2.05-6.90$\mathrm{\mu m}$)
 [$L_\odot$] & 0.13 \\
mid IR excess ($\lambda=$6.90-29.3$\mathrm{\mu m}$)
 [$L_\odot$] & 0.10 \\
far IR excess ($\lambda=$29.3-970.$\mathrm{\mu m}$)
 [$L_\odot$] & 0.03 \\
\hline
\end{tabular}

\vspace{0.5cm}\centering\begin{tabular}{lrccccr}
\hline & & \multicolumn{2}{c}{line flux $[\mathrm{W\,m^{-2}}$]} & \multicolumn{2}{c}{FWHM $[\mathrm{km\,s^{-1}}]$} &  \\
species & $\lambda\,$[$\mathrm{\mu m}]$ & observed & model & observed & model & size[AU] \\
\hline
$\mathrm{CO}$ & 1300.403 & 8.7$\pm$2.0(-21) & 1.1(-20) & 4.2$\pm$0.4 & 2.9 & 125.0 \\
$\mathrm{^{13}CO}$ & 1360.227 & 1.1$\pm$0.5(-21) & 1.5(-21) &  & 3.8 & 76.2 \\
$\mathrm{CN}$ & 1321.390 & 2.4$\pm$0.5(-21) & 2.8(-22) &  & 2.6 & 143.4 \\
$\mathrm{CO}$ & 4.633 & 1.7$\pm$0.3(-17) & 1.2(-17) & 65.0$\pm$4.0 & 87.0 & 0.6 \\
$\mathrm{CO}$ & 4.477 & 1.9$\pm$0.2(-17) & 1.3(-17) & 70.0$\pm$3.0 & 88.4 & 0.3 \\
$\mathrm{OI}$ & 63.183 & 9.5$\pm$2.7(-18) & 2.7(-17) &  & 2.7 & 140.2 \\
$\mathrm{o\!-\!H_2O}$ & 63.323 & $<$9.3(-18) & 5.9(-19) &  & 27.4 & 1.2 \\
\hline
\end{tabular}

\label{tab:BPTau}
\end{table}

\noindent{\sffamily\bfseries\underline{11.\ \ BP\,Tau}}\ \ The SED of
BP\,Tau (Fig.\,\ref{fig:SEDs}) is characterized by a strong near-IR
excess and a large amplitude 10\,$\mu$m silicate emission
feature, followed by a steep and steady decline of the flux towards
about 300\,$\mu$m, where the SED eventually kinks downward with a
modest millimeter-slope of about 2.1.

The model fits these properties, and a number of gas line observations
in the millimeter, far-IR and near-IR regions, by assuming a tenuous
but vertically highly extended inner disk (extending radially to
1.3\,AU) that casts a shadow onto the outer massive disk which is flat
and strongly settled (Fig.~\ref{fig:BPTau-1}). The fit includes
high-resolution Keck/NIRSPEC observations of fundamental
ro-vibrational CO emission around 4.6\,$\mu$m \citep{Najita2003} as
shown in Fig.~\ref{fig:BPTau-2}. We used the new Fast Line Tracer
\citep[FLiTs,][]{Woitke2018} to calculate the CO
spectrum from this disk. A good fit with the ProDiMo $\to$ FLiTs model
was found {\sl only after} viscous heating was taken into account, and
the inner disk zone was assumed to to be tenuous and vertically highly
extended, which creates sufficient hot gas (Table~\ref{tab:BPTau}).
The fit of the [O\,{\sc i}]\,63\,$\mu$m line is not quite
satisfactory, a factor of about 3 too bright, but that factor would be
even larger if there was no tall inner disk assumed, which provides
some shielding from the stellar UV (BP\,Tau is a strong UV and X-ray
source).

\clearpage
\newpage

\begin{figure}[!t]
\centering
\vspace*{2mm}
{\large\sffamily\bfseries\underline{DM\,Tau}}\\[2mm]
$M_{\rm gas} = 1.6\times 10^{-2}\,M_\odot$, 
$M_{\rm dust} = 1.6\times 10^{-4}\,M_\odot$, 
dust settling $\alpha_{\rm settle} = 6.4\times10^{-5}$\\
\hspace*{-4mm}
\begin{tabular}{cc}
\begin{minipage}{75mm}
\includegraphics[width=79mm]{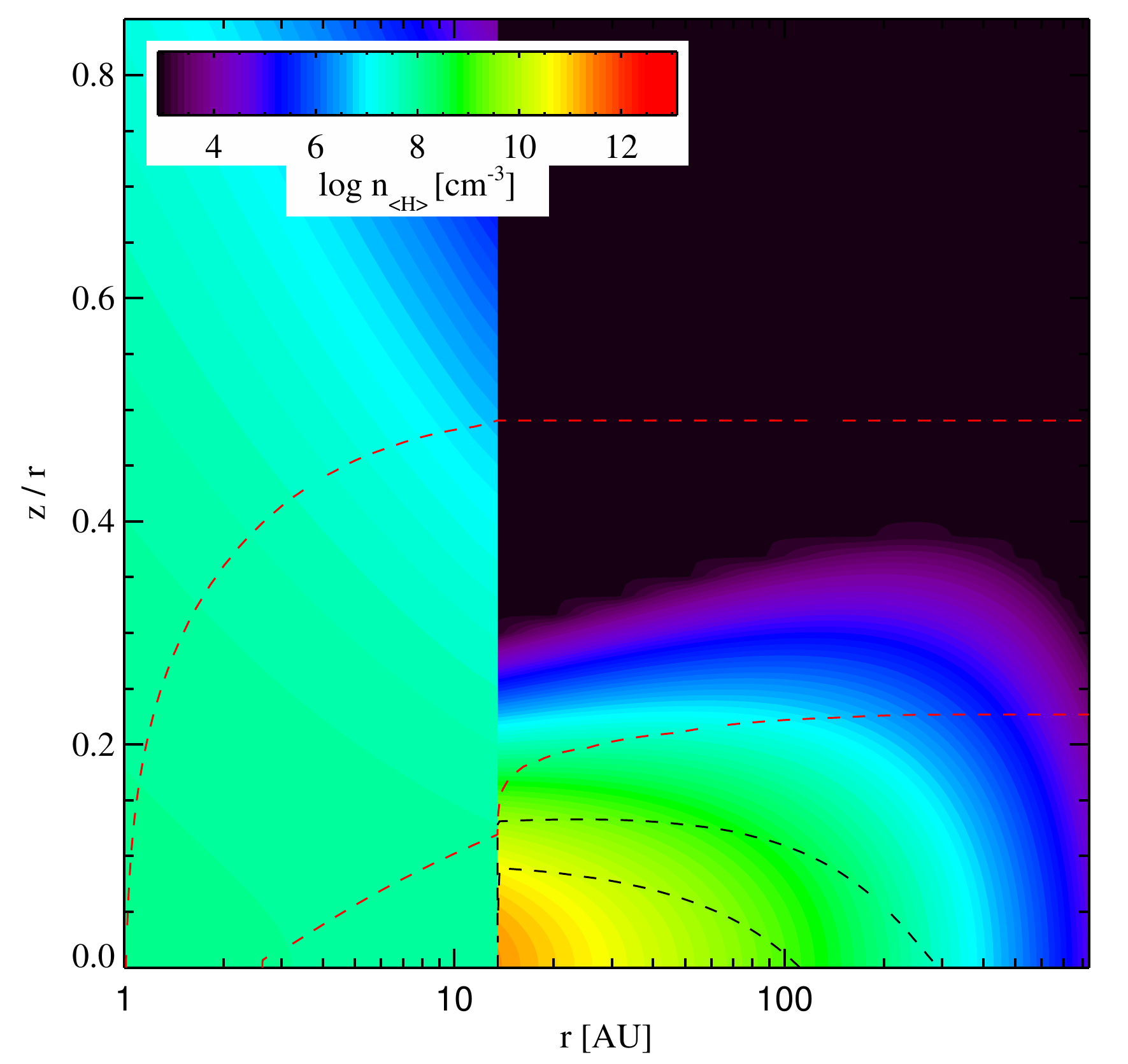}
\end{minipage}
&
\begin{minipage}{87mm}
\vspace*{1mm}
\includegraphics[width=87mm]{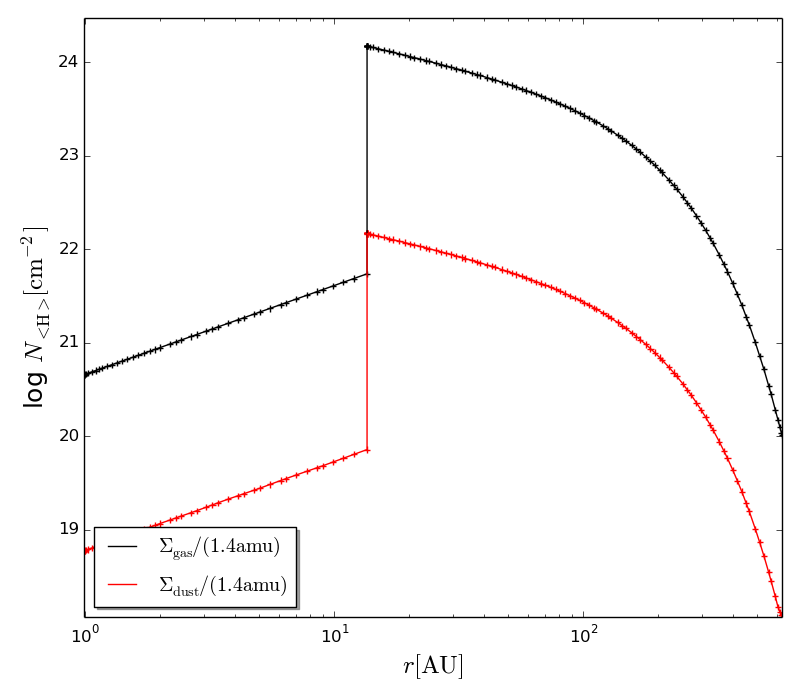}
\end{minipage}
\end{tabular}\\[-1mm]
\caption{DM\,Tau density and surface density plots. For details see Fig.\,\ref{fig:ABAur-1}.}
\label{fig:DMTau-1}
\end{figure}

\begin{figure}[!h]
\vspace*{0mm}\hspace*{-9mm}
\begin{tabular}{cc}
\begin{minipage}{85mm}
\includegraphics[width=85mm,trim=0 0 0 19, clip]{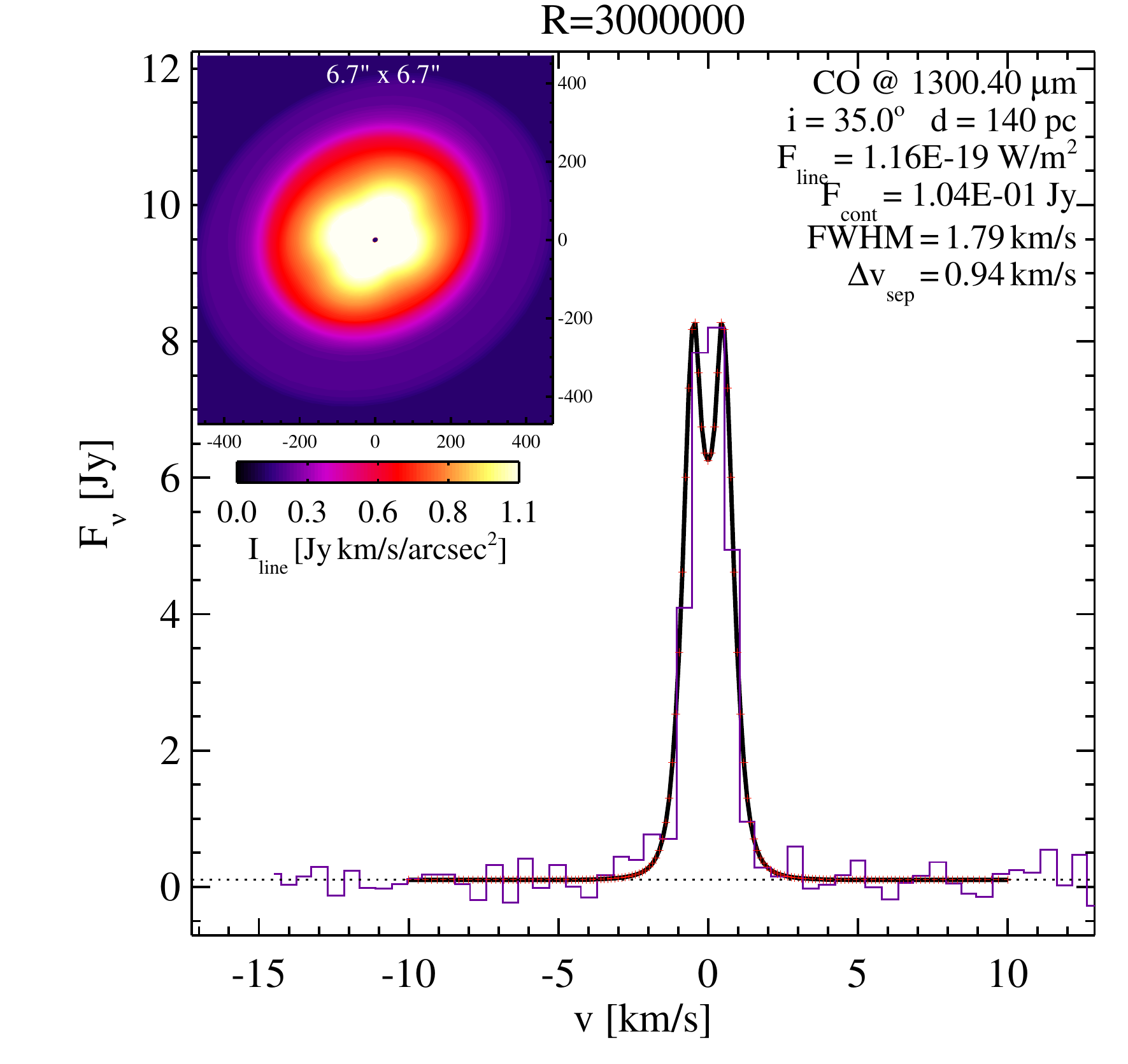}
\end{minipage} &
\hspace*{-6mm}
\begin{minipage}{85mm}
\includegraphics[width=85mm,trim=0 0 0 19, clip]{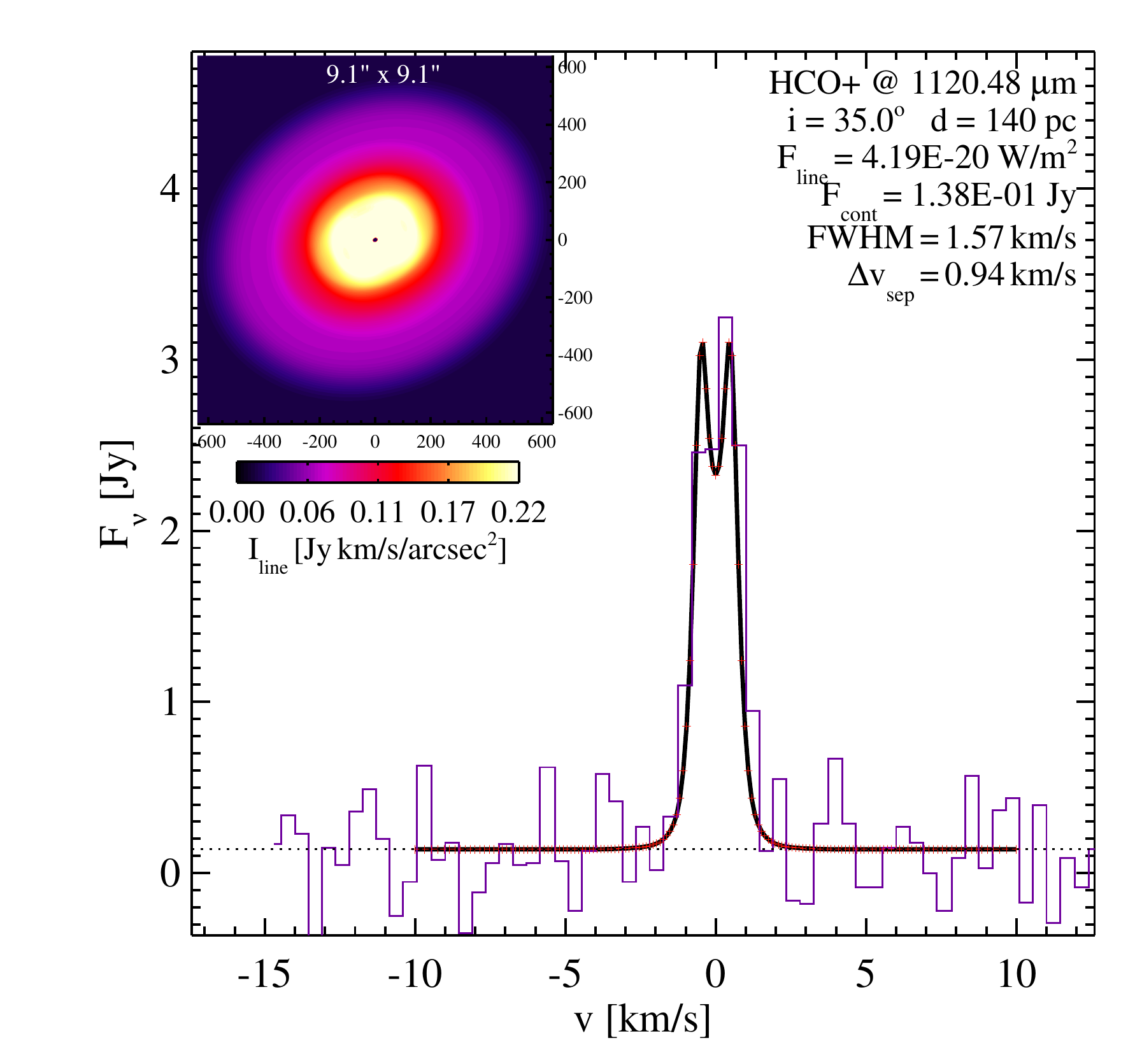}
\end{minipage}
\end{tabular}\\*[-3mm]
\caption{{\bf Left:} CO 2-1 line flux and velocity profile in comparison to SMA data.
{\bf Right:} Same for HCO$^+$ 3-2 line at 1.12~mm.}
\label{fig:DMTau-2}
\vspace*{0mm}
\end{figure}

\begin{table}
\caption{Model properties and comparison of computed spectral line
  properties with observations from {\sf DMTau.properties}. }
\centering\begin{tabular}{rr}
\hline
DIANA standard fit model properties &  \\
\hline
minimum dust temperature [K] & 5 \\
maximum dust temperature [K] & 242 \\
mass-mean dust temperature [K] & 9 \\
minimum  gas temperature [K] & 5 \\
maximum  gas temperature [K] & 15978 \\
mass-mean  gas temperature [K] & 13 \\
mm-opacity-slope (0.85-1.3)mm & 1.0 \\
cm-opacity-slope (5-10)mm & 1.3 \\
mm-SED-slope (0.85-1.3)mm & 2.1 \\
cm-SED-slope (5-10)mm & 3.1 \\
10$\mathrm{\mu m}$ silicate emission amplitude & 2.0 \\
naked star luminosity [$L_\odot$] & 0.3 \\
bolometric luminosity [$L_\odot$] & 0.4 \\
near IR excess ($\lambda=$2.02-6.72$\mathrm{\mu m}$)
 [$L_\odot$] & 0.00 \\
mid IR excess ($\lambda=$6.72-29.5$\mathrm{\mu m}$)
 [$L_\odot$] & 0.02 \\
far IR excess ($\lambda=$29.5-991.$\mathrm{\mu m}$)
 [$L_\odot$] & 0.04 \\
\hline
\end{tabular}

\vspace{0.5cm}\centering\begin{tabular}{lrccccr}
\hline & & \multicolumn{2}{c}{line flux $[\mathrm{W\,m^{-2}}$]} & \multicolumn{2}{c}{FWHM $[\mathrm{km\,s^{-1}}]$} &  \\
species & $\lambda\,$[$\mathrm{\mu m}]$ & observed & model & observed & model & size[AU] \\
\hline
$\mathrm{Ne^+}$ & 12.814 & 5.5$\pm$1.1(-18) & 3.3(-17) &  & 15.3 & 7.4 \\
$\mathrm{OI}$ & 63.183 & 7.0$\pm$2.0(-18) & 4.6(-17) &  & 3.0 & 331.6 \\
$\mathrm{OI}$ & 145.525 & $<$3.6(-18) & 1.2(-18) &  & 8.5 & 164.3 \\
$\mathrm{CII}$ & 157.740 & $<$3.6(-18) & 4.5(-18) &  & 1.5 & 565.0 \\
$\mathrm{o\!-\!H_2O}$ & 63.323 & $<$5.4(-18) & 8.8(-19) &  & 9.3 & 11.9 \\
$\mathrm{CO}$ & 866.963 & 1.9$\pm$0.1(-19) & 3.7(-19) &  & 1.8 & 448.9 \\
$\mathrm{CO}$ & 1300.404 & 1.1$\pm$0.3(-19) & 1.2(-19) &  & 1.8 & 443.9 \\
$\mathrm{^{13}CO}$ & 1360.227 & 4.0$\pm$0.5(-20) & 1.8(-20) &  & 2.2 & 324.3 \\
$\mathrm{C^{18}O}$ & 1365.421 & 5.0$\pm$0.5(-21) & 3.8(-21) &  & 2.4 & 289.5 \\
$\mathrm{CN}$ & 881.097 & 6.7$\pm$0.7(-20) & 1.6(-19) &  & 1.7 & 470.5 \\
$\mathrm{CN}$ & 1321.390 & 3.6$\pm$0.3(-20) & 6.5(-20) &  & 1.6 & 482.0 \\
$\mathrm{HCN}$ & 845.663 & 1.9$\pm$0.2(-20) & 1.2(-19) &  & 1.8 & 415.3 \\
$\mathrm{HCN}$ & 1127.520 & 2.6$\pm$0.5(-20) & 7.6(-20) &  & 1.7 & 440.5 \\
$\mathrm{HCO^+}$ & 840.380 & 7.5$\pm$0.4(-20) & 7.2(-20) &  & 1.8 & 439.8 \\
$\mathrm{HCO^+}$ & 1120.478 & 4.8$\pm$0.1(-20) & 4.4(-20) &  & 1.7 & 461.8 \\
$\mathrm{N_2H^+}$ & 804.439 & $<$6.7(-21) & 1.2(-20) &  & 2.0 & 273.2 \\
$\mathrm{N_2H^+}$ & 1072.557 & 9.3$\pm$0.9(-21) & 7.1(-21) &  & 2.0 & 289.6 \\
\hline
\end{tabular}

\label{tab:DMTau}
\end{table}

\noindent{\sffamily\bfseries\underline{12.\ \ DM\,Tau}}\ \ This is a
simple model on top of the SED-fit, so it does not necessarily provide
any further insight about the disk mass and gas/dust ratio. DM Tau is
one of the largest, brightest, and best-studied T Tauri disks. Its SED
(see Fig.\,\ref{fig:SEDs}) is similar to TW Hya and GM Aur, showing
the typical features of transitional disks, where the near-IR excess
is mostly lacking. For the inner disk we find an increasing surface
density profile. The massive outer disk starts at 13.5\,AU in the
model (Fig.~\ref{fig:DMTau-1}). Most (sub-)mm lines fit well,
including HCO$^+$ (Fig. 21) and N$_2$H$^+$ lines. The fit of the HCN
and CN lines is less convincing. The CO 2-1 isotopologue line show
that the model may be a bit too extended. The [O\,{\sc i}]\,63\,$\mu$m
line is too strong by a factor of seven, although the outer disk is
quite flat and already partly shielded by a tall inner
disk. Table~\ref{tab:DMTau} provides an overview of all line
results. The very faint near-IR excess does not allow to have much
more shielding by the inner disk (as in the case of CY Tau or GM Aur)
to further reduce the UV and X-ray irradiation of the outer disk,
which would weaken the [O\,{\sc i}]\,63\,$\mu$m line. This issue might
be related with the different X-ray luminosities reported in the
literature (factor of ten, see Sect.~\ref{sec:stellar} for
details). But we have not tested the older lower X-ray luminosity with
our model, furthermore a strong variability in X-rays is also a
possible scenario.

\citet{McClure2016} derived a disk gas mass in the range of
$1.0-4.7\times10^{-2}\,M_\mathrm{\odot}$ based on measurements of the
HD$\,J\!=1\!-\!0$ spectral line and suggests a lower CO gas phase
abundance of up to a factor of 5 compared to the canonical abundance
of $\approx10^{-4}$ (relative to molecular hydrogen), but the case of
no CO depletion is also possible. This is consistent with our model
with a disk gas mass of $1.6\times10^{-2}\,M_\mathrm{\odot}$ and no CO
depletion additionally to freeze-out. We note that the HD line was not
included in our modeling.

As the disk of DM Tau is large and bright it is a popular target for
molecular line observations and disk chemistry studies
\citep[e.g.][]{Dutrey1997b,Oberg2010c,Teague2015,Loomis2015,Semenov2018}. One
interesting example is the detection of C$_2$H
\citep{Henning2010n,Bergin2016} in DM Tau. The authors suggest that
the X-ray/UV radiation and dust evolution play a crucial role for the
C$_2$H abundance. Our model and the collected data provides additional
constraints on the disk physical structure (e.g. radiation fields) and
is therefore a ideal test-bed for further chemical studies although a
more elaborate chemical network is likely required.

\clearpage
\newpage

\begin{figure}[!t]
\centering
\vspace*{2mm}
{\large\sffamily\bfseries\underline{CY\,Tau}}\\[2mm]
$M_{\rm gas} = 0.12\,M_\odot$, 
$M_{\rm dust} = 1.2\times 10^{-3}\,M_\odot$, 
dust settling $\alpha_{\rm settle} = 6.4\times10^{-5}$\\
\hspace*{-4mm}
\begin{tabular}{cc}
\begin{minipage}{75mm}
\includegraphics[width=79mm]{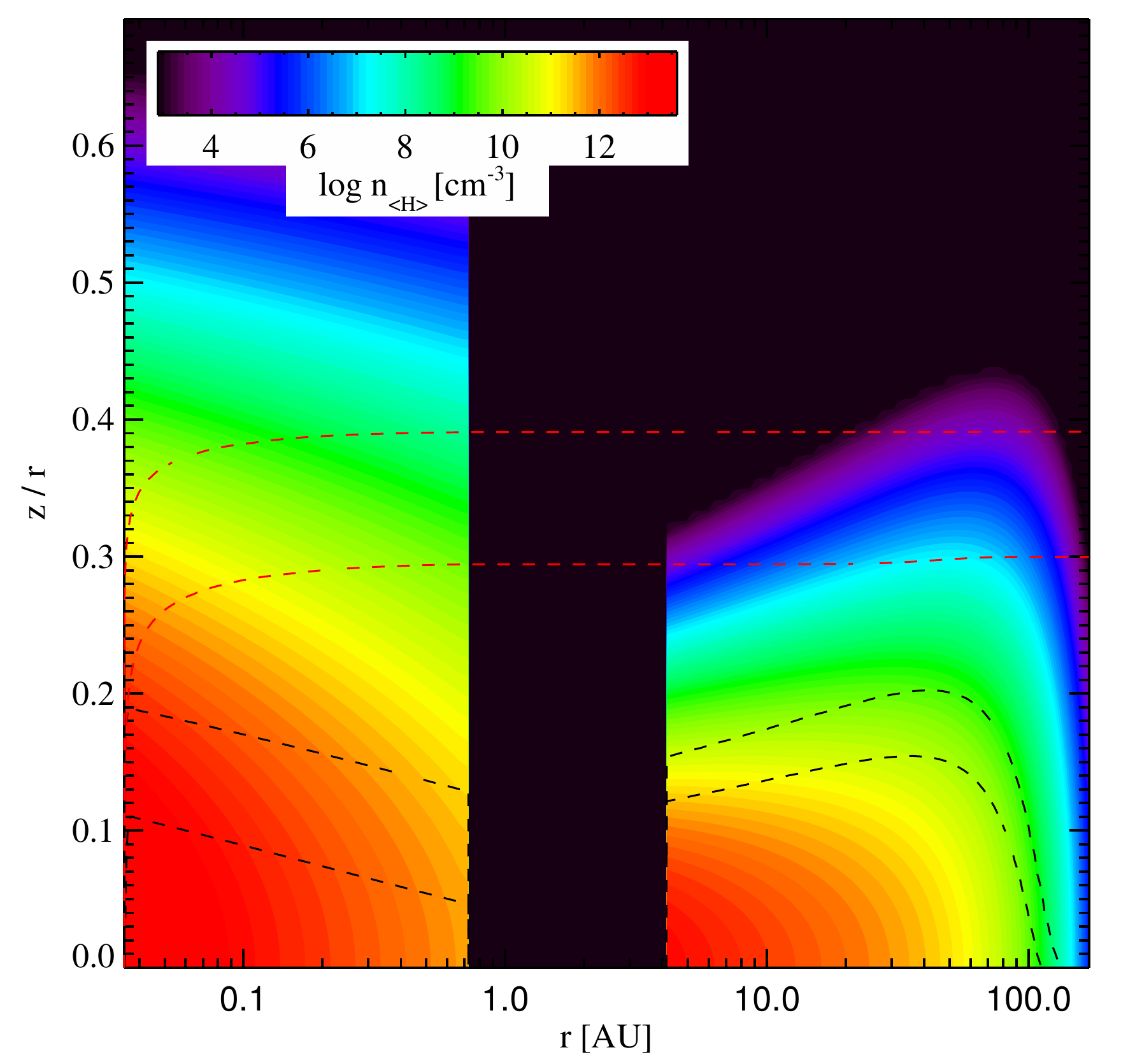}
\end{minipage}
&
\begin{minipage}{87mm}
\vspace*{1mm}
\includegraphics[width=87mm]{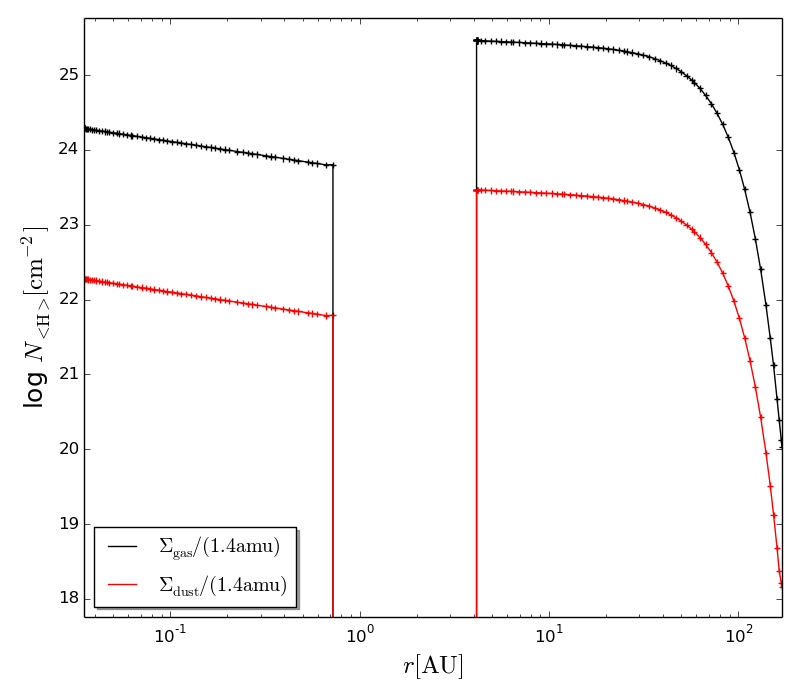}
\end{minipage}
\end{tabular}\\[-1mm]
\caption{CY\,Tau density and surface density plots. For details see Fig.\,\ref{fig:ABAur-1}.}
\label{fig:CYTau-1}
\end{figure}

\begin{figure}[!h]
\centering
\includegraphics[width=150mm]{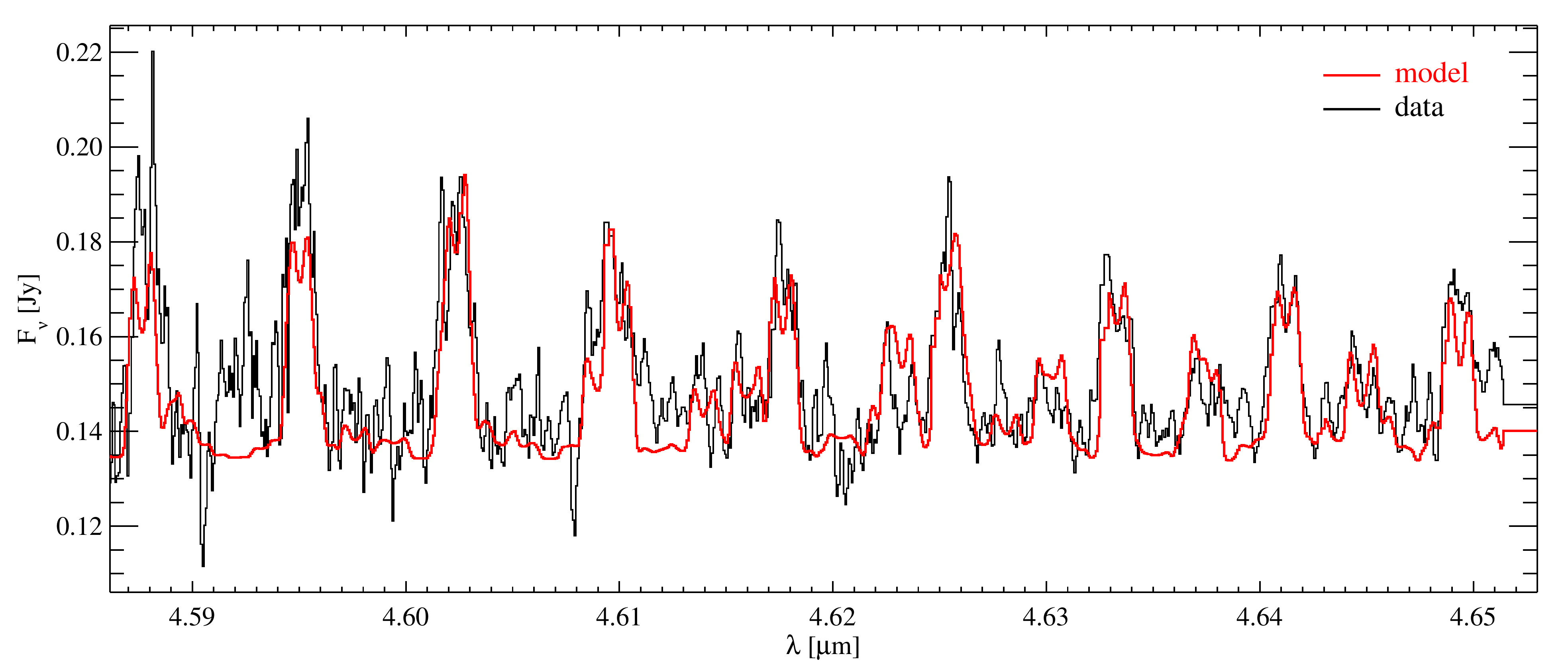}
\vspace*{-2mm}
\caption{Comparison of the observed and modeled CO ro-vibrational spectrum with continuum.}
\label{fig:CYTau-2}
\end{figure}

\begin{table}
\caption{Model properties and comparison of computed spectral line
  properties with observations from {\sf CYTau.properties}. }
\centering\begin{tabular}{rr}
\hline
DIANA standard fit model properties &  \\
\hline
minimum dust temperature [K] & 5 \\
maximum dust temperature [K] & 1683 \\
mass-mean dust temperature [K] & 9 \\
minimum  gas temperature [K] & 5 \\
maximum  gas temperature [K] & 40000 \\
mass-mean  gas temperature [K] & 12 \\
mm-opacity-slope (0.85-1.3)mm & 1.3 \\
cm-opacity-slope (5-10)mm & 1.7 \\
mm-SED-slope (0.85-1.3)mm & 1.6 \\
cm-SED-slope (5-10)mm & 3.0 \\
10$\mathrm{\mu m}$ silicate emission amplitude & 1.6 \\
naked star luminosity [$L_\odot$] & 0.4 \\
bolometric luminosity [$L_\odot$] & 0.5 \\
near IR excess ($\lambda=$2.07-6.96$\mathrm{\mu m}$)
 [$L_\odot$] & 0.06 \\
mid IR excess ($\lambda=$6.96-29.5$\mathrm{\mu m}$)
 [$L_\odot$] & 0.03 \\
far IR excess ($\lambda=$29.5-973.$\mathrm{\mu m}$)
 [$L_\odot$] & 0.01 \\
\hline
\end{tabular}

\vspace{0.5cm}\centering\begin{tabular}{lrccccr}
\hline & & \multicolumn{2}{c}{line flux $[\mathrm{W\,m^{-2}}$]} & \multicolumn{2}{c}{FWHM $[\mathrm{km\,s^{-1}}]$} &  \\
species & $\lambda\,$[$\mathrm{\mu m}]$ & observed & model & observed & model & size[AU] \\
\hline
$\mathrm{CO}$ & 1300.403 & 1.6$\pm$0.3(-20) & 2.0(-20) & 2.4$\pm$0.2 & 2.3 & 199.7 \\
$\mathrm{^{13}CO}$ & 1360.227 & 7.8$\pm$0.9(-21) & 6.0(-21) & 2.5$\pm$0.2 & 2.6 & 199.3 \\
$\mathrm{OI}$ & 63.183 & 1.2$\pm$0.1(-17) & 5.6(-18) &  & 3.2 & 143.2 \\
$\mathrm{OI}$ & 0.630 & 7.8$\pm$0.5(-18) & 9.4(-18) & 42.0$\pm$3.0 & 59.0 & 0.4 \\
$\mathrm{o\!-\!H_2O}$ & 63.323 & $<$1.2(-17) & 1.4(-19) &  & 29.8 & 0.7 \\
$\mathrm{CO}$ & 4.633 & 6.0$\pm$2.0(-18) & 6.6(-18) & 75.0$\pm$11.0 & 14.0 & 4.1 \\
$\mathrm{CO}$ & 4.920 & 5.0$\pm$1.0(-18) & 4.2(-18) & 86.0$\pm$9.0 & 92.1 & 0.1 \\
\hline
\end{tabular}

\label{tab:CYTau}
\end{table}

\noindent{\sffamily\bfseries\underline{13.\ \ CY\,Tau}}\ \ has an SED
(see Fig.\,\ref{fig:SEDs}) with almost no 10\,$\mu$m and 20\,$\mu$m
silicate emission features, that is steeply declining in the mid-IR,
and has only a modest far-IR excess. At longer wavelengths this excess
drops steadily all the way up to about 850\,$\mu$m. The mm-fluxes are
strong.  The model explains this particular SED-shape by a very cold
yet massive and strongly settled outer disk which is located in the
shadow of a tenuous, tall inner disk (Fig.~\ref{fig:CYTau-1}).
 
The gas/dust ratio was fixed at 100 during the model fitting. The
total disk mass of $0.12\,M_\odot$ is exceptionally large for this
$M_\star\!=\!0.43\,M_\odot$ T\,Tauri star, even larger than the value
$0.10\,M_\odot$ obtained from the pure SED-fit, see
Table~\ref{tab:dustmass}. All observed emission lines are predicted
well (Table~\ref{tab:CYTau}). The CO $J\!=$2-1 to $^{13}$CO $J\!=$2-1
line ratio is observed to be quite small, only $\sim\,$2. The model
manages to explain this peculiar line ratio by an almost sharp outer
edge with $\epsilon\!=\!0.11$ and $\gamma\!=\!-0.34$.

Figure~\ref{fig:CYTau-2} shows a FLiTs spectrum for
the $R$-branch of fundamental $\rm CO\ v\!=\!1\!-\!0$, with a very good
fit of a number of individual lines including some $\rm
CO\ v\!=\!2\!-\!1$ and o-H$_2$O lines, which was only obtained after
lowering the column densities in the inner disk zone while
simultaneously increasing the scale heights. 

\clearpage
\newpage

\begin{figure}[!t]
\vspace*{2mm}
\centering
{\large\sffamily\bfseries\underline{RECX\,15}}\\[2mm]
$M_{\rm gas} = 1.2\times 10^{-4}\,M_\odot$,
$M_{\rm dust} = 3.5\times 10^{-8}\,M_\odot$, 
dust settling $\alpha_{\rm settle} = 1.0\times 10^{-4}$\\
\hspace*{-4mm}
\begin{tabular}{cc}
\begin{minipage}{75mm}
\includegraphics[width=79mm]{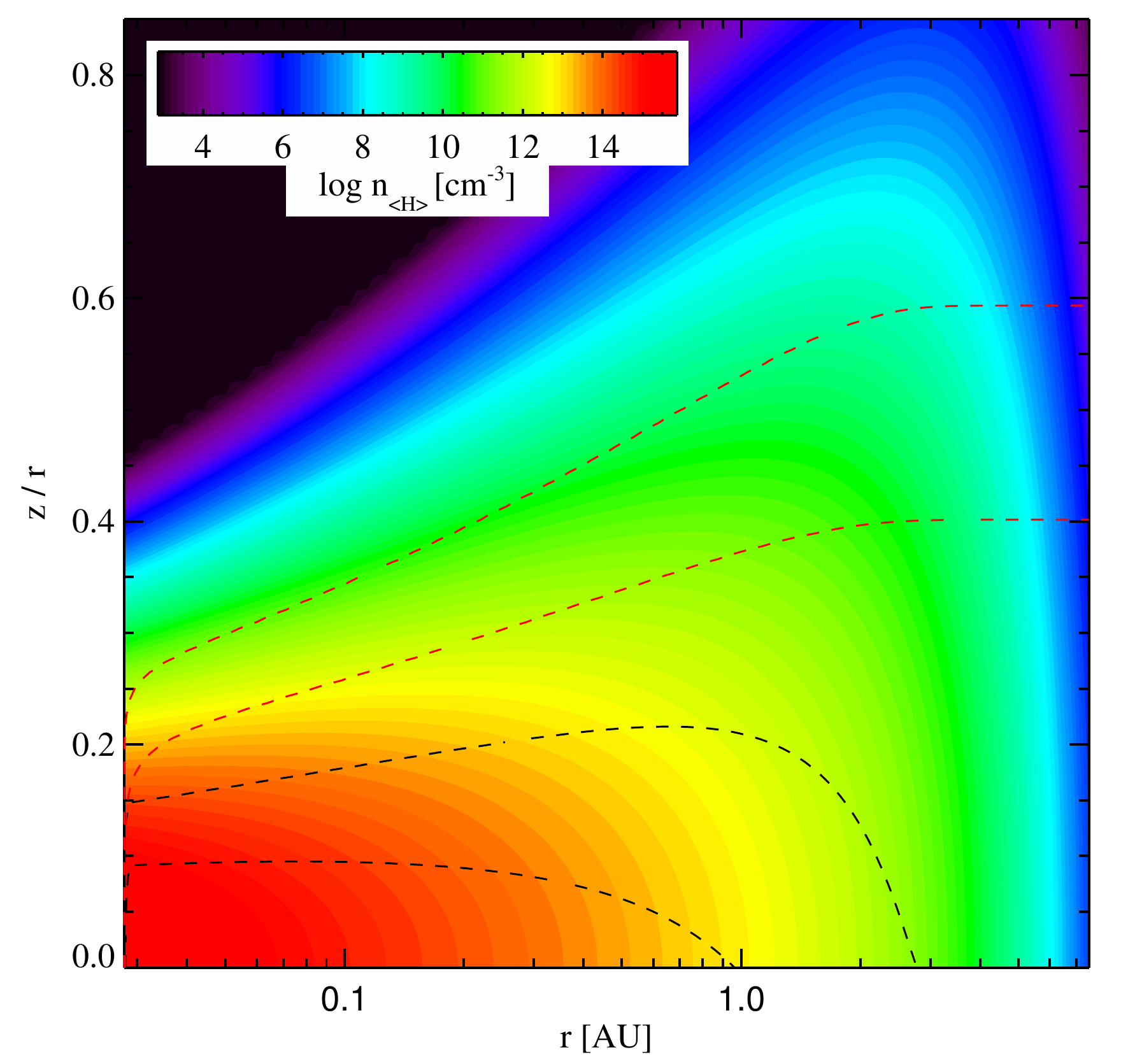}
\end{minipage}
&
\begin{minipage}{87mm}
\vspace*{1mm}
\includegraphics[width=87mm]{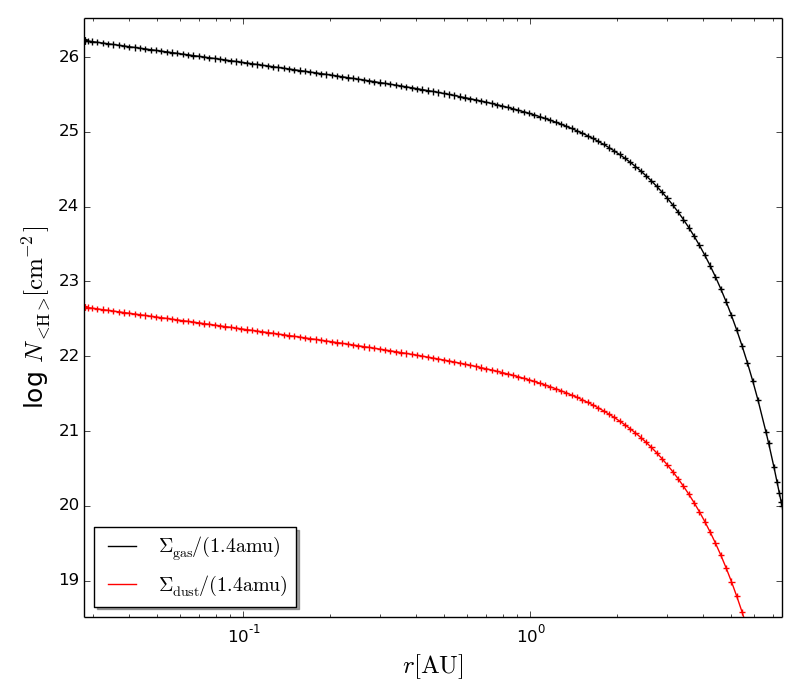}
\end{minipage}
\end{tabular}\\[-1mm]
\caption{RECX\,15 density and surface density plots. For details see Fig.\,\ref{fig:ABAur-1}.}
\label{fig:RECX15-1}
\end{figure}

\begin{figure}[!h]
\centering
\includegraphics[width=85mm]{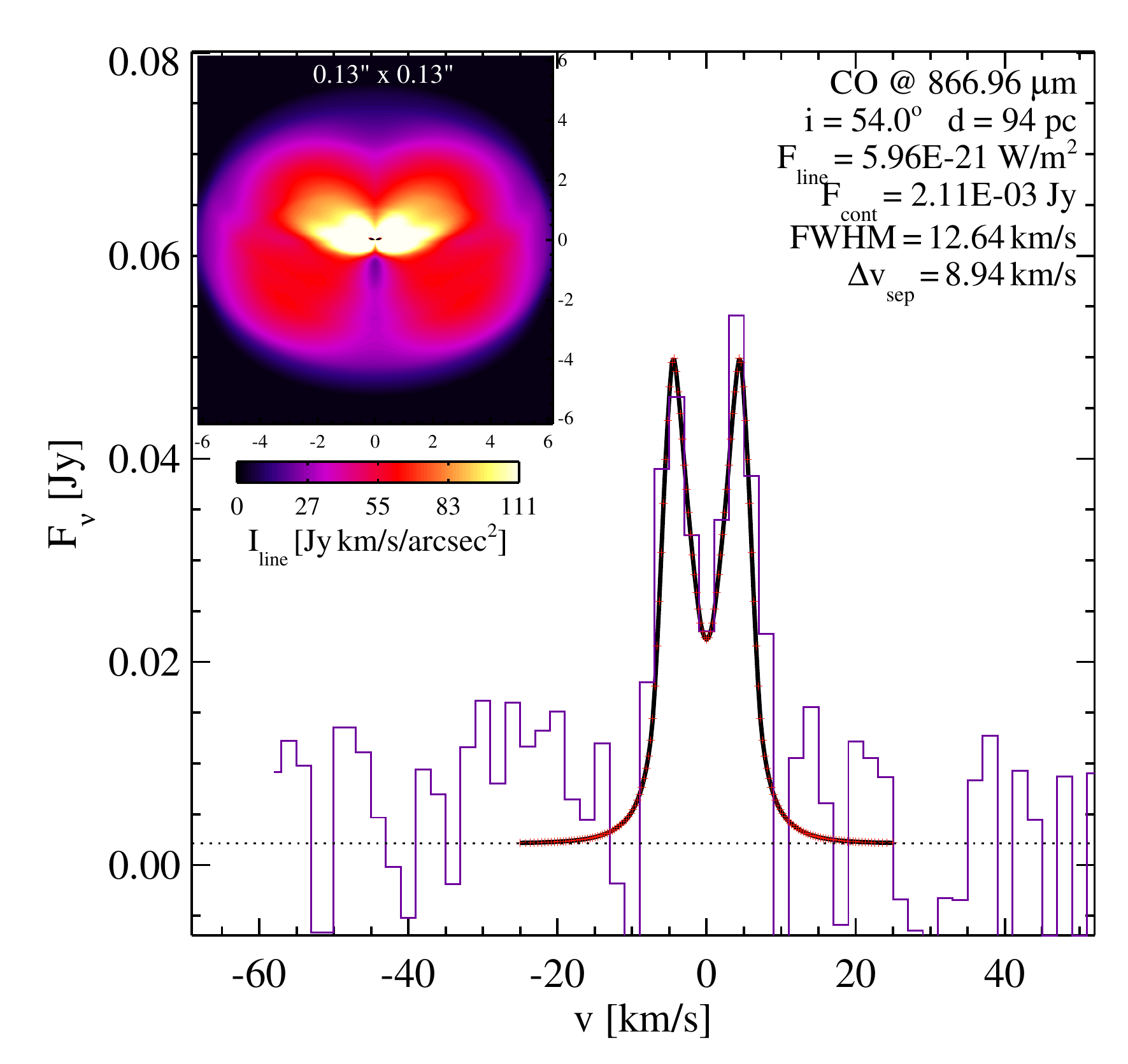}
\vspace*{-2mm}
\caption{Model fit to our CO 3-2 ALMA cycle-0 observations of RECX\,15.}
\label{fig:RECX15-2}
\end{figure}

\begin{table}
\caption{Model properties and comparison of computed spectral
      line properties with observations from {\sf
        RECX15.properties}. }
\centering\begin{tabular}{rr}
\hline
DIANA standard fit model properties &  \\
\hline
minimum dust temperature [K] & 32 \\
maximum dust temperature [K] & 1396 \\
mass-mean dust temperature [K] & 86 \\
minimum  gas temperature [K] & 44 \\
maximum  gas temperature [K] & 26982 \\
mass-mean  gas temperature [K] & 116 \\
mm-opacity-slope (0.85-1.3)mm & 0.8 \\
cm-opacity-slope (5-10)mm & 1.6 \\
mm-SED-slope (0.85-1.3)mm & 2.5 \\
cm-SED-slope (5-10)mm & 3.5 \\
10$\mathrm{\mu m}$ silicate emission amplitude & 1.4 \\
naked star luminosity [$L_\odot$] & 0.1 \\
bolometric luminosity [$L_\odot$] & 0.2 \\
near IR excess ($\lambda=$2.02-6.72$\mathrm{\mu m}$)
 [$L_\odot$] & 0.02 \\
mid IR excess ($\lambda=$6.72-29.5$\mathrm{\mu m}$)
 [$L_\odot$] & 0.02 \\
far IR excess ($\lambda=$29.5-991.$\mathrm{\mu m}$)
 [$L_\odot$] & 0.01 \\
\hline
\end{tabular}

\vspace{0.5cm}\centering\begin{tabular}{lrccccr}
\hline & & \multicolumn{2}{c}{line flux $[\mathrm{W\,m^{-2}}$]} & \multicolumn{2}{c}{FWHM $[\mathrm{km\,s^{-1}}]$} &  \\
species & $\lambda\,$[$\mathrm{\mu m}]$ & observed & model & observed & model & size[AU] \\
\hline
$\mathrm{CO}$ & 866.963 & 8.2$\pm$0.6(-21) & 6.0(-21) & 14.9$\pm$2.3 & 12.6 & 5.4 \\
$\mathrm{OI}$ & 63.183 & 3.0$\pm$0.3(-17) & 2.6(-17) &  & 13.8 & 6.0 \\
$\mathrm{OI}$ & 0.630 & 6.5$\pm$2.5(-17) & 1.3(-16) & 38.0$\pm$10.0 & 20.8 & 3.4 \\
$\mathrm{o\!-\!H_2}$ & 2.121 & 2.5$\pm$0.1(-18) & 1.7(-19) & 18.0$\pm$1.2 & 66.9 & 2.0 \\
$\mathrm{OI}$ & 145.525 & $<$6.0(-18) & 1.4(-18) &  & 13.2 & 5.5 \\
$\mathrm{CII}$ & 157.740 & $<$9.0(-18) & 1.8(-19) &  & 14.0 & 6.0 \\
$\mathrm{CO}$ & 90.162 & $<$9.6(-18) & 7.3(-19) &  & 19.9 & 3.7 \\
$\mathrm{CO}$ & 79.359 & $<$2.4(-17) & 5.6(-19) &  & 25.0 & 2.3 \\
$\mathrm{CO}$ & 72.842 & $<$8.1(-18) & 4.7(-19) &  & 28.7 & 2.1 \\
$\mathrm{o\!-\!H_2O}$ & 180.488 & $<$5.1(-18) & 9.2(-20) &  & 24.2 & 4.3 \\
$\mathrm{o\!-\!H_2O}$ & 179.526 & $<$5.1(-18) & 1.2(-19) &  & 20.1 & 4.8 \\
$\mathrm{o\!-\!H_2O}$ & 78.742 & $<$3.0(-17) & 1.2(-18) &  & 21.3 & 4.8 \\
$\mathrm{p\!-\!H_2O}$ & 89.988 & $<$9.6(-18) & 6.8(-19) &  & 24.2 & 4.4 \\
\hline
\end{tabular}

\label{tab:RECX15}
\end{table}

\noindent{\sffamily\bfseries\underline{14.\ \ RECX\,15}}\ \ is an
exceptional case, the only gas-rich protoplanetary disk in an
otherwise rather old star formation cluster, which seems to be as
small as 5\,AU in radius \citep{Woitke2011, Woitke2013}.
Figure~\ref{fig:RECX15-1} shows the density structure and surface
density profiles (gas and dust) for this object. RECX\,15 must have
lost its outer disk for some reason, possibly due to a close
encounter. Only few gas lines have been detected, among them [O\,{\sc
    i}]\,63\,$\mu$m, [O\,{\sc i}]\,6300\,\AA\ and o-H$_2$
2.121\,$\mu$m. Our ALMA cycle-0 data shows a faint, very broad CO
$J\!=$3-2 line (FWHM\,$\approx\!15\,$km/s) which only a tiny Keplerian
disk can explain (Fig.~\ref{fig:RECX15-2}).  There is no spatially
resolved data, not even with ALMA.  The model manages to fit the SED
(Fig.\,\ref{fig:SEDs}) and the emission lines with a strongly flared,
tall disk, strengthening our general conclusion that the inner disks
of T\,Tauri stars are vertically much more extended than previously
thought. The o-H$_2$ 2.121\,$\mu$m line is underpredicted by a factor
15, though. The gas/dust mass ratio is found to be $\sim$\,3500, an
very unusual value for outer disks, but maybe not so atypical for the
inner disks of transitional disks (Table~\ref{tab:dustmass}). All line
fits are summarized in Table~\ref{tab:RECX15}.

\clearpage
\newpage

\subsection{Systematic line flux deviations}

\noindent We conclude this study by looking out for possible
systematic weaknesses in our models, for example lines that are always
predicted to be too strong or too weak. This would indicate some
principle problem in our modeling assumptions or techniques, for
example that certain molecules are always underabundant or
overabundant with respect to observations due to an issue in the
chemistry.  Figures~\ref{fig:Lines1} and \ref{fig:Lines2} show the
ratios of predicted to observed line fluxes, for a sample of
frequently observed emission lines. We note here again that all lines have
been simultaneously fitted by one model for each object, which at the
same time fits the continuum observations (SED and some images) as
good as possible.

These figures do not reveal any severe weaknesses. The CO 3-2 and CO
2-1 isotopologue lines are typically well-fitted within a factor of
two or better. The [O\,{\sc i}]\,6300\,\AA\ line maybe somewhat
underpredicted for luminous stars, whereas the fits are fine for the
T\,Tauri stars.  The [O\,{\sc i}]\,63\,$\mu$m line proves to be quite
difficult to fit, we arrive at deviations within a factor 4 or less,
with a slight tendency to overpredict. The other lines shown in
Fig.\,\ref{fig:Lines2} show no obvious trends either. The frequently
observed HCO$^+$ 3-2 line usually fits fine, as well as the N$_2$H$^+$
line, whereas the situation is more diverse for the HCN 3-2 and CN
lines. An exception is AB\,Aur where the line data might be confused
with foreground cloud absorption. The CO $J\!=$18-17 line tends to be
somewhat underpredicted by our models for the T\,Tauri stars.

\begin{figure}[!t]
  \centering
  \hspace*{-5mm}\includegraphics[width=100mm]{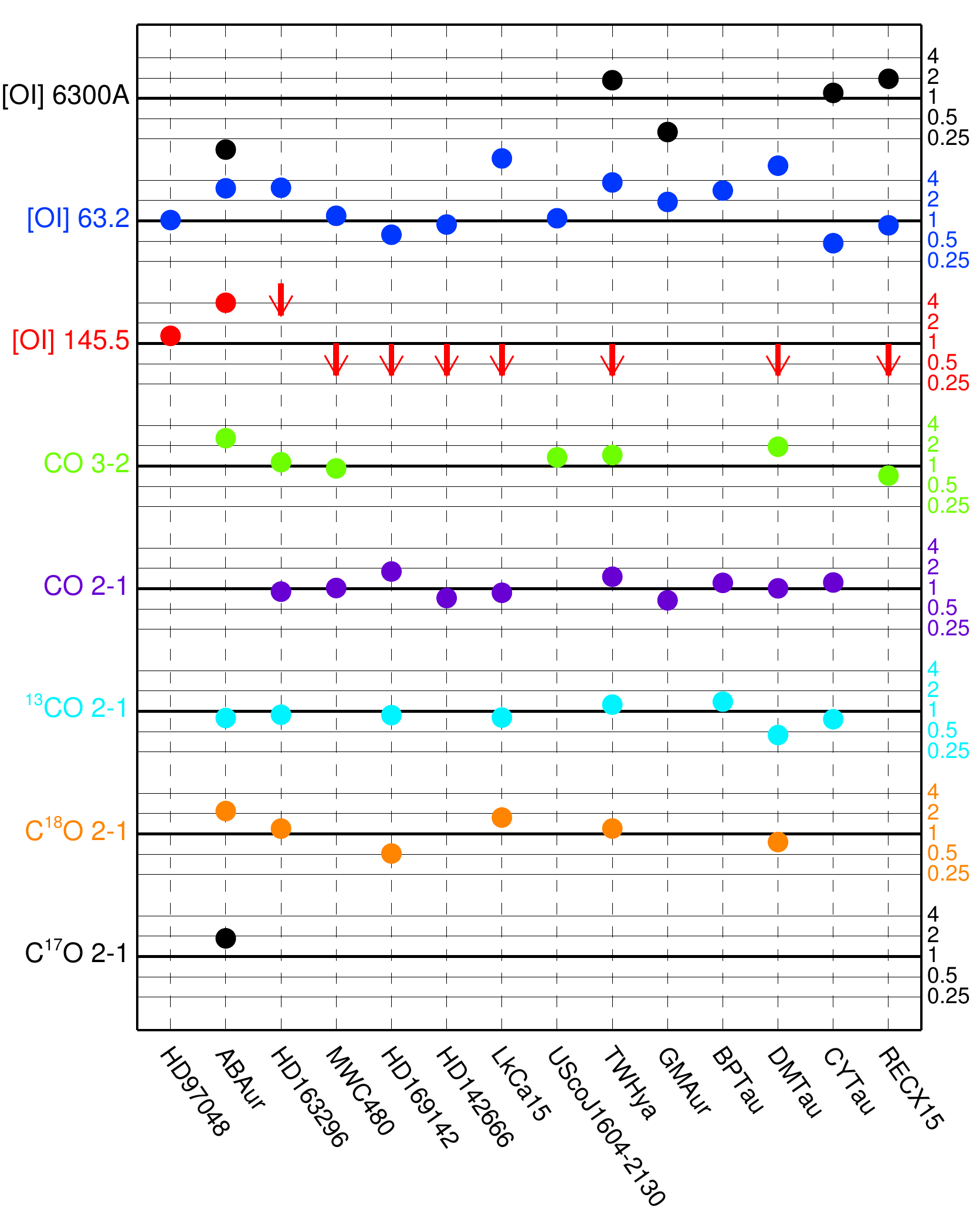}\\[-2mm]
  \caption{Summary of line fitting results from the {\sf
      DIANA}-standard models. The $y$-axis shows the line fluxes as
    predicted by the models divided by the observed line fluxes $F_{\rm
      line}^{\rm mod}/F_{\rm line}^{\rm obs}$ on log-scalings as
    indicated on the right. If a dot is positioned at 0.5, for
    example, it means that the model underpredicts the line flux by a
    factor of 2. An arrow indicates that the observation is an upper
    limit. An arrow starting at 1 indicates a good match -- the model
    predicts a line flux lower or equal to the observational
    $3\,\sigma$ upper limit. If, however, the arrow starts at 2, it
    means that the model predicts a line flux that is a factor 2
    larger than the $3\,\sigma$ upper limit. Systematic errors in the
    observations are typically of order $(10-30)$\%.}
  \label{fig:Lines1}
\end{figure}

\begin{figure}
  \centering
  \hspace*{-5mm}\includegraphics[width=110mm]{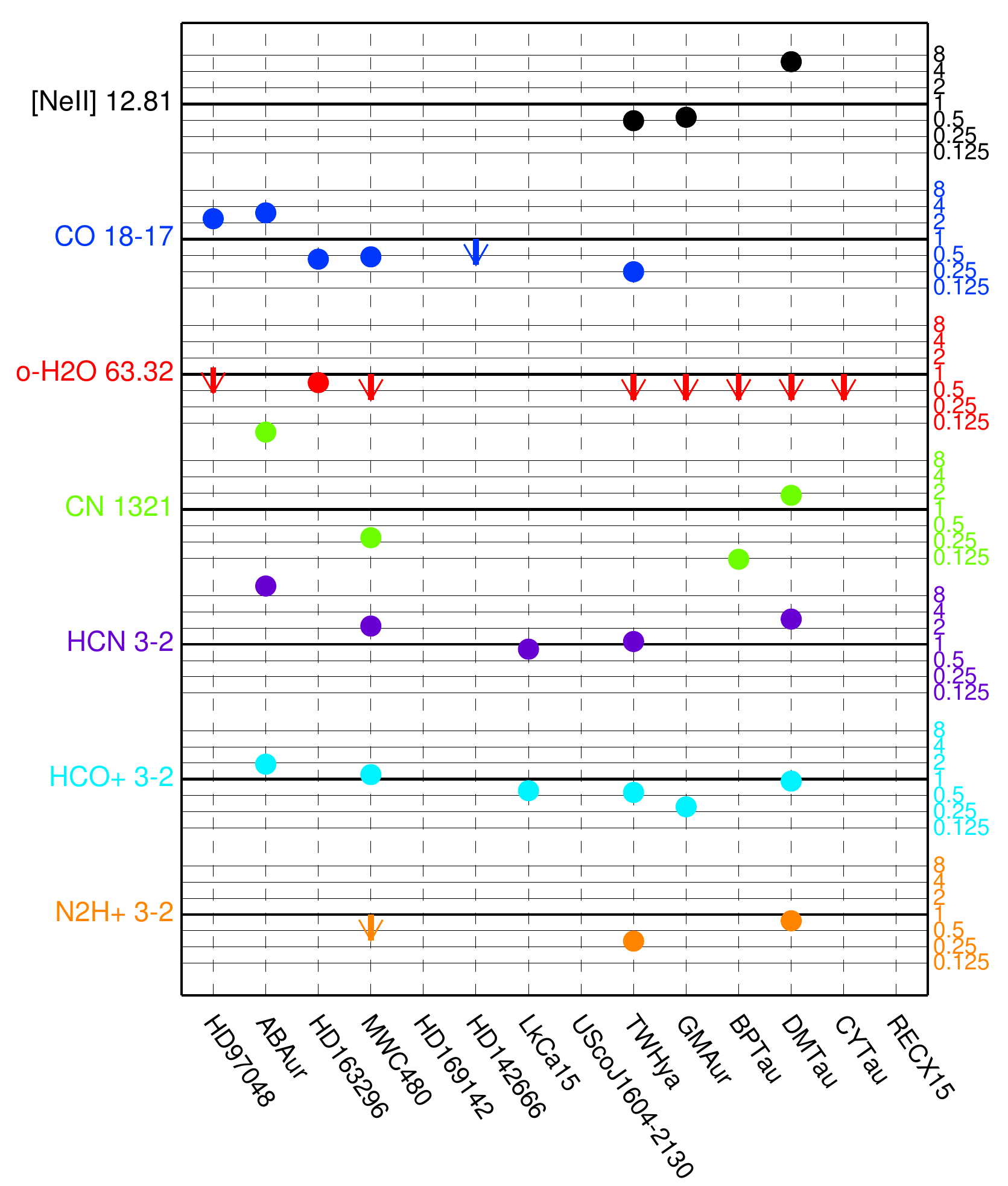}\\[-3mm]
 \caption{continued from Fig.~\ref{fig:Lines1}.}
  \label{fig:Lines2}
\end{figure}

\section{Summary and Outlook}

\noindent The European FP7 project {\sf DIANA} has performed a
coherent analysis of a large set of observational data of
protoplanetary disks by means of state-of-the-art thermo-chemical disk
models.  We used multi-wavelength, multi-instrument, mostly archival
data comprising photometry, low-resolution IR to far-IR spectra,
continuum images and line observations of different kinds and
different quality.  {Our goal was to fit all collected
  observational data simultaneously by means of a single disk model,
  separately for each object in our target list. Our aim was} to
conclude about the disk shape, the masses of the disks and their
physical parameters, the properties of the dust grains, the internal
gas and dust temperature structures in the disks, and the chemical
composition.  The driving question of the project was {\sl ``does this
  work?''} Can we fit all available observational data with a
standardized modeling approach, using 2D thermo-chemical disk models,
only by varying the disk mass and shape parameters, the gas/dust ratio
and the dust size and opacity parameters?
 
The answer is a surprisingly clear yes.  In reality, disks are very
complicated objects, individual, time-dependent and not strictly
axi-symmetric. Yet, by allowing just for two disk zones, an inner and
an outer disk, we could find parameter combinations for our models,
which predict observations that resemble most of the continuum
and line data we could find, simultaneously. In particular, we did not
need individual adjustments of element abundances, but achieved all our
fits by using standard ISM element abundances with strongly depleted
heavy elements.

Our data analysis was performed in three steps, (i) finding the
stellar parameters including detailed UV and X-ray properties, (ii)
using fast Monte-Carlo radiative transfer models to fit the SED in
order to roughly determine the disk shape and dust opacity parameters,
and (iii) using self-consistent radiation thermo-chemical models in
application to an enlarged set of observational data, including all
line and image data, to complete the fit. During the last modeling
phase, the various disk shape, dust settling and opacity parameters
were not frozen, but were continued to be varied for finding the
best-fitting parameter values. This procedure distinguishes our work
from most previous studies.

For most objects, however, we found at least one observation which we
could not fit at all together with the other observations. This could
be caused, for example, by the variability of an object or simply due
to issues with foreground clouds or secondary sources in the field of
view observed with different instruments. It could also be caused, of
course, by some missing physics or chemistry in our models, but as
Figs.~\ref{fig:Lines1} and \ref{fig:Lines2} show, we could not find
any systematic problem.  In case of such unclear or un-fittable
data, the only practical way forward was to exclude such data or to
artificially increase the respective measurement errors, otherwise the
$\chi^2$-minimization desperately tries to improve the fit of exactly
those data which we trust the least. Thus, our data selection and
definition of $\chi^2$ was not mathematically sound; it was based on
and required human judgment.  We therefore do not claim to have found
a unique solution of the disk structure of our targets.

An MCMC analysis might have revealed some interesting parameter
degeneracies and credibility intervals, but was judged to be
computationally unfeasible.  A single thermo-chemical disk model takes
about 10 CPU hours, the number of free parameters is about 20, and so
hundreds of thousands of disk models would have needed to be
calculated to determine those errorbars, which would have taken about 
500 CPU-years for a single object. 

Therefore, the intention of this project was not to determine all disk
parameters with errorbars\footnote{Appendix~\ref{app:uncertainties}
  contains a rough estimate of parameter uncertainties by probing the
  curvature in the local $\chi^2$ minimum.}.  Instead, we have
found some re-occurring patterns and trends in the models that helped
us to better fit the data, which offers new ways to understand and
explain disk observations in general, new clues for data
interpretation, and useful starting points for future modeling
purposes. We summarize these findings below.

\vspace{1.5mm}\noindent {\bf Dust properties:}\ \ The key to arrive at
our simultaneous continuum and line fits often was to vary the dust
size and opacity parameters that have an important influence on how
deep the stellar UV photons penetrate into the disk, causing gas
heating and line emission.  We used the DIANA dust opacities for disks
\citep{Min2016}, based on a power-law dust size distribution with an
effective mixture of laboratory silicate and amorphous carbon. Since
our size distribution is typically extended to a few millimeters, our
dust is much more transparent in the UV than standard interstellar
dust, but rather opaque at 850\,$\mu$m, $\kappa^{\rm abs}_{850}\approx
6.3^{+3.5}_{-2.3}\rm\,cm^2/g(dust)$, which is somewhat larger than the
frequently used value of 3.5\,cm$^2$/g(dust) originally proposed by
\citet{Beckwith1990} as order-of-magnitude estimate of
10\,cm$^2$/g(dust) at 1000\,GHz, and later scaled to 850\,$\mu$m using
$\kappa^{\rm abs}_\nu\!\propto\!1/\lambda$.

\vspace{1.5mm}\noindent {\bf Dust settling:}\ \ Another important
degree of freedom for the fitting was the dust settling, which has
opposite effects on continuum and line fluxes in the mid and far-IR,
and can hence be used to break some degeneracies with disk flaring as
known from pure SED-fitting. Our results suggest that dust settling is
rather strong in these disks, and hence the turbulence rather weak,
$\log_{10}\alpha_{\rm set}=-2.9\pm0.9$ as compared to the standard
value of $10^{-2}$. This is in agreement with the analysis by
\citet{Pinte2016} of the HL\,Tau rings as seen with ALMA.

\vspace{1.5mm}\noindent {\bf PAH properties:}\ \ Our simultaneous fits
of the PAH properties in protoplanetary disks show an underabundance
of PAHs $f_{\rm PAH} \approx 0.005 - 0.8$ with respect to the standard
abundance in the interstellar medium \citep[$10^{-6.52}$ PAH
  molecules/H-nucleus][]{Tielens2008}, with large individual scatter.
The PAH abundance is relevant to the model, in particular, by
converting blue and soft UV photon energies into heat, which leads to
additional line emission. We have solely considered small PAHs
(circumcononene) with 54 carbon atoms, and our fits to a few Herbig~Ae
stars, where multiple mid-IR PAH emission features have been detected,
show that these PAHs are mostly charged (60\%-98\%).  We did not
consider PAHs in disk gaps as discussed by \citep{Maaskant2014}.

\vspace{1.5mm}\noindent {\bf Two-zone disks:}\ \ In 9 out of 27 cases,
we decided to switch from a one-zone disk setup to a two-zone setup
already during the SED-fitting phase. This was found to be necessary
to fit certain properties in the photometric and low-resolution
spectroscopic data including the PAH emission features. Some of these
objects are well-known transitional disks, but for others this is not
so clear, for example CY\,Tau, where the SED points to a massive yet
very cold outer disk. Such disk properties can be produced by setting
up a semi-transparent tall inner disk that casts a shadow onto the
outer disk.

\vspace{1.5mm}\noindent {\bf Gas-rich, tall inner disks:}\ \ The
two-zone scenario was often found to be useful to explain objects
with very faint far-IR lines. In fact, 12 out of our 14 completed
DIANA standard models used a two-zone setup. These tall inner
disks shield the outer disk from the UV irradiation by the star, hence
lead to less disk heating and line emission. Such inner disks do not
obey the condition of hydrostatic equilibrium, in particular those of
the T\,Tauri stars, so their nature can be disputed. However, we could
not find any observation that could rule out such a scenario.  On the
contrary, having that tenuous, tall inner disk definitely helps to
explain some of the strong optical lines (such as [O\,{\sc
    i}]\,6300\,\AA) and strong near-IR lines (such as CO fundamental),
which are preferentially emitted by these inner disks. Interestingly,
the gas/dust ratio in these inner disk zones was often found to be
large, up to 90000 for HD\,163296, but never smaller than the
standard value of 100.

\vspace{1.5mm}\noindent 
{\bf Dust masses and cold disks:}\ \ We found that the classical dust
mass-determination method according to Eq.\,(\ref{eq1}) seems not
entirely justified.  According to our results, disks can be very cold
in the midplane $<\!10\,$K, and since a large fraction of the disk
mass resides in those cold midplane areas, emission at 850\,$\mu$m is
generally far less intense than expected from the Rayleigh-Jeans
limit. This finding is supported by \citet{Guilloteau2016} who found
temperatures as low as $5-7\,$K for the large grains in the Flying Saucer.
We find the mean disk temperatures to vary by an order of magnitude among
individual disks. Disks can be optically thick even at mm-wavelengths.
We therefore found disk masses that are often much larger than stated
elsewhere, see Table~\ref{tab:discpara}. The determination of dust
sizes from the SED millimeter-slope is equally affected by these
physical effects. We found that the dust mm-opacity-slope is only weakly
correlated to the resulting SED-slope.
 
\vspace{1.5mm}\noindent
{\bf Dust/gas ratio:}\ \ Concerning the gas/dust ratio in the outer
disk, our results are inconclusive.  Disregarding those DIANA standard
models where $\rm gas/dust=100$ was enforced, the remaining 8 models
have $\rm gas/dust=120^{+170}_{-70}$.

\vspace{3mm}\noindent We offer all collected observational data at
\url{http://www.univie.ac.at/diana} and provide all modeling results
at \url{http://www-star.st-and.ac.uk/~pw31/DIANA/DIANAstandard}.  The
2D physico-chemical structures of our objects can be downloaded from
these internet portals for further inspection and analysis.  All model
parameters are offered in an easy-to-use format, see
App.~\ref{app:uploaded-data}, and all users of {\sf MCFOST}, {\sf
  MCMax} or {\sf ProDiMo} can download setup files to re-run our disk
models. This ensures the transparency of our modeling work, and their
re-use in future applications.

The use of the {\sf DIANA} disk models beyond the project has already
started. \citet{Stolker2017} used the {\sf DIANA} SED model for the
interpretation of multi-epoch VLT/SPHERE scattered light images of
HD\,135344B. They use our model to illustrate that a small
misalignment between the inner and the outer disk ($2^{\circ}.6$) can
explain the observed broad, quasi-stationary shadowing in
north-northwest direction.  The effects of inclined inner disks on
line fluxes would be worth studying in the future.  As a second
example, \citet{Muro-Arena2018} used the {\sf DIANA} SED model of
HD\,163296 to analyze the combined VLT/SPHERE and ALMA continuum
dataset for this object. Even though clear ring structures are seen
and the model is eventually refined to show a surface density
modulation, the average surface density profile stays the same as in
the {\sf DIANA} SED model. This illustrates that in some cases, the
detailed disk substructures are minor modulations of the global 2D
disk models. These substructures are key for the planet formation
processes, but have only little effect on global observables such SEDs
and unresolved line observations that arise from an extended radial
region of the disk.  Lastly, the MWC\,480 and Lk\,Ca\,15 disk models
can serve as an interesting starting point for the analysis of the
full ALMA spectral scan (ALMA community proposal 2015.1.00657.S
searching for complex molecules, led by Karin {\"O}berg et al.).
These examples illustrate the potential of the 2D disk models presented 
in this paper for future data analysis and interpretation, even in the
realm of observed 3D disk substructures. 


\bigskip\noindent {\bf Acknowledgements}\\[1mm] The authors
acknowledge funding from the EU FP7-2011 under Grant Agreement
no. 284405. JDI gratefully acknowledges support from the DISCSIM
project, grant agreement 341137, funded by the European Research
Council under ERC-2013-ADG. Astrophysics at Queen's University Belfast
is supported by a grant from the STFC (ST/P000312/1).  OD acknowledges
the support of the Austrian Research Promotion Agency (FFG) to the
project JetPro, funded within the framework of Austrian Space
Applications Program (ASAP) FFG-854025.  This paper makes use of the
following ALMA data: RECX\,15: ADS/JAO.ALMA\#2011.0.00133.S (RECX\,15)
and ADS/JAO.ALMA\#2011.0.00629.S (DM\,Tau and MWC\,480). ALMA is a
partnership of ESO (representing its member states), NSF (USA) and
NINS (Japan), together with NRC (Canada) and NSC and ASIAA (Taiwan)
and KASI (Republic of Korea), in cooperation with the Republic of
Chile. The Joint ALMA Observatory is operated by ESO, AUI/NRAO and
NAOJ.

\bibliographystyle{aa}
\bibliography{references}

\begin{thebibliography}{143}
\expandafter\ifx\csname natexlab\endcsname\relax\def\natexlab#1{#1}\fi

\bibitem[{{Acke} {et~al.}(2010){Acke}, {Bouwman}, {Juh{\'a}sz}, {Henning}, {van
  den Ancker}, {Meeus}, {Tielens}, \& {Waters}}]{Acke2010}
{Acke}, B., {Bouwman}, J., {Juh{\'a}sz}, A., {et~al.} 2010, \apj, 718, 558

\bibitem[{{Andrews} {et~al.}(2013){Andrews}, {Rosenfeld}, {Kraus}, \&
  {Wilner}}]{Andrews2013}
{Andrews}, S.~M., {Rosenfeld}, K.~A., {Kraus}, A.~L., \& {Wilner}, D.~J. 2013,
  \apj, 771, 129

\bibitem[{{Andrews} {et~al.}(2011){Andrews}, {Rosenfeld}, {Wilner}, \&
  {Bremer}}]{Andrews2011}
{Andrews}, S.~M., {Rosenfeld}, K.~A., {Wilner}, D.~J., \& {Bremer}, M. 2011,
  \apjl, 742, L5

\bibitem[{{Andrews} \& {Williams}(2005)}]{Andrews2005}
{Andrews}, S.~M. \& {Williams}, J.~P. 2005, \apj, 631, 1134

\bibitem[{{Andrews} \& {Williams}(2007)}]{Andrews2007}
{Andrews}, S.~M. \& {Williams}, J.~P. 2007, \apj, 659, 705

\bibitem[{{Andrews} {et~al.}(2016){Andrews}, {Wilner}, {Zhu}, {Birnstiel},
  {Carpenter}, {P{\'e}rez}, {Bai}, {{\"O}berg}, {Hughes}, {Isella}, \&
  {Ricci}}]{Andrews2016}
{Andrews}, S.~M., {Wilner}, D.~J., {Zhu}, Z., {et~al.} 2016, \apjl, 820, L40

\bibitem[{{Aresu} {et~al.}(2011){Aresu}, {Kamp}, {Meijerink}, {Woitke}, {Thi},
  \& {Spaans}}]{Aresu2011}
{Aresu}, G., {Kamp}, I., {Meijerink}, R., {et~al.} 2011, \aap, 526, A163

\bibitem[{Banzatti {et~al.}(2012)Banzatti, Meyer, Bruderer, Geers, Pascucci,
  Lahuis, Juh{\'a}sz, Henning, \& {\'A}brah{\'a}m}]{Banzatti2012}
Banzatti, A., Meyer, M.~R., Bruderer, S., {et~al.} 2012, \apj, 745, 90

\bibitem[{{Beckwith} {et~al.}(1990){Beckwith}, {Sargent}, {Chini}, \&
  {Guesten}}]{Beckwith1990}
{Beckwith}, S.~V.~W., {Sargent}, A.~I., {Chini}, R.~S., \& {Guesten}, R. 1990,
  \aj, 99, 924

\bibitem[{{Bergin} {et~al.}(2013){Bergin}, {Cleeves}, {Gorti}, {Zhang},
  {Blake}, {Green}, {Andrews}, {Evans}, {Henning}, {{\"O}berg}, {Pontoppidan},
  {Qi}, {Salyk}, \& {van Dishoeck}}]{Bergin2013}
{Bergin}, E.~A., {Cleeves}, L.~I., {Gorti}, U., {et~al.} 2013, \nat, 493, 644

\bibitem[{{Bergin} {et~al.}(2016){Bergin}, {Du}, {Cleeves}, {Blake}, {Schwarz},
  {Visser}, \& {Zhang}}]{Bergin2016}
{Bergin}, E.~A., {Du}, F., {Cleeves}, L.~I., {et~al.} 2016, \apj, 831, 101

\bibitem[{{Bergner} {et~al.}(2018){Bergner}, {Guzm{\'a}n}, {{\"O}berg},
  {Loomis}, \& {Pegues}}]{Bergner2018}
{Bergner}, J.~B., {Guzm{\'a}n}, V.~G., {{\"O}berg}, K.~I., {Loomis}, R.~A., \&
  {Pegues}, J. 2018, \apj, 857, 69

\bibitem[{{Blevins} {et~al.}(2016){Blevins}, {Pontoppidan}, {Banzatti},
  {Zhang}, {Najita}, {Carr}, {Salyk}, \& {Blake}}]{Blevins2016}
{Blevins}, S.~M., {Pontoppidan}, K.~M., {Banzatti}, A., {et~al.} 2016, \apj,
  818, 22

\bibitem[{{Boneberg} {et~al.}(2016){Boneberg}, {Pani{\'c}}, {Haworth},
  {Clarke}, \& {Min}}]{Boneberg2016}
{Boneberg}, D.~M., {Pani{\'c}}, O., {Haworth}, T.~J., {Clarke}, C.~J., \&
  {Min}, M. 2016, \mnras, 461, 385

\bibitem[{{Brinch} \& {Hogerheijde}(2011)}]{Brinch2011}
{Brinch}, C. \& {Hogerheijde}, M.~R. 2011, {LIME: Flexible, Non-LTE Line
  Excitation and Radiation Transfer Method for Millimeter and Far-infrared
  Wavelengths}, Astrophysics Source Code Library

\bibitem[{{Brott} \& {Hauschildt}(2005)}]{Brott2005}
{Brott}, I. \& {Hauschildt}, P.~H. 2005, in ESA Special Publication, Vol. 576,
  The Three-Dimensional Universe with Gaia, ed. C.~{Turon}, K.~S. {O'Flaherty},
  \& M.~A.~C. {Perryman}, 565--+

\bibitem[{{Bruderer}(2013)}]{Bruderer2013}
{Bruderer}, S. 2013, \aap, 559, A46

\bibitem[{{Bruderer} {et~al.}(2009){Bruderer}, {Doty}, \&
  {Benz}}]{Bruderer2009}
{Bruderer}, S., {Doty}, S.~D., \& {Benz}, A.~O. 2009, \apjs, 183, 179

\bibitem[{{Bruderer} {et~al.}(2012){Bruderer}, {van Dishoeck}, {Doty}, \&
  {Herczeg}}]{Bruderer2012}
{Bruderer}, S., {van Dishoeck}, E.~F., {Doty}, S.~D., \& {Herczeg}, G.~J. 2012,
  \aap, 541, A91

\bibitem[{{Calvet} {et~al.}(2002){Calvet}, {D'Alessio}, {Hartmann}, {Wilner},
  {Walsh}, \& {Sitko}}]{Calvet2002}
{Calvet}, N., {D'Alessio}, P., {Hartmann}, L., {et~al.} 2002, \apj, 568, 1008

\bibitem[{{Carmona} {et~al.}(2014){Carmona}, {Pinte}, {Thi}, {Benisty},
  {M{\'e}nard}, {Grady}, {Kamp}, {Woitke}, {Olofsson}, {Roberge}, {Brittain},
  {Duch{\^e}ne}, {Meeus}, {Martin-Za{\"i}di}, {Dent}, {Le Bouquin}, \&
  {Berger}}]{Carmona2014}
{Carmona}, A., {Pinte}, C., {Thi}, W.~F., {et~al.} 2014, \aap, 567, A51

\bibitem[{{Carmona} {et~al.}(2017){Carmona}, {Thi}, {Kamp}, {Baruteau},
  {Matter}, {van den Ancker}, {Pinte}, {K{\'o}sp{\'a}l}, {Audard}, {Liebhart},
  {Sicilia-Aguilar}, {Pinilla}, {Reg{\'a}ly}, {G{\"u}del}, {Henning}, {Cieza},
  {Baldovin-Saavedra}, {Meeus}, \& {Eiroa}}]{Carmona2017}
{Carmona}, A., {Thi}, W.~F., {Kamp}, I., {et~al.} 2017, \aap, 598, A118

\bibitem[{{Cazaux} \& {Tielens}(2004)}]{Cazaux2004}
{Cazaux}, S. \& {Tielens}, A.~G.~G.~M. 2004, \apj, 604, 222

\bibitem[{{Cazaux} \& {Tielens}(2010)}]{Cazaux2010}
{Cazaux}, S. \& {Tielens}, A.~G.~G.~M. 2010, \apj, 715, 698

\bibitem[{{Cazzoletti} {et~al.}(2018){Cazzoletti}, {van Dishoeck}, {Visser},
  {Facchini}, \& {Bruderer}}]{Cazzoletti2018}
{Cazzoletti}, P., {van Dishoeck}, E.~F., {Visser}, R., {Facchini}, S., \&
  {Bruderer}, S. 2018, \aap, 609, A93

\bibitem[{{Chapillon} {et~al.}(2012{\natexlab{a}}){Chapillon}, {Dutrey},
  {Guilloteau}, {Pi{\'e}tu}, {Wakelam}, {Hersant}, {Gueth}, {Henning},
  {Launhardt}, {Schreyer}, \& {Semenov}}]{Chapillon2012a}
{Chapillon}, E., {Dutrey}, A., {Guilloteau}, S., {et~al.} 2012{\natexlab{a}},
  \apj, 756, 58

\bibitem[{{Chapillon} {et~al.}(2012{\natexlab{b}}){Chapillon}, {Guilloteau},
  {Dutrey}, {Pi{\'e}tu}, \& {Gu{\'e}lin}}]{Chapillon2012}
{Chapillon}, E., {Guilloteau}, S., {Dutrey}, A., {Pi{\'e}tu}, V., \&
  {Gu{\'e}lin}, M. 2012{\natexlab{b}}, \aap, 537, A60

\bibitem[{{Cleeves} {et~al.}(2015){Cleeves}, {Bergin}, {Qi}, {Adams}, \&
  {{\"O}berg}}]{Cleeves2015a}
{Cleeves}, L.~I., {Bergin}, E.~A., {Qi}, C., {Adams}, F.~C., \& {{\"O}berg},
  K.~I. 2015, \apj, 799, 204

\bibitem[{{Cleeves} {et~al.}(2016){Cleeves}, {{\"O}berg}, {Wilner}, {Huang},
  {Loomis}, {Andrews}, \& {Czekala}}]{Cleeves2016}
{Cleeves}, L.~I., {{\"O}berg}, K.~I., {Wilner}, D.~J., {et~al.} 2016, \apj,
  832, 110

\bibitem[{{D'Alessio} {et~al.}(1998){D'Alessio}, {Canto}, {Calvet}, \&
  {Lizano}}]{Dalessio1998}
{D'Alessio}, P., {Canto}, J., {Calvet}, N., \& {Lizano}, S. 1998, \apj, 500,
  411

\bibitem[{Davies {et~al.}(2018)Davies, Kraus, Harries, Kreplin, Monnier,
  Labdon, Kloppenborg, Acreman, Baron, Millan-Gabet, Sturmann, Sturmann, \&
  Ten~Brummelaar}]{Davies2018}
Davies, C.~L., Kraus, S., Harries, T.~J., {et~al.} 2018, arXiv.org,
  arXiv:1808.10762

\bibitem[{{de Gregorio-Monsalvo} {et~al.}(2013){de Gregorio-Monsalvo},
  {M{\'e}nard}, {Dent}, {Pinte}, {L{\'o}pez}, {Klaassen}, {Hales},
  {Cort{\'e}s}, {Rawlings}, {Tachihara}, {Testi}, {Takahashi}, {Chapillon},
  {Mathews}, {Juhasz}, {Akiyama}, {Higuchi}, {Saito}, {Nyman}, {Phillips},
  {Rod{\'o}n}, {Corder}, \& {Van Kempen}}]{deGregorio2013}
{de Gregorio-Monsalvo}, I., {M{\'e}nard}, F., {Dent}, W., {et~al.} 2013, \aap,
  557, A133

\bibitem[{{Dionatos et al.\ submitted}(2018)}]{Dionatos2018}
{Dionatos et al.\ submitted}, O. 2018, ArXiv e-prints

\bibitem[{{Drabek-Maunder} {et~al.}(2016){Drabek-Maunder}, {Mohanty},
  {Greaves}, {Kamp}, {Meijerink}, {Spaans}, {Thi}, \& {Woitke}}]{Drabek2016}
{Drabek-Maunder}, E., {Mohanty}, S., {Greaves}, J., {et~al.} 2016, \apj, 833,
  260

\bibitem[{{Du} \& {Bergin}(2014)}]{Du2014}
{Du}, F. \& {Bergin}, E.~A. 2014, \apj, 792, 2

\bibitem[{{Du} {et~al.}(2017){Du}, {Bergin}, {Hogerheijde}, {van Dishoeck},
  {Blake}, {Bruderer}, {Cleeves}, {Dominik}, {Fedele}, {Lis}, {Melnick},
  {Neufeld}, {Pearson}, \& {Y{\i}ld{\i}z}}]{Du2017}
{Du}, F., {Bergin}, E.~A., {Hogerheijde}, M., {et~al.} 2017, \apj, 842, 98

\bibitem[{{Du} {et~al.}(2015){Du}, {Bergin}, \& {Hogerheijde}}]{Du2015}
{Du}, F., {Bergin}, E.~A., \& {Hogerheijde}, M.~R. 2015, \apjl, 807, L32

\bibitem[{{Dubrulle} {et~al.}(1995){Dubrulle}, {Morfill}, \&
  {Sterzik}}]{Dubrulle1995}
{Dubrulle}, B., {Morfill}, G., \& {Sterzik}, M. 1995, \icarus, 114, 237

\bibitem[{{Dullemond} \& {Dominik}(2004)}]{Dullemond2004}
{Dullemond}, C.~P. \& {Dominik}, C. 2004, \aap, 417, 159

\bibitem[{{Dutrey} {et~al.}(1997){Dutrey}, {Guilloteau}, \&
  {Guelin}}]{Dutrey1997b}
{Dutrey}, A., {Guilloteau}, S., \& {Guelin}, M. 1997, \aap, 317, L55

\bibitem[{{Dutrey} {et~al.}(2007){Dutrey}, {Henning}, {Guilloteau}, {Semenov},
  {Pi{\'e}tu}, {Schreyer}, {Bacmann}, {Launhardt}, {Pety}, \&
  {Gueth}}]{Dutrey2007}
{Dutrey}, A., {Henning}, T., {Guilloteau}, S., {et~al.} 2007, \aap, 464, 615

\bibitem[{{Facchini} {et~al.}(2017){Facchini}, {Birnstiel}, {Bruderer}, \& {van
  Dishoeck}}]{Facchini2017}
{Facchini}, S., {Birnstiel}, T., {Bruderer}, S., \& {van Dishoeck}, E.~F. 2017,
  \aap, 605, A16

\bibitem[{{Favre} {et~al.}(2013){Favre}, {Cleeves}, {Bergin}, {Qi}, \&
  {Blake}}]{Favre2013}
{Favre}, C., {Cleeves}, L.~I., {Bergin}, E.~A., {Qi}, C., \& {Blake}, G.~A.
  2013, \apjl, 776, L38

\bibitem[{{Fedele} {et~al.}(2013){Fedele}, {Bruderer}, {van Dishoeck}, {Carr},
  {Herczeg}, {Salyk}, {Evans}, {Bouwman}, {Meeus}, {Henning}, {Green},
  {Najita}, \& {G{\"u}del}}]{Fedele2013}
{Fedele}, D., {Bruderer}, S., {van Dishoeck}, E.~F., {et~al.} 2013, \aap, 559,
  A77

\bibitem[{{Fedele} {et~al.}(2017){Fedele}, {Carney}, {Hogerheijde}, {Walsh},
  {Miotello}, {Klaassen}, {Bruderer}, {Henning}, \& {van
  Dishoeck}}]{Fedele2017}
{Fedele}, D., {Carney}, M., {Hogerheijde}, M.~R., {et~al.} 2017, \aap, 600, A72

\bibitem[{{Fernandes} {et~al.}(2018){Fernandes}, {Long}, {Pikhartova}, {Sitko},
  {Grady}, {Russell}, {Luria}, {Tyler}, {Bayyari}, {Danchi}, \&
  {Wisniewski}}]{Fernandes2018}
{Fernandes}, R.~B., {Long}, Z.~C., {Pikhartova}, M., {et~al.} 2018, \apj, 856,
  103

\bibitem[{{Fitzpatrick}(1999)}]{Fitzpatrick1999}
{Fitzpatrick}, E.~L. 1999, \pasp, 111, 63

\bibitem[{{Fuente} {et~al.}(2010){Fuente}, {Cernicharo}, {Ag{\'u}ndez},
  {Bern{\'e}}, {Goicoechea}, {Alonso-Albi}, \& {Marcelino}}]{Fuente2010}
{Fuente}, A., {Cernicharo}, J., {Ag{\'u}ndez}, M., {et~al.} 2010, \aap, 524,
  A19

\bibitem[{{Fukagawa} {et~al.}(2004){Fukagawa}, {Hayashi}, {Tamura}, {Itoh},
  {Hayashi}, {Oasa}, {Takeuchi}, {Morino}, {Murakawa}, {Oya}, {Yamashita},
  {Suto}, {Mayama}, {Naoi}, {Ishii}, {Pyo}, {Nishikawa}, {Takato}, {Usuda},
  {Ando}, {Iye}, {Miyama}, \& {Kaifu}}]{Fukagawa2004}
{Fukagawa}, M., {Hayashi}, M., {Tamura}, M., {et~al.} 2004, \apjl, 605, L53

\bibitem[{Garufi {et~al.}(2017)Garufi, Meeus, Benisty, Quanz, Banzatti, Kama,
  Canovas, Eiroa, Schmid, Stolker, Pohl, Rigliaco, M{\'e}nard, Meyer, van
  Boekel, \& Dominik}]{Garufi2017}
Garufi, A., Meeus, G., Benisty, M., {et~al.} 2017, arXiv.org, A21

\bibitem[{{Geers} {et~al.}(2009){Geers}, {van Dishoeck}, {Pontoppidan},
  {Lahuis}, {Crapsi}, {Dullemond}, \& {Blake}}]{Geers2009}
{Geers}, V.~C., {van Dishoeck}, E.~F., {Pontoppidan}, K.~M., {et~al.} 2009,
  \aap, 495, 837

\bibitem[{{Geers} {et~al.}(2007){Geers}, {van Dishoeck}, {Visser},
  {Pontoppidan}, {Augereau}, {Habart}, \& {Lagrange}}]{Geers2007}
{Geers}, V.~C., {van Dishoeck}, E.~F., {Visser}, R., {et~al.} 2007, \aap, 476,
  279

\bibitem[{{Gorti} {et~al.}(2011){Gorti}, {Hollenbach}, {Najita}, \&
  {Pascucci}}]{Gorti2011}
{Gorti}, U., {Hollenbach}, D., {Najita}, J., \& {Pascucci}, I. 2011, \apj, 735,
  90

\bibitem[{{Grady} {et~al.}(2010){Grady}, {Hamaguchi}, {Schneider}, {Stecklum},
  {Woodgate}, {McCleary}, {Williger}, {Sitko}, {M{\'e}nard}, {Henning},
  {Brittain}, {Troutmann}, {Donehew}, {Hines}, {Wisniewski}, {Lynch},
  {Russell}, {Rudy}, {Day}, {Shenoy}, {Wilner}, {Silverstone}, {Bouret},
  {Meusinger}, {Clampin}, {Kim}, {Petre}, {Sahu}, {Endres}, \&
  {Collins}}]{Grady2010b}
{Grady}, C.~A., {Hamaguchi}, K., {Schneider}, G., {et~al.} 2010, \apj, 719,
  1565

\bibitem[{{G{\"u}del} {et~al.}(2007){G{\"u}del}, {Briggs}, {Arzner}, {Audard},
  {Bouvier}, {Feigelson}, {Franciosini}, {Glauser}, {Grosso}, {Micela},
  {Monin}, {Montmerle}, {Padgett}, {Palla}, {Pillitteri}, {Rebull}, {Scelsi},
  {Silva}, {Skinner}, {Stelzer}, \& {Telleschi}}]{Guedel2007}
{G{\"u}del}, M., {Briggs}, K.~R., {Arzner}, K., {et~al.} 2007, \aap, 468, 353

\bibitem[{{G{\"u}del} {et~al.}(2010){G{\"u}del}, {Lahuis}, {Briggs}, {Carr},
  {Glassgold}, {Henning}, {Najita}, {van Boekel}, \& {van
  Dishoeck}}]{Guedel2010}
{G{\"u}del}, M., {Lahuis}, F., {Briggs}, K.~R., {et~al.} 2010, \aap, 519, A113

\bibitem[{{Guilloteau} {et~al.}(2013){Guilloteau}, {Di Folco}, {Dutrey},
  {Simon}, {Grosso}, \& {Pi{\'e}tu}}]{Guilloteau2013}
{Guilloteau}, S., {Di Folco}, E., {Dutrey}, A., {et~al.} 2013, \aap, 549, A92

\bibitem[{{Guilloteau} {et~al.}(2011){Guilloteau}, {Dutrey}, {Pi{\'e}tu}, \&
  {Boehler}}]{Guilloteau2011}
{Guilloteau}, S., {Dutrey}, A., {Pi{\'e}tu}, V., \& {Boehler}, Y. 2011, \aap,
  529, A105

\bibitem[{{Guilloteau} {et~al.}(2016){Guilloteau}, {Pi{\'e}tu}, {Chapillon},
  {Di Folco}, {Dutrey}, {Henning}, {Semenov}, {Birnstiel}, \&
  {Grosso}}]{Guilloteau2016}
{Guilloteau}, S., {Pi{\'e}tu}, V., {Chapillon}, E., {et~al.} 2016, \aap, 586,
  L1

\bibitem[{{Guzm{\'a}n} {et~al.}(2015){Guzm{\'a}n}, {{\"O}berg}, {Loomis}, \&
  {Qi}}]{Guzman2015}
{Guzm{\'a}n}, V.~V., {{\"O}berg}, K.~I., {Loomis}, R., \& {Qi}, C. 2015, \apj,
  814, 53

\bibitem[{{Hales} {et~al.}(2014){Hales}, {De Gregorio-Monsalvo}, {Montesinos},
  {Casassus}, {Dent}, {Dougados}, {Eiroa}, {Hughes}, {Garay}, {Mardones},
  {M{\'e}nard}, {Palau}, {P{\'e}rez}, {Phillips}, {Torrelles}, \&
  {Wilner}}]{Hales2014}
{Hales}, A.~S., {De Gregorio-Monsalvo}, I., {Montesinos}, B., {et~al.} 2014,
  \aj, 148, 47

\bibitem[{{Harries}(2000)}]{Harries2000}
{Harries}, T.~J. 2000, \mnras, 315, 722

\bibitem[{{Henning} {et~al.}(2010){Henning}, {Semenov}, {Guilloteau}, {Dutrey},
  {Hersant}, {Wakelam}, {Chapillon}, {Launhardt}, {Pi{\'e}tu}, \&
  {Schreyer}}]{Henning2010n}
{Henning}, T., {Semenov}, D., {Guilloteau}, S., {et~al.} 2010, \apj, 714, 1511

\bibitem[{{Hillenbrand} \& {White}(2004)}]{Hillenbrand2004}
{Hillenbrand}, L.~A. \& {White}, R.~J. 2004, \apj, 604, 741

\bibitem[{{Honda} {et~al.}(2010){Honda}, {Inoue}, {Okamoto}, {Kataza},
  {Fukagawa}, {Yamashita}, {Fujiyoshi}, {Tamura}, {Hashimoto}, {Miyata},
  {Sako}, {Sakon}, {Fujiwara}, {Kamizuka}, \& {Onaka}}]{Honda2010}
{Honda}, M., {Inoue}, A.~K., {Okamoto}, Y.~K., {et~al.} 2010, \apjl, 718, L199

\bibitem[{{Huang} {et~al.}(2017){Huang}, {{\"O}berg}, {Qi}, {Aikawa},
  {Andrews}, {Furuya}, {Guzm{\'a}n}, {Loomis}, {van Dishoeck}, \&
  {Wilner}}]{Huang2017}
{Huang}, J., {{\"O}berg}, K.~I., {Qi}, C., {et~al.} 2017, \apj, 835, 231

\bibitem[{{Isella} {et~al.}(2016){Isella}, {Guidi}, {Testi}, {Liu}, {Li}, {Li},
  {Weaver}, {Boehler}, {Carperter}, {De Gregorio-Monsalvo}, {Manara}, {Natta},
  {P{\'e}rez}, {Ricci}, {Sargent}, {Tazzari}, \& {Turner}}]{Isella2016}
{Isella}, A., {Guidi}, G., {Testi}, L., {et~al.} 2016, PRL, 117, 251101

\bibitem[{{Jamialahmadi} {et~al.}(2018){Jamialahmadi}, {Lopez}, {Berio},
  {Matter}, {Flament}, {Fathivavsari}, {Ratzka}, {Sitko}, {Spang}, \&
  {Russell}}]{Jamialahmadi2018}
{Jamialahmadi}, N., {Lopez}, B., {Berio}, P., {et~al.} 2018, \mnras, 473, 3147

\bibitem[{{Kama} {et~al.}(2016{\natexlab{a}}){Kama}, {Bruderer}, {Carney},
  {Hogerheijde}, {van Dishoeck}, {Fedele}, {Baryshev}, {Boland}, {G{\"u}sten},
  {Aikutalp}, {Choi}, {Endo}, {Frieswijk}, {Karska}, {Klaassen}, {Koumpia},
  {Kristensen}, {Leurini}, {Nagy}, {Perez Beaupuits}, {Risacher}, {van der
  Marel}, {van Kempen}, {van Weeren}, {Wyrowski}, \&
  {Y{\i}ld{\i}z}}]{Kama2016b}
{Kama}, M., {Bruderer}, S., {Carney}, M., {et~al.} 2016{\natexlab{a}}, \aap,
  588, A108

\bibitem[{{Kama} {et~al.}(2016{\natexlab{b}}){Kama}, {Bruderer}, {van
  Dishoeck}, {Hogerheijde}, {Folsom}, {Miotello}, {Fedele}, {Belloche},
  {G{\"u}sten}, \& {Wyrowski}}]{Kama2016}
{Kama}, M., {Bruderer}, S., {van Dishoeck}, E.~F., {et~al.} 2016{\natexlab{b}},
  \aap, 592, A83

\bibitem[{{Kamp} {et~al.}(2013){Kamp}, {Thi}, {Meeus}, {Woitke}, {Pinte},
  {Meijerink}, {Spaans}, {Pascucci}, {Aresu}, \& {Dent}}]{Kamp2013}
{Kamp}, I., {Thi}, W.-F., {Meeus}, G., {et~al.} 2013, \aap, 559, A24

\bibitem[{{Kamp} {et~al.}(2017){Kamp}, {Thi}, {Woitke}, {Rab}, {Bouma}, \&
  {M{\'e}nard}}]{Kamp2017}
{Kamp}, I., {Thi}, W.-F., {Woitke}, P., {et~al.} 2017, \aap, 607, A41

\bibitem[{{Kamp} {et~al.}(2010){Kamp}, {Tilling}, {Woitke}, {Thi}, \&
  {Hogerheijde}}]{Kamp2010}
{Kamp}, I., {Tilling}, I., {Woitke}, P., {Thi}, W., \& {Hogerheijde}, M. 2010,
  \aap, 510, A260000+

\bibitem[{{Kamp} {et~al.}(2011){Kamp}, {Woitke}, {Pinte}, {Tilling}, {Thi},
  {Menard}, {Duchene}, \& {Augereau}}]{Kamp2011}
{Kamp}, I., {Woitke}, P., {Pinte}, C., {et~al.} 2011, \aap, 532, A85

\bibitem[{{Kraus} \& {Ireland}(2012)}]{Kraus2012}
{Kraus}, A.~L. \& {Ireland}, M.~J. 2012, \apj, 745, 5

\bibitem[{{Lagage} {et~al.}(2006){Lagage}, {Doucet}, {Pantin}, {Habart},
  {Duch{\^e}ne}, {M{\'e}nard}, {Pinte}, {Charnoz}, \& {Pel}}]{Lagage2006}
{Lagage}, P.-O., {Doucet}, C., {Pantin}, E., {et~al.} 2006, Science, 314, 621

\bibitem[{{Loomis} {et~al.}(2015){Loomis}, {Cleeves}, {{\"O}berg}, {Guzman}, \&
  {Andrews}}]{Loomis2015}
{Loomis}, R.~A., {Cleeves}, L.~I., {{\"O}berg}, K.~I., {Guzman}, V.~V., \&
  {Andrews}, S.~M. 2015, \apj, 809, L25

\bibitem[{{Maaskant} {et~al.}(2013){Maaskant}, {Honda}, {Waters}, {Tielens},
  {Dominik}, {Min}, {Verhoeff}, {Meeus}, \& {van den Ancker}}]{Maaskant2013}
{Maaskant}, K.~M., {Honda}, M., {Waters}, L.~B.~F.~M., {et~al.} 2013, \aap,
  555, A64

\bibitem[{{Maaskant} {et~al.}(2014){Maaskant}, {Min}, {Waters}, \&
  {Tielens}}]{Maaskant2014}
{Maaskant}, K.~M., {Min}, M., {Waters}, L.~B.~F.~M., \& {Tielens}, A.~G.~G.~M.
  2014, \aap, 563, A78

\bibitem[{{Mannings} \& {Sargent}(1997)}]{Mannings1997b}
{Mannings}, V. \& {Sargent}, A.~I. 1997, \apj, 490, 792

\bibitem[{{Mathews} {et~al.}(2013){Mathews}, {Klaassen}, {Juh{\'a}sz},
  {Harsono}, {Chapillon}, {van Dishoeck}, {Espada}, {de Gregorio-Monsalvo},
  {Hales}, {Hogerheijde}, {Mottram}, {Rawlings}, {Takahashi}, \&
  {Testi}}]{Mathews2013}
{Mathews}, G.~S., {Klaassen}, P.~D., {Juh{\'a}sz}, A., {et~al.} 2013, \aap,
  557, A132

\bibitem[{{Mathews} {et~al.}(2012){Mathews}, {Williams}, \&
  {M{\'e}nard}}]{mathews2012b}
{Mathews}, G.~S., {Williams}, J.~P., \& {M{\'e}nard}, F. 2012, \apj, 753, 59

\bibitem[{{McClure} {et~al.}(2016){McClure}, {Bergin}, {Cleeves}, {van
  Dishoeck}, {Blake}, {Evans}, {Green}, {Henning}, {{\"O}berg}, {Pontoppidan},
  \& {Salyk}}]{McClure2016}
{McClure}, M.~K., {Bergin}, E.~A., {Cleeves}, L.~I., {et~al.} 2016, \apj, 831,
  167

\bibitem[{{Meeus} {et~al.}(2012){Meeus}, {Montesinos}, {Mendigut{\'{\i}}a},
  {Kamp}, {Thi}, {Eiroa}, {Grady}, {Mathews}, {Sandell}, {Martin-Za{\"i}di},
  {Brittain}, {Dent}, {Howard}, {M{\'e}nard}, {Pinte}, {Roberge},
  {Vandenbussche}, \& {Williams}}]{Meeus2012}
{Meeus}, G., {Montesinos}, B., {Mendigut{\'{\i}}a}, I., {et~al.} 2012, \aap,
  544, A78

\bibitem[{{Meeus} {et~al.}(2001){Meeus}, {Waters}, {Bouwman}, {van den Ancker},
  {Waelkens}, \& {Malfait}}]{Meeus2001}
{Meeus}, G., {Waters}, L.~B.~F.~M., {Bouwman}, J., {et~al.} 2001, \aap, 365,
  476

\bibitem[{{Menu} {et~al.}(2014){Menu}, {van Boekel}, {Henning}, {Chandler},
  {Linz}, {Benisty}, {Lacour}, {Min}, {Waelkens}, {Andrews}, {Calvet},
  {Carpenter}, {Corder}, {Deller}, {Greaves}, {Harris}, {Isella}, {Kwon},
  {Lazio}, {Le Bouquin}, {M{\'e}nard}, {Mundy}, {P{\'e}rez}, {Ricci},
  {Sargent}, {Storm}, {Testi}, \& {Wilner}}]{Menu2014}
{Menu}, J., {van Boekel}, R., {Henning}, T., {et~al.} 2014, \aap, 564, A93

\bibitem[{Menu {et~al.}(2015)Menu, van Boekel, Henning, Leinert, Waelkens, \&
  Waters}]{Menu2015}
Menu, J., van Boekel, R., Henning, T., {et~al.} 2015, A\&A, 581, A107

\bibitem[{{Min} {et~al.}(2016{\natexlab{a}}){Min}, {Bouwman}, {Dominik},
  {Waters}, {Pontoppidan}, {Hony}, {Mulders}, {Henning}, {van Dishoeck},
  {Woitke}, {Evans}, \& {Digit Team}}]{Min2016a}
{Min}, M., {Bouwman}, J., {Dominik}, C., {et~al.} 2016{\natexlab{a}}, \aap,
  593, A11

\bibitem[{{Min} {et~al.}(2009){Min}, {Dullemond}, {Dominik}, {de Koter}, \&
  {Hovenier}}]{Min2009}
{Min}, M., {Dullemond}, C.~P., {Dominik}, C., {de Koter}, A., \& {Hovenier},
  J.~W. 2009, \aap, 497, 155

\bibitem[{{Min} {et~al.}(2016{\natexlab{b}}){Min}, {Rab}, {Woitke}, {Dominik},
  \& {M{\'e}nard}}]{Min2016}
{Min}, M., {Rab}, C., {Woitke}, P., {Dominik}, C., \& {M{\'e}nard}, F.
  2016{\natexlab{b}}, \aap, 585, A13

\bibitem[{{Miotello} {et~al.}(2016){Miotello}, {van Dishoeck}, {Kama}, \&
  {Bruderer}}]{Miotello2016}
{Miotello}, A., {van Dishoeck}, E.~F., {Kama}, M., \& {Bruderer}, S. 2016,
  \aap, 594, A85

\bibitem[{{Mohanty} {et~al.}(2013){Mohanty}, {Greaves}, {Mortlock}, {Pascucci},
  {Scholz}, {Thompson}, {Apai}, {Lodato}, \& {Looper}}]{Mohanty2013}
{Mohanty}, S., {Greaves}, J., {Mortlock}, D., {et~al.} 2013, \apj, 773, 168

\bibitem[{{Muro-Arena} {et~al.}(2018){Muro-Arena}, {Dominik}, {Waters}, {Min},
  {Klarmann}, {Ginski}, {Isella}, {Benisty}, {Pohl}, {Garufi}, {Hagelberg},
  {Langlois}, {Menard}, {Pinte}, {Sezestre}, {van der Plas}, {Villenave},
  {Delboulb{\'e}}, {Magnard}, {M{\"o}ller-Nilsson}, {Pragt}, {Rabou}, \&
  {Roelfsema}}]{Muro-Arena2018}
{Muro-Arena}, G.~A., {Dominik}, C., {Waters}, L.~B.~F.~M., {et~al.} 2018, \aap,
  614, A24

\bibitem[{{Najita} {et~al.}(2003){Najita}, {Carr}, \& {Mathieu}}]{Najita2003}
{Najita}, J., {Carr}, J.~S., \& {Mathieu}, R.~D. 2003, \apj, 589, 931

\bibitem[{{Natta} {et~al.}(2007){Natta}, {Testi}, {Calvet}, {Henning},
  {Waters}, \& {Wilner}}]{Natta2007}
{Natta}, A., {Testi}, L., {Calvet}, N., {et~al.} 2007, Protostars and Planets,
  V, 767

\bibitem[{{{\"O}berg} {et~al.}(2015){{\"O}berg}, {Guzm{\'a}n}, {Furuya}, {Qi},
  {Aikawa}, {Andrews}, {Loomis}, \& {Wilner}}]{Oeberg2015}
{{\"O}berg}, K.~I., {Guzm{\'a}n}, V.~V., {Furuya}, K., {et~al.} 2015, \nat,
  520, 198

\bibitem[{{{\"O}berg} {et~al.}(2010){{\"O}berg}, {Qi}, {Fogel}, {Bergin},
  {Andrews}, {Espaillat}, {van Kempen}, {Wilner}, \& {Pascucci}}]{Oberg2010c}
{{\"O}berg}, K.~I., {Qi}, C., {Fogel}, J.~K.~J., {et~al.} 2010, \apj, 720, 480

\bibitem[{{Pascucci} {et~al.}(2016){Pascucci}, {Testi}, {Herczeg}, {Long},
  {Manara}, {Hendler}, {Mulders}, {Krijt}, {Ciesla}, {Henning}, {Mohanty},
  {Drabek-Maunder}, {Apai}, {Sz{\'{u}}cs}, {Sacco}, \&
  {Olofsson}}]{Pascucci2016}
{Pascucci}, I., {Testi}, L., {Herczeg}, G.~J., {et~al.} 2016, \apj, 831, 125

\bibitem[{{Perrin} {et~al.}(2009){Perrin}, {Schneider}, {Duchene}, {Pinte},
  {Grady}, {Wisniewski}, \& {Hines}}]{Perrin2009}
{Perrin}, M.~D., {Schneider}, G., {Duchene}, G., {et~al.} 2009, \apjl, 707,
  L132

\bibitem[{{Pi{\'e}tu} {et~al.}(2007){Pi{\'e}tu}, {Dutrey}, \&
  {Guilloteau}}]{Pietu2007}
{Pi{\'e}tu}, V., {Dutrey}, A., \& {Guilloteau}, S. 2007, \aap, 467, 163

\bibitem[{{Pinte} {et~al.}(2016){Pinte}, {Dent}, {M{\'e}nard}, {Hales}, {Hill},
  {Cortes}, \& {de Gregorio-Monsalvo}}]{Pinte2016}
{Pinte}, C., {Dent}, W.~R.~F., {M{\'e}nard}, F., {et~al.} 2016, \apj, 816, 25

\bibitem[{{Pinte} {et~al.}(2009){Pinte}, {Harries}, {Min}, {Watson},
  {Dullemond}, {Woitke}, {M{\'e}nard}, \& {Dur{\'a}n-Rojas}}]{Pinte2009}
{Pinte}, C., {Harries}, T.~J., {Min}, M., {et~al.} 2009, \aap, 498, 967

\bibitem[{{Pinte} {et~al.}(2006){Pinte}, {M{\'e}nard}, {Duch{\^e}ne}, \&
  {Bastien}}]{Pinte2006}
{Pinte}, C., {M{\'e}nard}, F., {Duch{\^e}ne}, G., \& {Bastien}, P. 2006, \aap,
  459, 797

\bibitem[{{Pinte} {et~al.}(2008){Pinte}, {Padgett}, {M{\'e}nard},
  {Stapelfeldt}, {Schneider}, \& et~al.}]{Pinte2008}
{Pinte}, C., {Padgett}, D.~L., {M{\'e}nard}, F., {et~al.} 2008, \aap, 489, 633

\bibitem[{{Pinte} {et~al.}(2018){Pinte}, {Price}, {M{\'e}nard}, {Duch{\^e}ne},
  {Dent}, {Hill}, {de Gregorio- Monsalvo}, {Hales}, \& {Mentiplay}}]{Pinte2018}
{Pinte}, C., {Price}, D.~J., {M{\'e}nard}, F., {et~al.} 2018, \apj, 860, L13

\bibitem[{{Pinte} {et~al.}(2010){Pinte}, {Woitke}, {M{\'e}nard}, {Duch{\^e}ne},
  {Kamp}, {Meeus}, {Mathews}, {Howard}, {Grady}, {Thi}, {Tilling}, {Augereau},
  {Dent}, {Alacid}, {Andrews}, {Ardila}, {Aresu}, {Barrado}, {Brittain},
  {Ciardi}, {Danchi}, {Eiroa}, {Fedele}, {de Gregorio-Monsalvo}, {Heras},
  {Huelamo}, {Krivov}, {Lebreton}, {Liseau}, {Martin-Za{\"i}di},
  {Mendigut{\'{\i}}a}, {Montesinos}, {Mora}, {Morales-Calderon}, {Nomura},
  {Pantin}, {Pascucci}, {Phillips}, {Podio}, {Poelman}, {Ramsay}, {Riaz},
  {Rice}, {Riviere-Marichalar}, {Roberge}, {Sandell}, {Solano},
  {Vandenbussche}, {Walker}, {Williams}, {White}, \& {Wright}}]{Pinte2010}
{Pinte}, C., {Woitke}, P., {M{\'e}nard}, F., {et~al.} 2010, \aap, 518, L126

\bibitem[{{Pontoppidan} \& {Blevins}(2014)}]{Pontoppidan2014b}
{Pontoppidan}, K.~M. \& {Blevins}, S.~M. 2014, Faraday Discussions, 169, 49

\bibitem[{{Price} {et~al.}(2018){Price}, {Cuello}, {Pinte}, {Mentiplay},
  {Casassus}, {Christiaens}, {Kennedy}, {Cuadra}, {Sebastian Perez}, {Marino},
  {Armitage}, {Zurlo}, {Juhasz}, {Ragusa}, {Laibe}, \& {Lodato}}]{Price2018}
{Price}, D.~J., {Cuello}, N., {Pinte}, C., {et~al.} 2018, \mnras, 477, 1270

\bibitem[{{Quanz} {et~al.}(2013){Quanz}, {Avenhaus}, {Buenzli}, {Garufi},
  {Schmid}, \& {Wolf}}]{Quanz2013}
{Quanz}, S.~P., {Avenhaus}, H., {Buenzli}, E., {et~al.} 2013, \apjl, 766, L2

\bibitem[{{Rab} {et~al.}(2018){Rab}, {G{\"u}del}, {Woitke}, {Kamp}, {Thi},
  {Min}, {Aresu}, \& {Meijerink}}]{Rab2018}
{Rab}, C., {G{\"u}del}, M., {Woitke}, P., {et~al.} 2018, \aap, 609, A91

\bibitem[{Rechenberg(1994)}]{Rechenberg1994}
Rechenberg, I. 1994, {Evolutionsstrategie}, {Opti\-mie\-rung} technischer
  {Systeme} nach {Prinzipien} der biolo\-gi\-schen {Evolution} (Stuttgart:
  frommann-holzboog), 246--414

\bibitem[{Rubinstein {et~al.}(2018)Rubinstein, Mac{\'\i}as, Espaillat, Zhang,
  Calvet, \& Robinson}]{Rubinstein2018}
Rubinstein, A.~E., Mac{\'\i}as, E., Espaillat, C.~C., {et~al.} 2018, The
  Astrophysical Journal, 860, 7

\bibitem[{{Semenov} {et~al.}(2018){Semenov}, {Favre}, {Fedele}, {Guilloteau},
  {Teague}, {Henning}, {Dutrey}, {Chapillon}, {Hersant}, \&
  {Pi{\'e}tu}}]{Semenov2018}
{Semenov}, D., {Favre}, C., {Fedele}, D., {et~al.} 2018, ArXiv e-prints,
  arXiv:1806.07707

\bibitem[{{Siess} {et~al.}(2000){Siess}, {Dufour}, \& {Forestini}}]{Siess2000}
{Siess}, L., {Dufour}, E., \& {Forestini}, M. 2000, \aap, 358, 593

\bibitem[{{Sitko} {et~al.}(2008){Sitko}, {Carpenter}, {Kimes}, {Wilde},
  {Lynch}, {Russell}, {Rudy}, {Mazuk}, {Venturini}, {Puetter}, {Grady},
  {Polomski}, {Wisnewski}, {Brafford}, {Hammel}, \& {Perry}}]{Sitko2008a}
{Sitko}, M.~L., {Carpenter}, W.~J., {Kimes}, R.~L., {et~al.} 2008, \apj, 678,
  1070

\bibitem[{{Stolker} {et~al.}(2017){Stolker}, {Sitko}, {Lazareff}, {Benisty},
  {Dominik}, {Waters}, {Min}, {Perez}, {Milli}, {Garufi}, {de Boer}, {Ginski},
  {Kraus}, {Berger}, \& {Avenhaus}}]{Stolker2017}
{Stolker}, T., {Sitko}, M., {Lazareff}, B., {et~al.} 2017, \apj, 849, 143

\bibitem[{{Teague} {et~al.}(2015){Teague}, {Semenov}, {Guilloteau}, {Henning},
  {Dutrey}, {Wakelam}, {Chapillon}, \& {Pietu}}]{Teague2015}
{Teague}, R., {Semenov}, D., {Guilloteau}, S., {et~al.} 2015, \aap, 574, A137

\bibitem[{{Testi} {et~al.}(2014){Testi}, {Birnstiel}, {Ricci}, {Andrews},
  {Blum}, {Carpenter}, {Dominik}, {Isella}, {Natta}, {Williams}, \&
  {Wilner}}]{Testi2014}
{Testi}, L., {Birnstiel}, T., {Ricci}, L., {et~al.} 2014, Protostars and
  Planets, VI, 339

\bibitem[{{Thalmann} {et~al.}(2010){Thalmann}, {Grady}, {Goto}, {Wisniewski},
  {Janson}, {Henning}, {Fukagawa}, {Honda}, {Mulders}, {Min},
  {Moro-Mart{\'{\i}}n}, {McElwain}, {Hodapp}, {Carson}, {Abe}, {Brandner},
  {Egner}, {Feldt}, {Fukue}, {Golota}, {Guyon}, {Hashimoto}, {Hayano},
  {Hayashi}, {Hayashi}, {Ishii}, {Kandori}, {Knapp}, {Kudo}, {Kusakabe},
  {Kuzuhara}, {Matsuo}, {Miyama}, {Morino}, {Nishimura}, {Pyo}, {Serabyn},
  {Shibai}, {Suto}, {Suzuki}, {Takami}, {Takato}, {Terada}, {Tomono}, {Turner},
  {Watanabe}, {Yamada}, {Takami}, {Usuda}, \& {Tamura}}]{Thalmann2010}
{Thalmann}, C., {Grady}, C.~A., {Goto}, M., {et~al.} 2010, \apjl, 718, L87

\bibitem[{{Thi} {et~al.}(2010){Thi}, {Mathews}, {M{\'e}nard}, {Woitke},
  {Meeus}, {Riviere-Marichalar}, {Pinte}, {Howard}, {Roberge}, {Sandell},
  {Pascucci}, {Riaz}, {Grady}, {Dent}, {Kamp}, {Duch{\^e}ne}, {Augereau},
  {Pantin}, {Vandenbussche}, {Tilling}, {Williams}, {Eiroa}, {Barrado},
  {Alacid}, {Andrews}, {Ardila}, {Aresu}, {Brittain}, {Ciardi}, {Danchi},
  {Fedele}, {de Gregorio-Monsalvo}, {Heras}, {Huelamo}, {Krivov}, {Lebreton},
  {Liseau}, {Martin-Zaidi}, {Mendigut{\'{\i}}a}, {Montesinos}, {Mora},
  {Morales-Calderon}, {Nomura}, {Phillips}, {Podio}, {Poelman}, {Ramsay},
  {Rice}, {Solano}, {Walker}, {White}, \& {Wright}}]{Thi2010}
{Thi}, W.-F., {Mathews}, G., {M{\'e}nard}, F., {et~al.} 2010, \aap, 518, L125

\bibitem[{{Thi} {et~al.}(2014){Thi}, {Pinte}, {Pantin}, {Augereau}, {Meeus},
  {M{\'e}nard}, {Martin-Za{\"i}di}, {Woitke}, {Riviere-Marichalar}, {Kamp},
  {Carmona}, {Sandell}, {Eiroa}, {Dent}, {Montesinos}, {Aresu}, {Meijerink},
  {Spaans}, {White}, {Ardila}, {Lebreton}, {Mendigut{\'{\i}}a}, \&
  {Brittain}}]{Thi2014}
{Thi}, W.-F., {Pinte}, C., {Pantin}, E., {et~al.} 2014, \aap, 561, A50

\bibitem[{{Thi} {et~al.}(2011){Thi}, {Woitke}, \& {Kamp}}]{Thi2011}
{Thi}, W.-F., {Woitke}, P., \& {Kamp}, I. 2011, \mnras, 412, 711

\bibitem[{{Tielens}(2008)}]{Tielens2008}
{Tielens}, A.~G.~G.~M. 2008, \araa, 46, 289

\bibitem[{{Tilling} {et~al.}(2012){Tilling}, {Woitke}, {Meeus}, {Mora},
  {Montesinos}, {Riviere-Marichalar}, {Eiroa}, {Thi}, {Isella}, {Roberge},
  {Martin-Zaidi}, {Kamp}, {Pinte}, {Sandell}, {Vacca}, {M{\'e}nard},
  {Mendigut{\'{\i}}a}, {Duch{\^e}ne}, {Dent}, {Aresu}, {Meijerink}, \&
  {Spaans}}]{Tilling2012}
{Tilling}, I., {Woitke}, P., {Meeus}, G., {et~al.} 2012, \aap, 538, A20

\bibitem[{{Valenti} {et~al.}(2003){Valenti}, {Fallon}, \&
  {Johns-Krull}}]{Valenti2003}
{Valenti}, J.~A., {Fallon}, A.~A., \& {Johns-Krull}, C.~M. 2003, \apjs, 147,
  305

\bibitem[{{Valenti} {et~al.}(2000){Valenti}, {Johns-Krull}, \&
  {Linsky}}]{Valenti2000}
{Valenti}, J.~A., {Johns-Krull}, C.~M., \& {Linsky}, J.~L. 2000, \apjs, 129,
  399

\bibitem[{{van der Marel} {et~al.}(2016){van der Marel}, {van Dishoeck},
  {Bruderer}, {Andrews}, {Pontoppidan}, {Herczeg}, {van Kempen}, \&
  {Miotello}}]{vanderMarel2016}
{van der Marel}, N., {van Dishoeck}, E.~F., {Bruderer}, S., {et~al.} 2016,
  \aap, 585, A58

\bibitem[{{van der Plas} {et~al.}(2015){van der Plas}, {van den Ancker},
  {Waters}, \& {Dominik}}]{vanderPlas2015}
{van der Plas}, G., {van den Ancker}, M.~E., {Waters}, L.~B.~F.~M., \&
  {Dominik}, C. 2015, \aap, 574, A75

\bibitem[{{van der Plas} {et~al.}(2017){van der Plas}, {Wright}, {M{\'e}nard},
  {Casassus}, {Canovas}, {Pinte}, {Maddison}, {Maaskant}, {Avenhaus}, {Cieza},
  {Perez}, \& {Ubach}}]{vanderPlas2017}
{van der Plas}, G., {Wright}, C.~M., {M{\'e}nard}, F., {et~al.} 2017, \aap,
  597, A32

\bibitem[{{van der Wiel} {et~al.}(2014){van der Wiel}, {Naylor}, {Kamp},
  {M{\'e}nard}, {Thi}, {Woitke}, {Olofsson}, {Pontoppidan}, {Di Francesco},
  {Glauser}, {Greaves}, \& {Ivison}}]{vanderWiel2014}
{van der Wiel}, M.~H.~D., {Naylor}, D.~A., {Kamp}, I., {et~al.} 2014, \mnras,
  444, 3911

\bibitem[{{Vasyunin} {et~al.}(2008){Vasyunin}, {Semenov}, {Henning}, {Wakelam},
  {Herbst}, \& {Sobolev}}]{Vasyunin2008}
{Vasyunin}, A.~I., {Semenov}, D., {Henning}, T., {et~al.} 2008, \apj, 672, 629

\bibitem[{{Walsh} {et~al.}(2016){Walsh}, {Juh{\'a}sz}, {Meeus}, {Dent}, {Maud},
  {Aikawa}, {Millar}, \& {Nomura}}]{Walsh2016}
{Walsh}, C., {Juh{\'a}sz}, A., {Meeus}, G., {et~al.} 2016, \apj, 831, 200

\bibitem[{{Williams} \& {Best}(2014)}]{Williams2014dk}
{Williams}, J.~P. \& {Best}, W.~M.~J. 2014, \apj, 788, 59

\bibitem[{{Woitke} {et~al.}(2013){Woitke}, {Dent}, {Thi}, {Menard}, {Pinte},
  {Duchene}, {Sandell}, {Lawson}, \& {Kamp}}]{Woitke2013}
{Woitke}, P., {Dent}, W.~R.~F., {Thi}, W.-F., {et~al.} 2013, in Protostars and
  Planets VI

\bibitem[{{Woitke} {et~al.}(2009{\natexlab{a}}){Woitke}, {Kamp}, \&
  {Thi}}]{Woitke2009a}
{Woitke}, P., {Kamp}, I., \& {Thi}, W. 2009{\natexlab{a}}, \aap, 501, 383

\bibitem[{{Woitke} {et~al.}(2009{\natexlab{b}}){Woitke}, {Kamp}, \&
  {Thi}}]{Woitke2009}
{Woitke}, P., {Kamp}, I., \& {Thi}, W.-F. 2009{\natexlab{b}}, \aap, 501, 383

\bibitem[{{Woitke} {et~al.}(2016){Woitke}, {Min}, {Pinte}, {Thi}, {Kamp},
  {Rab}, {Anthonioz}, {Antonellini}, {Baldovin-Saavedra}, {Carmona}, {Dominik},
  {Dionatos}, {Greaves}, {G{\"u}del}, {Ilee}, {Liebhart}, {M{\'e}nard},
  {Rigon}, {Waters}, {Aresu}, {Meijerink}, \& {Spaans}}]{Woitke2016}
{Woitke}, P., {Min}, M., {Pinte}, C., {et~al.} 2016, \aap, 586, A103

\bibitem[{{Woitke} {et~al.}(2018){Woitke}, {Min}, {Thi}, {Roberts}, {Carmona},
  {Kamp}, {Menard}, \& {Pinte}}]{Woitke2018}
{Woitke}, P., {Min}, M., {Thi}, W.-F., {et~al.} 2018, ArXiv e-prints

\bibitem[{{Woitke} {et~al.}(2010){Woitke}, {Pinte}, {Tilling}, {M{\'e}nard},
  {Kamp}, {Thi}, {Duch{\^e}ne}, \& {Augereau}}]{Woitke2010}
{Woitke}, P., {Pinte}, C., {Tilling}, I., {et~al.} 2010, \mnras, 405, L26

\bibitem[{{Woitke} {et~al.}(2011){Woitke}, {Riaz}, {Duch{\^e}ne}, {Pascucci},
  {Lyo}, {Dent}, {Phillips}, {Thi}, {M{\'e}nard}, {Herczeg}, {Bergin}, {Brown},
  {Mora}, {Kamp}, {Aresu}, {Brittain}, {de Gregorio-Monsalvo}, \&
  {Sandell}}]{Woitke2011}
{Woitke}, P., {Riaz}, B., {Duch{\^e}ne}, G., {et~al.} 2011, \aap, 534, A44

\bibitem[{{Wolff} {et~al.}(2017){Wolff}, {Perrin}, {Stapelfeldt},
  {Duch{\^e}ne}, {M{\'e}nard}, {Padgett}, {Pinte}, {Pueyo}, \&
  {Fischer}}]{Wolff2017a}
{Wolff}, S.~G., {Perrin}, M.~D., {Stapelfeldt}, K., {et~al.} 2017, \apj, 851,
  56

\bibitem[{{Zhang} {et~al.}(2013){Zhang}, {Pontoppidan}, {Salyk}, \&
  {Blake}}]{Zhang2013}
{Zhang}, K., {Pontoppidan}, K.~M., {Salyk}, C., \& {Blake}, G.~A. 2013, \apj,
  766, 82

\bibitem[{{Zwintz} {et~al.}(2009){Zwintz}, {Kallinger}, {Guenther},
  {Gruberbauer}, {Huber}, {Rowe}, {Kuschnig}, {Weiss}, {Matthews}, {Moffat},
  {Rucinski}, {Sasselov}, {Walker}, \& {Casey}}]{Zwintz2009}
{Zwintz}, K., {Kallinger}, T., {Guenther}, D.~B., {et~al.} 2009, \aap, 494,
  1031

\end{thebibliography}

\newpage
\appendix

\section{Uploaded model data}
\label{app:uploaded-data}

\begin{figure}
\centering
\includegraphics[width=140mm]{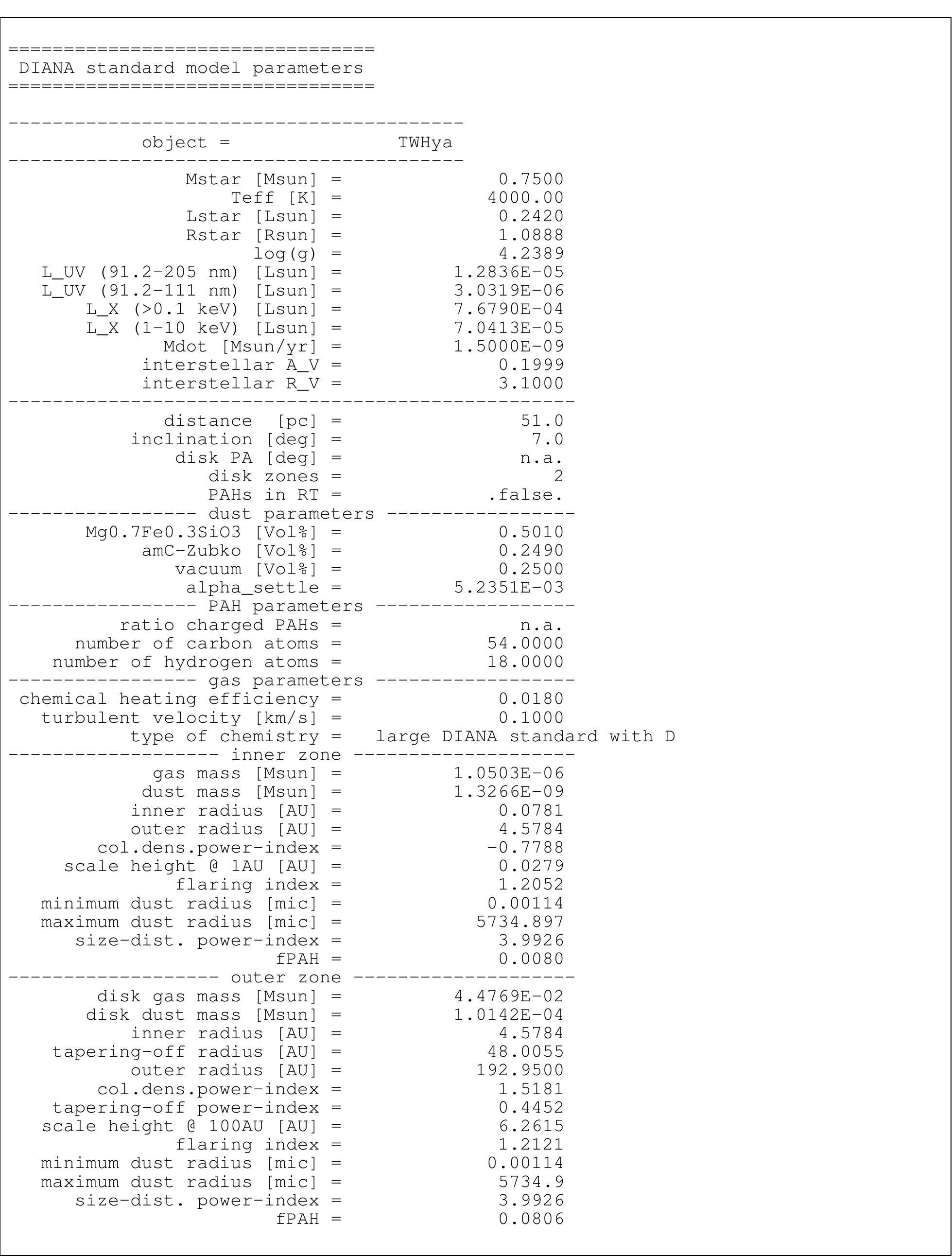}\\[-4mm]
\hspace*{0mm}\rule{140mm}{0.2pt}\\
\caption{Example file {\sf TWHya.para} which contains all model
  parameters of the TW\,Hya disk model in an easy to understand, 
  simple generic format.}
\label{fig:para}
\end{figure}

\begin{figure}
\hspace*{8mm}
\includegraphics[width=160mm,trim=0 370 0 0, clip]
                {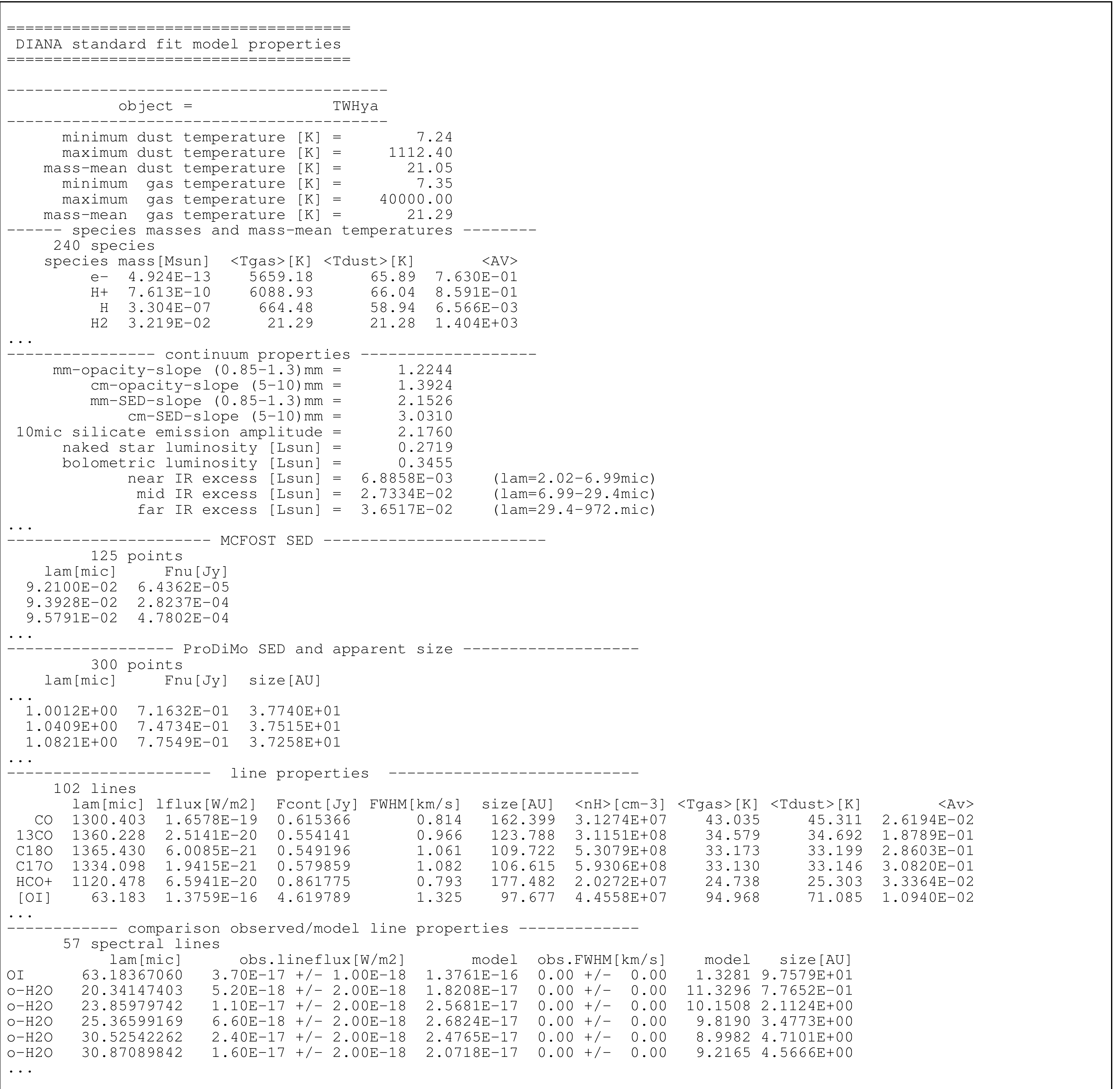}\\[-3.7mm]
\hspace*{9mm}\rule{158.5mm}{0.2pt}\\
\caption{Example file {\sf TWHya.properties} which contains a
  collection of important physico-chemical results and predictions of
  observational quantities from the model, such as continuum and line
  fluxes, apparent sizes, FWHM of lines, etc..  Note the ``...'' in
  the figure which means that a number of lines have been omitted here
  for clarity. Mean value $\langle.\rangle$ refer to averages over the
  specific line emission regions. The size\,[AU] is the radius of a
  circle in the image plane that contains 95\% of the flux.}
\label{fig:prop}
\end{figure}

\noindent The model data uploaded to the {\sf DIANA}
database\footnote{ \url{http://www.univie.ac.at/diana}, see also
  \url{http://www-star.st-and.ac.uk/~pw31/DIANA/SEDfit/SEDmodels_index.html}
  and \url{http://www-star.st-and.ac.uk/~pw31/DIANA/DIANAstandard} for
  direct access.} are organized in five sets
\begin{enumerate}
\item \sf \{object-name\}.para,
\item \sf \{object-name\}.properties,
\item \sf \{object-name\}\_DIANAfit.ps.gz,
\item \sf \{object-name\}\_ModelOutput.tgz, and
\item \sf \{object-name\}\_ModelSetup.tgz.
\end{enumerate}
Examples of the {\sf .para} and {\sf .properties} files are shown in
Figs.~\ref{fig:para} and \ref{fig:prop}, respectively.

\smallskip
\noindent{\sffamily\bfseries\underline{The .para files}}\ \ contain
the complete set of modeling parameters, as well as a few quantities
related to the stellar properties. These values are listed in a
generic form to make them readable even to end-users not working with
{\sf ProDiMo}, {\sf MCFOST} or {\sf MCMax}.  The different blocks of
the {\sf .para} files (see Fig.\,\ref{fig:para}) list the parameter
names and units of the stellar parameter, some observational
parameters like distance and inclination, two essential switches (one
or two-zone model? PAHs included in radiative transfer?), the dust
material and settling parameters, the PAH parameters, the gas
parameters, as well as the disk shape, mass, and dust size parameters
separately for the inner (for two-zone models) and the outer disk
zones. The meaning of all physical quantities is explained in
\citep{Woitke2016}.

\smallskip
\noindent{\sffamily\bfseries\underline{The .properties
    files}}\ \ contain a selection of important model output
quantities, as well as some predicted observations, see
Fig.\,\ref{fig:prop}. The first block contains minimum, maximum and
mean values of the gas and dust temperatures in the disk. The second
block lists total masses of chemicals, and mean values thereof, for
example
\begin{equation}
  \langle T_{\rm gas}^{\rm H_2}\rangle =
  \frac{\int T_{\rm gas}\,n_{\rm H_2}\,dV}{\int n_{\rm H_2}\,dV} \ ,
\end{equation}
where $n_{\rm H_2}$ is the particle density of molecular hydrogen,
$T_{\rm gas}$ the gas temperature, and $\int\,dV$ is the integration
in cylindrical coordinates over radius $r$ and height over midplane
$z$ in the disk. In the shown example, we find the mean gas
temperature of atomic hydrogen to be $\approx\!  660\,$K and the mean
gas temperature of molecular hydrogen to be $\approx\! 21\,$K.

The next block summarizes some predictions of directly observable
continuum properties, such as the millimeter and centimeter slopes,
the amplitude of the 10\,$\mu$m silicate emission feature, and
near-IR, mid-IR and far-IR excesses, followed by two blocks which
contain the model SED computed by {\sf MCFOST} and {\sf ProDiMo}, the
latter comes with an apparent size at all wavelengths, i.e.\ the
radius of a circle in the image plane which contains 95\,\% of the
spectral flux. The apparent sizes are derived without convolution with
any instrumental point spread function (PSF), and are computed without
direct star light.

The second block from the bottom shows predictions and properties of a
generic sample of 102 spectral emission lines. The name of the atom or
molecule, the line center wavelength, the total line flux, the
continuum flux, the FWHM of the line velocity-profile, the apparent
size of the line emitting region (as above), and averages of density,
gas and dust temperature and optical extinction over the line emitting
regions. Studying the output for the CO isotopologue lines around
1.3\,mm shows that the $^{12}$CO line comes from less dense but warmer
gas, whereas the $^{13}$CO, C$^{18}$O and C$^{17}$O lines originate in
successively cooler and deeper disk layers. Note that $\langle T_{\rm
  gas}\rangle\!\approx\!\langle T_{\rm dust}\rangle$ for all CO
isotopologue lines, whereas the [O\,{\sc i}]\,63\,$\mu$m line has
$\langle T_{\rm gas}\rangle\!>\!\langle T_{\rm dust}\rangle$. This
information is available for all 102 lines for all finished {\sf
  DIANA} standard models.

The last block compares observed with predicted line properties
including line fluxes, FWHMs and apparent sizes. In the case of
TW\,Hya, 57 line observations have been collected, and accounted for
in the fitting of the model parameters.

\smallskip
\noindent{\sffamily\bfseries\underline{The DIANAfit.ps.gz
    files}}\ \ contain graphical output pages of the following model
results: 2D-density structure (gas and dust), gas/dust ratio, mean
dust sizes, column density structure, scale height as function of
radius, UV field-strength $\chi$, radial optical depths for UV photons
and X-rays, X-ray ionization rate $\zeta_{\rm X}$, dust, gas and PAH
temperature structures, chemical and cooling relaxation timescales,
Rosseland-mean dust opacity, sound speed, the most important heating
and cooling processes, mean dust opacity as function of wavelength,
SED, magnification of mid-IR region, apparent size and continuum
emission regions as function of wavelength, enlargements of SED in
selected spectral windows with emission lines overplotted, continuum
images at near-IR to mm wavelengths, comparison of model to observed
intensity profiles after PSF-convolution, concentration of selected
chemicals as 2D-contour plots, 2D-plots of the line emission regions
that produce $>\!50\,\%$ of the line flux (available for all 102
generic emission lines), computed velocity profiles for the observed
lines in comparison to the data (where available), and line intensity
plots (after averaging over $\theta$ in concentric rings) as function
of $r$.
 
\smallskip
\noindent{\sffamily\bfseries\underline{The ModelOutput.tgz
    files}}\ \ contain all model output data as listed above (and
more) in numerical tables where each line corresponds to a
$(r,z)$-point in the model. These data can be used, for example, to
make your own plots, for example to look for the various ice-lines.

\smallskip
\noindent{\sffamily\bfseries\underline{The ModelSetup.tgz
    files}}\ \ contain all numerical input and data-files necessary to
recompute the model. These files are included in the format required
for {\sc ProDiMo}, {\sc MCFOST} and {\sc MCMax}. The data-files
include the observational data in the ``model-friendly'' input format
requested by {\sc ProDiMo}.

\clearpage

\begin{table}[!b]
\caption{Error estimations for three example disks, see text for 
         further explanations.}
\label{uncert}
\vspace*{2mm}\hspace*{-3mm}
\resizebox{163mm}{!}{
\begin{tabular}{|rl|r|r|r|}
\hline
&&&&\\[-2.3ex]
   && \sffamily\bfseries HD\,163296 & \sffamily\bfseries TW\,Hya &\sffamily\bfseries  BP\,Tau\\
\hline
\multicolumn{5}{|l|}{ }\\[-2.2ex]
\multicolumn{5}{|l|}{\sffamily\bfseries dust parameters}\\
\hline
minimum dust size                     & $a_{\rm min}$\,[$\mu$m] 
  & 0.020 $\pm$ 140\%   & 0.0011 $\pm$ 55\%       & 0.049   $\pm$ 14\%  \\
maximum dust size                     & $a_{\rm max}$\,[mm]     
  & 8.2   $\pm$  71\%   & 5.7    $\pm$ 105\%      & 3.1     $\pm$ 60\%  \\
dust size distribution powerlaw index & $a_{\rm pow}$           
  & 3.71  $\pm$ 0.08    & 3.99   $\pm$ 0.08       & 3.97    $\pm$ 0.05  \\
volume fraction of amorphous carbon   & amC\,[\%]              
  & 6.0   $\pm$  1.3 \  & 25 $\pm$ 3\hspace*{5.5mm} & 17    $\pm$ 4 \ \ \\
dust settling parameter               & $\alpha_{\rm settle}$   
  & 6.6(-3) $\pm$  87\% & 5.2(-3) $\pm$  120\%    & 6.0(-5) $\pm$ 20\%  \\
\hline
\multicolumn{5}{|l|}{ }\\[-2.2ex]
\multicolumn{5}{|l|}{\sffamily\bfseries PAH and gas parameters}\\
\hline
abundance of PAHs                     & $f_{\rm PAH}$           
  & 0.076 $\pm$ 0.06    & 0.08 $\pm$ 0.1 \ \ \    & 0.12    $\pm$ 0.28  \\
efficiency of chemical heating        & $\gamma_{\rm chem}$    
  & 0.19 $\pm$ 0.28     & 0.014 $\pm$ 0.025       & 0.13    $\pm$ 0.15  \\
\hline
\multicolumn{5}{|l|}{ }\\[-2.2ex]
\multicolumn{5}{|l|}{\sffamily\bfseries inner disk parameters}\\
\hline
gas mass                           & $M_{\rm gas}$\,[$M_\odot$]  
  & 1.3(-4) $\pm$ 58\%  & 1.1(-6) $\pm$ 31\%  & 7.1(-7) $\pm$ 17\%  \\
dust mass                          & $M_{\rm dust}$\,[$M_\odot$] 
  & 1.5(-9) $\pm$ 45\%  & 1.3(-9) $\pm$ 62\%  & 6.9(-9) $\pm$ 18\%  \\
inner radius                       & $R_{\rm in}$\,[AU]         
  & 0.41    $\pm$ 14\%  & 0.078   $\pm$ 46\%  & 0.060   $\pm$ 33\%  \\
column density exponent            & $\epsilon$                
  & 1.11    $\pm$ 0.39  & -0.78   $\pm$ 0.29  & 0.52    $\pm$ 0.06  \\
scale height at 1\,AU              & $H$\,[AU]                 
  & 0.077   $\pm$ 5\% \ & 0.028   $\pm$ 14\%  & 0.12    $\pm$ 6\% \ \\
flaring exponent                   & $\beta$                   
  & 1.01    $\pm$ 0.11  & 1.21    $\pm$ 0.13  & 1.22    $\pm$ 0.02  \\
\hline
\multicolumn{5}{|l|}{ }\\[-2.2ex]
\multicolumn{5}{|l|}{\sffamily\bfseries outer disk parameters}\\
\hline
gas mass                           & $M_{\rm gas}$\,[$M_\odot$]  
  & 5.8(-1) $\pm$ 60\%  & 4.5(-2) $\pm$ 45\%  & 6.4(-3) $\pm$ 50\%   \\
dust mass                          & $M_{\rm dust}$\,[$M_\odot$] 
  & 1.7(-3) $\pm$ 15\%  & 1.0(-4) $\pm$ 14\%  & 1.4(-4) $\pm$ 25\%   \\
inner radius                       & $R_{\rm in}$\,[AU]         
  & 3.7     $\pm$ 18\%  & 4.6     $\pm$ 9\% \ & 1.3     $\pm$  7\% \ \\
tapering-off radius                & $R_{\rm taper}$\,[AU]      
  & 130     $\pm$ 15\%  & 48      $\pm$ 9\% \ & 32      $\pm$ 55\%   \\
column density exponent            & $\epsilon$                
  & 0.95    $\pm$ 0.20  & 1.52    $\pm$ 0.13  & 0.65    $\pm$ 0.18   \\
tapering-off exponent              & $\gamma$                  
  & 0.20    $\pm$ 0.30  & 0.45    $\pm$ 0.23  & 0.67    $\pm$ 0.23   \\
scale height at 100\,AU            & $H$\,[AU]                 
  & 6.5     $\pm$ 0.2 \ \ & 6.3   $\pm$ 0.3 \ & 7.1     $\pm$ 0.3 \ \ \\
flaring exponent                   & $\beta$                  
  & 1.11    $\pm$ 0.03  & 1.21    $\pm$ 0.03  & 1.12    $\pm$ 0.02   \\
\hline
\end{tabular}}
\end{table}

\section{Uncertainties in disk parameter determination}
\label{app:uncertainties}

\noindent Our derived values of the model parameter concerning
disk shape, gas, dust and PAH parameters are subject to large
uncertainties, partly due to measurement errors and calibration
uncertainties of the astronomical instruments used, but partly also
due to well-known fitting degeneracies, in particular if the disk model is
only fitted to SED data.

Absolute measurement errors are typically 10\% to 30\%, for example
due to calibration uncertainties, usually larger than the quoted
measurement uncertainties. This is particularly relevant for
multi-instrument data as we consider in this paper. Another source for
uncertainties are numerical errors. For computational grid sizes $100
\times 100$ as used here, line flux predictions from the models are
uncertain by about 10\% \citep[][see Appendix~F therein]{Woitke2016}
due to numerical problems to properly resolve the sometimes very small
(i.e.\ very thin) line forming regions.

A proper resolution of all modeling uncertainties and degeneracies
would require a Bayesian analysis, which needs about
$10^6$ models for each object in a $\sim$\,20-dimensional parameter
space.  After careful consideration, the team decided {\em not} to
perform such a Bayesian analyses, because it would surpass our
already considerable numerical efforts in this project to determine the
$\chi^2$ minimums by a large factor.

However, to get a rough impression of our modeling uncertainties,
we consider here some deviations $dp_j$ from the best values $p_j^{\rm opt}$ 
of parameter $j$ in both directions from the local $\chi^2$-minimum as
\begin{eqnarray}
  \chi^2_{\rm min} &=& \chi^2(p_j^{\rm opt}) \\
      \chi^2_{j+} &=& \chi^2(p_j^{\rm opt} + \delta_{j,j'}\,dp_{j'}) \\
      \chi^2_{j-} &=& \chi^2(p_j^{\rm opt} - \delta_{j,j'}\,dp_{j'}) \ ,
\end{eqnarray}
and then fit a parabola to the three points $\{\chi^2_{j-}, \chi^2_{\rm
  min}, \chi^2_{j+}\}$ in each parameter dimension $j$ to (i)
check whether the best parameter value $p_j^{\rm opt}$ actually sits
in a multi-dimensional minimum as expected, and (ii) to determine the distance $\Delta
p_j$ by which parameter $j$ can be varied, until a considerable
worsening of the fit is obtained as
\begin{equation}
  \chi^2(p_j^{\rm opt} + \delta_{j,j'}\,\Delta p_{j'}) = \chi^2_{\rm min} + 0.1 \ .
  \label{eq:offset}
\end{equation}
Typically, our best fits have $\chi_{\rm min}\,\approx\,1\,...\,3$.
The constant offset of 0.1 is chosen ``by eye'', such that a human can
effortlessly identify the best fitting model among all three models
with parameter values $p_j^{\rm opt}-\Delta p_j$, $p_j^{\rm opt}$ and
$p_j^{\rm opt}+\Delta p_j$ by looking at all the fits to the various
continuum and line data.

Table~\ref{uncert} lists these local estimates $\Delta p_j$ of our parameter 
confidence intervals for three selected {\sf DIANA} standard models. The first two
objects HD\,163296 and TW\,Hya have rich continuum and line data, the
last object BP\,Tau has less data, with emphasis on the CO
ro-vibrational data which probes first and foremost the inner disk.

The formal conclusion from this exercise is that we can fit disk gas masses
by roughly a factor of two, and dust masses by about 20\%. However, this
disregards any degeneracies with e.g.\ the dust opacity
parameters. Some parameters like the PAH abundance (none of the
selected objects has clear PAH detections in the mid-IR) and the
chemical heating efficiencies can only be determined by order of
magnitude, power law indices can roughly be determined by 0.1 to 0.3,
but some disk shape parameters like the scale height and flaring
exponent can be determined more precisely, by about $5-15$\,\%.

\section{Detailed mid-IR spectra}
\label{app:midIR}

\noindent Figure~\ref{fig:midIR} shows enlargements of our SED-fitting
models in the mid-IR region in comparison to the Spitzer/IRS, ISO/SWS
and photometric data for 27 objects. These results are discussed in
Sect.~\ref{sec:SED}.

\begin{figure}
  \vspace{-0mm}\hspace*{-7mm}
  \begin{tabular}{rr}
  \hspace*{-5mm}\includegraphics[width=87mm,height=74mm]{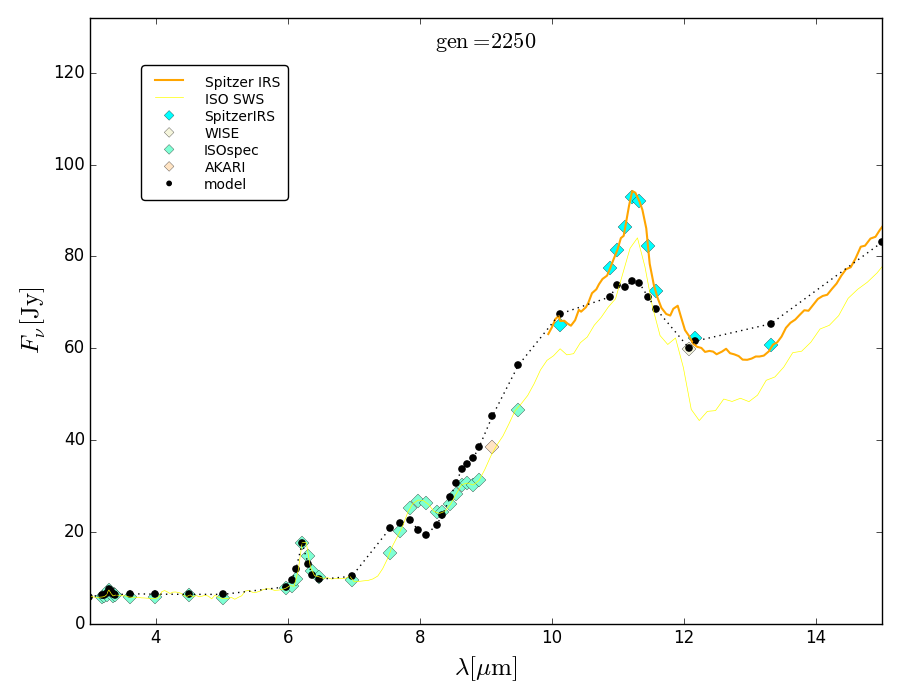} &
  \hspace*{-4mm}\includegraphics[width=87mm,height=74mm]{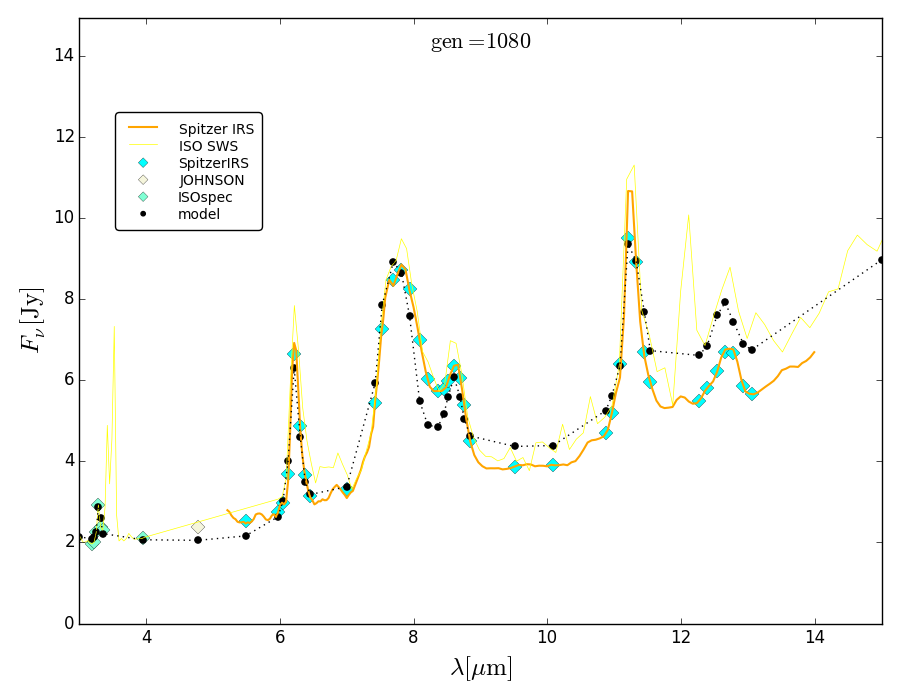}\\[-70mm]
  \sf HD\,100546\hspace*{3mm} & \sf HD\,97048\hspace*{3mm} \\[65mm]
  \hspace*{-5mm}\includegraphics[width=87mm,height=74mm]{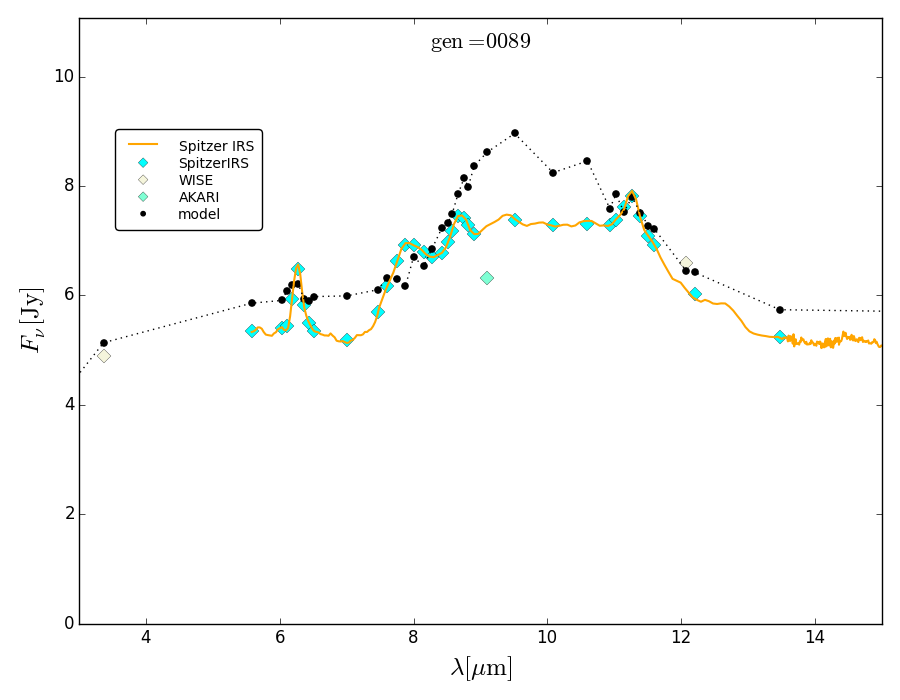} &
  \hspace*{-4mm}\includegraphics[width=87mm,height=74mm]{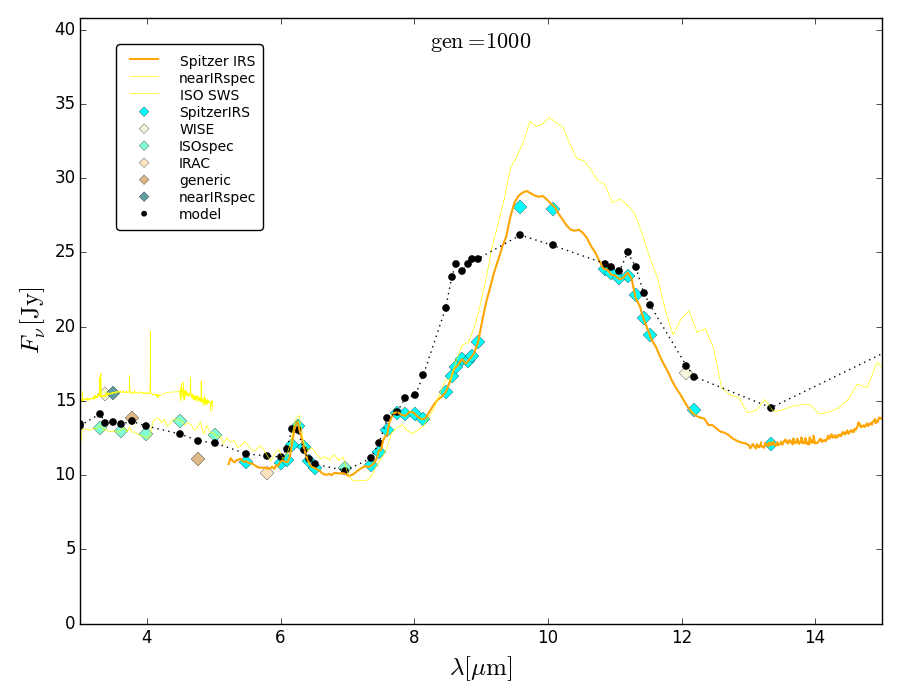} \\[-70mm]
 \sf HD\,95881\hspace*{3mm} & \sf AB\,Aur\hspace*{3mm} \\[65mm]
  \hspace*{-5mm}\includegraphics[width=87mm,height=74mm]{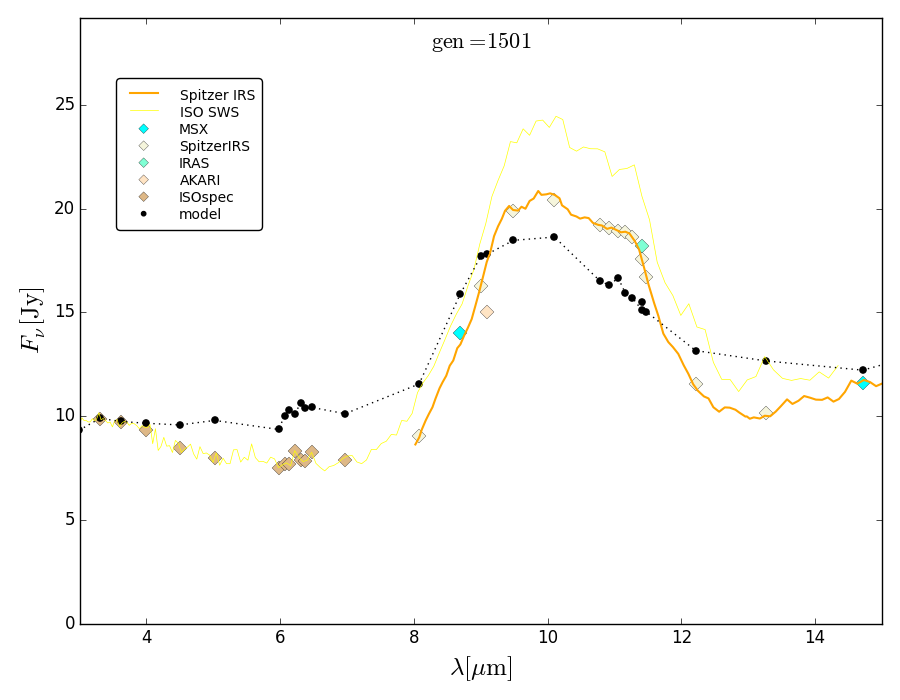} &
  \hspace*{-4mm}\includegraphics[width=87mm,height=74mm]{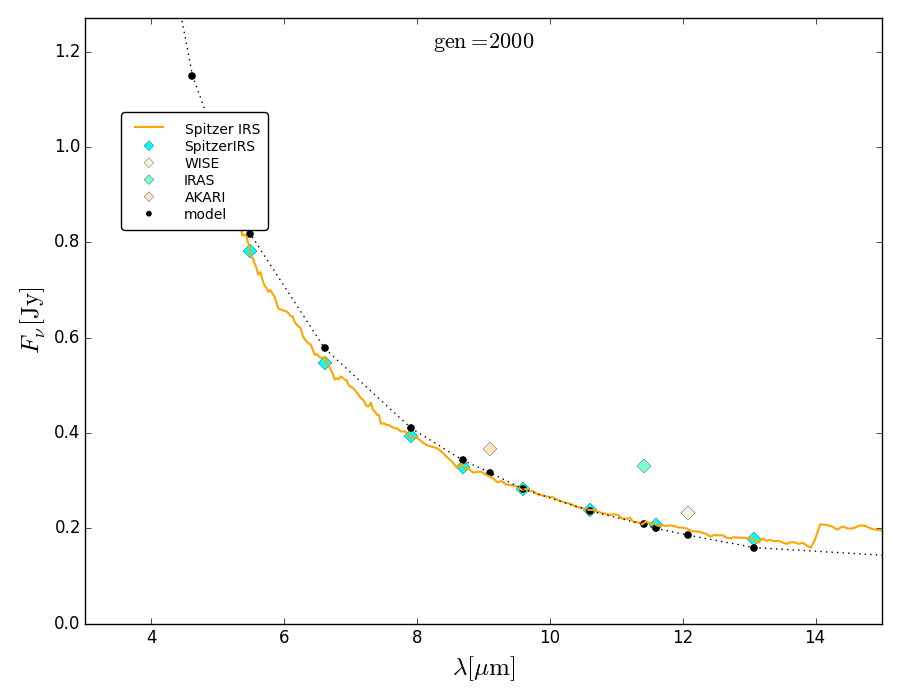}\\[-70mm] 
  \sf HD\,163296\hspace*{3mm} & \sf 49\,Cet\hspace*{3mm}\\[65mm]
  \end{tabular}
  \vspace*{-2mm}
 \caption{Enlargements of the SED-fits in the mid-IR region on linear
   scales, centered around the broad silicate emission feature at
   10\,$\mu$m and the PAH emission features at 3.3\,$\mu$m,
   6.25\,$\mu$m, 7.85\,$\mu$m, 8.7\,$\mu$m, 11.3\,$\mu$m, and
   12.7\,$\mu$m. All results have been obtained with the DIANA
   standard dust opacities, see text, where only the powerlaw
   parameters of the dust size-distribution and the volume fraction of
   amorphous carbon was varied for the fit.  Therefore, detailed fits
   of the spectral shapes of the features are not expected. Concerning
   the PAHs, only the abundance of the PAHs and the fraction of
   charged PAHs was varied for fitting.}
\label{fig:midIR}
\end{figure}

\addtocounter{figure}{-1}
\begin{figure}
  \small
  \vspace{-0mm}\hspace*{-7mm}
  \begin{tabular}{rr}
  \hspace*{-5mm}\includegraphics[width=87mm,height=74mm]{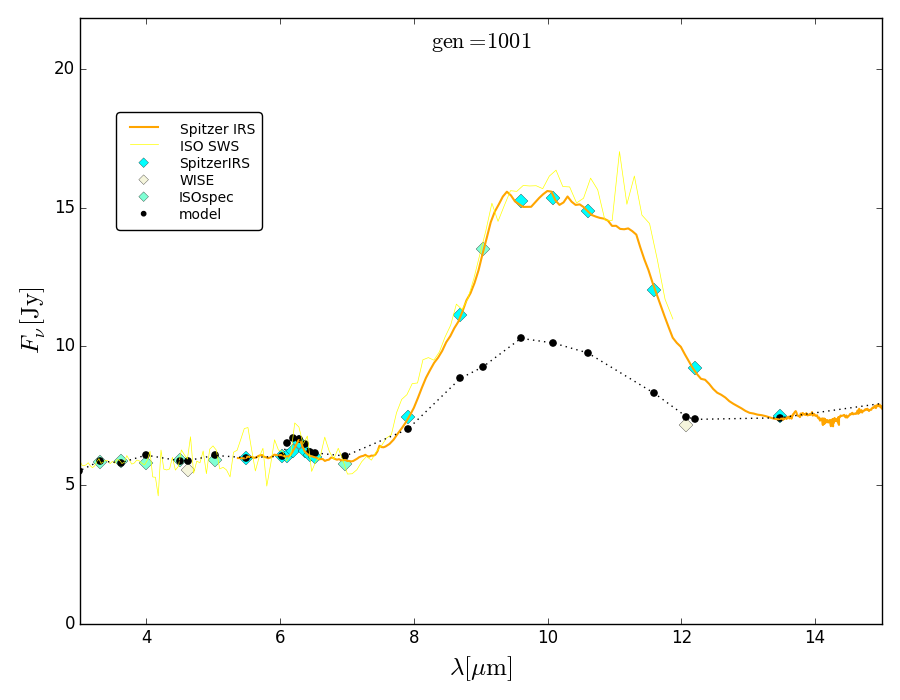} &
  \hspace*{-4mm}\includegraphics[width=87mm,height=74mm]{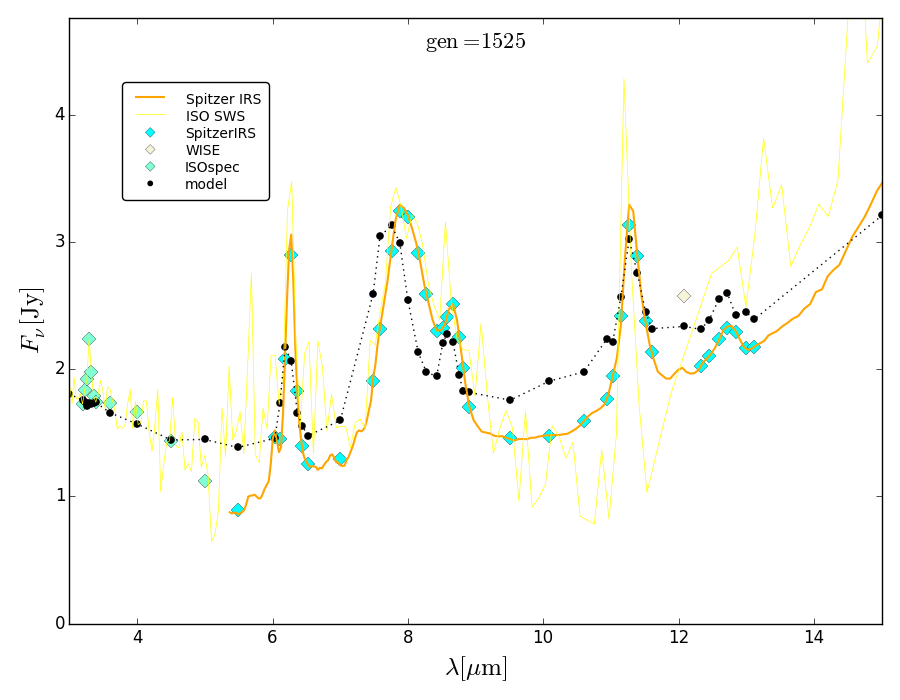} \\[-70mm]
  \sf MWC\,480\hspace*{3mm} & \sf HD\,169142\hspace*{3mm} \\[65mm]
  \hspace*{-5mm}\includegraphics[width=87mm,height=74mm]{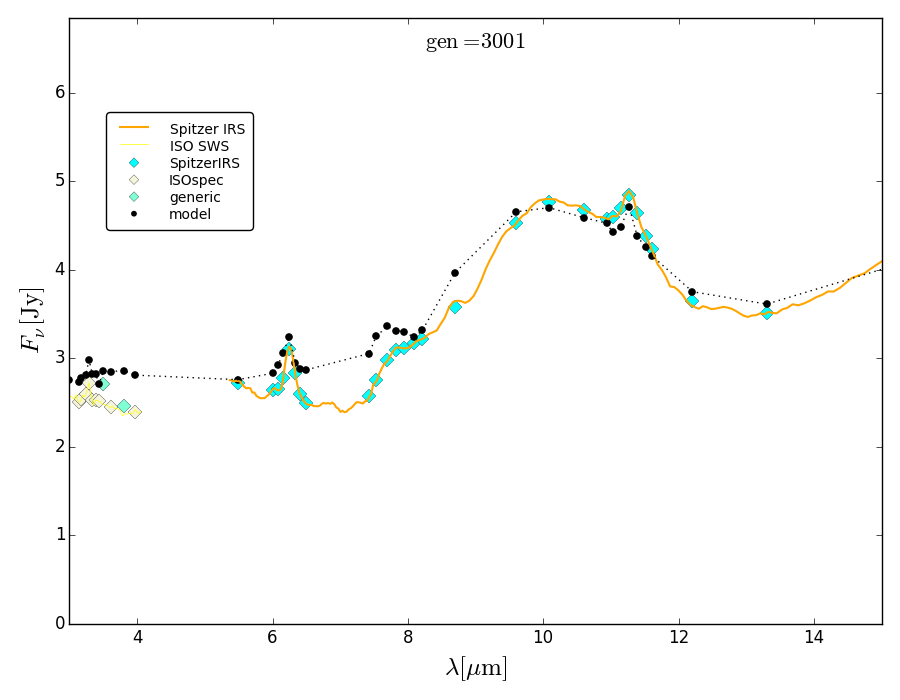} &
  \hspace*{-4mm}\includegraphics[width=87mm,height=74mm]{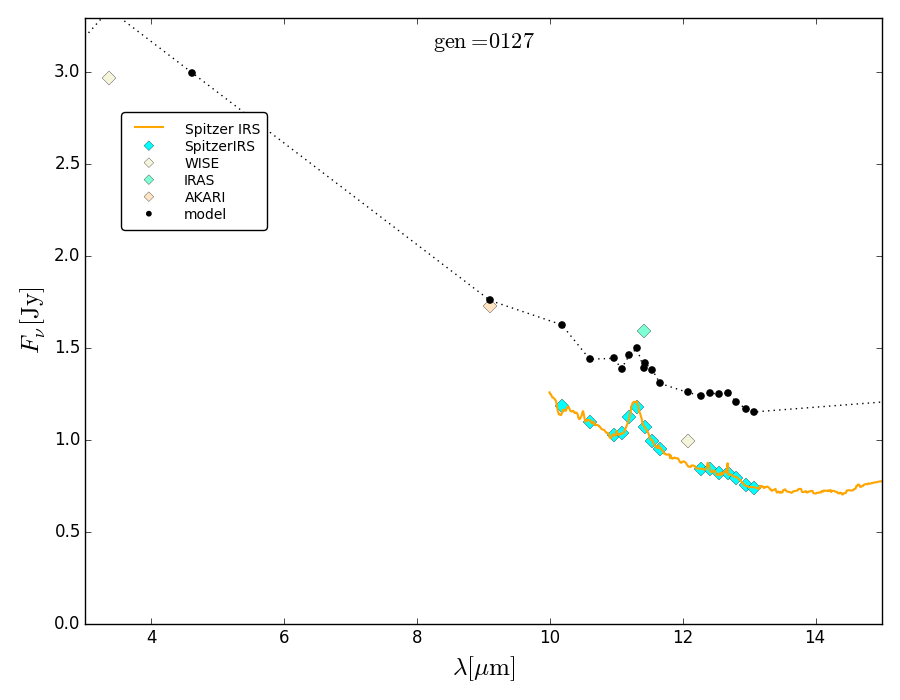} \\[-70mm]
  \sf HD\,142666\hspace*{3mm} & \sf HD\,135344B\hspace*{3mm} \\[65mm]
  \hspace*{-5mm}\includegraphics[width=87mm,height=74mm]{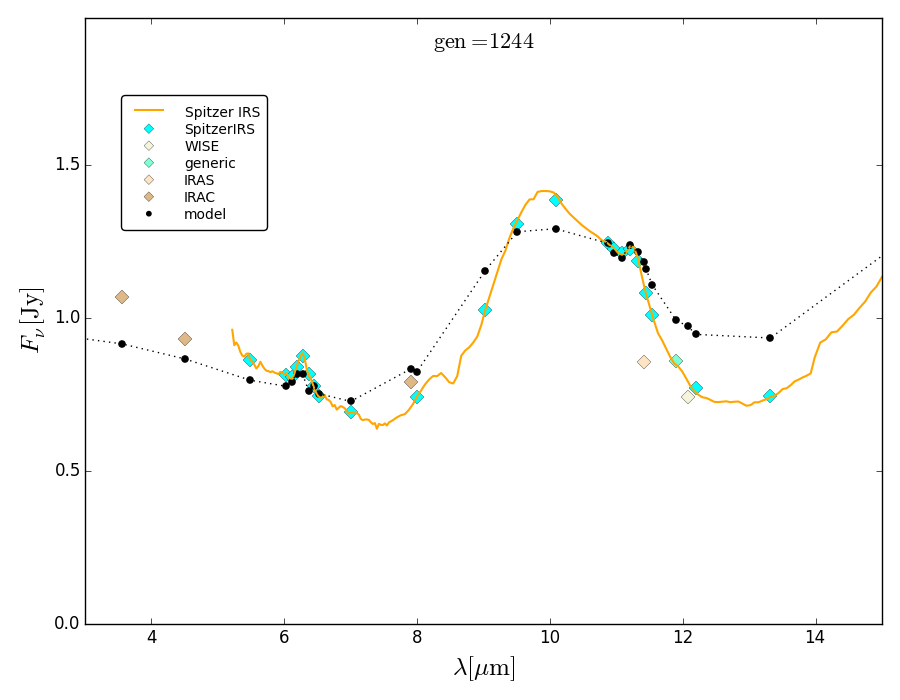} &
  \hspace*{-4mm}\includegraphics[width=87mm,height=74mm]{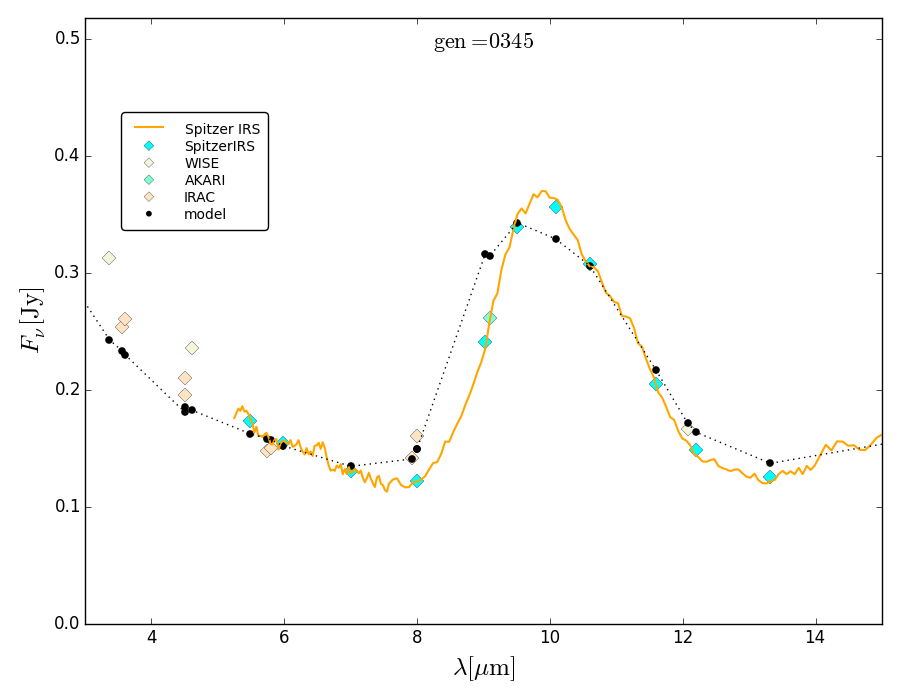}\\[-70mm]
  \sf V\,1149\,Sco\hspace*{3mm} & \sf Lk\,Ca\,15\hspace*{3mm}\\[65mm]
  \end{tabular}
  \vspace*{-2mm}
\caption{(continued)}
\end{figure}

\addtocounter{figure}{-1}
\begin{figure}
  \small
  \vspace{-0mm}\hspace*{-7mm}
  \begin{tabular}{rr}
  \hspace*{-5mm}\includegraphics[width=87mm,height=74mm]{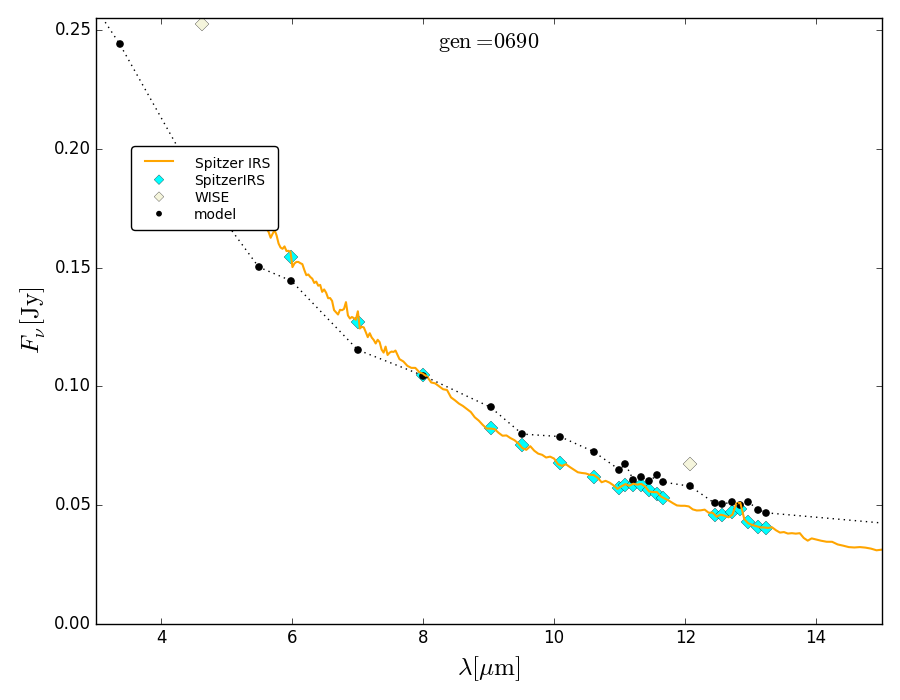} &
  \hspace*{-4mm}\includegraphics[width=87mm,height=74mm]{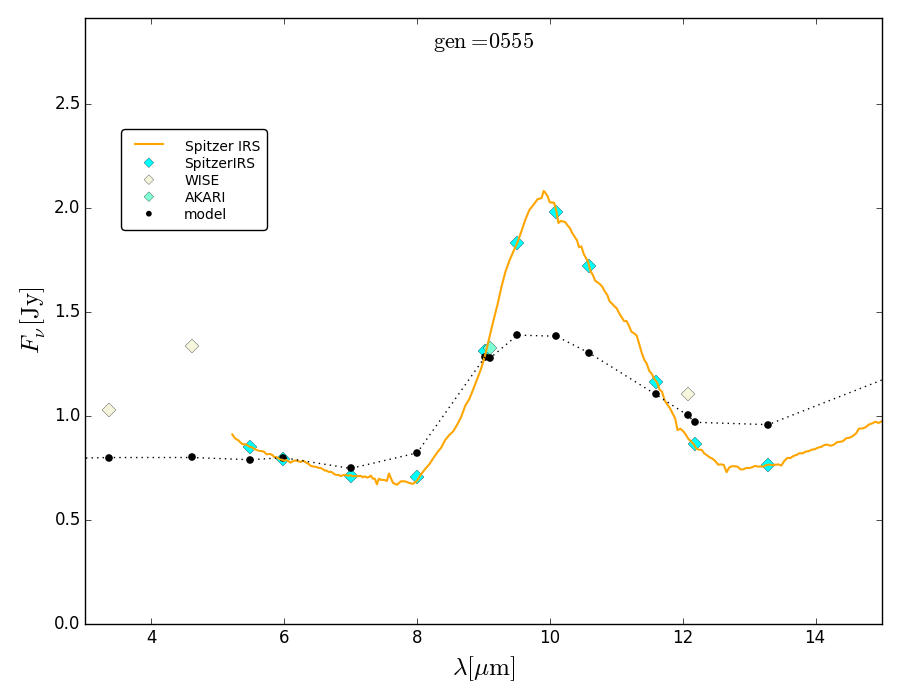} \\[-70mm]
  \sf UScoJ1604-2130\hspace*{3mm} & \sf RY\,Lup\hspace*{3mm} \\[65mm]
  \hspace*{-5mm}\includegraphics[width=87mm,height=74mm]{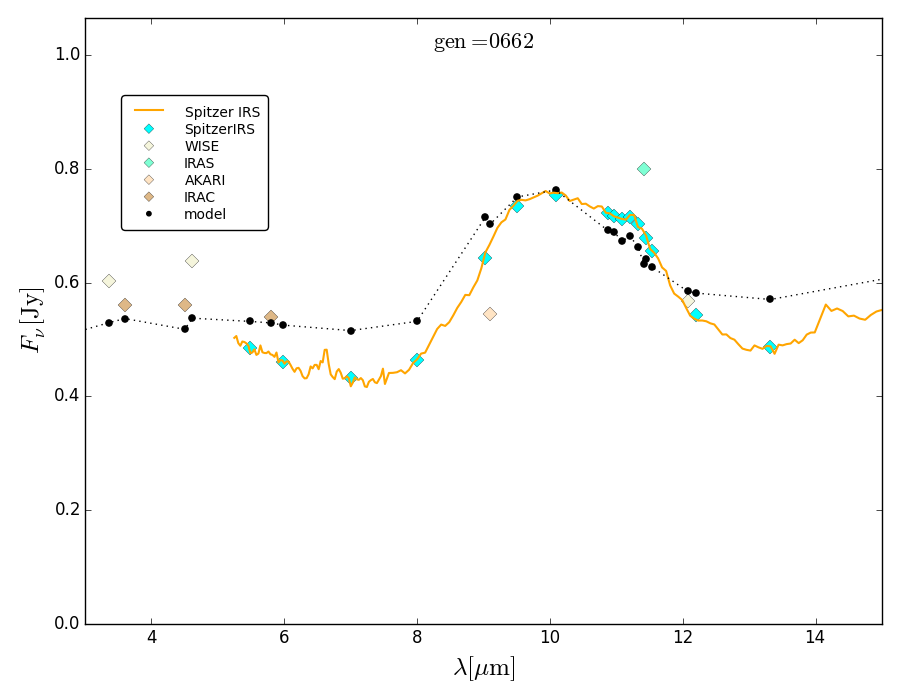} &
  \hspace*{-4mm}\includegraphics[width=87mm,height=74mm]{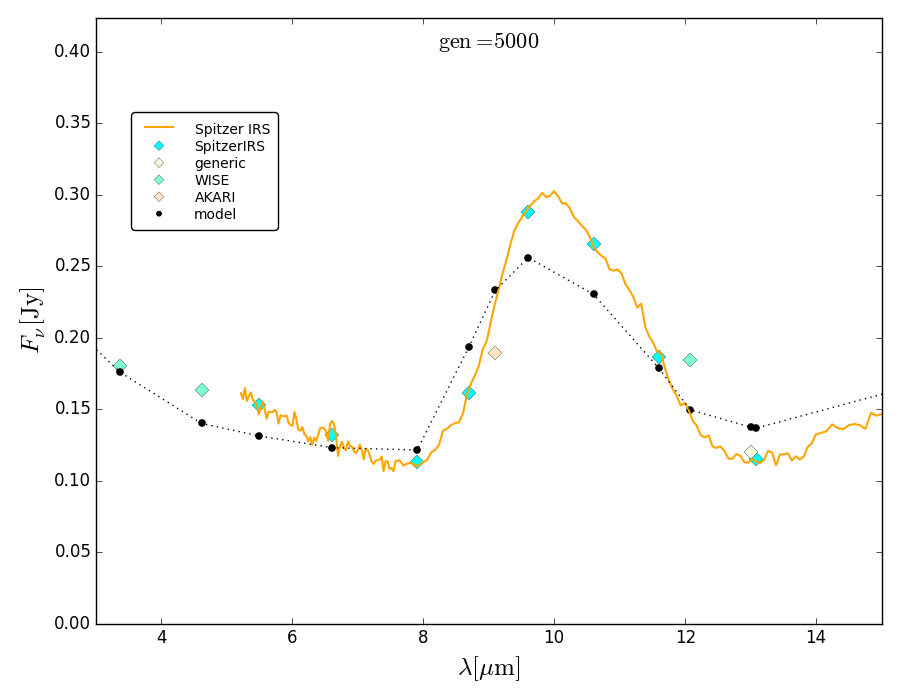} \\[-70mm]
  \sf CI\,Tau\hspace*{3mm} & \sf TW\,Cha\hspace*{3mm} \\[65mm]
  \hspace*{-5mm}\includegraphics[width=87mm,height=74mm]{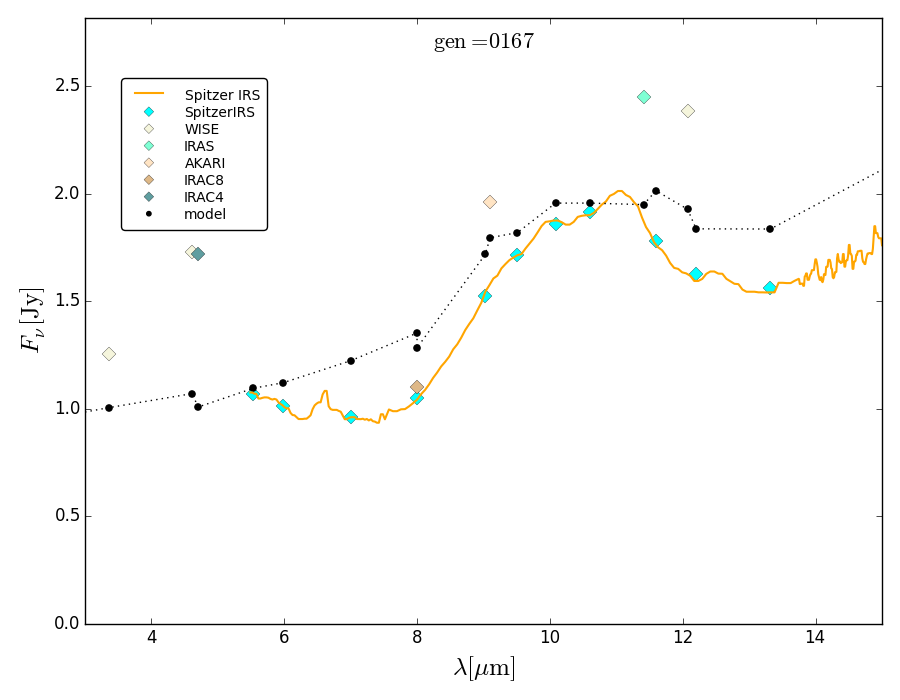} &
  \hspace*{-4mm}\includegraphics[width=87mm,height=74mm]{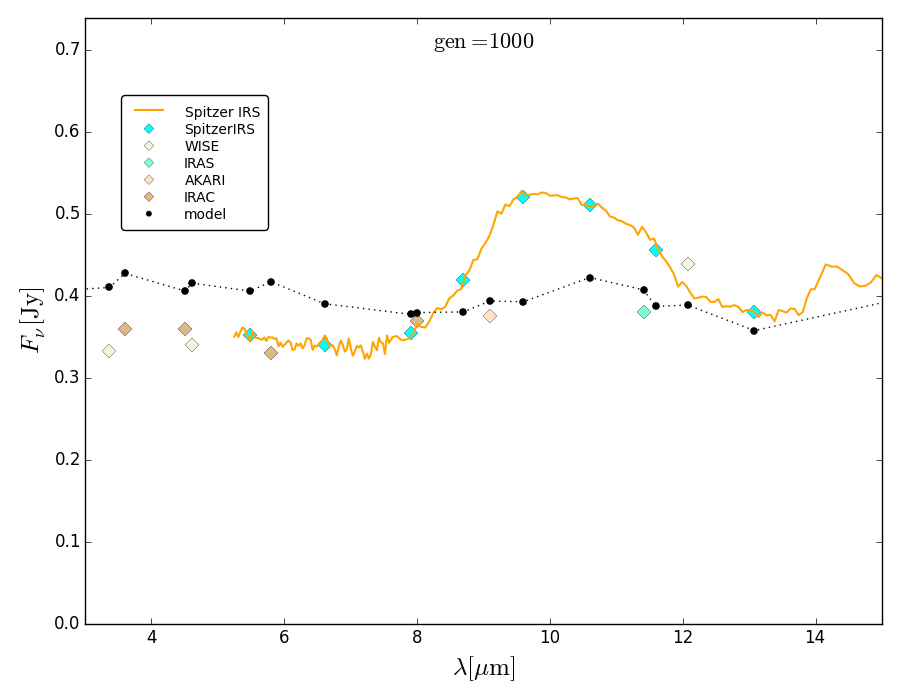} \\[-70mm]
  \sf RU\,Lup\hspace*{3mm} & \sf AA\,Tau\hspace*{3mm} \\[65mm]
  \end{tabular}
  \vspace*{-2mm}
\caption{(continued)}
\end{figure}

\addtocounter{figure}{-1}
\begin{figure}
  \small
  \vspace{-0mm}\hspace*{-7mm}
  \begin{tabular}{rr}
  \hspace*{-4mm}\includegraphics[width=87mm,height=74mm]{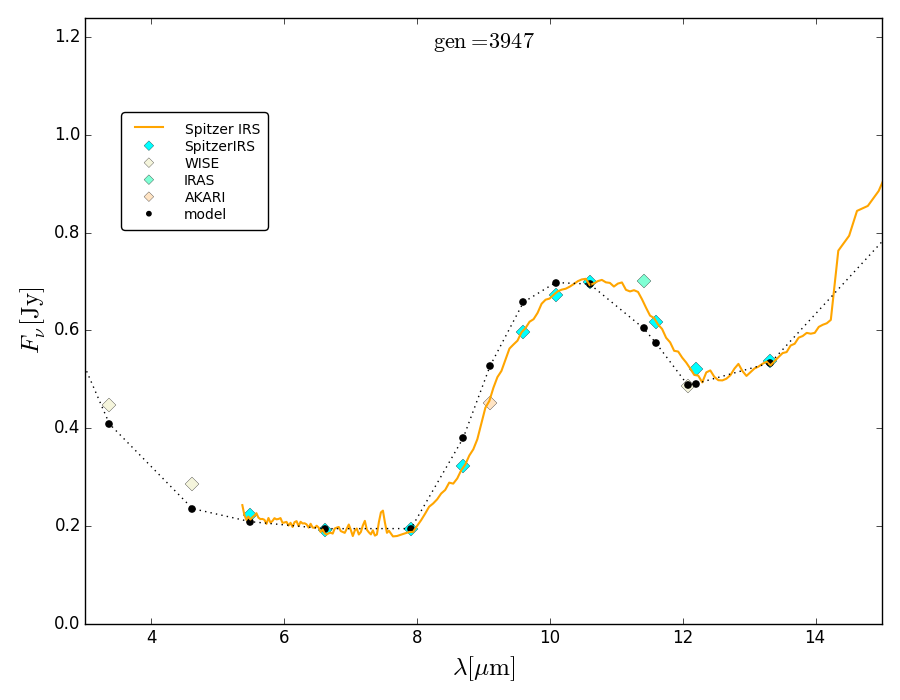} &
  \hspace*{-4mm}\includegraphics[width=87mm,height=74mm]{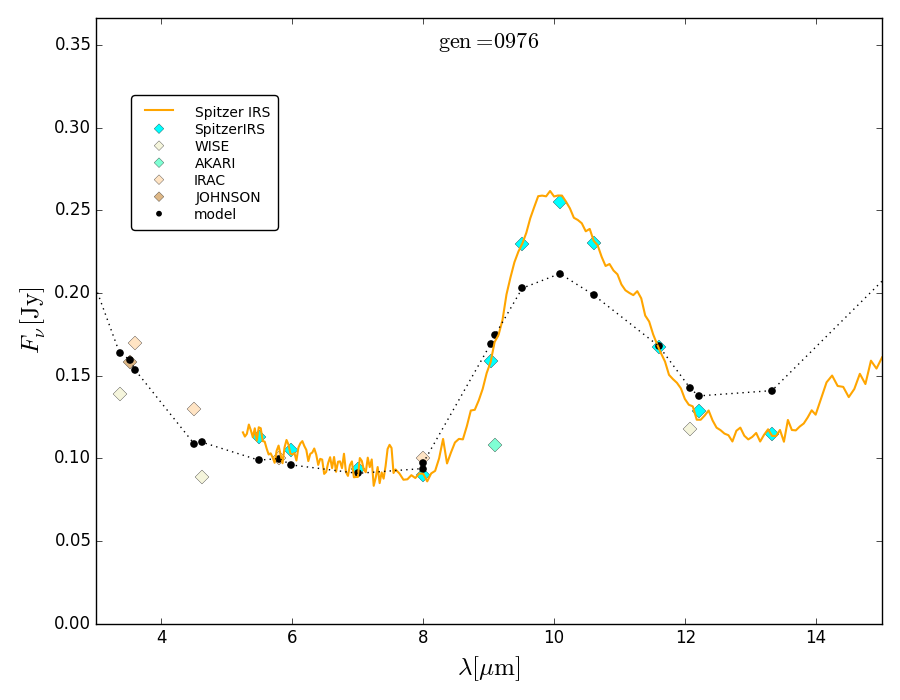}\\[-70mm]
  \sf TW\,Hya\hspace*{3mm} & \sf GM\,Aur\hspace*{3mm}\\[65mm]
  \hspace*{-5mm}\includegraphics[width=87mm,height=74mm]{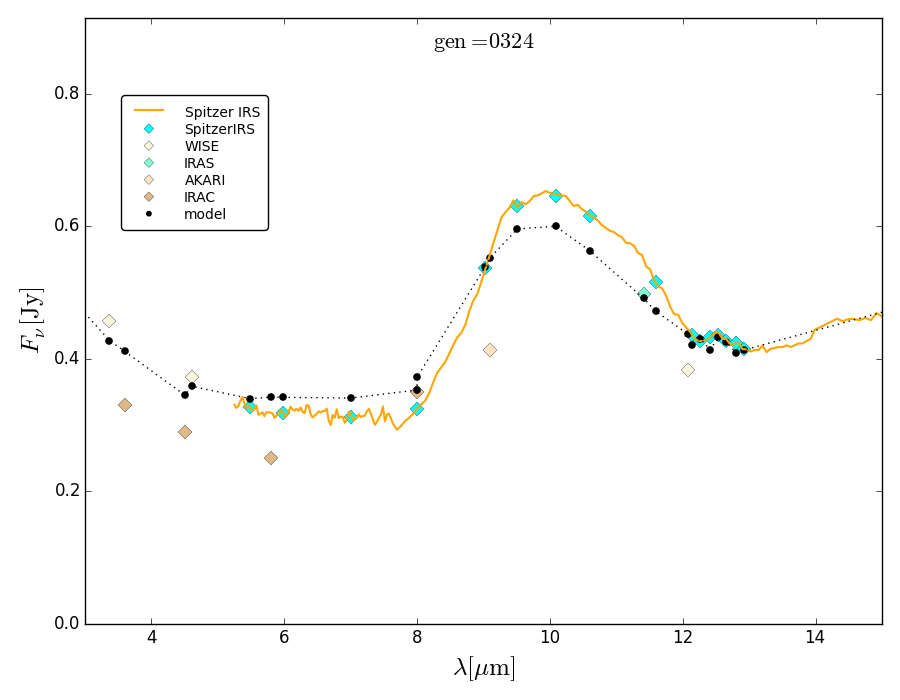} &
  \hspace*{-4mm}\includegraphics[width=87mm,height=74mm]{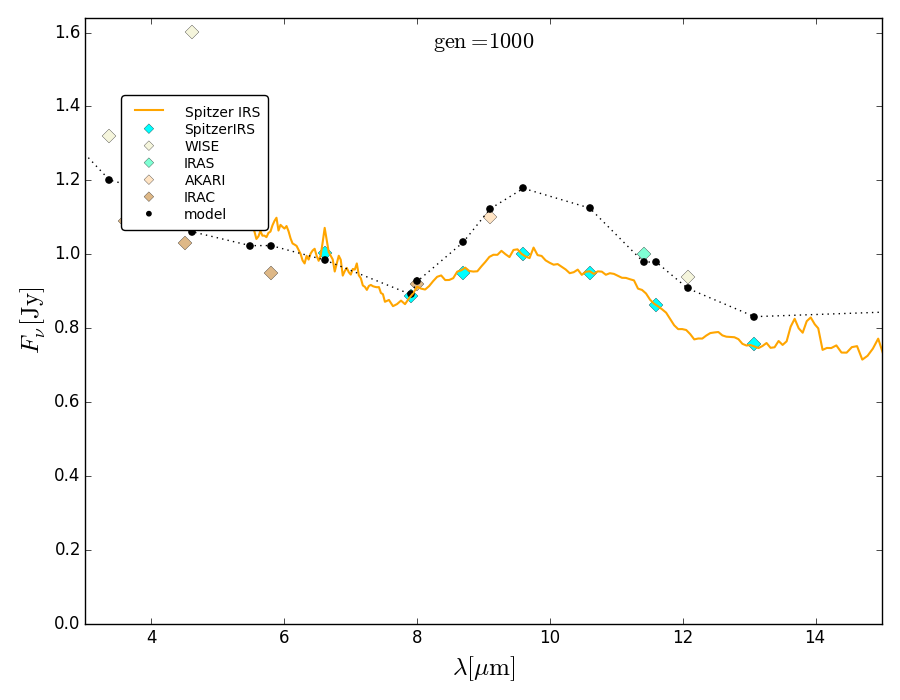} \\[-70mm]
  \sf BP\,Tau\hspace*{3mm} & \sf DF\,Tau\hspace*{3mm} \\[65mm]
  \hspace*{-5mm}\includegraphics[width=87mm,height=74mm]{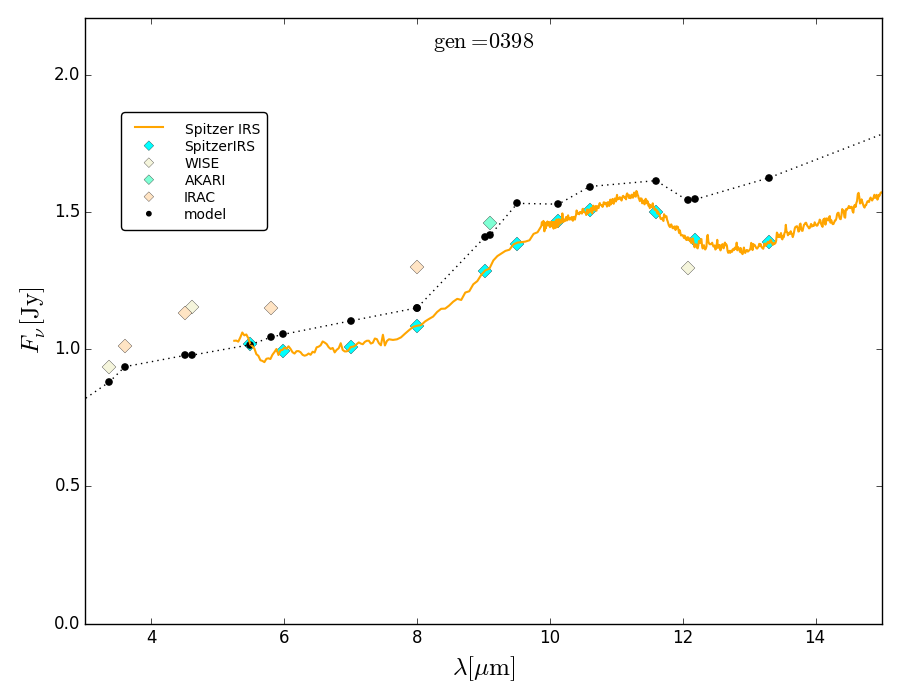} &
  \hspace*{-4mm}\includegraphics[width=87mm,height=74mm]{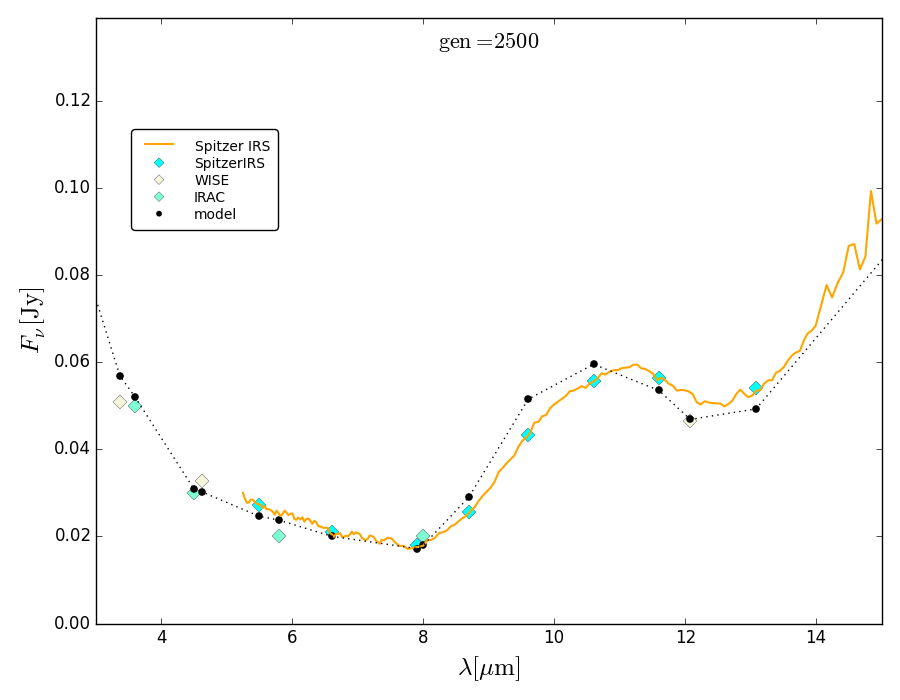} \\[-70mm]
  \sf DO\,Tau\hspace*{3mm} & \sf DM\,Tau\hspace*{3mm} \\[65mm]
  \end{tabular}
  \vspace*{-2mm}
\caption{(continued)}
\end{figure}

\addtocounter{figure}{-1}
\begin{figure}
  \small
  \vspace{-0mm}\hspace*{-7mm}
  \begin{tabular}{rr}
  \hspace*{-4mm}\includegraphics[width=87mm,height=74mm]{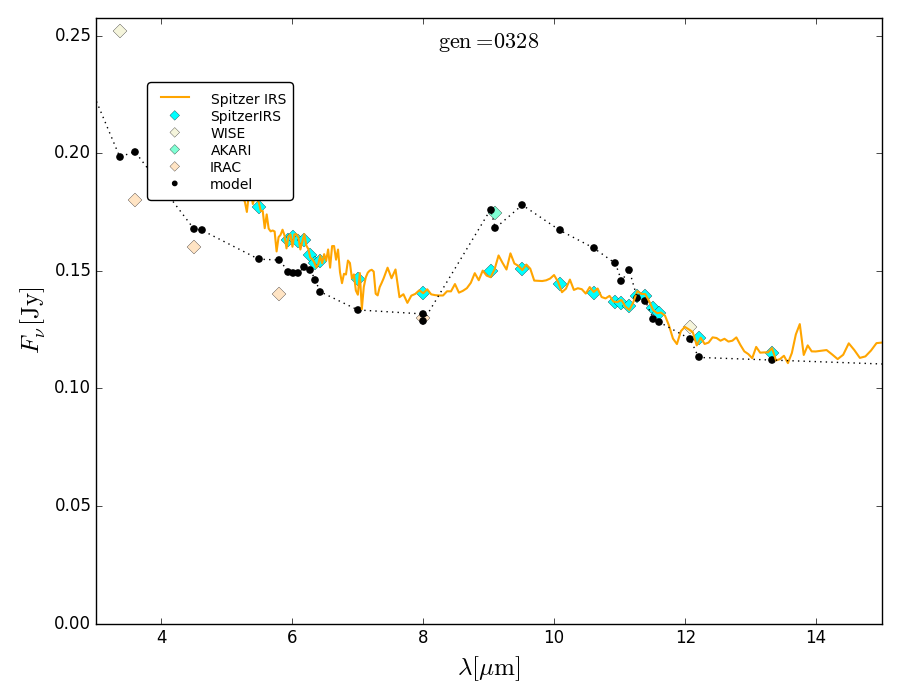} &
  \hspace*{-4mm}\includegraphics[width=87mm,height=74mm]{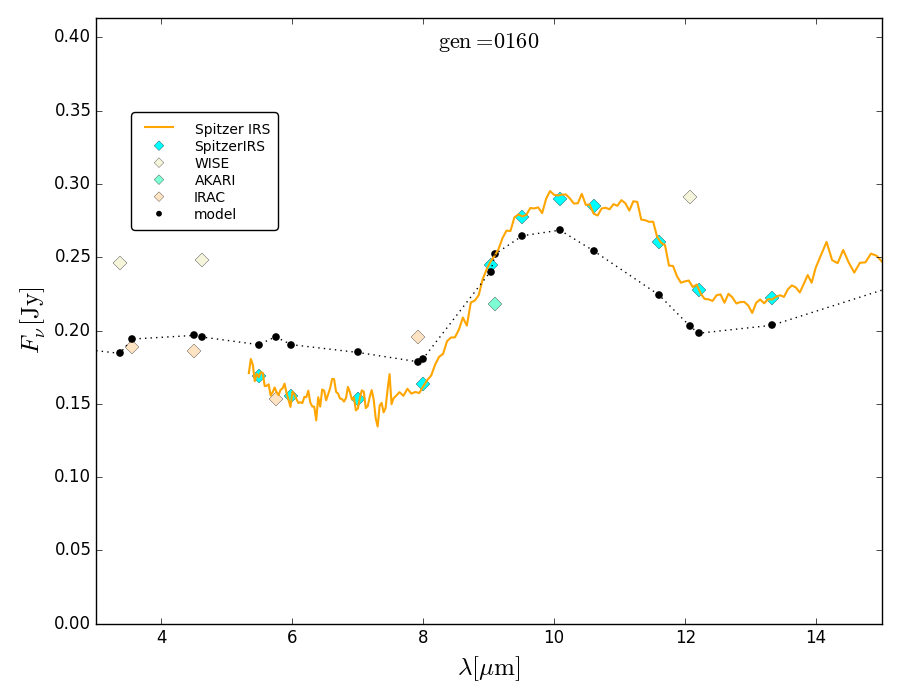}\\[-70mm]
  \sf CY\,Tau\hspace*{3mm} & \sf FT\,Tau\hspace*{3mm}\\[65mm]
  \hspace*{-4mm}\includegraphics[width=87mm,height=74mm]{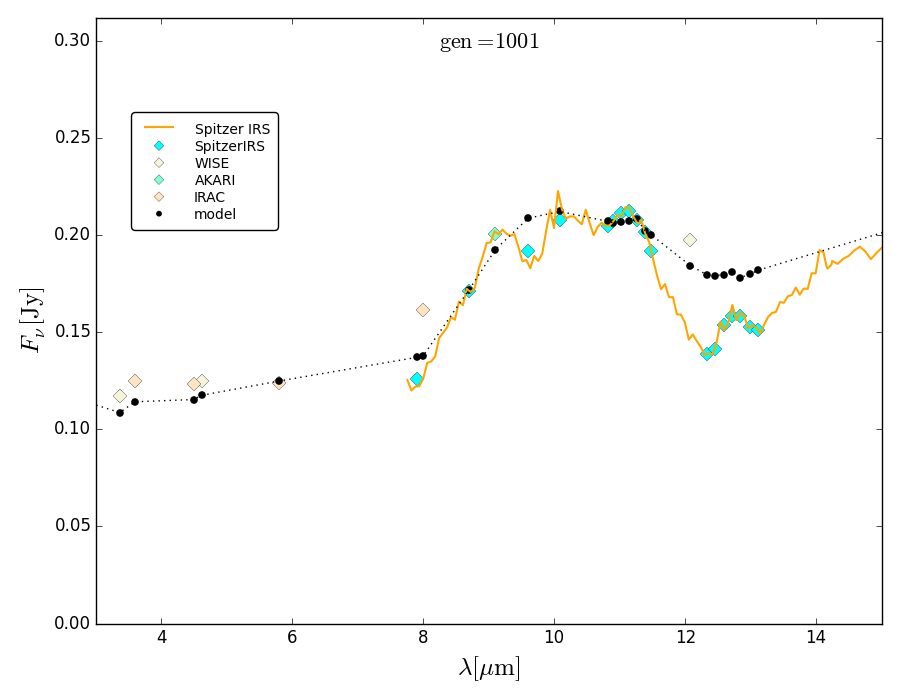}\\[-70mm]
  \sf RECX\,15\hspace*{3mm} \\[65mm]
  \end{tabular}
  \vspace*{-2mm}
\caption{(continued)}
\end{figure}

\end{document}